\documentclass[aps,prd,twocolumn,showpacs,superscriptaddress,nofootinbib,groupedaddress]{revtex4} 
\usepackage[a4paper, total={7in, 10in}]{geometry}
\usepackage{graphicx}
\usepackage{amssymb}
\usepackage{empheq}
\usepackage[dvipsnames]{xcolor}
\usepackage{pgf}
\usepackage{verbatim}
\usepackage{slashed}
\usepackage{MnSymbol}
\usepackage{float}
\usepackage{titlesec}
\usepackage[colorlinks=true,linkcolor=blue,urlcolor=black,bookmarksopen=true]{hyperref}

\newcommand{\qu}{q}
\newcommand{\qb}{\bar{q}}
\newcommand{\gl}{g}

\newcommand{\qtt}{\vec{q}}
\newcommand{\gtt}{\vec{g}}
\newcommand{\ptt}{\vec{p}_{b}}

\def\doubleunderline#1{\underline{\underline{#1}}}
\newcommand{\slt}{\underline{t}}
\newcommand{\slu}{\underline{u}}
\newcommand{\dlt}{\doubleunderline{t}}
\newcommand{\dlu}{\doubleunderline{u}}

\begin{document}
\title{Equilibration of weakly coupled QCD plasmas}
\author{Xiaojian Du}
\affiliation{Fakultät für Physik, Universität Bielefeld, D-33615 Bielefeld, Germany}
\author{Sören Schlichting}
\affiliation{Fakultät für Physik, Universität Bielefeld, D-33615 Bielefeld, Germany}
\date{\today}

\begin{abstract}
We employ a non-equilibrium Quantum Chromodynamics (QCD) kinetic description to study the kinetic and chemical equilibration of the Quark-Gluon Plasma (QGP) at weak coupling. Based on our numerical framework, which explicitly includes all leading order processes involving light flavor degrees of freedom, we investigate the thermalization process of homogeneous and isotropic plasmas far-from equilibrium and determine the relevant time scales for kinetic and chemical equilibration. We further simulate the longitudinally expanding pre-equilibrium plasma created in ultrarelativistic heavy-ion collisions at zero and non-zero density of the conserved charges and study its microscopic and macroscopic evolution towards equilibrium.

\end{abstract}
\maketitle
\tableofcontents

\section{Introduction}
\label{chap-introduction}

Non-equilibrium systems are ubiquitous in nature and of relevance to essentially all disciplines of modern physics. Despite the appearance of non-equilibrium phenomena in a variety of different contexts, there is a rather limited number of theoretical methods to study the real-time evolution of quantum systems, most of which rely on a set of approximations to study microscopic and macroscopic real-time properties of complex many-body systems. 

Specifically, for fundamental theories of nature, the question of understanding and describing non-equilibrium processes in the strongly interacting Quantum Chromodynamics sector of the standard model, has gained considerable attention in light of high-energy heavy-ion collision experiments at the Relativistic Heavy-Ion Collider (RHIC) and the Large Hadron Collider (LHC). Somewhat surprisingly, it turns out
that the complex space-time dynamics of high-energy heavy-ion collisions on space- and timescales $\sim 10 {\rm fm}$, can be rather well described by modern formulations of relativistic viscous hydrodynamics~\cite{Romatschke:2017ejr}, which has become the primary tool of heavy-ion phenomenology~\cite{Romatschke:2009im,Shen:2020mgh}. Nevertheless, due to the limited availability of theoretical approaches, it remains to some extent an open question how the macroscopic hydrodynamic behavior emerges from the underlying non-equilibrium dynamics of QCD, albeit significant progress in this direction has been achieved in recent years~\cite{Kurkela:2015qoa,Romatschke:2017vte,Kurkela:2018wud,Kurkela:2018vqr,Kurkela:2018xxd,Almaalol:2020rnu,Strickland:2017kux,Heller:2011ju,Heller:2016rtz,Kurkela:2019set,Blaizot:2017ucy,Blaizot:2019scw}.

Beyond high-energy heavy-ion collisions similar questions arise in Cosmology, where the non-equilibrium dynamics of QCD and QCD-like theories can certainly be expected to play a prominent role in producing a thermal abundance of standard model particles between the end of inflation and big bang nucleosynthesis. However, at the relevant energy scales, the field content of the early universe is not necessarily well constrained, and a detailed understanding of the thermalization of the early universe at least requires the knowledge of the coupling of the standard model degrees of freedom to the inflation sector, which makes this problem significantly more difficult. Nevertheless studies of the thermalization of the isolated QCD sector still bear relevance to this question, as some of the basic insights into the thermalization process of QCD or QCD-like plasmas can be adapted to Cosmological models, as recently discussed e.g. in Ref.~\cite{Harigaya:2013vwa,Mukaida:2015ria,Asaka:2003vt}.

Even though Quantum Chromodynamics (QCD) exhibits essentially non-perturbative phenomena such as confinement at low energies, strong interaction matter becomes weakly coupled at asymptotically high energies owing to the renowned property of asymptotic freedom. Specifically, for thermal QCD properties, it is established from first principles lattice QCD simulations, that above temperatures $T_{\rm pc} \sim 155 {\rm MeV}$~\cite{Aoki:2006br,Aoki:2009sc,Borsanyi:2010bp,Bazavov:2011nk} hadronic bound states dissolve into a Quark-Gluon plasma (QGP) and the approximate chiral symmetry of light-flavor QCD is restored. While (resummed) perturbative approaches to QCD are able to describe the most important static thermal properties of high-temperature QCD down to approximately $\sim 2 T_{\rm pc}$~\cite{Haque:2014rua}, the perturbative description appears to be worse for dynamical properties, where e.g. next-to-leading order calculations of transport coefficients \cite{Ghiglieri:2018dib,Ghiglieri:2018dgf} yield large corrections to the leading order results \cite{Arnold:2003zc,Arnold:2000dr}, indicating a poor convergence of the perturbative expansion. Nevertheless, it is conceivable that at energy scales corresponding to $\gtrsim 4T_{\rm pc}$, achieved during the early stages of high-energy heavy-ion collisions~\cite{Giacalone:2019ldn}, perturbative descriptions can provide useful insights into the early-time non-equilibrium dynamics of the system. Besides the potential relevance to early-universe Cosmology and Heavy-Ion phenomenology, it is also of genuine theoretical interest to understand the unique microscopic dynamics of thermalization processes in QCD or QCD-like plasmas.

During the past few year, significant progress in understanding thermalization and ``hydrodynamization", i.e. the onset of hydrodynamic behavior,  in high-energy heavy-ion collisions has been achieved, within the limiting cases of weakly coupled QCD~\cite{Kurkela:2015qoa,Kurkela:2018wud,Kurkela:2018vqr,Kurkela:2018xxd,Almaalol:2020rnu} and strongly-coupled holographic descriptions~\cite{Chesler:2008hg,Balasubramanian:2010ce,Heller:2011ju,Keegan:2015avk}. Despite clear microscopic differences, a common finding is that the evolution of macroscopic quantities, such as the energy momentum tensor, follows a hydrodynamic behavior well before the system reaches an approximate state of local thermal equilibrium.

Specifically, for weakly-coupled QCD plasmas, a detailed microscopic understanding of the thermalization process has also been established, as described e.g. in the recent reviews~\cite{Schlichting:2019abc,Berges:2020fwq}. Different weak-coupling thermalization scenarios based on parametric estimates~\cite{Baier:2000sb,Bodeker:2005nv,Kurkela:2011ti,Blaizot:2011xf}, distinguish between two broadly defined classes of non-equilibrium systems, commonly referred to as over-occupied or under-occupied~\cite{Schlichting:2019abc}, which undergo qualitatively different thermalization processes. While the thermalization of over-occupied QCD plasmas proceeds via a self-similar direct energy cascade~\cite{Kurkela:2012hp,Schlichting:2012es,Berges:2013eia,Berges:2013fga}, as is the case for many far-from equilibrium systems~\cite{Micha:2004bv,Berges:2014bba,Orioli:2015dxa}, under-occupied QCD plasmas undergo the so-called ``bottom-up'' scenario~\cite{Baier:2000sb} where thermalization proceeds via an inverse energy cascade, which is in many ways unique to QCD and QCD-like systems.
Earlier parametric estimates have now been supplemented with detailed simulations of the non-equilibrium dynamics based on classical-statistical lattice gauge theory~\cite{Kurkela:2012hp,Schlichting:2012es,Berges:2013fga,Berges:2013eia,Berges:2017igc} and effective kinetic theory~\cite{Kurkela:2014tea,Blaizot:2014jna,Scardina:2014gxa,Xu:2014ega,Kurkela:2018oqw}. However, with the exception of Ref.~\cite{Blaizot:2014jna,Kurkela:2018oqw}, all of the aforementioned studies have been performed for $SU(N_c)$ Yang-Mills theory, i.e. only taking into account the bosonic degrees of freedom and neglecting the effect of dynamical fermions.

Central objective of this paper is to extend the study of thermalization processes of weakly coupled non-abelian plasmas, to include all relevant quark and gluon degrees of freedom. Based the leading order effective kinetic theory of QCD~\cite{Arnold:2002zm}, we perform numerical simulations of the non-equilibrium dynamics of the QGP, to characterize the mechanisms and time scales for kinetic and chemical equilibration processes. By explicitly taking into account all light flavor degrees of freedom, i.e. gluons ($g$) as well as $u,\bar{u},d,\bar{d},s,\bar{s}$ quarks/anti-quarks, we further investigate the non-equilibrium dynamics of QCD plasmas at zero and non-zero values of the conserved $u,d,s$ charges.

We organize the discussion in this paper as follows. We begin with an brief explanation of the general setup in Sec.~\ref{sec-theory}, where we discuss the characterization of weakly coupled non-equilibrium QCD plasmas in Sec.~\ref{sec-theory-thermodynamics}, and outline their effective kinetic description in Sec.~\ref{sec-theory-transport}. Based on this framework, we study different thermalization mechanisms of the QGP, starting with the chemical equilibration of near-equilibrium systems in Sec.~\ref{sec-evol-gluonquark}.
Subsequently, in Sec.~\ref{sec:FarEq} we investigate kinetic and chemical equilibration processes in far-from equilibrium systems considering the two stereotypical examples of over-occupied systems in Sec.~\ref{sec-evol-overoccupied} and under-occupied systems in Sec.~\ref{sec-evol-underoccupied}. In Sec.~\ref{sec-evol-expansion} we continue with the study of longitudinally expanding QCD plasmas, which are relevant to describe the early time dynamics of high-energy heavy-ion collisions. Here, we mainly focus on the microscopic aspects underlying the isotropization of the pressure, and evolution of the QGP chemistry at zero and non-zero net-baryon density, and refer to our companion paper ~\cite{Du:2020pre} for additional discussions on the implications of our findings in the context of relativistic heavy-ion collisions. We conclude in Sec.~\ref{sec:conclusions} with a brief a summary of our most important findings and a discussion of possible future extensions. Several Appendices~\ref{sec-algorithm},~\ref{sec-discretization},~\ref{sec-algorithm3} contain additional details regarding the details of our numerical implementation of the QCD kinetic equations.  
\section{Non-equilibrium QCD}
\label{sec-theory}
Generally the description of non-equilibrium processes in Quantum Chromo Dynamics (QCD) represents a challenging task, and at present can only be achieved in limiting cases, such as the weak coupling limit. We employ a leading order kinetic description of QCD~\cite{Arnold:2002zm}, where the non-equilibrium evolution of the system is described in terms of the phase-space density $f(\vec{x},\vec{p},t)$ of on-shell quarks and gluons. We will focus on homogenous systems, for which the phase-space density $f(\vec{x},\vec{p},t)=f(\vec{p},t)$ only depends on momenta and time, and investigate the non-equilibrium dynamics of the QGP, based on numerical solutions of the QCD kinetic equations. Below we provide an overview of the relevant ingredients, with additional details on the numerical implementation provided in Appendices~\ref{sec-algorithm},~\ref{sec-discretization},~\ref{sec-algorithm3}.

\subsection{Non-equilibrium properties of the Quark-Gluon Plasma}
\label{sec-theory-thermodynamics}
Before we address the details of the QCD kinetic description, we briefly introduce a few relevant quantities that will be used to characterize static properties and interactions in non-equilibrium systems. We first note that both equilibrium, as well as non-equilibrium systems can be characterized in terms of their conserved charges, which for the light flavor degrees of freedom of QCD correspond to the conserved energy density $e$, and the conserved net-charge densities $\Delta n_{u}, \Delta n_{d},\Delta n_{s}$ of up,down and strange quarks. Evidently in thermal equilibrium, the maximal entropy principle uniquely determines the phase-space distribution of gluons and quarks 
\begin{eqnarray}
\nonumber
f_g^{\rm eq}(p,T)&&=\frac{1}{e^{p/T}-1},\\
\nonumber
f_{q_{f}}^{\rm eq}(p,T,\mu_f)&&=\frac{1}{e^{(p-\mu_f)/T}+1},\\
f_{\bar{q}_{f}}^{\rm eq}(p,T,\mu_f)&&=\frac{1}{e^{(p+\mu_f)/T}+1}.
\end{eqnarray}
with well-defined temperature $T_{\rm eq}$ and chemical potential $\mu_{f,\rm eq}$ determined by the values of the densities of the charges according to 
\begin{eqnarray}
\label{eq-energydensity}
\nonumber
e&&=\int \frac{d^3p}{(2\pi)^3}p\left[\nu_g f_{g}(\vec{p})+\nu_q \sum_f\left(f_{q_{f}}(\vec{p})+f_{\bar{q}_{f}}(\vec{p})\right)\right]\\
\nonumber
&&\overset{(eq)}{=}\left[\nu_g\frac{\pi^2}{30}
+\nu_q\frac{\pi^2}{120}\sum_f\left(7+\frac{30}{\pi^2}z_f^2+\frac{15}{\pi^4}z_f^4\right)\right]T_{\rm eq}^4\\
\nonumber
\Delta n_{f}&&=(n_{q}-n_{\bar{q}})_{f}=\int \frac{d^3p}{(2\pi)^3}\left[\nu_q\left(f_{q_{f}}(\vec{p})-f_{\bar{q}_{f}}(\vec{p})\right)\right]\\
&&\overset{(eq)}{=}
\frac{\nu_q}{6}\left[z_f+\frac{1}{\pi^2}z_f^3\right]T_{\rm eq}^3.
\end{eqnarray}
where we denote $z_f=\frac{{f,\rm eq}}{T_{\rm eq}}$ for the three light flavors $f=u,d,s$, which we will treat as massless throughout this work. Even though a non-equilibrium system can no longer be characterized uniquely in terms of its conserved charges, it is nevertheless useful to associate effective temperatures $T_{ldm}$ and chemical potentials $\mu_{f,ldm}$ with the system, which can be determined via the so called Landau matching procedure of determining $T_{ldm}, \mu_{f,ldm}$ from the conserved charges according to the relations in Eq.~(\ref{eq-energydensity}). Specifically for systems with conserved energy and charge densities, $T_{ldm}$ and $\mu_{f,ldm}$ will ultimately determine the equilibrium temperature $T_{\rm eq}=T_{ldm}$ and chemical potential $\mu_{f,\rm eq}=\mu_{f,ldm}$ once the system has thermalized.

Besides the densities of the conserved quantities, there is another set of important quantities relevant to describe the interactions in  non-equilibrium QCD plasmas~\cite{Arnold:2002zm}. Specifically, this includes the in-medium screening masses of quarks and gluons, which in the case of the gluon can be expressed as in terms of the Debye mass
\begin{eqnarray}
\label{eq-debyemass}
m_D^2&&=\frac{4g^2}{d_A}\int \frac{d^3p}{(2\pi)^3}\frac{1}{2p}\left[\nu_g C_Af_g(\vec{p})\right.\\
\nonumber
&&\left.+\nu_q C_F\sum_f\left(f_{q_{f}}(\vec{p})+f_{\bar{q}_{f}}(\vec{p})\right)\right]\\
\nonumber
&&\overset{(eq)}{=}\frac{g^2}{6d_A}\left[\nu_g C_A+\nu_q C_F\sum_f\left(1+\frac{3}{\pi^2}z_f^2\right)\right]T_{\rm eq}^2
\end{eqnarray}
with non-equilibrium gluon and quark distributions $f_{g}(\vec{p})$, $f_{q}(\vec{p})$, $f_{\bar{q}}(\vec{p})$. Similarly, the thermal quark masses $m_{Q_{f}}^2$ for $f=u,d,s$ quarks also enter in the kinetic description and can be expressed as 
\begin{eqnarray}
\label{eq-quarkmass}
\nonumber
m_{Q_{f}}^2&&=g^2C_F\int \frac{d^3p}{(2\pi)^3}\frac{1}{2p}\left[2f_{g}(\vec{p})+\left(f_{q_{f}}(\vec{p})+f_{\bar{q}_{f}}(\vec{p})\right)\right]\\
&&\overset{(eq)}{=}\frac{g^2}{8}C_F\left[1+\frac{1}{\pi^2}z_f^2\right]T_{\rm eq}^2.
\end{eqnarray}
While the screening masses $m_{D}^2$ and $m_{Q_{f}}^2$ determine the elastic scattering matrix elements, the calculation of the effective rates for inelastic processes also requires the asymptotic masses of quarks and gluons, $m_{\infty,a}^2$ which to leading order in perturbation theory can be related to the respective screening masses according to $m_{\infty,g}^{2}=m_{D}^{2}/2$ and $m_{\infty,Q_{f}}^{2}=2 m_{Q_{f}}^{2}$. Since inelastic interactions are induced by elastic collisions, their effective in-medium rates are also sensitive to the density of elastic interaction partners
\begin{eqnarray}
\label{eq-effetivetemperature}
&&m_D^2 T^{*}=\frac{g^2}{d_A}\int \frac{d^3p}{(2\pi)^3}\bigg\{\nu_g C_Af_g(\vec{p})(1+f_{g}(\vec{p}))\\
\nonumber
&&+\nu_q C_F\sum_f\left[f_{q_{f}}(\vec{p})(1-f_{q_{f}}(\vec{p}))+f_{\bar{q}_{f}}(\vec{p})(1-f_{\bar{q}_{f}}(\vec{p}))\right]\bigg\}
\end{eqnarray}
which receives the usual Bose enhancement $f_{g}(\vec{p})(1+f_{g}(\vec{p}))$ and Fermi blocking $f_{q/\bar{q}}(\vec{p})(1-f_{q/\bar{q}}(\vec{p}))$ factors. Since we will frequently characterize the non-equilibrium evolution of the QGP in terms of the above dynamical scales, we further note that the quantity $g^2 T^{*}$ characterizes the rate of small angle scatterings in the plasma, with $T^{*}$ defined such that in equilibrium $T^{*}\overset{(\rm eq)}{=}T_{\rm eq}$ corresponds to the equilibrium temperature.

\subsection{Effective Kinetic Theory of Quark-Gluon Plasma}
\label{sec-theory-transport}
We adopt an effective kinetic description of the QGP, which at leading order includes both ``$2\leftrightarrow2$'' elastic processes as well as effective ``$1\leftrightarrow2$'' collinear inelastic processes. Specifically for a spatially homogeneous system, the time evolution of the phase-space density of quarks and gluons is then described by the Boltzmann equation for QCD light particles ``$a={g,u,\bar{u},d,\bar{d},s,\bar{s}}$''
\begin{eqnarray}
\label{eq-bolzmann}
\frac{\partial}{\partial t}f_a(\vec{p},t)
=-C^{{2\leftrightarrow2}}_a[f](\vec{p},t)-C^{{1\leftrightarrow2}}_a[f](\vec{p},t)
\end{eqnarray}
where $C^{{2\leftrightarrow2}}_a[f](\vec{p},t)$ is the $2\leftrightarrow2$ elastic collision term and $C^{{1\leftrightarrow2}}_a[f](\vec{p},t)$ is the $1\leftrightarrow2$ inelastic collision term.

With regards to the numerical implementation, we follow previous works and solve the QCD Boltzmann equation directly as an integro-differential equation using pseudo spectral methods~\cite{York:2014wja,Keegan:2015avk}. Our numerical implementation of the non-equilibrium dynamics rely on a discretized form of the Boltzmann equation
\begin{eqnarray}
\label{eq-fton0}
n(i_{p},j_{\theta},k_{\phi},t)&&= \int \frac{d^3p}{(2\pi)^3} W_{(i_{p},j_{\theta},k_{\phi})}(\vec{p})f(\vec{p},t)\\
\nonumber
&&=\int \frac{d^3p}{(2\pi)^3}
w^{(p)}_{i}(p) w^{(\theta)}_{j}(\theta) w^{(\phi)}_{k}(\phi)f(\vec{p},t)
\end{eqnarray}
based on a weight function algorithm ~\cite{Lu:2015frp}, which is described in detail in Appendices~\ref{sec-algorithm},~\ref{sec-discretization}. Based on Eq.~(\ref{eq-fton0}), the discretized form of the Boltzmann equation for species ``a'' can be written as
\begin{eqnarray}
&&\frac{\partial}{\partial t}n_a(i_{p},j_{\theta},k_{\phi},t)\\
\nonumber
&&=-C^{{2\leftrightarrow2}}_a[n](i_{p},j_{\theta},k_{\phi},t)-C^{{1\leftrightarrow2}}_a[n](i_{p},j_{\theta},k_{\phi},t)
\end{eqnarray}
where in accordance with Eq.~(\ref{eq-fton0}), $C^{{2\leftrightarrow2}}_a[n](i_{p},j_{\theta},k_{\phi},t)$ and $C^{{1\leftrightarrow2}}_a[n](i_{p},j_{\theta},k_{\phi},t)$ correspond to discretized moments of the collision integral. Based on a suitable choice of the weight functions $w^{(p)}_{i}(p)$,$w^{(\theta)}_{j}(\theta)$ and $w^{(\phi)}_{k}(\phi)$, the discretization of the collision integrals is performed such that it ensures an exact conservation of the particle number for elastic collision, as well as an exact conservation of energy for both elastic and inelastic collisions. 

\subsubsection{Elastic Collisions}
\label{subsec-elastic}
Within our effective kinetic description, we include all leading order elastic scattering processes between quarks and gluons, where following previous works~\cite{Kurkela:2014tea,Kurkela:2015qoa,Kurkela:2018vqr,Kurkela:2018oqw} the relevant in-medium scattering matrix elements are determined based on an effective isotropic screening assumption.
\paragraph{Collision Integral}
We follow the notation of Arnold, Moore and Yaffe (AMY)~\cite{Arnold:2002zm}, where the elastic collision integrals for particle $a$ with momentum $\vec{p}_1$ participating in scattering process $a,b \rightarrow c,d$ with $p_{1},p_{2}\leftrightarrow p_{3},p_{4}$ takes the form
\begin{eqnarray}
\label{eq-cint-elastic}
&&C^{{2\leftrightarrow2}}_a[f](\vec{p}_1)
=\frac{1}{2 \nu_{a}}\frac{1}{2 E_{p_1}}\\
\nonumber
&&{\sum_{cd}}
\int d\Pi_{2\leftrightarrow2}
|\mathcal{M}_{cd}^{ab}(\vec{p}_1,\vec{p}_2|\vec{p}_3,\vec{p}_4)|^2F_{cd}^{ab}(\vec{p}_1,\vec{p}_2|\vec{p}_3,\vec{p}_4)
\end{eqnarray}
with $d\Pi_{2\leftrightarrow2}$ denoting the measure
\begin{eqnarray}
\nonumber
&&d\Pi_{2\leftrightarrow2} = \frac{d^3p_2}{(2\pi)^3} \frac{1}{2E_{p_2}} \frac{d^3p_3}{(2\pi)^3} \frac{1}{2E_{p_3}} \frac{d^3p_4}{(2\pi)^3} \frac{1}{2E_{p_4}}\\
&&\times(2\pi)^4 \delta^{(4)}({p}_1+{p}_2-{p}_3-{p}_4)
\end{eqnarray}
and $\nu_G=2(N_c^2-1)=16, \nu_Q=2N_c=6$ denoting the number of gluon and quark degrees of freedom. By $|\mathcal{M}_{cd}^{ab}(\vec{p}_1,\vec{p}_2|\vec{p}_3,\vec{p}_4)|^2$ we denote the square matrix element for the process ``$a,b\leftrightarrow c,d$'' summed over spin and color for all particles, while $F_{cd}^{ab}(\vec{p}_1,\vec{p}_2|\vec{p}_3,\vec{p}_4)$ is the statistical factor for the ``$a,b \leftrightarrow c,d$'' scattering process
\begin{eqnarray}
\nonumber
&&F_{cd}^{ab}(\vec{p}_1,\vec{p}_2|\vec{p}_3,\vec{p}_4) =f_a(\vec{p}_1)f_b(\vec{p}_2)(1\pm f_c(\vec{p}_3))(1\pm f_d(\vec{p}_4))\\
&&-f_c(\vec{p}_3)f_d(\vec{p}_4)(1\pm f_a(\vec{p}_1))(1\pm f_b(\vec{p}_2))
\label{eq:StatFacFDef}
\end{eqnarray}
where ``$\pm$" provides a Bose enhancement $(+)$ for gluons and Fermi blocking $(-)$ for quarks, such that the first term in Eq.(\ref{eq:StatFacFDef}) represents a loss term, whereas the second term in Eq.(\ref{eq:StatFacFDef}) corresponds to a gain term associated with the inverse process.

\paragraph{Scattering matrix elements}
\begin{center}
	\begin{table*}
		\begin{tabular}{|c|c|}
			\hline
			Processes $a,b \leftrightarrow c,d$ & Matrix element $|\mathcal{M}_{cd}^{ab}(\vec{p}_1,\vec{p}_2|\vec{p}_3,\vec{p}_4)|^2$\\ 
			\hline
			$gg \leftrightarrow gg$ & $4g^4d_{A}C_{A}^2\left(9+\frac{(s-u)^2}{\slt^2}+\frac{(s-t)^2}{\slu^2}+\frac{(t-u)^2}{s^2}\right)$\\ 
			\hline
			$qg \leftrightarrow qg$, $\bar{q}g \leftrightarrow \bar{q}g$ & $-8 g^4 d_F C_F^2 \left(\frac{u}{s} +\frac{s}{\dlu}\right) + 8 g^4 d_F C_F C_A\left(\frac{s^2+u^2}{\slt^2}\right)$\\ 
			\hline
			$q\bar{q} \leftrightarrow gg$ & $8 g^4 d_F C_F^2 \left(\frac{u}{\dlt} +\frac{t}{\dlu}\right) - 8 g^4 d_F C_F C_A\left(\frac{t^2+u^2}{s^2}\right)$\\
			\hline
			$qq \leftrightarrow qq$, $\bar{q}\bar{q} \leftrightarrow \bar{q}\bar{q}$ & $8 g^4 \frac{d_F^2 C_F^2}{d_A} \left(\frac{t^2+s^2}{\slu^2}+\frac{u^2+s^2}{\slt^2}\right)+
			16 g^4 d_F C_F \left(C_F-\frac{C_A}{2}\right)
			\left(\frac{s^2}{\slt\slu}\right)$\\ 
			\hline
			$q\bar{q} \leftrightarrow q\bar{q}$ & $8 g^4 \frac{d_F^2 C_F^2}{d_A} \left(\frac{s^2+u^2}{\slt^2}+\frac{t^2+u^2}{s^2}\right)+
			16 g^4 d_F C_F \left(C_F-\frac{C_A}{2}\right)
			\left(\frac{u^2}{s\slt}\right)$\\ 
			\hline
			$q_1q_2 \leftrightarrow q_1q_2$ & \\ 
			$q_1\bar{q}_2 \leftrightarrow q_1\bar{q}_2$ & $8 g^4 \frac{d_F^2 C_F^2}{d_A} \left(\frac{s^2+u^2}{\slt^2}\right)$\\
			$\bar{q}_1\bar{q}_2 \leftrightarrow \bar{q}_1\bar{q}_2$ & \\ 
			\hline
			$q_1\bar{q}_1 \leftrightarrow q_2\bar{q}_2$ & $8 g^4 \frac{d_F^2 C_F^2}{d_A} \left(\frac{t^2+u^2}{s^2}\right)$\\ 
			\hline
		\end{tabular}
		\caption{Summary of $2\leftrightarrow2$ elastic processes and their leading order pQCD scattering amplitudes.}
		\label{tb-pQCD}
	\end{table*}
\end{center}
Elastic scattering matrix elements for the various $2\leftrightarrow 2$ processes can be calculated in perturbative QCD (pQCD)~\cite{Arnold:2002zm}, with the corresponding leading order matrix elements listed in Table~(\ref{tb-pQCD}), where $g$ is the gauge coupling, $s=(p_1+p_2)^2$, $t=(p_1-p_3)^3$ and $u=(p_1-p_4)^2$ denote the usual Mandelstam variables and $C_F=\frac{N_c^2-1}{2N_c}=\frac{4}{3}$, $C_A=N_c=3$, $d_F=N_c=3$, $d_A=N_c^2-1=8$ denote the group theoretical factors. However, due to the enhancement of soft $t,u$-channel gluon and quark exchanges, the vacuum matrix elements in Table~\ref{tb-pQCD} give rise to divergent scattering rates, which inside the medium are regulated by incorporating screening effects via the insertions of the Hard-Thermal loop (HTL) self-energies,
as discussed in detail in~\cite{Arnold:2002zm}. Even though it should in principle be possible to include the full HTL self-energies in the calculation of the in-medium elastic scattering matrix elements (at least for homogenous and isotropic systems), this would represent yet another significant complication as the corresponding expressions would have to be re-evaluated numerically at each time step and we did not pursue this further. Instead, we follow previous works~\cite{Kurkela:2014tea,Kurkela:2015qoa,Kurkela:2018vqr,Kurkela:2018oqw}, and incorporate an effective isotropic screening, where soft $t-$ and $u-$ channel exchanges are regulated by screening masses $m_D^2$ and $m_{Q_{f}}^2$ for different species of internal exchange particles, by replacing $t$ and $u$ in the singly and doubly underlined expressions in Table~\ref{tb-pQCD} with
\begin{eqnarray}
\nonumber
&&\slt= t\left(1+\frac{\xi_g^2m_g^2}{\vec{q}^2_{t}}\right),~~~~~
\slu= u\left(1+\frac{\xi_g^2m_g^2}{\vec{q}^2_{u}}\right),\\
&&\dlt= t\left(1+\frac{\xi_q^2m_q^2}{\vec{q}^2_{t}}\right),~~~~~
\dlu= u\left(1+\frac{\xi_q^2m_q^2}{\vec{q}^2_{u}}\right)
\end{eqnarray}
where $\vec{q}_{t}=\vec{p}_1-\vec{p}_{3}$, $\vec{q}_{u}=\vec{p}_1-\vec{p}_{4}$ is the spatial momentum of the exchanged particle, and the parameters $\xi_g=\frac{e^{\frac{5}{6}}}{2\sqrt{2}}$, $\xi_q=\frac{e}{\sqrt{2}}$ have been determined in~\cite{Kurkela:2018oqw} by matching to leading order HTL results. Based on the above expressions for the collision integrals and scattering matrix elements, the corresponding integrals for the discretized moments $C^{{2\leftrightarrow2}}_a[n](i_{p},j_{\theta},k_{\phi},t)$, is then calculated at each time step by performing a Monte-Carlo sampling described in detail in Appendix~\ref{sec-discretization-elastic}.

\subsubsection{Inelastic Collisions}
\label{subsec-inelastic}
Within our effective kinetic description, we include all leading order inelastic scattering processes between quarks and gluons, where following previous works~\cite{Kurkela:2014tea,Kurkela:2015qoa,Kurkela:2018vqr,Kurkela:2018oqw} the relevant in-medium scattering matrix elements are determined within the formalism of Arnold, Moore and Yaffe~\cite{Arnold:2002zm}, by solving an integro-differential equation for the effective collinear emission/absorption rates to take into account account coherence effects associated with the Landau-Pomeranchuk-Migdal (LPM) effect~\cite{Landau:1953gr,Landau:1953um,Migdal:1955nv}.

\paragraph{Collision Integral}
\label{subsec-inelastic-collision}
Generally, the inelastic collision integral for particle ``a'' with momentum $\vec{p}_1$ participating in the splitting process $a \rightarrow b,c$ ($p_{1}\leftrightarrow p_{2},p_{3}$) and the inverse joining process $a,b \rightarrow c$ ($p_{1},p_{2}\leftrightarrow p_{3}$) takes the form
\begin{eqnarray}
\label{eq-cint-inelastic}
&&C^{{1\leftrightarrow2}}_a[f](\vec{p}_1)\\
\nonumber
&&=\frac{1}{2 \nu_{a}}\frac{1}{2 E_{p_1}}\sum_{bc}\int d\Pi_{1\leftrightarrow 2}^{a\leftrightarrow bc}
|\mathcal{M}_{bc}^{a}(\vec{p}_1|\vec{p}_2,\vec{p}_3)|^2F_{bc}^{a}(\vec{p}_1|\vec{p}_2,\vec{p}_3)\\
\nonumber
&&+\frac{1}{\nu_{a}}\frac{1}{2 E_{p_1}}\int d\Pi_{1\leftrightarrow 2}^{ab\leftrightarrow c}
|\mathcal{M}_{c}^{ab}(\vec{p}_1,\vec{p}_2|\vec{p}_3)|^2F_{c}^{ab}(\vec{p}_1,\vec{p}_2|\vec{p}_3)\\
\nonumber
&&=\frac{1}{2 \nu_{a}}\frac{1}{2 E_{p_1}}\sum_{bc}\int d\Pi_{1\leftrightarrow 2}^{a\leftrightarrow bc}
|\mathcal{M}_{bc}^{a}(\vec{p}_1|\vec{p}_2,\vec{p}_3)|^2F_{bc}^{a}(\vec{p}_1|\vec{p}_2,\vec{p}_3)\\
\nonumber
&&-\frac{1}{\nu_{a}}\frac{1}{2 E_{p_1}}\int d\Pi_{1\leftrightarrow 2}^{ab\leftrightarrow c}
|\mathcal{M}_{ab}^{c}(\vec{p}_3|\vec{p}_1,\vec{p}_2)|^2F_{ab}^{c}(\vec{p}_3|\vec{p}_1,\vec{p}_2)
\end{eqnarray}
where $d\Pi_{1\leftrightarrow2}^{a\leftrightarrow bc}$ and $d\Pi_{1\leftrightarrow2}^{ab\leftrightarrow c}$ denote the measures
\begin{eqnarray}
\nonumber
&&\int d\Pi_{1\leftrightarrow2}^{a\leftrightarrow bc}= \int\frac{d^3p_2}{(2\pi)^3} \frac{1}{2E_{p_2}} \int\frac{d^3p_3}{(2\pi)^3} \frac{1}{2E_{p_3}}\\
\nonumber
&&\times(2\pi)^4\delta^{(4)}({p}_1-{p}_2-{p}_3).\\
\nonumber
&&\int d\Pi_{1\leftrightarrow2}^{ab\leftrightarrow c}= \int\frac{d^3p_2}{(2\pi)^3} \frac{1}{2E_{p_2}} \int\frac{d^3p_3}{(2\pi)^3} \frac{1}{2E_{p_3}}\\
&&\times(2\pi)^4\delta^{(4)}({p}_1+{p}_2-{p}_3).
\end{eqnarray}
$|\mathcal{M}_{bc}^{a}(\vec{p}_1|\vec{p}_2,\vec{p}_3)|^2$ and $|\mathcal{M}_{c}^{ab}(\vec{p}_1,\vec{p}_2|\vec{p}_3)|^2$ are the matrix element squares for process ``$a\leftrightarrow b,c$'' and ``$a,b\leftrightarrow c$''. $F_{cd}^{a}(\vec{p}_1|\vec{p}_3,\vec{p}_4)$ and $F_{c}^{ab}(\vec{p}_1,\vec{p}_2|\vec{p}_3)$
are the statistical factors
\begin{eqnarray}
\nonumber
&&F_{bc}^{a}(\vec{p}_1|\vec{p}_2,\vec{p}_3)
=f_a(\vec{p}_1)(1\pm f_b(\vec{p}_2))(1\pm f_c(\vec{p}_3))\\
\nonumber
&&-f_b(\vec{p}_2)f_c(\vec{p}_3)(1\pm f_a(\vec{p}_1))\\
\nonumber
&&F_{c}^{ab}(\vec{p}_1,\vec{p}_2|\vec{p}_3) =f_a(\vec{p}_1)f_b(\vec{p}_2)(1\pm f_c(\vec{p}_3))\\
&&-f_c(\vec{p}_3)(1\pm f_a(\vec{p}_1))(1\pm f_b(\vec{p}_2))
\end{eqnarray}
where again ``$\pm$" provides a Bose enhancement ($+$) for gluon  and Fermi blocking ($-$) for quarks.

Since for ultra-relativistic particles, the $1\leftrightarrow 2$ processes require collinearity to be kinematically allowed, the collision integral in Eq.(\ref{eq-cint-inelastic}) can be recast into an effectively one-dimensional collinear process
\begin{eqnarray}
\label{eq-ccint-inelastic-cast}
\nonumber
&&C^{{1\leftrightarrow2}}_a[f](p)
=\frac{1}{2\nu_{a}}\int_0^1dz
\left[
\sum_{bc}\frac{d\Gamma_{bc}^{a}}{dz}\big(p,z\big)\nu_{a}F_{bc}^{a}(p|zp,\bar{z}p)\right.\\
\nonumber
&&\left.-\frac{1}{z^3}\frac{d\Gamma_{ab}^{c}}{dz}\big(\frac{p}{z},z\big)\nu_{c}F_{ab}^{c}(\frac{p}{z}|p,\frac{\bar{z}}{z}p)\right.\\
&&\left.-\frac{1}{\bar{z}^3}\frac{d\Gamma_{ab}^{c}}{dz}\big(\frac{p}{\bar{z}},\bar{z}\big)\nu_{c}F_{ab}^{c}(\frac{p}{\bar{z}}|p,\frac{z}{\bar{z}}p)
\right]
\end{eqnarray}
where $z$ and $\bar{z}=1-z$ are the collinear splitting/joining fractions, and the effective inelastic rate $\frac{d\Gamma_{bc}^{a}}{dz}$ in Eq.~(\ref{eq-ccint-inelastic-cast}) is obtained within the framework of AMY~\cite{Arnold:2002zm}, by considering the overall probability of a single radiative emission/absorption over the course of multiple successive elastic interactions, reduced to an effective collinear rate by integrating over the parametrically small transverse momentum accumulated during the emission/absorption process.\footnote{We note that several different notations for the rate $\frac{d\Gamma_{bc}^{a}}{dz}$ exist in the literature, and we refer to the Appendix of Ref.~\cite{Arnold:2008iy} for a comparison of different notations.} 

\paragraph{Effective Inelastic Rate}
\label{subsec-inelastic-rate}
Based on the formalism of AMY~\cite{Arnold:2002zm}, the effective inelastic rate can be expressed in the following factorized form
\begin{eqnarray}
\frac{d\Gamma_{bc}^{a}}{dz}(p,z)=\frac{\alpha_s P_{bc}^{a}(z)}{[2pz(1-z)]^2}\int\frac{d^2 \ptt}{(2\pi)^2}{\rm Re}\left[2\ptt\cdot \gtt_{(p,z)}(\ptt)\right]\;, \nonumber \\
\end{eqnarray}
where the matrix element for the collinear splitting is expressed in terms of the leading-order QCD splitting functions (DGLAP)~\cite{Dokshitzer:1977sg,Gribov:1972ri,Altarelli:1977zs},
\begin{eqnarray}
&&P_{gg}^{g}(z)=\frac{2C_A(1-z(1-z))^2}{z(1-z)}, \nonumber\\ 
&&P_{qq}^{g}(z)=\frac{z^2+(1-z)^2}{2}, \nonumber\\
&&P_{gq}^{q}(z)=\frac{C_F(1+(1-z)^2)}{z}\;, \nonumber \\
&&P_{qg}^{q}(z)=\frac{C_F(1+z^2)}{1-z}\;.
\end{eqnarray}
Secondly, the factor $\int\frac{d^2 \ptt}{(2\pi)^2}{\rm Re}\left[2\ptt\cdot \gtt_{(p,z)}(\ptt)\right]$ encodes the relevant aspects of the current-current correlation function inside the medium, and satisfies the following integral equation for particles $a,b,c$ carrying momentum fraction $1,z,1-z$
\begin{eqnarray}
\label{eq-integralequation}
&&2\ptt=i\delta E_{(p,z)}(\ptt)\times \gtt_{(p,z)}(\ptt)\\
\nonumber
&&+\frac{1}{2}\int\frac{d^2\qtt}{(2\pi)^2}~\frac{d\bar{\Gamma}^{\rm el}}{d^2\qtt}\times\\
\nonumber
&&\left\{\left(C^{R}_b+C^{R}_c-C^{R}_a\right)
\left[\gtt_{(p,z)}(\ptt)-\gtt_{(p,z)}(\ptt-\qtt)\right]\right.\\
\nonumber
&&+\left.\left(C^{R}_c+C^{R}_a-C^{R}_b\right)
\left[\gtt_{(z,P)}(\ptt)-\gtt_{(p,z)}(\ptt -z\qtt)\right]\right.\\
\nonumber
&&\left.+\left(C^{R}_a+C^{R}_b-C^{R}_c\right)
\left[\gtt_{(p,z)}(\ptt)- \gtt_{(p,z)}(\ptt-(1-z)\qtt)\right]\right\},
\end{eqnarray}
with 
\begin{eqnarray}
&&\delta E_{(p,z)}(\ptt)=\frac{\ptt^2}{2pz(1-z)} +\frac{m^2_{\infty,b}}{2zp}  + \frac{m^2_{\infty,c}}{2(1-z)p} -\frac{m^2_{\infty,a}}{2p}\;, \nonumber \\
\end{eqnarray}
where $m_{\infty,a}^2,m_{\infty,b}^2,m_{\infty,c}^2$ denote the asymptotic masses of particles $a,b,c$, i.e $m_{\infty,g}^2=\frac{m_D^2}{2}$ for gluons and $m_{\infty,q_{f}}^2=2m_{Q_{f}}^2$ for quarks, $C^{R}_a,C^{R}_b, C^{R}_c$ denote the Casimir of the representation of, i.e. $C^{R}_q=C_{F}$ for quarks and $C^{R}_g=C_{A}$ for gluons, and $\frac{d\bar{\Gamma}^{\rm el}}{d^2\qtt}$ denotes the differential elastic scattering rate stripped by its color factor, which is given by
\begin{eqnarray}
\frac{d\bar{\Gamma}^{\rm el}}{d^2\qtt}= g^2 T^{*}~\frac{ m_D^2}{\qtt^2 (\qtt^2+m_D^2)}
\end{eqnarray}
at leading order~\cite{Arnold:2002zm}. 

Self-consistent solutions to Eq.~(\ref{eq-integralequation}) can be efficiently constructed in impact parameter space, i.e. by performing a Fourier transformation w.r.t. to $\ptt$ (see e.g. \cite{Anisimov:2010gy}), and resum the effects of multiple elastic scatterings during a single emission/absorption. Since the effective inelastic rates depend on the kinematic variables $p,z$ as well as the time dependent medium parameters $T^{*}(t), m_{D}^2(t), m_{Q_{f}}^2(t)$, in practice we tabulate the rates as a function of $p,z$ for different values of $T^{*}, m_{D}^2, m_{Q_{f}}^2$ around their momentary values $T^{*}(t), m_{D}^2(t), m_{Q_{f}}^2(t)$, such that for small variations $T^{*}, m_{D}^2, m_{Q_{f}}^2$ which occur in every time step we interpolate between neighboring points, whereas for larger variations of $T^{*}, m_{D}^2, m_{Q_{f}}^2$ which occur over the course of many time steps the entire database gets updated. Similar to the elastic scattering processes, the discretized versions $C^{{1\leftrightarrow2}}_a[n](i_{p},j_{\theta},k_{\phi},t)$ of the inelastic collision integrals in Eq.~(\ref{eq-ccint-inelastic-cast}) are then calculated using a Monte-Carlo sampling, as described in more detail in Appendix~\ref{sec-discretization-inelastic}.

Even though we will always employ Eq.~(\ref{eq-integralequation}) to calculate the effective inelastic rates in our numerical studies, it proves insightful to briefly consider the two limiting cases where the formation time $t_{\rm form} \sim 1/\delta E_{(p,z)}(\ptt)$ of the splitting is small or large compared to inverse of the (small angle) elastic scattering rate $1/\Gamma_{el}\sim 1/(g^2 T^{*})$ and closed analytic expressions for the effective inelastic rates can be obtained. We first consider the limit of small formation times, commonly referred to as the Bethe-Heitler regime~\cite{Bethe:1934za}, where radiative emissions/adsorptions are induced by a single elastic scattering and Eq.~(\ref{eq-integralequation}) can be solved perturbatively (see e.g. \cite{Schlichting:2019abc}), yielding
\begin{eqnarray}
\label{eq-BHRate}
\left.\frac{d\Gamma_{bc}^{a}}{dz}\right|_{\rm BH}(p,z)  = 2\alpha_s P^{a}_{bc}(z)~g^2T^{*}~\mathcal{I}^{a,{\rm BH}}_{bc}(z)\;,
\end{eqnarray}
where
\begin{eqnarray}
&& \mathcal{I}^{a,{\rm BH}}_{bc}(z)= \int \frac{d^2\ptt}{(2\pi)^2}\int \frac{d^2\qtt}{(2\pi)^2} \frac{m_D^2}{\qtt^2(\qtt^2+m_{D}^2)} \frac{\ptt}{\ptt^2+\mu^{a}_{bc}(z)} \nonumber\\
&&\left\{\left(C^{R}_b+C^{R}_c-C^{R}_a\right)
\left[\frac{\ptt}{\ptt^2+\mu^{a}_{bc}(z)} -\frac{\ptt-\qtt}{(\ptt-\qtt)^2+\mu^{a}_{bc}(z)}\right]\right. \nonumber\\
&&+\left.\left(C^{R}_c+C^{R}_a-C^{R}_b\right)
\left[\frac{\ptt}{\ptt^2+\mu^{a}_{bc}(z)} -\frac{\ptt-z\qtt}{(\ptt-z\qtt)^2+\mu^{a}_{bc}(z)}\right]\right. \nonumber\\
\nonumber
&&\left.+\left(C^{R}_a+C^{R}_b-C^{R}_c\right)
\left[\frac{\ptt}{\ptt^2+\mu^{a}_{bc}(z)} -\frac{\ptt-\bar{z}\qtt}{(\ptt-\bar{z}\qtt)^2+\mu^{a}_{bc}(z)}\right]\right\}\;, \nonumber\\
\end{eqnarray}
with $\mu^{a}_{bc}(z)=(1-z)m_{\infty,b}^2+zm_{\infty,c}^2-z(1-z)m_{\infty,a}^2$, such that the effective inelastic rate is essentially determined by the small angle elastic scattering rate $(\sim g^2 T^{*})$. 

Since the typical transverse momentum acquired in a single scattering is $\ptt^2 \sim m_{D}^2$ the validity of this approximations requires the formation time $t_{\rm form}\sim \frac{2p z (1-z)}{m_{D}^2}$ to be small compared to the mean free time between small angle scatterings $t_{\rm mfp}\sim \Gamma_{el}^{-1}\sim 1/g^2 T^{*}$, giving rise to a characteristic energy scale $\omega_{\rm BH}=\frac{m_D^2}{g^2T^{*}}$ such that for $2p z (1-z) \lesssim \omega_{\rm BH}$ radiative emissions/adsorptions typically occur due to a single elastic scattering. Conversely, for $2p z (1-z) \gtrsim \omega_{\rm BH}$ the radiative emission/adsorption occurs coherently over the course of many elastic scatterings, leading to the famous Landau-Pomeranchuk-Migdal suppression~\cite{Landau:1953gr,Landau:1953um,Migdal:1955nv} of the effective inelastic interaction rate. Specifically, in the high-energy limit $2p z (1-z) \gg \omega_{\rm BH}$, the effective rate can be approximated as ~\cite{Arnold:2008iy,Arnold:2008zu,Schlichting:2019abc}
\begin{eqnarray}
\nonumber
&&\left.\frac{d\Gamma_{bc}^{a}}{dz}\right|_{HO}(p,z)=
\frac{\alpha_s}{2\pi}P_{bc}^{a}(z)
\sqrt{\frac{\hat{\bar{q}}}{p}}\\
&&\times\sqrt{\frac{zC^{R}_c+(1-z)C^{R}_b-z(1-z)C^{R}_a}{z(1-z)}},\label{eq-HORate}
\end{eqnarray}
with $\hat{\bar{q}}=g^2T^{*}\frac{m_D^2}{2\pi}$, where in contrast to Eq.~(\ref{eq-BHRate}) the effective rate is determined by the formation time $t^{-1}_{\rm form} \sim \sqrt{\frac{\hat{\bar{q}}}{2pz(1-z)}}$ of the splitting/merging rather than the elastic scattering rate.
\section{Chemical equilibration of near-equilibrium Systems}
\label{sec-evol-gluonquark}

Before we address kinetic and chemical equilibration of non-abelian plasmas which are initially far-from equilibrium, we will address the conceptually simpler case of studying the chemical equilibration of systems, where initially there is only one species of particles present. While it is conceivable that such kind of states could be created in a cosmological environment, whenever the QCD sector is selectively populated via the coupling to e.g. the standard model Higgs or other BSM particles, our primary goal is to understand and characterize the dynamics underlying chemical equilibration of the QGP, and we do not claim relevance to any particular physics application. We will for simplicity assume that, e.g. due to the interaction with other non-QCD particles, the particle species that is present initially is already in thermal equilibrium at a given temperature $T_{0}$ and chemical potential $\mu_{0}$, such that over the course of the chemical equilibration process the energy of the dominant species needs to be redistributed among all QCD degrees of freedom, until eventually the final equilibrium state with a different temperature $T_{\rm eq}$ and chemical potential $\mu_{\rm eq}$ is reached.

Since the leading order kinetic description of massless QCD degrees of freedom is manifestly scale invariant, we can express the relevant momentum and time scales in terms of an arbitrary chosen unit. Naturally, for this kind of investigation, we will express our results in terms of the final equilibrium temperature $T_{\rm eq}$ and chemical potential $\mu_{\rm eq}$, such that the corresponding estimates of the physical time scales can be obtained by evaluating the expressions for the relevant temperatures and densities. Even though we employ a leading order weak-coupling description, we will investigate the behavior for different values of the QCD coupling strength \footnote{While for pure Yang-Mills theory, one can show that the results do not separately depend on $g$ and $N_c$, but only on the combination $g^2 N_c$~\cite{York:2014wja}, in non-abelian gauge theories with fundamental fermions the general dependence on the gauge coupling $g$, the number of colors $N_c$ and the number of flavors $N_f$ are more complicated and we only consider the case $N_c=3$ and $N_{f}=3$ relevant to QCD at currently available collider energies.}, typically denoted by the t'Hooft coupling $\lambda=g^2N_c$, and frequently express the dependence on the coupling strength in terms of macroscopic quantities, such as the shear-viscosity to entropy density ratio $\eta/s \sim 1/g^4$~\cite{Arnold:2000dr,Arnold:2003zc}. 

\subsection{Chemical equilibration at zero density}
We first consider the case of chemical equilibration at zero (net-) density of the conserved $u,d,s$ charges, where the systems features equivalent amounts of quarks and antiquarks, resulting in zero chemical potentials for all quark flavors. We distinguish two cases, where in the first case the system initially features a thermal distribution of gluons, without any quarks or antiquarks present at initial time, whereas in the second case the system is initially described by the same distribution of quarks/antiquark for all flavors, without gluons present in the system. Specifically, for the first case with thermal gluons only, we have
\begin{eqnarray}
\label{eq-TMGQ-INITIAL-G0}
\nonumber
&&f_{g}(p,t=0)=\frac{1}{e^{p/T_0}-1},\\
\nonumber
&&f_{q_{f}}(p,t=0)=0,\\
&&f_{\bar{q}_{f}}(p,t=0)=0,
\end{eqnarray}
where due to energy conservation, the initial parameter $T_0$ can be related to thermal equilibrium temperature $T_{\rm eq}$ by $\nu_g\frac{\pi^2}{30}T_0^4=\left(4\nu_g+7\nu_q N_f\right)\frac{\pi^2}{120}T_{\rm eq}^4$ according to Eq.~(\ref{eq-energydensity}). 
Similarly, for the second case where only quarks/antiquarks are initially present in the system, we have
\begin{eqnarray}
\label{eq-TMGQ-INITIAL-Q0}
\nonumber
&&f_{g}(p,t=0)=0,\\
\nonumber
&&f_{q_{f}}(p,t=0)=\frac{1}{e^{p/T_0}+1},\\
&&f_{\bar{q}_{f}}(p,t=0)=\frac{1}{e^{p/T_0}+1}.
\end{eqnarray}
and the initial parameter $T_0$ has the following relation to final equilibrium temperature $T_{\rm eq}$ by
$\nu_q N_f\frac{7\pi^2}{120}T_0^4=\left(4\nu_g+7\nu_q N_f\right)\frac{\pi^2}{120}T_{\rm eq}^4$ according to Eq.~(\ref{eq-energydensity}). 

Since the final equilibrium temperature $T_{\rm eq}$ is a constant scale, it is then natural to express other scales in terms of $T_{\rm eq}$, or alternatively in terms their corresponding equilibrium values, such as $m_{D}^2(T_{\rm eq}),m_{Q}^2(T_{\rm eq}),\cdots$. Besides providing a reference scale for static equilibrium quantities, the inverse of the equilibrium temperature $\sim\frac{1}{T_{\rm eq}}$ also provides a natural time scale for the evolution of the system, and it is convenient to express the time evolution in units of the near-equilibrium kinetic relaxation time
\begin{eqnarray}
\label{eq-relaxation}
\tau_R=\frac{4\pi\eta/s}{T_{\rm eq}}
\end{eqnarray}
where $\eta/s$ is the constant shear viscosity to entropy density ratios, with $\eta/s\simeq~1900,~35,~1$ for t'Hooft couplings $\lambda=g^2N_c=4\pi\alpha_sN_c= ~0.1,~1,~10$ ~\cite{Kurkela:2018oqw}.\footnote{We have performed independent extractions of $\eta/s$ within our implementation of QCD kinetic theory, which  -- as discussed in Sec.~\ref{sec-evol-expansion} -- agree with the results previously obtained by Kurkela and Mazeliauskas in \cite{Kurkela:2018oqw}.}

\subsubsection{Spectra Evolution}
\begin{figure}[!tb]
		\centering
		\includegraphics[width=0.48\textwidth]{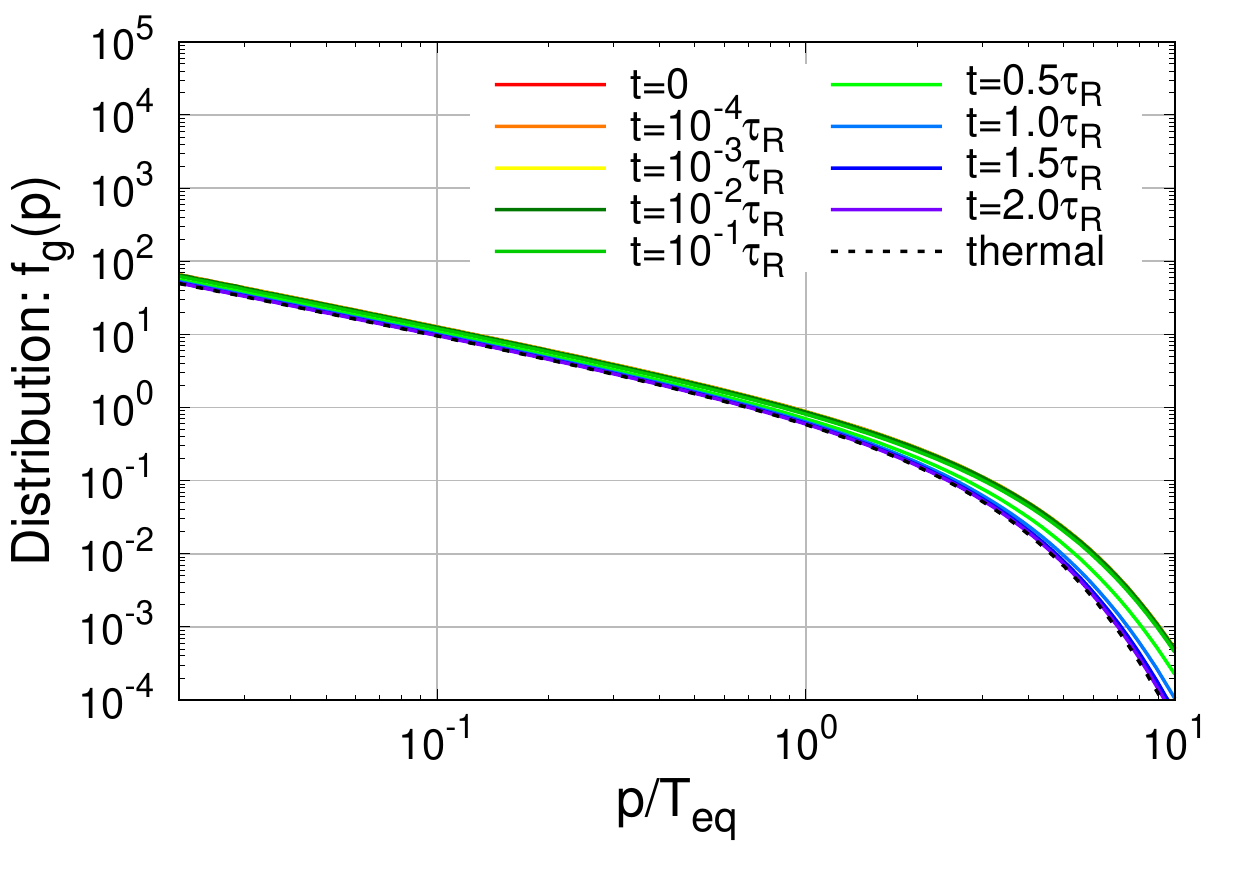}
		\centering
		\includegraphics[width=0.48\textwidth]{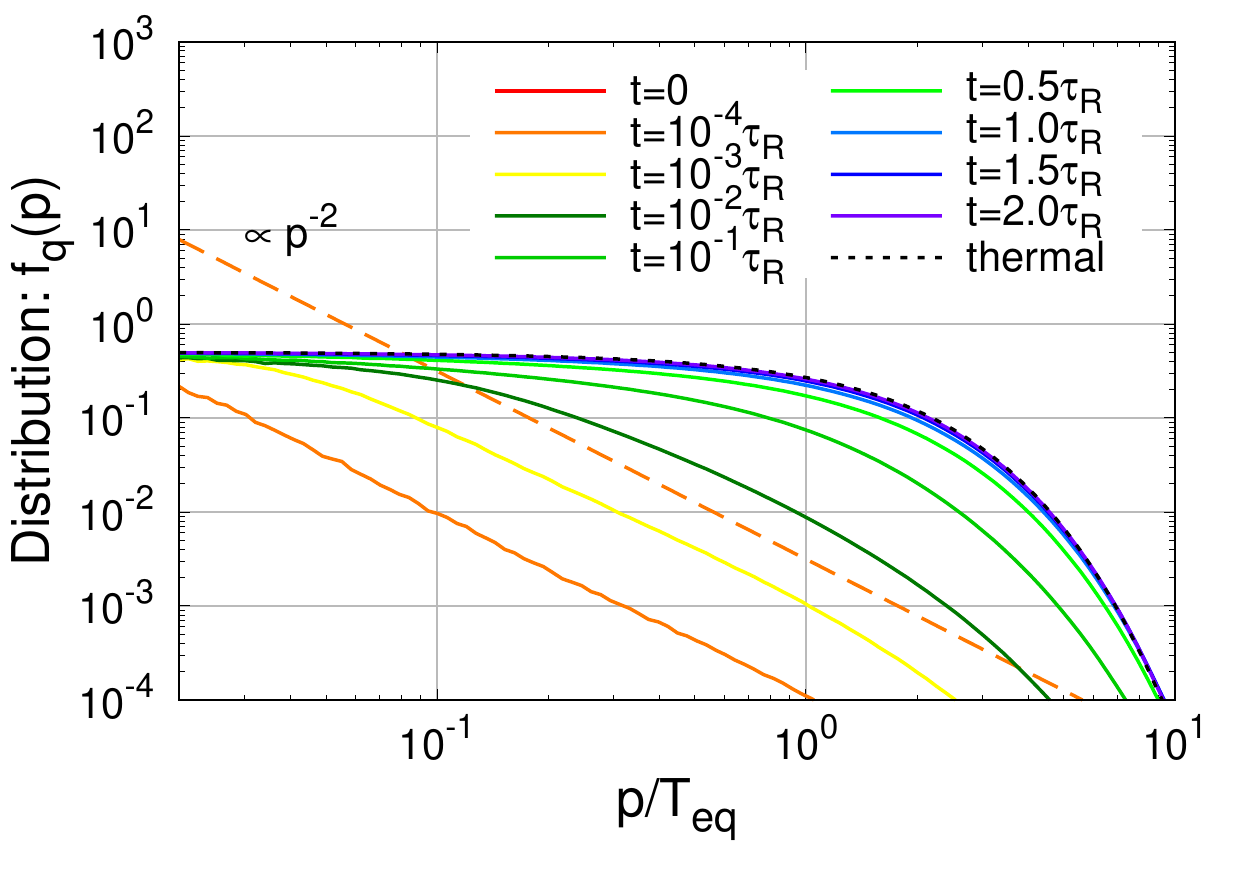}
	\caption{Evolution of gluon $f_{g}(t,p)$ and quark $f_{q}(t,p)$ distribution for \emph{gluon dominated initial conditions} $(\lambda=1)$ at different times $0\leq t \leq 2 \tau_{R}$ expressed in units of the equilibrium relaxation time $\tau_{R}$ in Eq.~(\ref{eq-relaxation}). Spectra of anti-quarks $f_{\bar{q}}(t,p)$ are identical to the spectra of quarks $f_{\bar{q}}(p)$ at zero density and not depicted in the figure.}
	\label{fig-TMGQ-MU0F_G0}
\end{figure}

\begin{figure}
		\centering
		\includegraphics[width=0.48\textwidth]{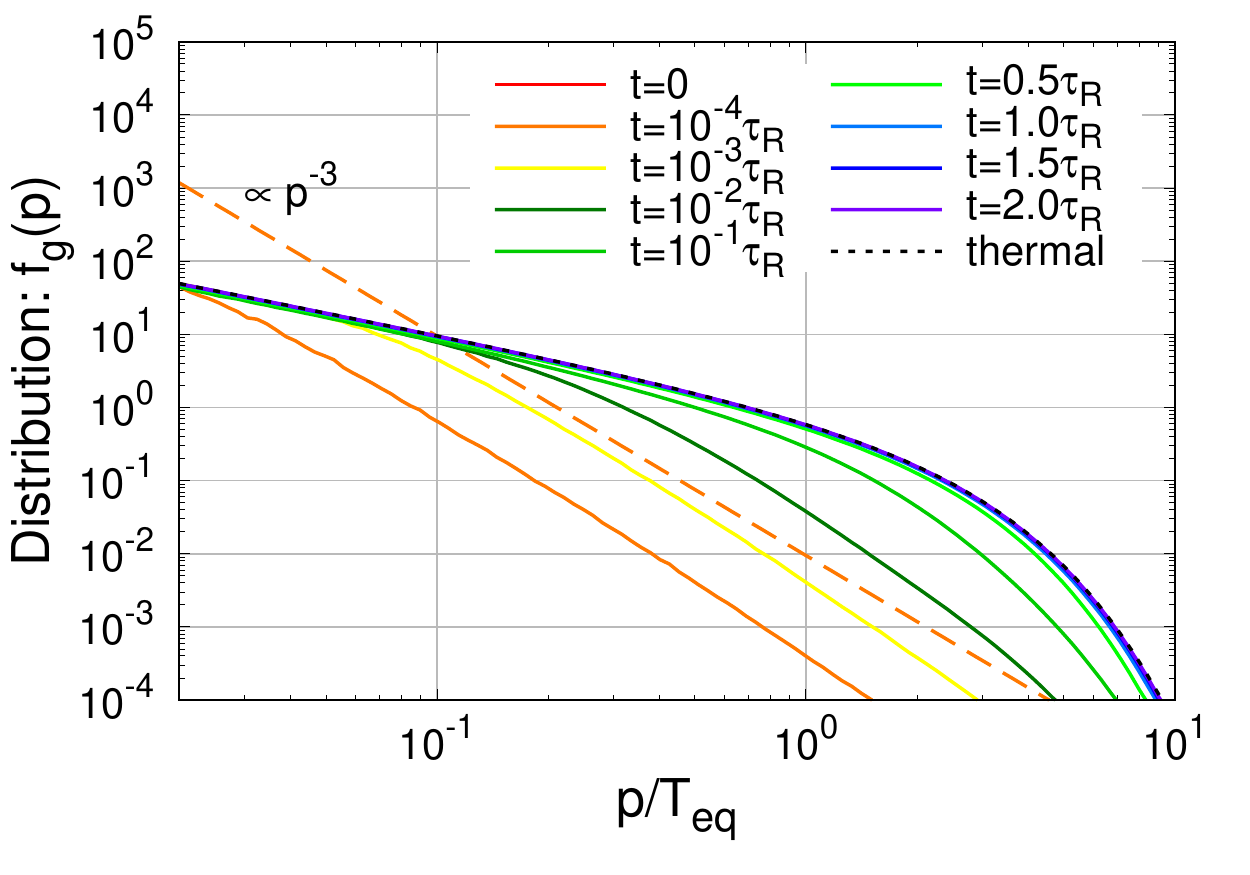}
		\centering
		\includegraphics[width=0.48\textwidth]{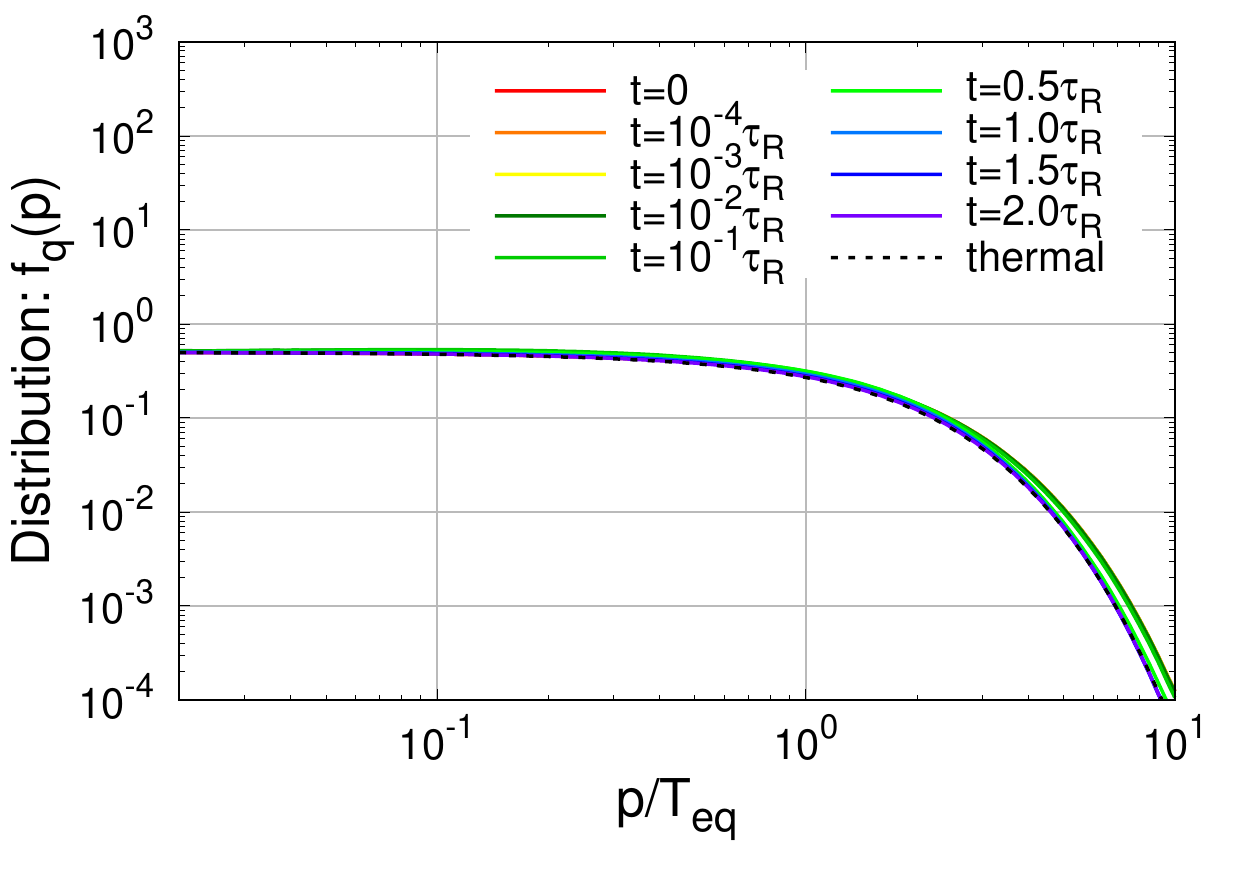}
	\caption{Evolution of gluon $f_{g}(t,p)$ and quark $f_{q}(t,p)$ distribution for \emph{quark/anti-quark dominated initial conditions} $(\lambda=1)$ at different times $0\leq t \leq 2 \tau_{R}$ expressed in units of the equilibrium relaxation time $\tau_{R}$ in Eq.~(\ref{eq-relaxation}). Spectra of anti-quarks $f_{\bar{q}}(t,p)$ are identical to the spectra of quarks $f_{\bar{q}}(p)$ at zero density and not depicted in the figure.}
	\label{fig-TMGQ-MU0F_Q0}
\end{figure}

We first investigate the evolution of the phase-space distribution of quarks and gluons over the course of the chemical equilibration of the QGP. We present our results in Figs.~\ref{fig-TMGQ-MU0F_G0} and \ref{fig-TMGQ-MU0F_Q0}, where we depict the evolution of the spectra of quarks and gluons for initially gluon (Fig.~\ref{fig-TMGQ-MU0F_G0}) and quark (Fig.~\ref{fig-TMGQ-MU0F_Q0}) dominated systems.
 
Starting with the evolution of the gluon dominated system in Fig.~\ref{fig-TMGQ-MU0F_G0}, one observes that the gluon spectrum only varies modestly over the course of the chemical equilibration of the system, such that throughout the evolution the spectrum can be rather well described by an effectively thermal distribution $f_g(\vec{p},t)\simeq\frac{1}{{\rm exp}(p/T_{g}(t))-1}$, with a time dependent temperature $T_{g}(t)$, decreasing monotonically from the initial value $T_{g}(t=0)=T_0$ to the final equilibrium temperature $T_{g}(t\to \infty)=T_{\rm eq}$. Due to soft gluon splittings $g\rightarrow q\bar{q}$ and elastic quark/gluon conversion $gg\rightarrow q\bar{q}$, the quark/antiquark spectra quickly built up at soft scales $p\lesssim T_{\rm eq}$, as can be seen from the spectra at early times $(t \ll \tau_{R})$ in the bottom panel of Fig.~\ref{fig-TMGQ-MU0F_G0}. The quark/antiquark follows a power-law spectrum $f_{q/\bar{q}}(\vec{p},t) \propto 1/p^2$ associated with Bethe-Heitler spectrum. While the production of quark/antiquark at low momentum continues throughout the early stages of the evolution, the momentum of previously produced quarks/antiquarks increases due to elastic interactions, primarily $q g \leftrightarrow q g$ and $\bar{q} g \leftrightarrow \bar{q} g$ scattering, such that by the time $t \simeq 0.5 \tau_R$ the spectrum of produced quarks/antiquarks extends all the way to the temperature scale $p\sim T_{\rm eq}$ and eventually approaches equilibrium on a time scale on the order one to two times the kinetic relaxation time $\tau_R$.

Similar behavior can be observed for the quark/antiquark dominated scenario, which is depicted in Fig.~\ref{fig-TMGQ-MU0F_Q0}. While quarks/antiquarks feature approximately thermal spectra $f_{q/\bar{q}}(\vec{p},t)\simeq\frac{1}{{\rm exp}(p/T_{q}(t))+1}$, gluons are initially produced at low momentum mainly due to the emission of soft gluon radiation $q\rightarrow gq$, which at early times ($t \ll \tau_{R}$) gives rise to a power law spectrum $f_g(\vec{p},t) \propto 1/p^{3}$ associated with the Bethe-Heitler spectrum. Subsequently, elastic and inelastic processes lead to a production of gluons with momenta $p \sim T_{\rm eq}$ until the system approaches equilibrium on a time scale on the order of the kinetic relaxation time $\tau_R$.

\subsubsection{Collision Rates}
\begin{center}
\begin{figure*}
		\centering
		\includegraphics[width=0.95\textwidth]{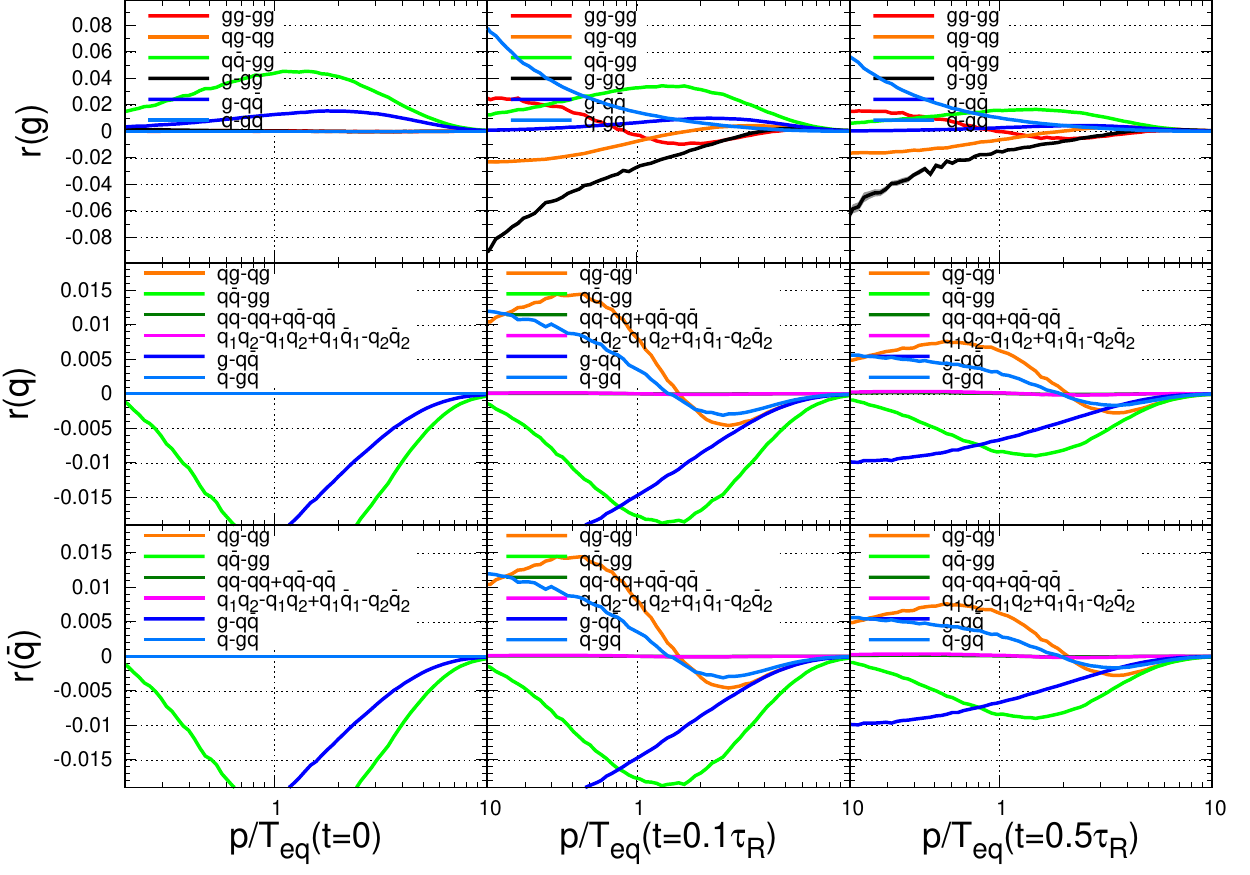}
	\caption{Collision rates $r$ for gluons (upper), quarks (middle) and antiquarks (lower) defined in Eq.~(\ref{eq:CollisionRatesR}) for \emph{gluon dominated initial conditions} at initial time $t$=0 (left), intermediate time $t$=0.1$\tau_R$ (middle) and during the approach towards equilibrium $t$=0.5$\tau_R$ (right) .}
	\label{fig-TMGQ-Cg0}
\end{figure*}
\end{center}

\begin{center}
	\begin{figure*}
		\centering
		\includegraphics[width=0.95\textwidth]{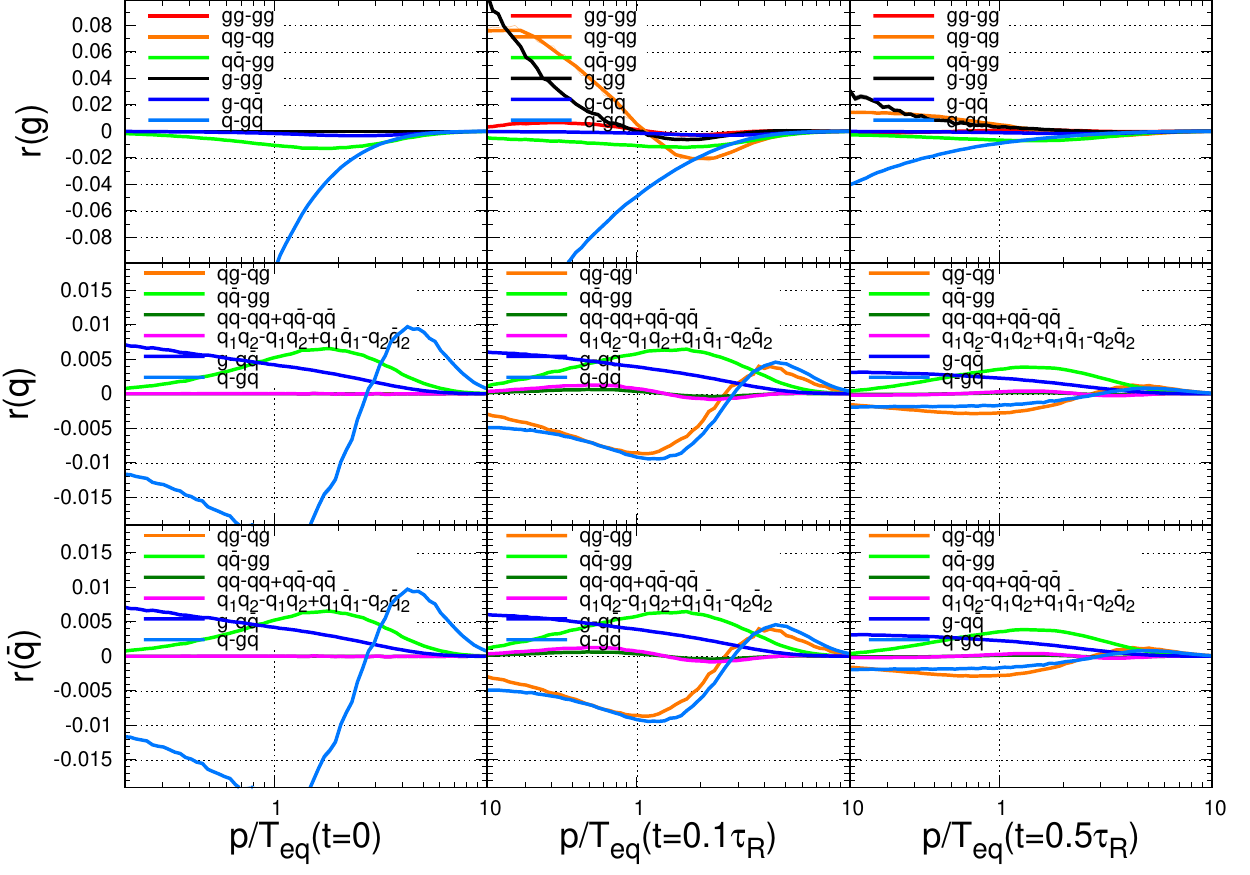}
	\caption{Collision rates $r$ for gluons (upper), quarks (middle) and antiquarks (lower) defined in Eq.~(\ref{eq:CollisionRatesR}) for \emph{quark/anti-quark dominated initial conditions} at initial time $t$=0 (left), intermediate time $t$=0.1$\tau_R$ (middle) and during the approach towards equilibrium $t$=0.5$\tau_R$ (right).}
		\label{fig-TMGQ-Cq0}
	\end{figure*}
\end{center}

While the evolution of the spectra in Figs.~\ref{fig-TMGQ-MU0F_G0} and \ref{fig-TMGQ-MU0F_Q0} provides an overview over the chemical equilibration process, we will now investigate how the individual QCD processes contribute to the evolutions of the gluon and quark/antiquark spectra in Figs.~\ref{fig-TMGQ-MU0F_G0} and \ref{fig-TMGQ-MU0F_Q0}. We provide a compact summary of our findings in Figs.~\ref{fig-TMGQ-Cg0} and \ref{fig-TMGQ-Cq0}, where we present result for the collision rates
\begin{eqnarray}
\label{eq:CollisionRatesR}
r_{g}=\frac{\nu_g}{\lambda^2T_{\rm eq}^3}p^2C_{g}[f]\;, \quad
r_{q/\bar{q}}=\frac{N_{f}\nu_q}{\lambda^2T_{\rm eq}^3}p^2C_{q/\bar{q}}[f]
\end{eqnarray} 
for initially gluon dominated (Fig.~\ref{fig-TMGQ-Cg0}) and initially quark dominated scenarios (Fig.~\ref{fig-TMGQ-Cq0}). Different columns in Figs.~\ref{fig-TMGQ-Cg0} and \ref{fig-TMGQ-Cq0} show the collision rates of individual processes at the initial time $t=0$, at an intermediate time $t=0.1\tau_R$ and near-equilibrium at time $t=0.5\tau_R$. We note that due to the zero net-density of $u,d,s$ quarks, the quark and antiquark collision rates in Figs.~\ref{fig-TMGQ-Cg0} and \ref{fig-TMGQ-Cq0} are identical and briefly remind the reader, that according to our convention in Eq.~(\ref{eq-bolzmann}), positive contributions to the collision rate represent a number loss and negative collision rates exhibit a number gain for the specific particle. 

\paragraph{Gluon dominated scenario}

Starting with the collision rates for the gluon dominated scenario in Fig.~\ref{fig-TMGQ-Cg0}, one observes that at initial time $t=0$, the gluon splitting process $g\rightarrow q\bar{q}$ shown by the dark blue curve is dominating the production of quarks/antiquarks. By comparing the collision rates for quarks and gluons, one finds that gluons with momenta $p\simeq 1-2\,T_{\rm eq}$ copiously produce soft quarks/antiquarks at low momenta $p\ll T_{\rm eq}$. Since the individual splittings are typically asymmetric with $z(1-z)\ll 1$, the energy of thermal gluons is re-distributed to soft quarks/antiquarks, and the splittings fall into the Bethe-Heitler regime as typically $p z(1-z) \lesssim \omega_{BH} \sim T_{\rm eq}$. In addition to the inelastic splitting, elastic conversion processes $gg\rightarrow q\bar{q}$ shown as a lime curve evenly re-distribute the energy of gluons with momenta $p \simeq T_{\rm eq}$ into quarks/antiquarks at an intermediate scale $p\simeq T_{\rm eq}$. Due to the absence of quarks and antiquarks at initial time, the contributions of all other processes involving quarks/antiquarks in the initial state vanish identically at initial time, as do the collision rates for processes involving only gluons due to the detaily balanced in the gluon sector.

Subsequently, as quarks/antiquarks are produced at low momenta, additional scattering processes involving quarks/antiquarks in the initial state become increasingly important, as can be seen from the second column of Fig.~\ref{fig-TMGQ-Cg0}, where we present the collision rates at $t=0.1\tau_R$. While the rate of the initial quark/antiquark production processes $g\rightarrow q\bar{q}$, $gg\rightarrow q\bar{q}$ decrease, as the corresponding inverse processes $q\bar{q}\rightarrow g$, $q\bar{q}\rightarrow gg$ start to become important, elastic scattering of quarks and gluons $qg\rightarrow qg$ (orange curve) and gluon absorption $gq\rightarrow q$ (light blue curve) become of comparable importance. Specifically, in each of these processes, the previously produced quarks/antiquarks at low momentum $p \ll T_{\rm eq}$ gain energy via elastic scattering or absorption of a gluons, resulting in an increase of the spectrum for $p\gtrsim 1.5 T_{\rm eq}$. By inspecting the collision rates for gluons in the top panel of Fig.~\ref{fig-TMGQ-Cg0}, one observes that the depletion of soft gluons $(p \ll T_{\rm eq})$ due to gluon absorption by quarks $gq\rightarrow q$ is primarily compensated by the emission of soft gluon radiation due to the $g\rightarrow gg$ process (black curve). Beside the aforementioned process, the elastic scattering of gluons $gg \to gg$ (red curve) also plays an equally important role in re-distributing energy among gluons, clearly indicating that over the course of the chemical equilibration process the gluon distribution also falls out of kinetic equilibrium.


Eventually, the chemical equilibration process proceeds in essentially the same way, until close to equilibrium the collision rates of all processes decrease as the corresponding inverse processes start to become of equal importance, as seen in the right column of Fig.~\ref{fig-TMGQ-Cg0} where we present the collision rates at $t=0.5\tau_R$. By the time $t\simeq 2\tau_R$, which is no longer shown in Fig.~\ref{fig-TMGQ-Cg0}, all the collision rates decrease by at least one order of magnitude as the the system gradually approaches the detailed balanced chemical and kinetic equilibrium state.

\paragraph{Quark/antiquark dominated scenario}
Next we will analyze the collision rates in the quark/antiquark dominated scenario shown in Fig.~\ref{fig-TMGQ-Cq0} and compare the underlying dynamics to the gluon dominated scenario in Fig.~\ref{fig-TMGQ-Cg0}.

Starting from the collision rates at initial time shown again in the left panel, one finds that in addition to quark/antiquark annihilation via elastic $q\bar{q}\rightarrow gg$ (lime) and inelastic $q\bar{q}\rightarrow g$ (dark blue) processes, soft gluons are copiously produced by $q \rightarrow gq$ Bremsstrahlungs processes initiated by hard quarks/antiquarks with momenta $ p\gtrsim 3 T_{\rm eq}$. Noteably the $q \rightarrow gq$ process also leads to the re-distribution of the energy of quarks/antiquarks from momenta $ p\gtrsim 3 T_{\rm eq}$, to lower momenta $ p\lesssim 3 T_{\rm eq}$; however the negative collision rate for the $q \rightarrow gq$ process partially cancel against the positive contribution from $q\bar{q}\rightarrow g$ processes, such that there is effectively no increase/decrease of the quark/antiquark distributions at very low momenta $p \ll T_{\rm eq}$.   Similar to the processes involving only gluons at $t=0$ in Fig.~\ref{fig-TMGQ-Cg0}, processes involving only quarks and antiquarks (green, pink) in Fig.~\ref{fig-TMGQ-Cq0} vanish identically at $t=0$ due to cancellations of gain and loss terms in the statistical factor, while other processes $gg\rightarrow gg$, $qg\rightarrow qg$, $g\rightarrow gg$ are exactly zero due to the absence of gluons in the initial state. By comparing the collision integrals for quarks and gluons in Figs.~\ref{fig-TMGQ-Cg0} and \ref{fig-TMGQ-Cq0}, one also observes that inelastic processes are initially much more dominant for the quark/antiquark dominated scenario in Fig.~\ref{fig-TMGQ-Cg0} as compared to the gluon dominated scenario in Fig.~\ref{fig-TMGQ-Cq0}.

Similarly to the evolution in the gluon dominated scenario, the energy of the soft gluons produced in the previous stage increases through successive elastic and inelastic interactions, as can be seen from the middle column of  Fig.~\ref{fig-TMGQ-Cq0}, where we present the collision rates at the intermediate time $t=0.1\tau_R$ for the quark dominated case. By inspecting the collision rates for gluon in more detail, one finds that quark-gluon scattering $qg\rightarrow qg$ (orange) as well as  $gg \rightarrow g$ (black) are the dominant processes that increase the number of hard $p \gtrsim T_{\rm eq}$ gluons. Elastic scattering between gluons $gg\rightarrow gg$ (red) plays a less prominent role for the evolution of the gluons, as do elastic (green) $q\bar{q}\rightarrow gg$ and inelastic (dark-blue) $q\bar{q}\rightarrow g$ conversions. With regards to the collision rates for quarks and antiquarks, one finds that elastic $q\bar{q}\rightarrow gg$ (lime) and inelastic $q\bar{q}\rightarrow g$ (dark blue) annihilation processes as well as $q \rightarrow gq$ Bremsstrahlungs processes continue to deplete the number of hard quarks/antiquarks. However at this stage of the evolution, $qg\rightarrow qg$ scattering processes (orange) also lead to an efficient energy transfer from quarks to gluons, depleting the number of hard quarks $p\gtrsim 2 T_{\rm eq}$ in the system. While the non-vanishing quark/antiquark scattering rates (green, pink) reveal slight deviations of quark/antiquark spectra from kinetic equilibrium, these processes clearly have a subleading effect.

Subsequently, the evolution continues along the same lines as illustrated in the right column for $t=0.5\tau_R$, with the collision rates of all processes decreasing as the system approaches kinetic and chemical equilibrium.

\begin{figure}
	\centering
	\includegraphics[width=0.75\textwidth]{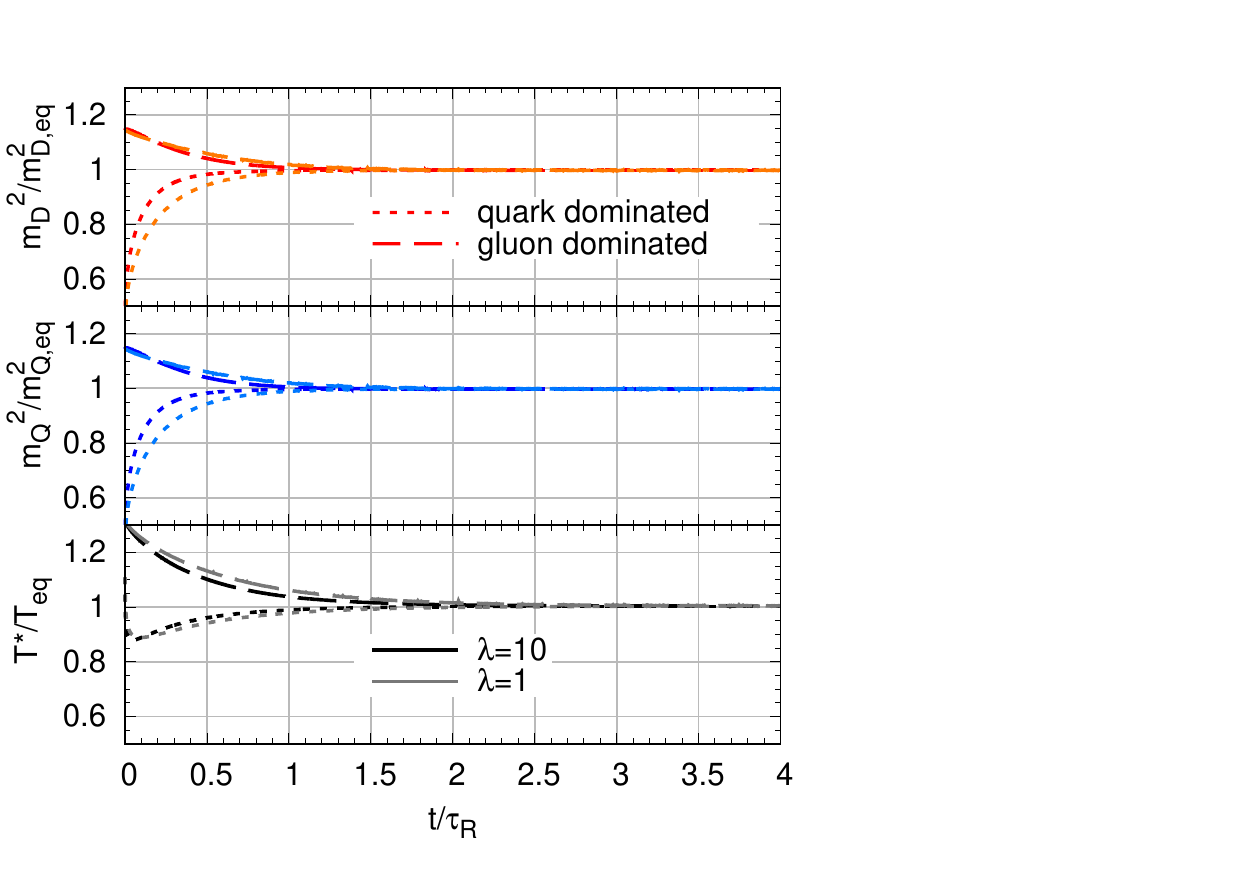}
	\caption{Evolution of the characteristic scales $m_D^2(t)$ (red), $m_Q^2(t)$ (blue) and $T^{*}(t)$ (black) during the chemical equilibration process for the quark dominated (fine dashed) and gluon dominated (long dashed) scenarios at two different coupling strengths $\lambda=1$ (lighter colors) and $\lambda=10$ (darker colors). Scales are normalized to their respective equilibrium values, while the evolution time $t$ is normalized to the equilibrium relaxation time $\tau_R$ in Eq.~(\ref{eq-relaxation}) in order to take into account the leading coupling dependence.}
	\label{fig-TMGQ-MU0S}
\end{figure}

\begin{figure}
	\centering
	\includegraphics[width=0.48\textwidth]{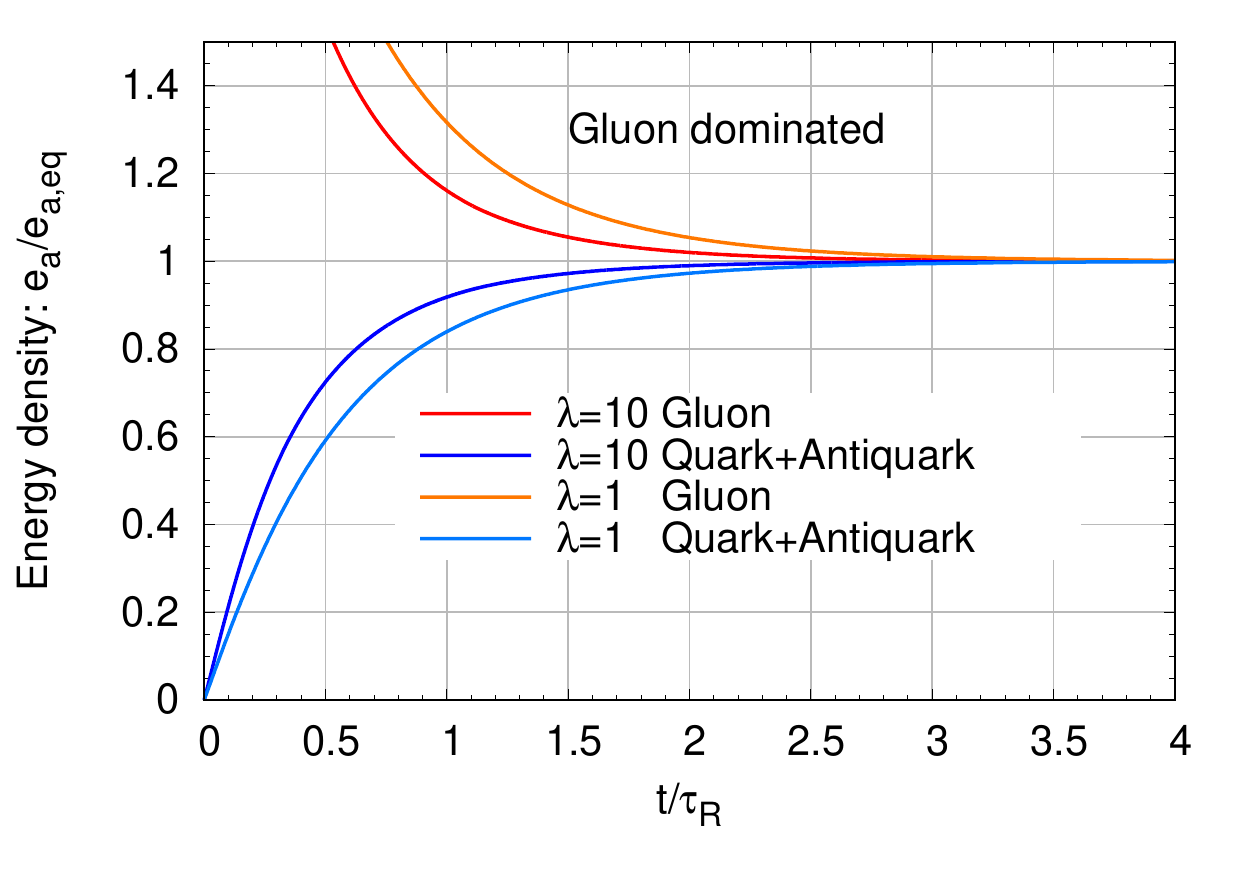}
	\centering
	\includegraphics[width=0.48\textwidth]{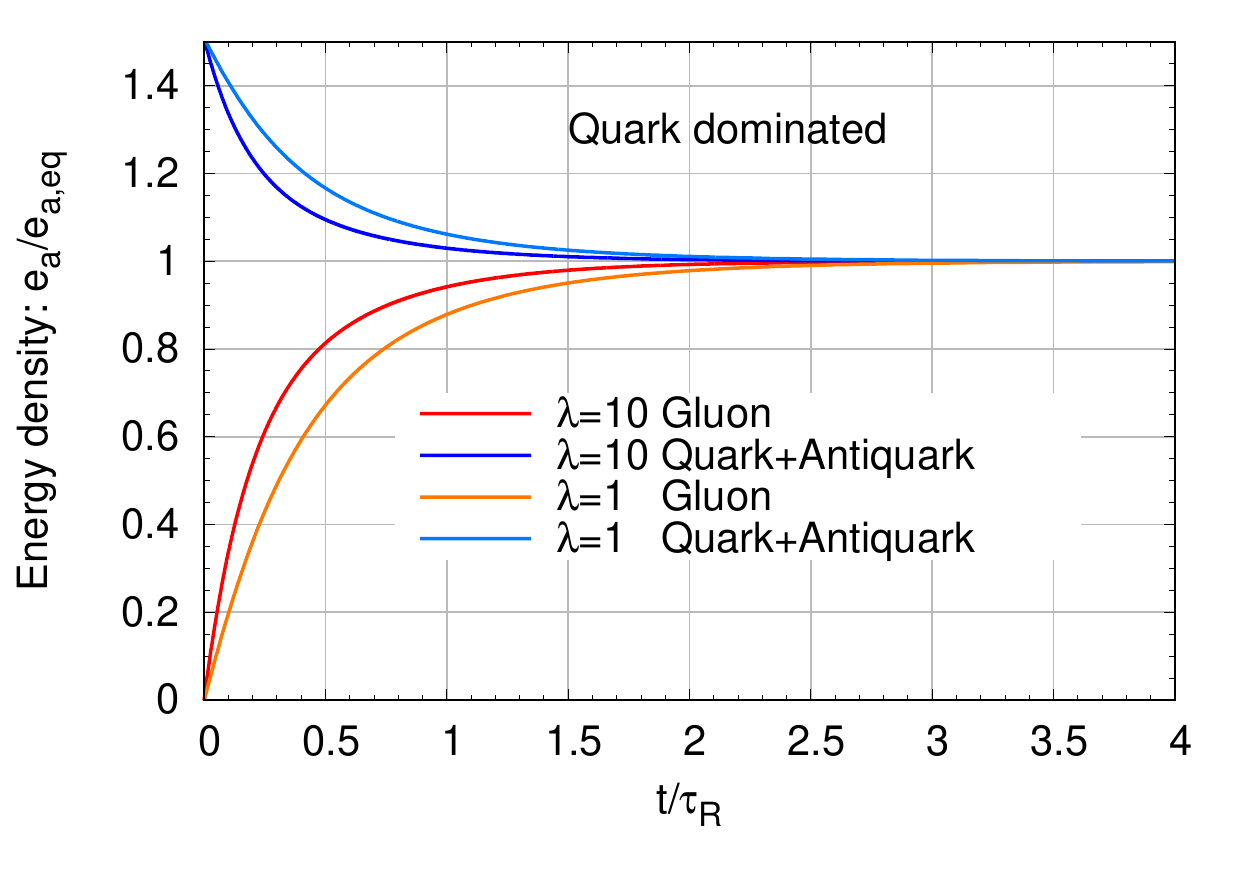}
	\caption{Evolution of the energy densities of gluons $e_{g}$ (red) and quarks plus anti-quarks $\sum_{f} e_{q_{f}}+e_{\bar{q}_{f}}$ (blue) for the quark dominated (bottom) and gluon dominated (top) scenarios at different couplings strength $\lambda$=10 (darker colors), $\lambda$=1 (lighter colors). Energies of each species are normalized to their equilibrium values, while the evolution time $t$ is normalized to the equilibrium relaxation time $\tau_R$ in Eq.~(\ref{eq-relaxation}) in order to take into account the leading coupling dependence.}
	\label{fig-TMGQ-MU0E}
\end{figure}

\subsubsection{Scale Evolutions}

Beyond the characterization of the microscopic processes in terms of spectra and collision rates, it is instructive to investigate the evolution of the characteristic scales $m_D^2$, $m_Q^2$ and $T^{*}$ defined in Sec.~\ref{sec-theory-thermodynamics}, which further provides a compact way to compare the time scales of the chemical equilibration process at different coupling strength. Corresponding results are presented in Fig.~\ref{fig-TMGQ-MU0S}, where we compare the evolution of the various scales for quark and gluon dominated initial condition at two different coupling strengths $\lambda=1,10$.  By taking into account the corresponding change in the equilibrium relaxation rate $\tau_{R}$ (c.f. Eq.~(\ref{eq-relaxation})), one finds that the time evolutions of the various scales are quite similar, and rather insensitive to the coupling strength, such that by the time $t=1\sim2\,\tau_{R}$ all relevant dynamical scales are within a few percent of their equilibrium values.

During the earlier stages, $t\lesssim \tau_{R}$, some interesting patterns emerge in the evolution of $m_D^2$, $m_Q^2$ and $T^{*}$, which can be readily understood from considering the evolution of the spectra in Figs.~\ref{fig-TMGQ-MU0F_G0} and \ref{fig-TMGQ-MU0F_Q0}, along with different effects that quarks and gluons have on each of these quantities. Since the occupancy of soft quarks is always limited to below unity, soft gluons contribute more significantly to in-medium screening, such that in the gluon dominated case the screening masses $m_{D}^2$ and $m_{Q}^2$ in Fig.~\ref{fig-TMGQ-MU0S} decrease monotonically, whereas in the quark dominated case on observes a monotonic increase of the same quantities. The effective temperature $T^{*}$ which characterizes the strength of elastic interactions inside the medium, drops throughout the chemical equilibration process for the gluon dominated case, whereas for quark dominated initial conditions, the evolution of $T^{*}$ shows a non-monotonic behavior featuring a rapid initial drop followed by a gradual increase of $T^{*}$ towards its equilibrium value. By careful inspection of the spectra in Fig.~\ref{fig-TMGQ-MU0F_Q0}, one finds that this rather subtle effect should be attributed to the effects of Bose enhancement and Fermi suppression in the determination of $T^{*}$.

Besides the evolution of the characteristic scales $m_D^2$, $m_Q^2$ and $T^{*}$, it is also important to understand how the overall energy is shared and transferred between quark and gluon degrees of freedom over the course of the evolution. A compact overview of the energy transfer during the chemical equilibration process is provided in Fig.~\ref{fig-TMGQ-MU0E}, where we show the evolution of the energy density of gluons as well as quarks and antiquark for the two scenarios. Starting from a rapid energy transfer at early times, the flattening of the individual energy densities towards later times eventually indicates the approach towards chemical equilibrium. 
Even though the evaluation of an exact chemical equilibration time depends on the quantitative criterion for how close the energy densities (or other scales) are compared their equilibrium values are, the figures still speak for themselves indicating the occurrence of chemical equilibration roughly at the same time scale as kinetic equilibration, with
\begin{eqnarray}
\label{eq:eqTimes-chemical}
t_{\rm chem}^{\rm eq} \simeq t_{\rm kin}^{\rm eq}\simeq 1-2 \times \frac{4\pi \eta/s}{T_{\rm eq}},
\end{eqnarray}
subject to mild variations for the two different coupling strengths. 

\begin{figure}
	\centering
	\includegraphics[width=0.48\textwidth]{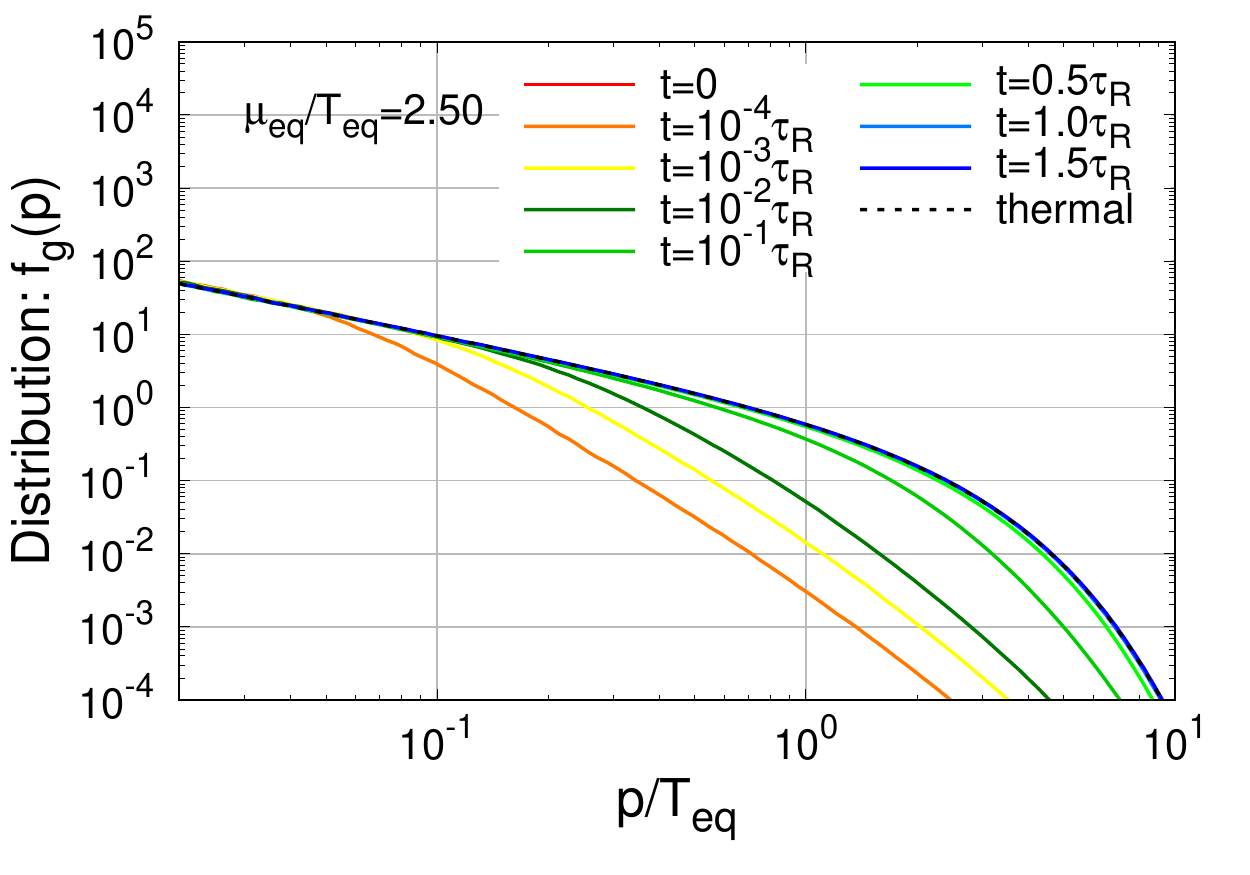}
	\centering
	\includegraphics[width=0.48\textwidth]{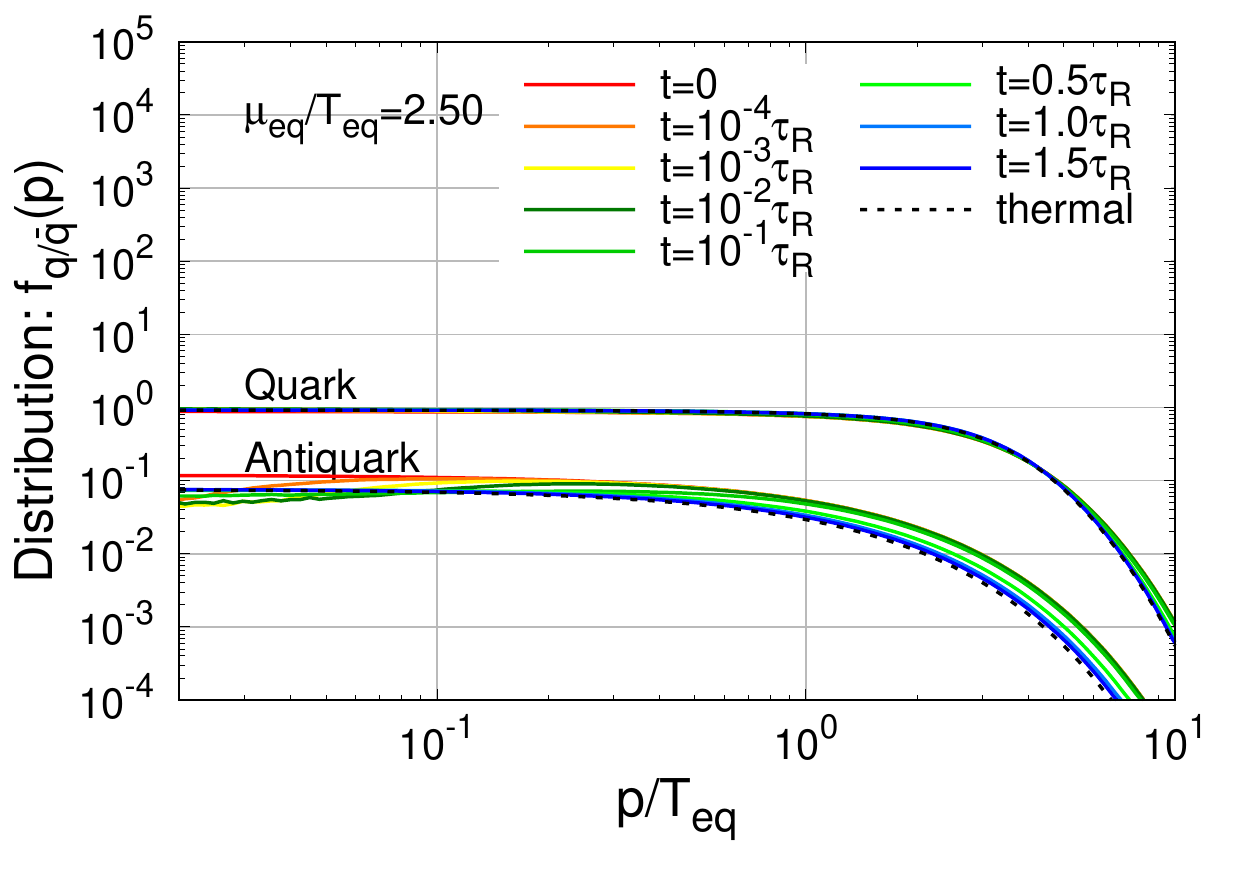}
	\caption{Evolution of the gluon $f_{g}(t,p)$, quark $f_{q}(t,p)$ and anti-quark $f_{\bar{q}}(t,p)$ distributions for \emph{quark dominated initial conditions} with large chemical potential $\mu_{f}/T_{\rm eq}=2.5$ at different times $0\leq t \leq 2 \tau_{R}$ expressed in units of the equilibrium relaxation time $\tau_{R}$ in Eq.~(\ref{eq-relaxation-finite}) for $\lambda=1$.}
	\label{fig-TMGQ-MUF}
\end{figure}

\begin{center}
	\begin{figure*}
		\begin{minipage}[b]{0.98\linewidth}
			\centering
			\includegraphics[width=0.95\textwidth]{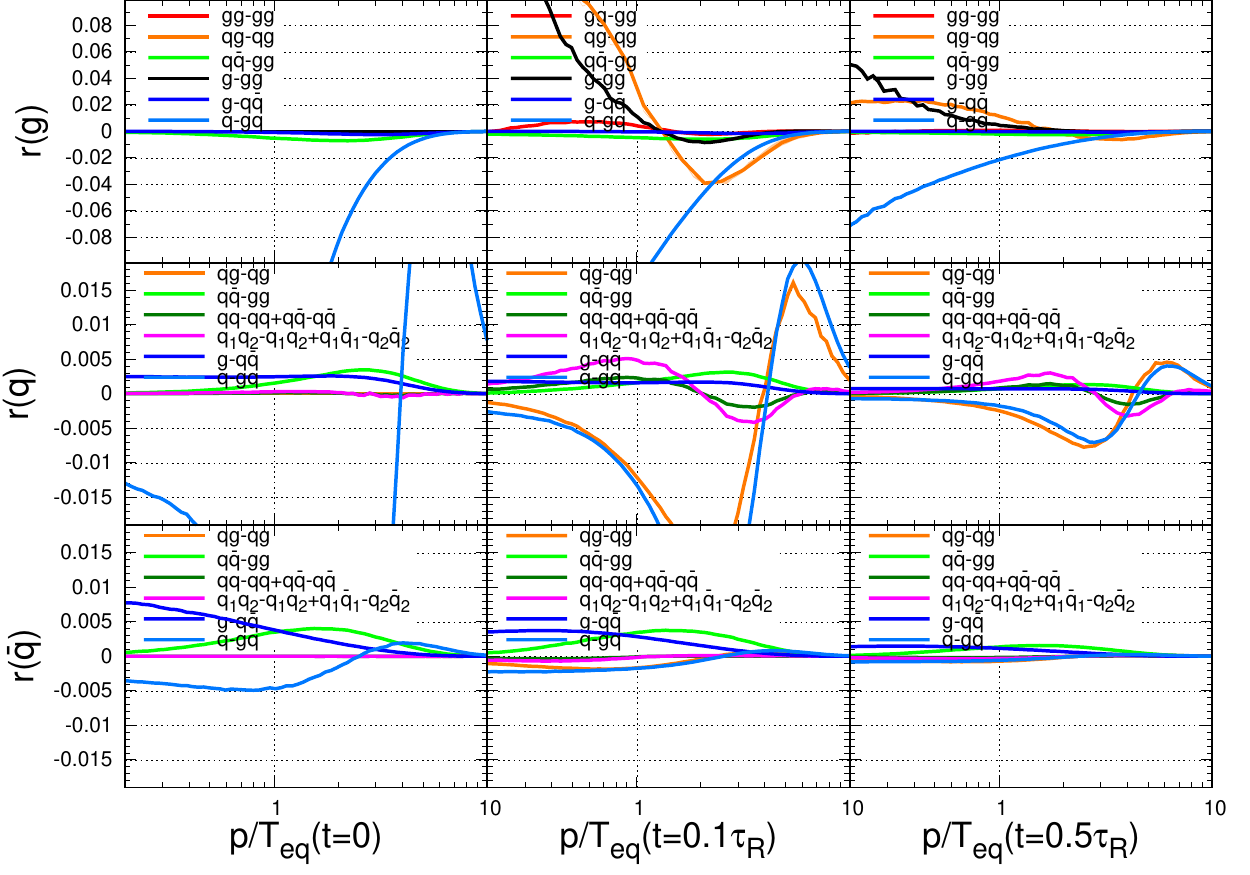}
		\end{minipage}
		\caption{Collision rates $r$ for gluons (upper), quarks (middle) and antiquarks (lower) defined in Eq.~(\ref{eq:CollisionRatesR}) for \emph{quark dominated initial conditions} with large chemical potential $\mu_{f}/T_{\rm eq}=2.5$ at initial time $t$=0 (left), intermediate time $t$=0.1$\tau_R$ (middle) and during the approach towards equilibrium $t$=0.5$\tau_R$ (right).}
		\label{fig-TMGQ-Cq2}
	\end{figure*}
\end{center}

\subsection{Chemical equilibration of finite density systems}
So far we have investigated the chemical equilibration of charge neutral QCD plasmas, and we will now proceed to study the chemical equilibration process of QCD plasmas at finite density of the conserved $u,d,s$ charges, featuring an excess of quarks to antiquarks (or vice versa). Since a finite net charge density of the system can only be realized in the presence of quarks/anit-quarks, we will focus on quark dominated initial conditions and modify the corresponding initial conditions as
\begin{eqnarray}
\label{eq-TMGQ-INITIAL-Q2}
\nonumber
&&f_g(p,t=0)=0,\\
\nonumber
&&f_{q_{f}}(p,t=0)=\frac{1}{e^{(p-\mu_f)/T_0}+1},\\
&&f_{\bar{q}_{f}}(p,t=0)=\frac{1}{e^{(p+\mu_f)/T_0}+1}.
\end{eqnarray}
where for simplicity, we consider equal densities of $u,d$ and $s$ quarks. Similar to Eqs.~(\ref{eq-TMGQ-INITIAL-G0}) and (\ref{eq-TMGQ-INITIAL-Q0}), the initial parameters $T_0$, $\mu_0$ can be related to corresponding equilibrium temperature $T_{\rm eq}$, and equilibrium chemical potential $\mu_{\rm eq}$ via the Landau matching procedure in Eq.~(\ref{eq-energydensity}). Due to energy and charge conservation, $T_{\rm eq}$ and $\mu_{\rm eq}$, then determine the final equilibrium state of the system, and we will characterize the different amounts of net charge in the system in terms of the ratio $\mu_{\rm eq}/T_{\rm eq}$, with $\mu_{\rm eq}/T_{\rm eq}=0$ corresponding to the charge neutral plasma considered in the previous section. 

When comparing the evolutions at different coupling strengths, we follow the same procedure as discussed above and express the evolution time in units of the kinetic relaxation time
\begin{eqnarray}
\label{eq-relaxation-finite}
&&\tau_R(T,\mu)=\frac{4\pi}{T_{\rm eff}}\left(\frac{\eta(T,\mu) T_{\rm eff}}{e+p}\right)
\overset{\mu=0}{\simeq}\frac{4\pi}{T_{\rm eq}}\left(\frac{\eta(T,\mu=0)}{s}\right)\;, \nonumber \\
\end{eqnarray}
which in accordance with the last equality reduces to the same expression for a charge neutral system ($\mu=0$) in Eq.~(\ref{eq-relaxation}). 
The effective temperature is evaluated as $T_{\rm eff}=\left(\frac{30e}{\pi^2\nu_{\rm eff}}\right)^{\frac{1}{4}}$ with effective degree of freedom $\nu_{\rm eff}=\nu_{G}+\frac{7}{4}\nu_{Q}N_f$, so that $T_{\rm eff}\overset{\mu=0}=T_{\rm eq}$.
Since we did not explicitly determine the dependence of the shear-viscosity $\eta(T,\mu)$ on the chemical potential $\mu$ for all coupling strengths $\lambda$, we will approximate $\frac{\eta(T,\mu) T}{e+p} \approx \left.\frac{\eta(T,\mu) T}{e+p}\right|_{\mu=0}$ by the corresponding value of $\frac{\eta(T,\mu=0)}{s}$ at vanishing density of the conserved charges, which are quoted below Eq.~(\ref{eq-relaxation}). 

\subsubsection{Spectra Evolutions}
We follow the same logic as in the charge neutral case and first investigate the evolution of the spectra of quarks, antiquarks and gluons, which is presented in Fig.~\ref{fig-TMGQ-MUF} for the chemical equilibration of a system with quark chemical potentials $\mu_{\rm eq}/T_{\rm eq}=2.5$. Similar to the quark dominated scenario at zero density, we find that the spectra for quarks and antiquarks are always close to a thermal distribution with the expected moderate deviation at intermediate times. Specifically, the antiquark spectra in the low momentum sector $p \lesssim 0.3 T_{\rm eq}$ are depleted at intermediate times $t \lesssim 0.5 \tau_{R}$, due to elastic $q \bar{q} \rightarrow gg$ and inelastic $q \bar{q} \rightarrow g$ conversions. Besides quark/antiquark annihilations, the radiative emission of gluons due to $q \rightarrow qg$ and $\bar{q} \rightarrow \bar{q} g$ processes leads to a rapid population of the soft gluon sector seen in the top panel of Fig.~\ref{fig-TMGQ-MUF}. By comparing the results in Figs.~\ref{fig-TMGQ-MU0F_Q0} and \ref{fig-TMGQ-MUF}, one finds that the soft gluon sector builds up even more rapidly at finite density as compared to zero density, such that already by the time $t=10^{-3} \tau_{R}$, the gluon distribution at low momentum $p \lesssim 0.1 T_{\rm eq}$ features a quasi-thermal spectrum $f(p \ll T_{\rm eq}) \simeq T_{\rm eq}/p$, whereas the high momentum tail is yet to be populated. Eventually on a time scale of $t\simeq 1.5 \tau_{R}$, a sufficiently large number of hard gluons has been produced and the spectra of all particle species relax towards equilibrium, such that significant deviations from the thermal distributions are no longer visible for $t=1.5 \tau_{R}$ in Fig.~\ref{fig-TMGQ-MUF}.

\subsubsection{Collision Rates}
Beyond the evolution of the of the spectra, it again proves insightful to investigate the collision rates in Fig.~\ref{fig-TMGQ-Cq2} in order to identify the microscopic processes that drive chemical and kinetic equilibration of gluons, quarks and antiquarks at different stages. 

\begin{figure}
	\centering
	\includegraphics[width=0.75\textwidth]{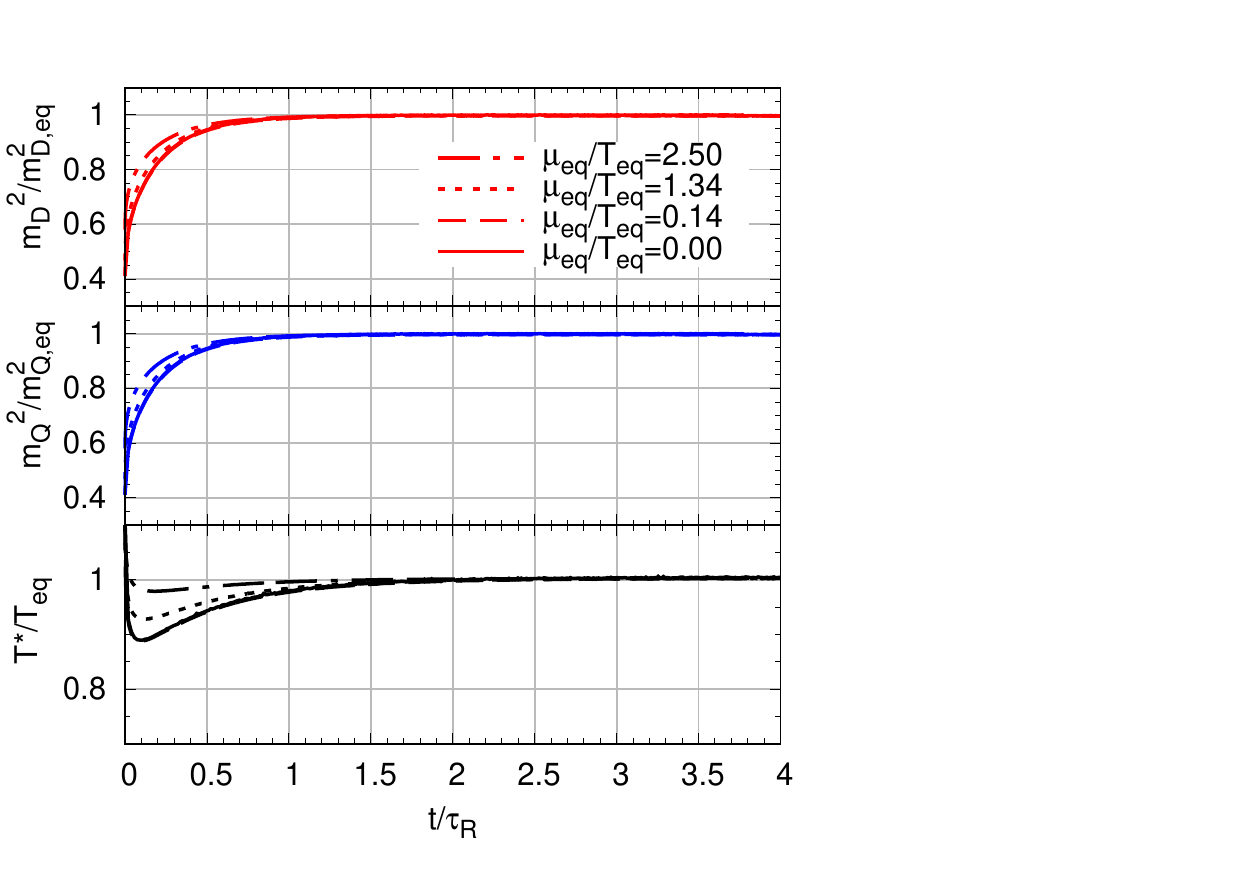}
		\caption{Evolution of the characteristic scales $m_D^2(t)$ (red), $m_Q^2(t)$ (blue) and $T^{*}(t)$ (black) during the chemical equilibration process for different chemical potentials $\mu_{\rm eq}/T_{\rm eq}$=0,0.14,1.34,2.5 (solid,dashed,dotted,dotted dashed). Scales are normalized to their respective equilibrium values, while the evolution time $t$ is normalized to the equilibrium relaxation time $\tau_R$ in Eq.~(\ref{eq-relaxation-finite}) for $\lambda=1$.}
	\label{fig-TMGQ-MUS}
\end{figure}

Similar to the results for the charge neutral case in Fig.~\ref{fig-TMGQ-Cq0}, the initial gluon production in Fig.~\ref{fig-TMGQ-Cq2} is still dominated by soft radiation $q\rightarrow gq + \bar{q} \rightarrow \bar{q}g$ (light blue), with even more substantial contributions due to the larger abundancies of quarks. Conversely, the gluon production from elastic  $q\bar{q}\rightarrow gg$ (lime) and inelastic $q\bar{q}\rightarrow g$ (dark blue) quark/antiquark annihilation processes is markedly suppressed due to the shortage of antiquarks. Similar differences between the evolution at zero and finite density can also be observed in the collision rates for quarks and antiquarks, where in the case of the quark the emission of gluon radiation leads to a depletion of the hard sector $p \gtrsim 3 T_{\rm eq}$, along with an increase of the population of softer quarks with typical momenta $p \sim T_{\rm eq}$. While elastic $q\bar{q}\rightarrow gg$ (lime) and inelastic $q\bar{q}\rightarrow g$ (dark blue) processes initially contribute at a much smaller rate, such that the inelastic $q 
\rightarrow q g$ process dominates the evolution of the quarks, a manifestly different picture emerges for the collision rates of antiquarks. 
Due to the large abundancies of quarks, elastic $q\bar{q}\rightarrow gg$ (lime) and inelastic $q\bar{q}\rightarrow g$ (dark blue) quark/antiquark annihilation initially occur at essentially the same rate as gluon radiation off antiquarks $\bar{q} \rightarrow \bar{q} g$ (light blue), resulting in a net-depletion of the antiquark sector across the entire range of momenta. Besides the aforementioned processes, the collision rates of all other processes vanish identically at initial time for all particle species due to cancellations of the statistical factors.

\begin{figure}
	\centering
	\includegraphics[width=0.48\textwidth]{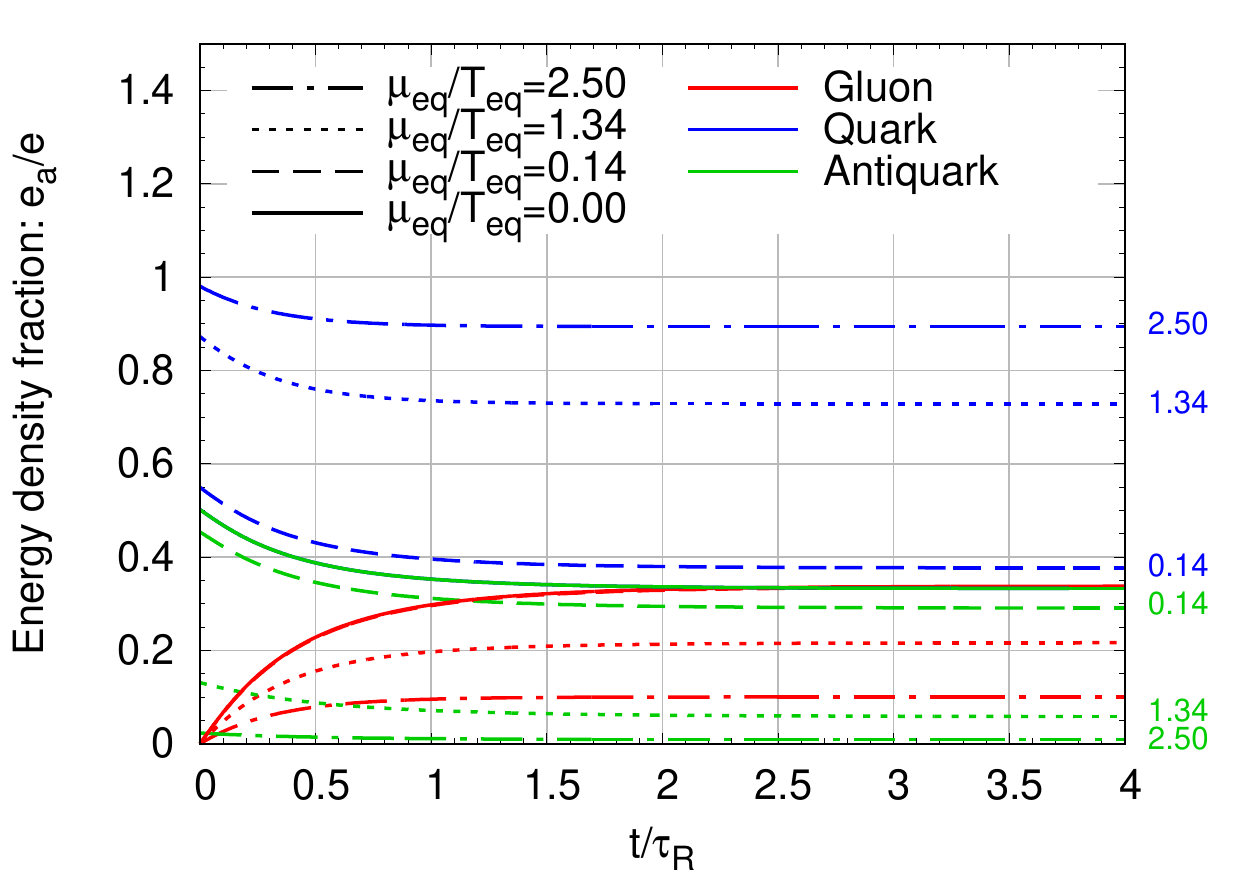}
	\centering
	\includegraphics[width=0.48\textwidth]{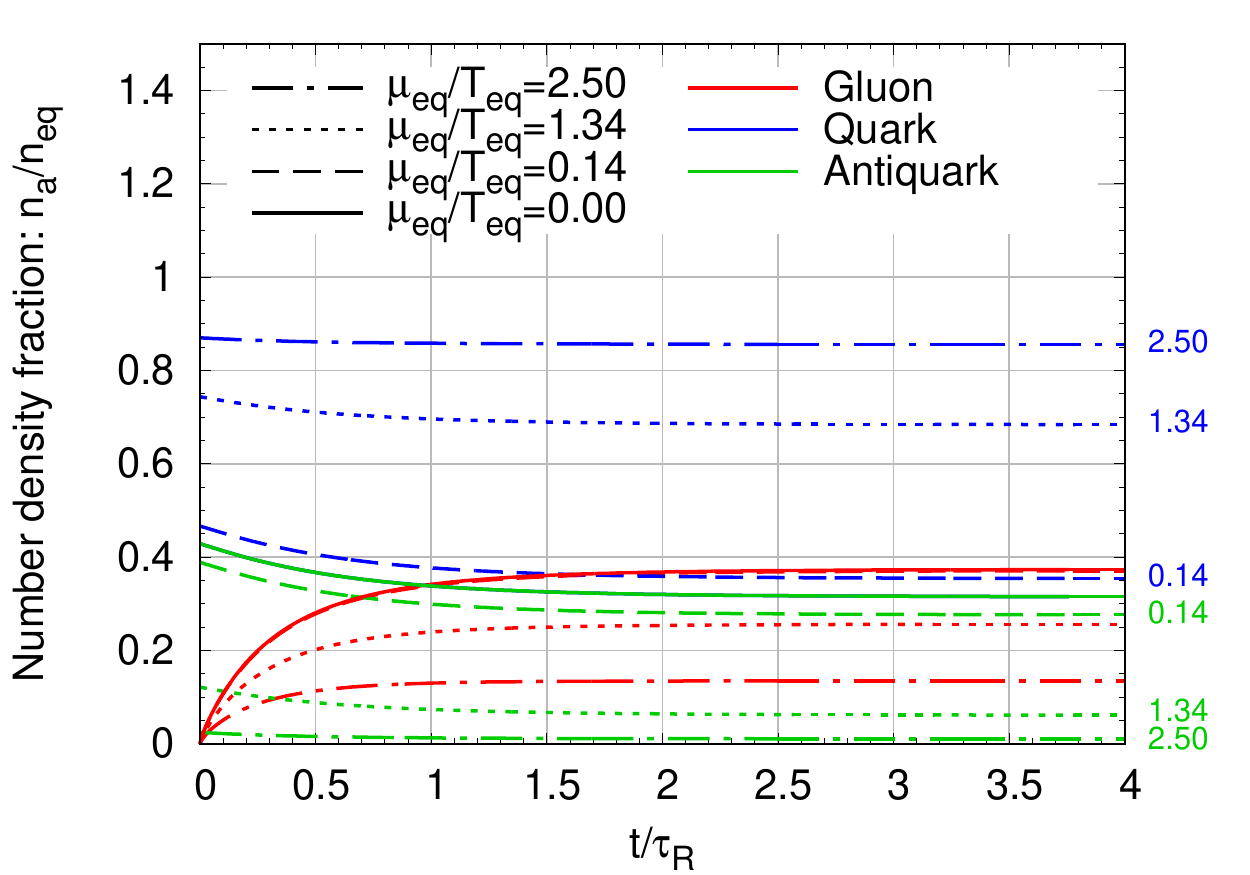}
	\caption{Evolution of the energy (top) and particle number (bottom) densities of gluons (red), quarks (blue) and anti-quarks (green) for various chemical potentials $\mu_{\rm eq}/T_{\rm eq}$=0,0.14,1.34,2.5 (solid,dashed,dotted,dotted dashed). Energy and particle number densities are normalized to the total equilibrium values $e_{\rm eq}$ and $n_{\rm eq}$ of all species, while the evolution time $t$ is normalized to the equilibrium relaxation time $\tau_R$ in Eq.~(\ref{eq-relaxation-finite}) for $\lambda=1$.}
	\label{fig-TMGQ-MUE}
\end{figure}

Subsequently, for $t=0.1\tau_R$ depicted in the central column of Fig.~\ref{fig-TMGQ-Cq2} a variety of different processes becomes relevant as soft gluons have been copiously produced during the previous evolution. Besides the processes involving quark-gluon interactions $q\rightarrow gq$ (light blue), $qg\rightarrow qg$ (orange), inelastic absorptions of soft gluons $gg\rightarrow g$ (black) also have an important effect for the thermalization of the gluon sector, whereas elastic scattering of gluons $gg\rightarrow gg$ (red) as well as elastic  $q\bar{q}\rightarrow gg$ (lime) and inelastic $q\bar{q}\rightarrow g$ (dark blue) quark/antiquark annihilation processes are clearly subleading. By comparing the results at zero and finite density in Figs.~\ref{fig-TMGQ-Cq0} and \ref{fig-TMGQ-Cq2}, one further notices an increment of the $gg \rightarrow g$ collision rates, indicating a more rapid gluon production from quarks at finite density, consistent with the observations of the spectra in Figs.~\ref{fig-TMGQ-MU0F_Q0} and \ref{fig-TMGQ-MUF}. Due to the fact that at finite density there are more quarks present in the system, the collision rates for quarks are generally larger compared to the zero density case. Nevertheless, the underlying dynamics remains essentially the same as compared to the zero density case, with gluon radiation $q\rightarrow gq$ (light blue) and quark-gluon scattering $qg\rightarrow gq$ providing the dominant mechanisms to transfer energy from hard quarks to softer gluons. Due to the larger abundance of quarks at finite density, elastic scattering processes $qq \rightarrow qq$ involving quarks of the same (green) and different flavors (pink), also play a more prominent role in restoring kinetic equilibrium in the quark sector, while they were more or less negligible at zero density. Surprisingly small changes appear in the collision rates for antiquarks between the initial time $t=0$ and $t=0.1 \tau_{R}$, where at later times the inelastic $\bar{q} \rightarrow \bar{q}g$ process becomes suppressed due to the fact the inverse process of absorbing a soft gluon $\bar{q}g \rightarrow  \bar{q}$ becomes increasingly likely. Similarly, elastic scattering processes $\bar{q}g \rightarrow  \bar{q}$ (orange) between antiquarks and gluons also contribute to the energy transfer from the antiquark to the gluon sector.

Eventually for $t=0.5 \tau_{R}$, the energy transfer from quarks to gluons due to elastic $qg \rightarrow qg$ (orange) and inelastic $q\rightarrow gg$ (light blue) becomes smaller and smaller, so do the collision rates for inelastic gluon absorptions $g \rightarrow gg$ (black) and elastic scatterings between quarks/antiquarks (pink and green), which are primarily responsible for restoring kinetic equilibrium in the gluon and quark sectors. Beyond the time scales shown in Fig.~\ref{fig-TMGQ-Cq2}, the evolution of the system continues in essentially the same way, with continuously collision rates decreasing until eventually gluons, quarks and antiquarks all approach their respective equilibrium distribution.

\subsubsection{Scale Evolutions}
Now that we have established the microscopic processes underlying the chemical equilibration of finite density, we again turn to the evolution of the dynamical scales $m_D^2$, $m_Q^2$ and $T^{*}$, which serve as a reference to determine the progress of kinetic and chemical equilibration. We present our results in Fig.~\ref{fig-TMGQ-MUS}, where we compare the evolution of the dynamical scales in systems with a different amount of net-charge density, as characterized by the ratio $\mu_{\rm eq}/T_{\rm eq}=0,0.14,1.34,2.5$ of the equilibrium chemical potential over the equilibrium temperature. By comparing the evolution of the various quantities in Fig.~\ref{fig-TMGQ-MUS}, one observes that for larger chemical potentials $m_D^2$, $m_Q^2$ as well as $T^{*}$ are generally closer to their final equilibrium values over the course of the entire evolution. While the smaller deviations of $m_D^2$, $m_Q^2$ and $T^{*}$ can partly be attributed to the fact that in the finite density system the initial values for these quantities are already closer to the final equilibrium value, it also appears that the ultimate approach towards equilibrium occurs on a slightly shorter time scale. We attribute this to the fact that for larger values of $\mu_{\rm eq}/T_{\rm eq}$ at a fixed temperature, the system features a larger energy density (c.f. Eq.~(\ref{eq-energydensity})), which should effectively speed up the various collision processes. 

Similar phenomena can also be observed in Fig.~\ref{fig-TMGQ-MUE}, where we present the evolution of the energy and number density of gluons, quarks and antiquarks over the course of the chemical equilibration process at different densities $\mu_{\rm eq}/T_{\rm eq}=0,0.14,1.34,2.5$. While initially there is always a rapid production and energy transfer to the gluon sector, the flattening of the curve at later times show the  relaxation towards chemical equilibrium, which occurs roughly on the same time scale as the kinetic equilibration of the dynamical scales $m_D^2$, $m_Q^2$ and $T^{*}$. By comparing the results for different $\mu_{\rm eq}/T_{\rm eq}$, one again observes that the chemical equilibration happens slightly earlier for larger chemical potential, consistent with the observations from spectra in Fig.~\ref{fig-TMGQ-MU0F_Q0}, Fig.~\ref{fig-TMGQ-MUF}, collision rates in Fig.~\ref{fig-TMGQ-Cq0}, Fig.~\ref{fig-TMGQ-Cq2} and from the scale evolutions in Fig.~\ref{fig-TMGQ-MUS}. Nevertheless, we believe that at least for the range of $\mu_{\rm eq}/T_{\rm eq}$ considered in Fig.~\ref{fig-TMGQ-MUS}, our estimate of the kinetic and chemical equilibration time scales in Eq.~(\ref{eq:eqTimes-chemical}), remains valid also at finite density.

\section{Equilibration of Far-From-Equilibrium Systems}
\label{sec:FarEq}
We will now analyze the equilibration process of QCD systems which are initially far from equilibrium. By focusing on systems which are spatially homogeneous and isotropic in momentum space, we can distinguish two broad classes of far-from equilibrium initial states which following \cite{Kurkela:2011ti,Schlichting:2019abc} can be conveniently characterized by considering the initial average energy per particle $\langle p \rangle_0$ in relation to the equilibrium temperature $T_{\rm eq}$ of the system. Specifically, for far-from equilibrium initial states, we can consider a situation where the average energy per particle is initially much smaller than the equilibrium temperature, i.e. $\langle p \rangle_0 \ll T_{\rm eq}$, such that the energy is initially carried by a large number $f_{0} \gg 1$ of low momentum gluons. Such \emph{over-occupied} initial states typically appear as a consequence of plasma instabilities \cite{Nielsen:1978rm,Kurkela:2011ti,Berges:2013fga} and they also bear some resemblance with the saturated ``Glasma'' initial state created in high-energy collisions of heavy nuclei~\cite{Krasnitz:2002mn,Lappi:2003bi,Lappi:2006fp,Blaizot:2010kh,Berges:2013eia,Berges:2014yta,Gelis:2013rba}, although the detailed properties of this state are quite different as the system is highly anisotropic and rapidly expanding in the longitudinal direction as discussed in more detail in Sec.~\ref{sec-evol-expansion}. While for $\langle p \rangle_0 \sim T_{\rm eq}$, the system is in some sense close to equilibrium and one would naturally expect kinetic and chemical equilibration to occur on the time scales of the equilibrium relaxation time $\sim \tau_{R}$, there is a second important class of far-from equilibrium initial states corresponding to under-occupied states. In $\emph{under-occupied}$ systems the average energy per particle is initially much larger then the equilibrium temperature $\langle p \rangle_0 \gg T_{\rm eq}$, such that the energy is initially carried by a small number $f_{0} \ll 1$ of highly energetic particles, as is for instance the case for an ensemble of high-energy jets. While earlier works~\cite{Berges:2013eia,Kurkela:2014tea} have established the equilibration patterns of such systems for pure glue QCD, we provide an extension of these studies to full QCD with three light flavors, as previously done for over-occupied systems in \cite{Kurkela:2018oqw}.

\subsection{Equilibration of Over-occupied Systems}
\label{sec-evol-overoccupied}
We first consider over-occupied systems characterized by a large occupation number $f_{0} \gg 1$ of low-energy $\langle p \rangle_0 \ll T_{\rm eq}$ gluons,\footnote{Due to the fact that the phase-space occupancies of quark/antiquarks are limited to $f_{q/\bar{q}} \leq 1$ due to Fermi statistics, such over-occupied systems are inevitably gluon dominated.} and  we may estimate the energy density of the over-occupied system as $e_{0}\sim f_0 \left<p\right>_0^4$. Since the total energy density is conserved, we have $e_{\rm eq}= e_{0}$, such that with $e_{\rm eq}\sim T_{\rm eq}^4$ the final equilibrium temperature $T_{\rm eq}\sim f_0^{\frac{1}{4}} \left<p\right>_0 \gg \left<p\right>_0$ is much larger than the average initial momentum $\langle p \rangle_0$. Due to this separation of scales, energy needs to be re-distributed from low momentum to high momentum degrees of freedom, which as will be discussed shortly is achieved via a direct energy cascade from the infrared to ultraviolet in momentum space.

\subsubsection{Theoretical Aspects}
Due to the large population of low momentum gluons, interaction rates for elastic and inelastic processes are initially strongly enhanced, such that e.g. the large angle elastic scattering rate $\Gamma_{\rm el} \sim g^2 T^{*} \frac{m_{D}^2}{\langle p \rangle^2}$ is initially much larger than in equilibrium $\Gamma_{\rm el}^{0} \sim g^4 f_{0}^2 \langle p \rangle \gg \Gamma_{\rm el}^{\rm eq} \sim g^4 T_{\rm eq}$.  Even though the time scale for the actual equilibration process is eventually controlled by the equilibrium rate $\sim 1/\Gamma_{\rm el}^{\rm eq}$, the system will therefore encounter a rapid memory loss of the details of the initial conditions on a time scale $\sim 1/\Gamma_{\rm el}^{0}$, and subsequently spend a significant amount of time in a transient non-equilibrium state, where the energy transfer from the infrared towards the ultraviolet is accomplished. 

Since the dynamics remains gluon dominated with $f_g\gg 1 \ge f_{q,\bar{q}}$ all the way until the system eventually approaches equilibrium, one should expect that the evolution of the over-occupied Quark-Gluon plasma follows that of pure-glue QCD, where it has been established~\cite{Schlichting:2012es,Kurkela:2012hp,Berges:2013fga,York:2014wja,Berges:2017igc}, that for intermediate times $1/\Gamma_{\rm el}^{0} \ll t \ll 1/\Gamma_{\rm el}^{\rm eq}$, the evolution of the gluon spectrum follows a self-similar scaling behavior of the form
\begin{eqnarray}
\label{eq-scalingfunc}
&&f_g(p,t)=(t/t_0)^{\alpha} f_{0}~f_S\left((t/t_0)^{\beta}\frac{p}{\langle p \rangle_{0}}\right)
\end{eqnarray}
where $t_0\simeq1/\Gamma_{\rm el}^{0}$, $\langle p \rangle_0$ are the characteristic time and momentum scales, $f_{0}$ is the initial occupancy and $f_S(x)$ is a universal scaling function up to amplitude normalization and we adopt the normalization conditions $f_{S}(x=1)=f'_{S}(x=1)=1$. We note that the emergence of self-similar behavior as in Eq.~(\ref{eq-scalingfunc}), is by no means unique to QCD, and in fact constitutes a rather generic pattern in the equilibration of far-from-equilibrium quantum systems, with similar observations reported in the context of relativistic and non-relativistic scalar field theories~\cite{Micha:2004bv,Berges:2013lsa}. Specifically, the scaling exponents $\alpha=-4/7$, $\beta=-1/7$ follow directly from a dimensional analysis of the underlying kinetic equations~\cite{Kurkela:2011ti,Kurkela:2012hp,Berges:2013fga,Blaizot:2011xf}, and describe the energy transport from the infrared towards the ultra-violet due to a direct energy cascade~\cite{nazarenko2011wave}.

Based on Eq.~(\ref{eq-scalingfunc}), we can further estimate the evolutions of some physical quantities knowing that gluon are dominant $f_g\gg 1 \ge f_{q,\bar{q}}$ in the self-similar scaling regime. 
In particular, the average momentum $\langle p \rangle$ increases as a function of time according to
\begin{eqnarray}
\label{eq-scale-p}
&&\left<p\right>(t)
\simeq \frac{\int \frac{d^3p}{(2\pi)^3}pf_g(p,t)}{\int \frac{d^3p}{(2\pi)^3}f_g(p,t)}
\sim \langle p \rangle_{0} \left(\frac{t}{t_0}\right)^{\frac{1}{7}}
\end{eqnarray}
while the typical occupancies of hard gluons decrease as
\begin{eqnarray} 
f\Big(t,\langle p \rangle(t)\Big) \simeq f_{0} \left(\frac{t}{t_0}\right)^{\alpha} \sim f_{0}\left(\frac{t}{t_0}\right)^{-\frac{4}{7}}
\end{eqnarray}

\begin{figure}
	\centering
	\includegraphics[width=0.48\textwidth]{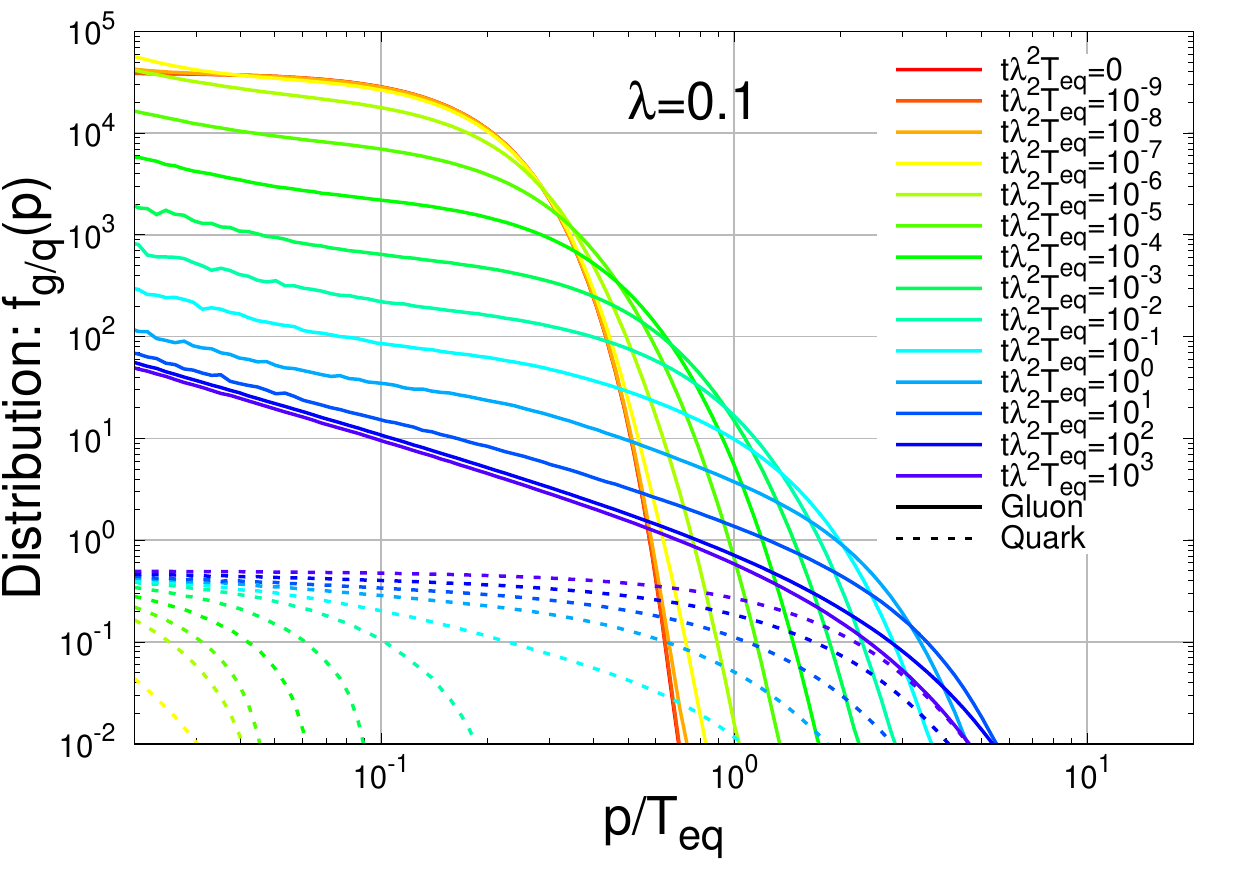}
	\centering
	\includegraphics[width=0.48\textwidth]{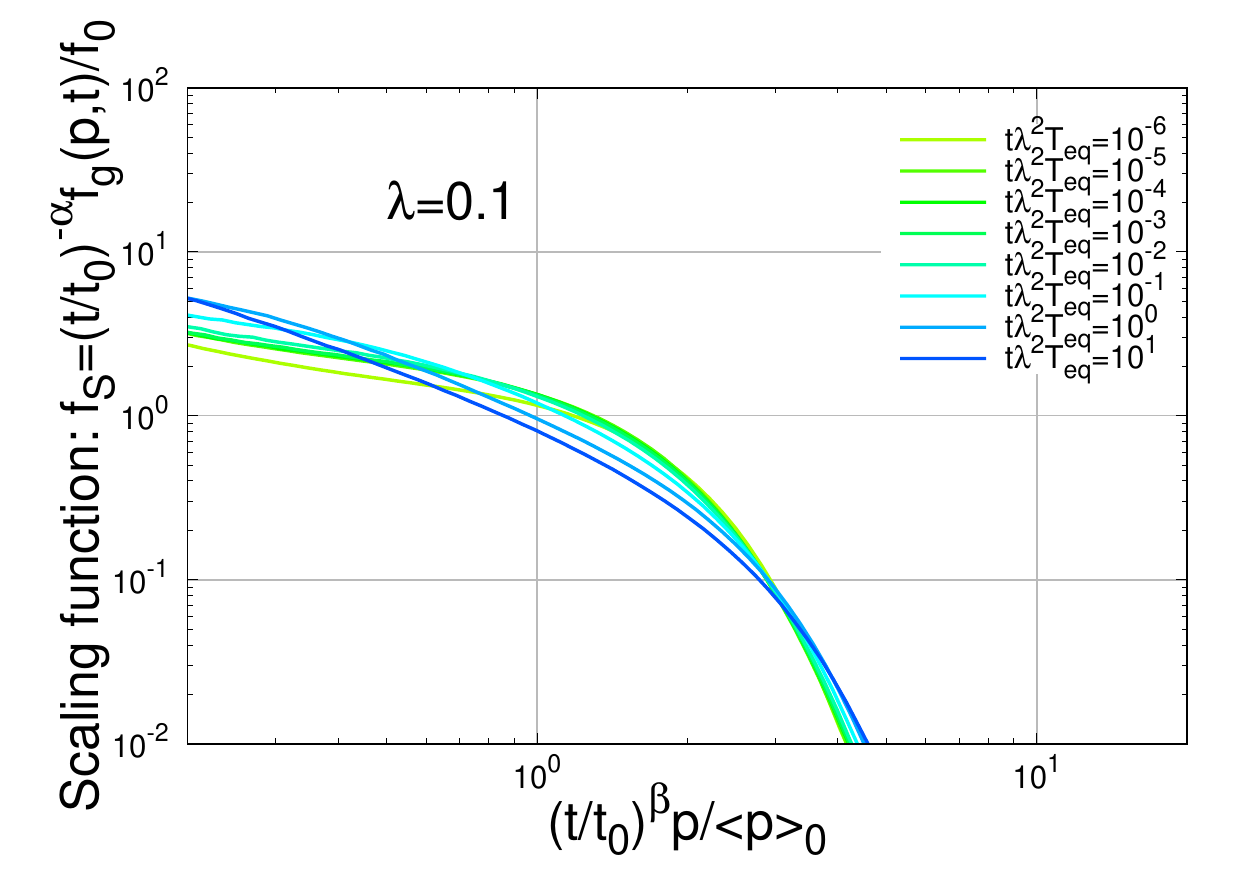}
	\caption{(top) Evolution of the phase-space distributions of gluons (solid) and quarks/anti-quarks (dashed) in an over-occupied QCD plasma with $\left<p\right>_0/T_{\rm eq}$=0.2 at weak coupling $\lambda$=0.1. (bottom) Scaling function $f_{s}$ obtained by re-scaling the gluon spectra in the top panel, with the exponents $\alpha$=-4/7, $\beta$=-1/7 and scale factors $t_0$=0.005$T_{\rm eq}^{-1}$ and $f_0$=1500.}
	\label{fig-OG-F}
\end{figure}

Similarly, one finds that the screening mass 
\begin{eqnarray}
\label{eq-scale-mD}
&&m_D^2(t)\simeq g^2 \int \frac{d^3p}{(2\pi)^3}\frac{1}{2p}f_g(p,t)
\sim g^2f_0 \langle p \rangle_{0}^{2} \left(\frac{t}{t_0}\right)^{-\frac{2}{7}}
\end{eqnarray}
decreases, such that the system dynamically establishes a separation between the soft ($\sim m_{D}$) and hard ($\sim \langle p \rangle$) scales over the course of the self-similar evolution~\cite{Kurkela:2011ti,Berges:2017igc}. 
Since the effective temperature also decreases according to ($f_g(p,t)\gg 1$)
\begin{eqnarray}
\label{eq-scale-T}
&&T^{*}(t)\simeq\frac{\nu_{g} C_{A}}{d_{A} m_D^2}\int\frac{d^3p}{(2\pi)^3}f_g^2(p,t)
\sim g^2f_0 \langle p \rangle_{0} \left(\frac{t}{t_0}\right)^{-\frac{3}{7}}.
\end{eqnarray}
the large-angle elastic scattering rate $\Gamma_{\rm el}(t) \sim g^2 T^{*} \frac{m_{D}^2}{\langle p \rangle^2}\sim g^4 f_{0}^{2} \langle p \rangle_{0} (t/t_0)^{-1}$ decreases over the course of the self-similar evolution and eventually becomes on the order of the equilibrium rate $\Gamma_{\rm el}(t) \sim g^4 T_{\rm eq}$ at the same time $t/t_0 \sim f_{0}^{7/4}$ when the occupancies of hard gluons $f\Big(t,\langle p \rangle(t)\Big)$ become of order unity, and the average momentum $\langle p \rangle(t)$ becomes on the order of the equilibrium temperature $T_{\rm eq} \sim \langle p \rangle_{0} f_{0}^{1/4}$ , indicating that the energy transfer towards the ultra-violet has been accomplished and gluons are no longer dominant for $t\gtrsim t_{0}f_{0}^{7/4} \sim g^{-4} f_{0}^{-1/4} \langle p \rangle_{0}^{-1} \sim g^{-4} T_{\rm eq}^{-1} $.

Beyond this time scale, the system can be considered as close to equilibrium, and should be expected to relax towards equilibrium on a time scale on the order of the kinetic relaxation time $\tau_{R}\sim g^{-4} T_{\rm eq}^{-1}$, which is parametrically of the same order as the time it takes to accomplish the energy transfer towards the ultra-violet.

\begin{figure}
	\centering
	\includegraphics[width=0.48\textwidth]{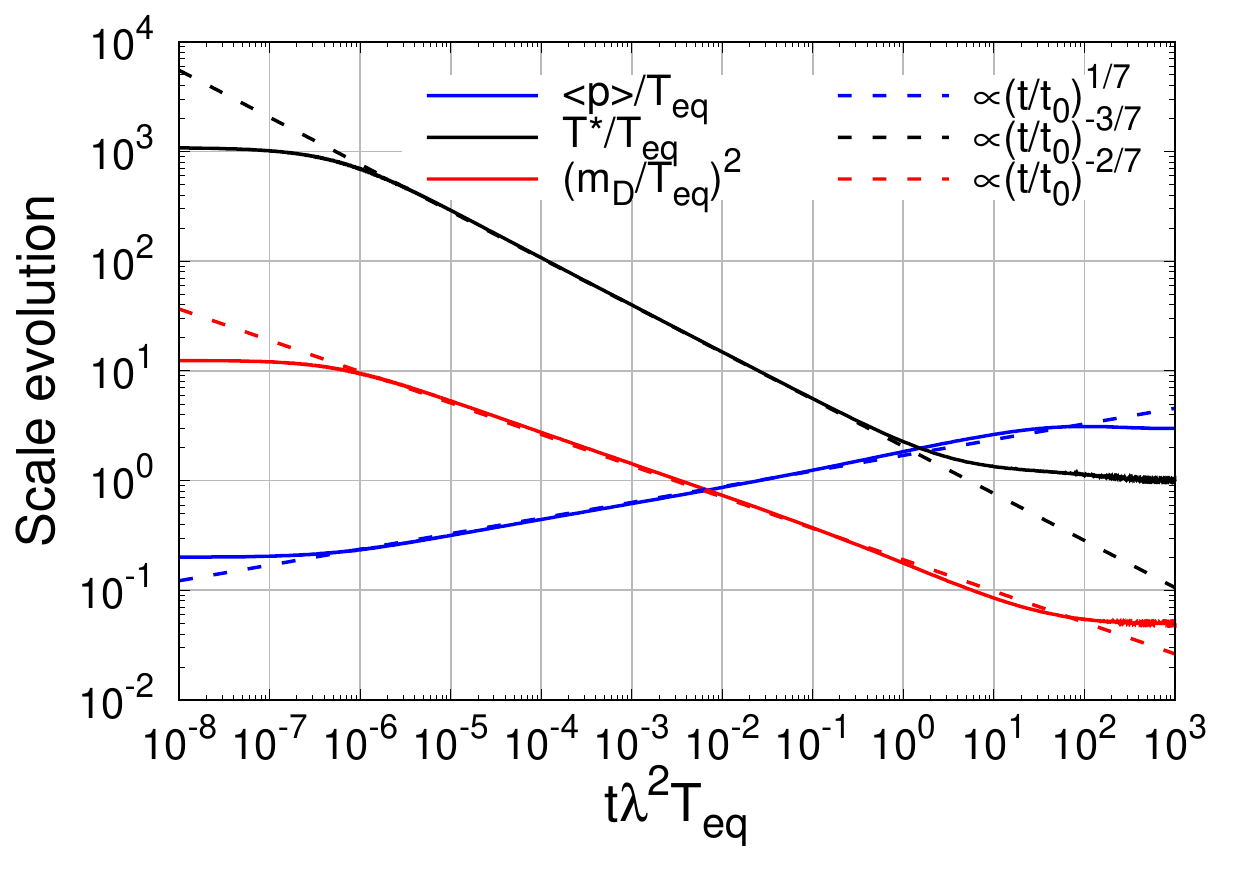}
	\centering
	\includegraphics[width=0.48\textwidth]{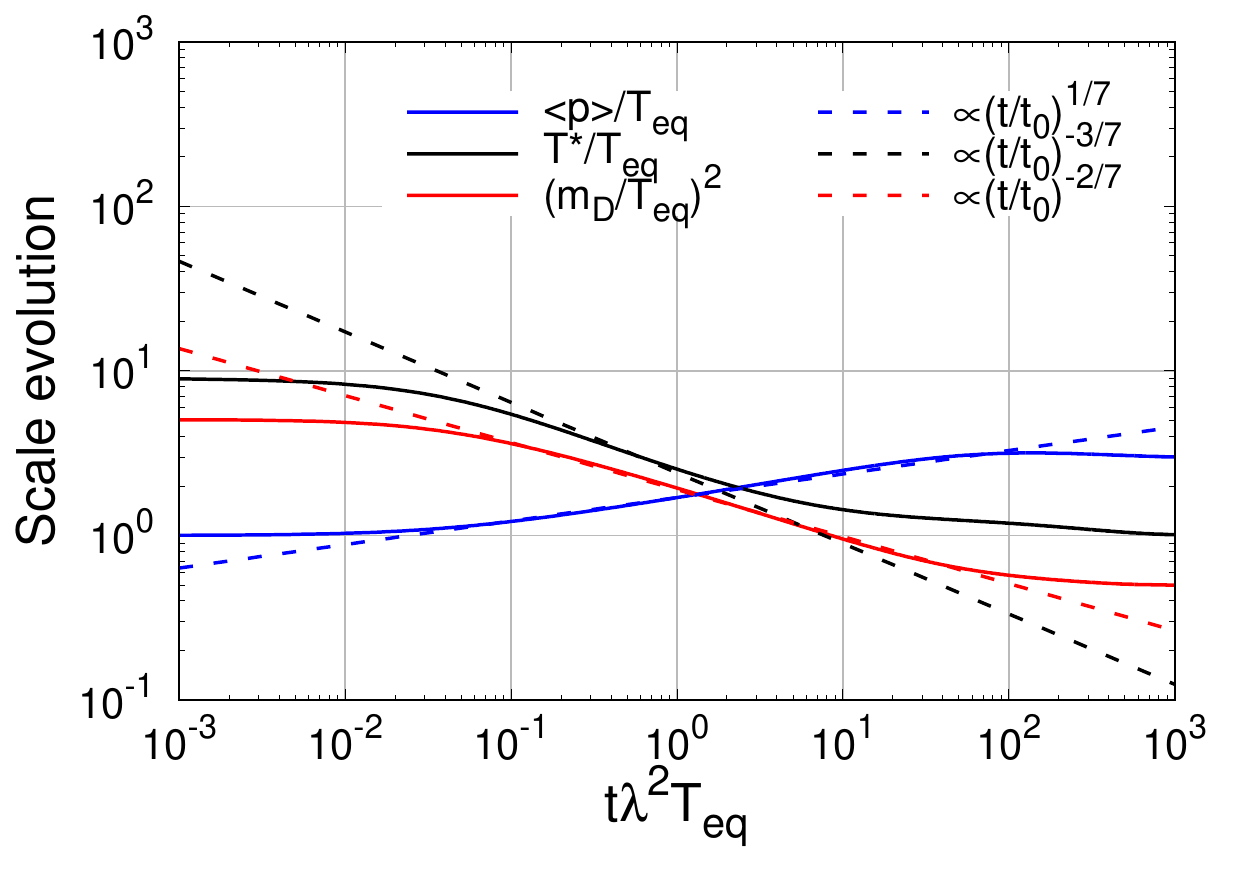}
	\centering
	\includegraphics[width=0.48\textwidth]{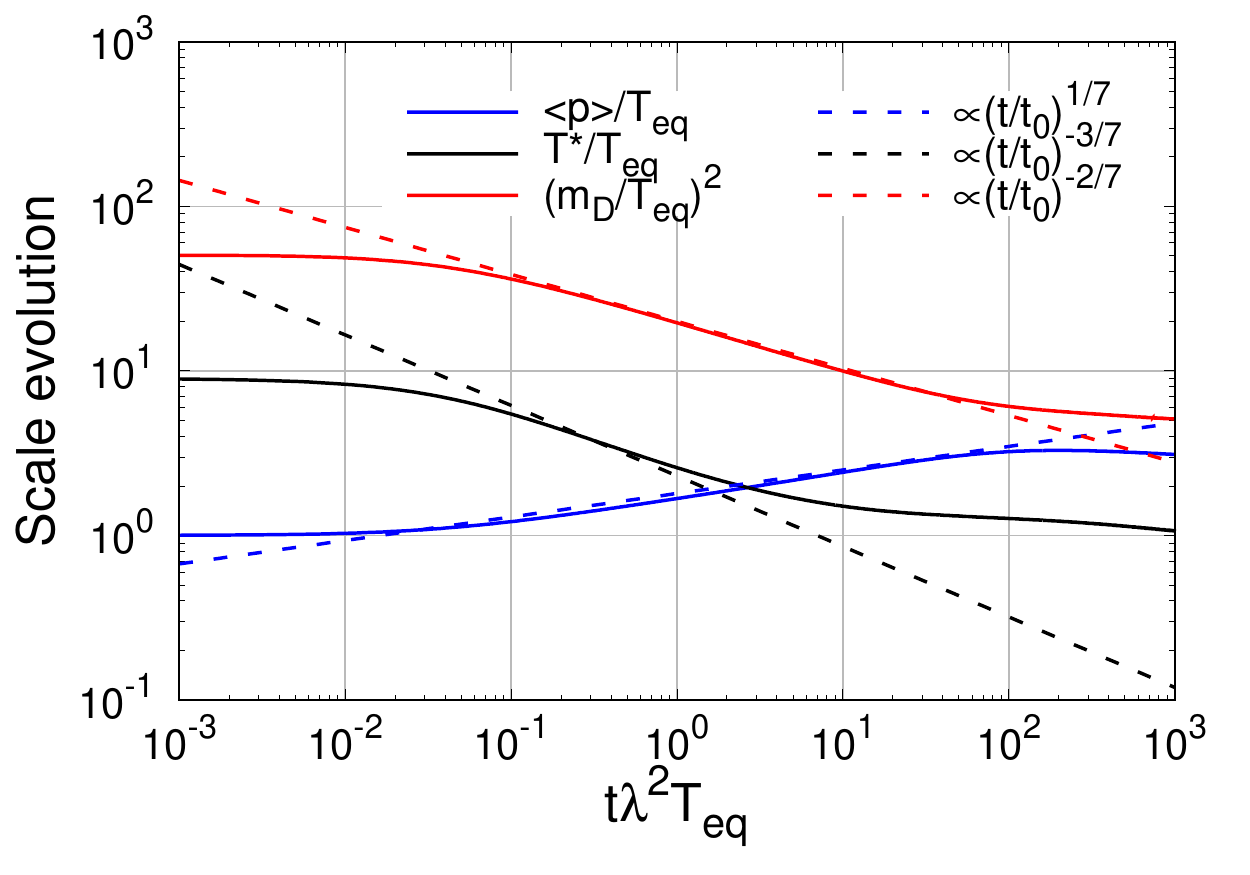}
	\caption{Evolution of the average momentum per particle $\left<p\right>$ (blue), effective temperature  $T^{*}$ (black) and screening mass $m_{D}^{2}$ in an over-occupied QCD plasma at weak coupling $\left<p\right>_0/T_{\rm eq}$=0.2, $\lambda$=0.1 (top), intermediate coupling $\left<p\right>_0/T_{\rm eq}$=1, $\lambda$=1 (middle) and at strong coupling $\left<p\right>_0/T_{\rm eq}$=1, $\lambda$=10 (bottom). Dashed lines show the characteristic power law behavior in the self-similar turbulent regime.}
	\label{fig-OG-S}
\end{figure}

\subsubsection{Numerical results}

We now turn to the results of effective kinetic theory simulation of the equilibration process in over-occupied QCD plasmas, extending earlier results in \cite{Kurkela:2018oqw}. We initialize the phase-space distributions as
\begin{eqnarray}
\label{eq-OG-INITIAL}
\nonumber
&&f_g(p,t=0)=e_{0}\frac{\pi^2}{4Q^4}e^{-\frac{p^2}{Q^2}},\\
\nonumber
&&f_q(p,t=0)=0,\\
&&f_{\bar{q}}(p,t=0)=0,
\end{eqnarray}
such that for $Q^{4} \ll e_{0}$ the system features a large occupancy $f_{0} \simeq e_{0}\frac{\pi^2}{4Q^4}$ of low-momentum gluons, with average momentum $\left<p\right>_0= \frac{2Q}{\sqrt{\pi}}$. Since the system in Eq.~(\ref{eq-OG-INITIAL}) is charge neutral, all species of quarks/antiquarks will be produced democratically over the course of the evolution.

\begin{figure}
	\centering
	\includegraphics[width=0.48\textwidth]{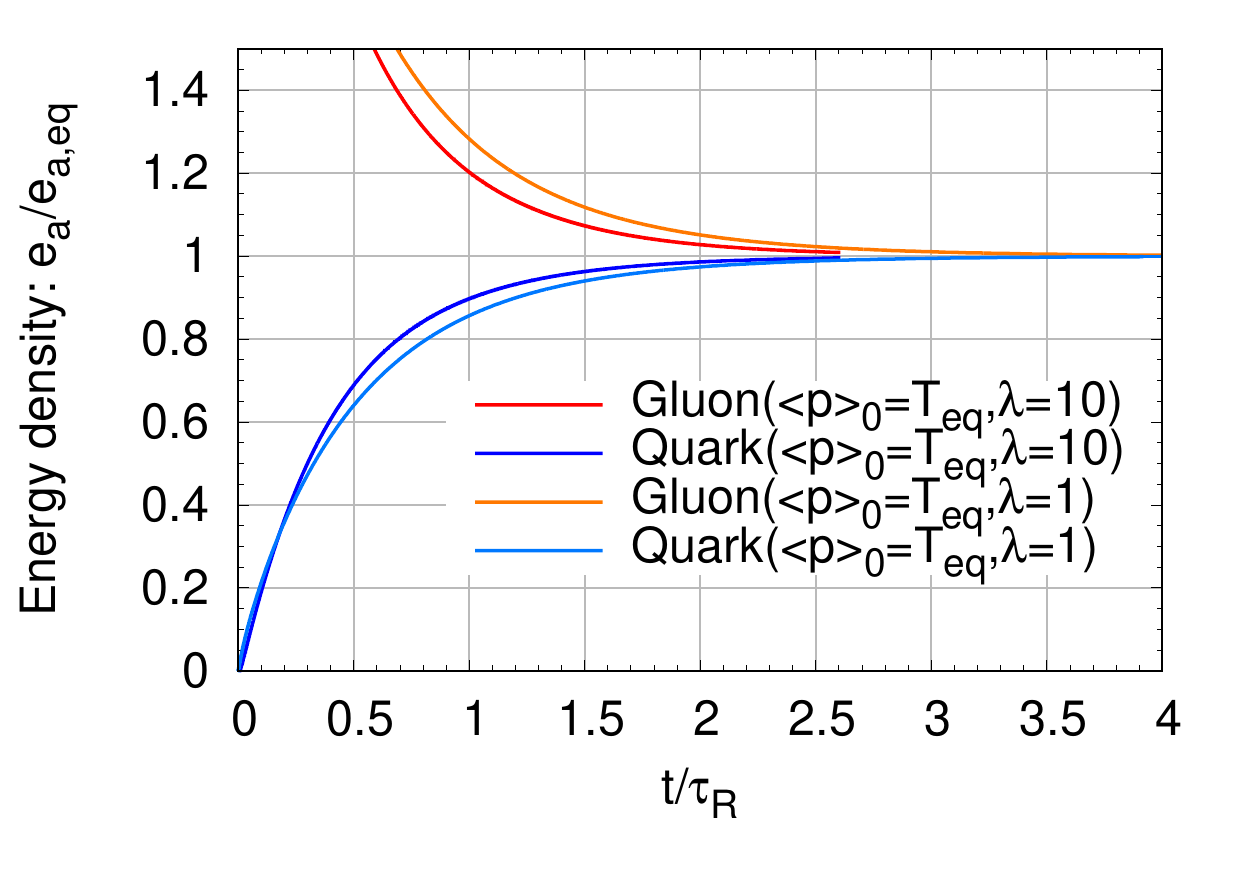}
	\centering
	\includegraphics[width=0.87\textwidth]{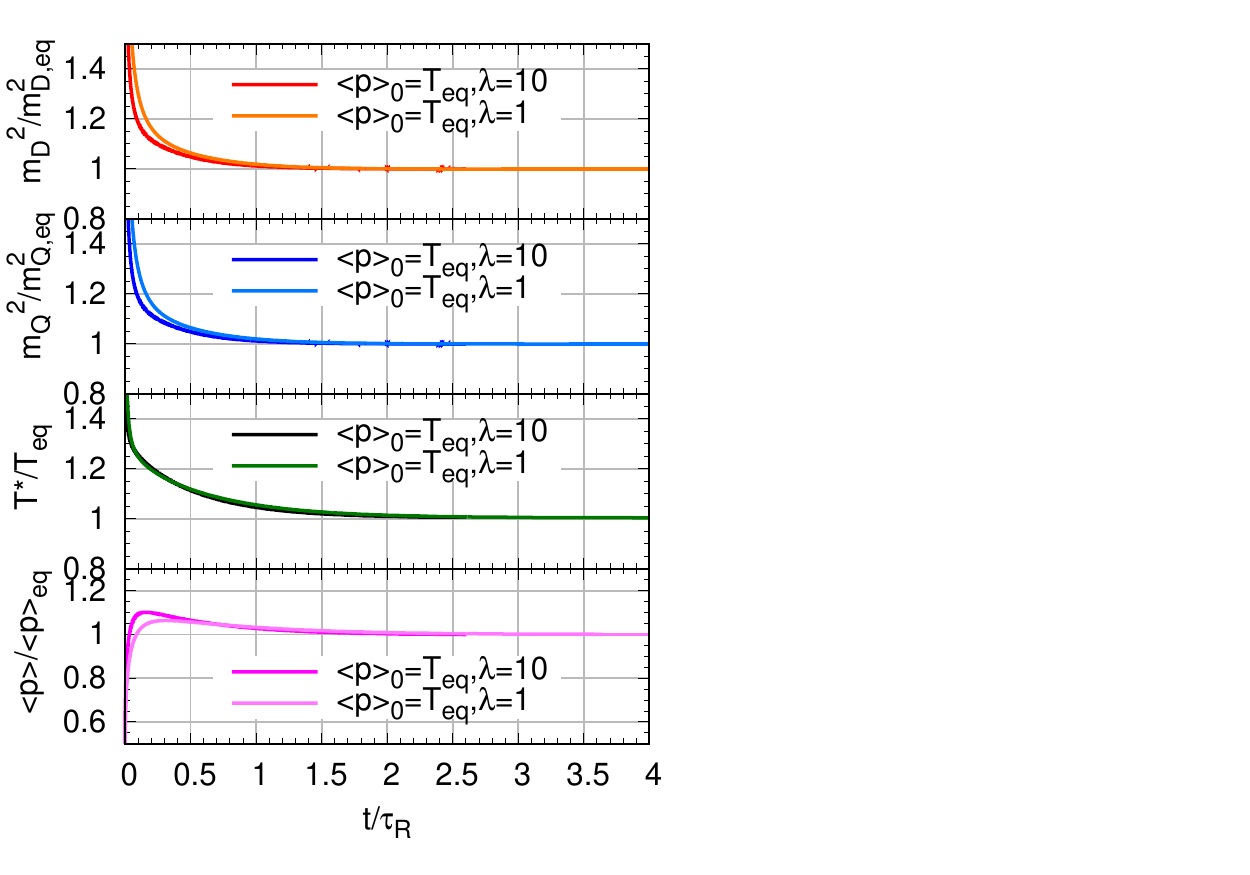}
	\caption{Evolution of the energy densities of quarks and gluons (top), the characteristic scales (bottom) $m_D^2(t)$ (red), $m_Q^2(t)$ (blue), $T^{*}(t)$ (green) and $\left<p\right>$ (pink) in an over-occupied QCD plasma at two different coupling strengths $\lambda=1$ (lighter colors) and $\lambda=10$ (darker colors). Scales are normalized to their respective equilibrium values, while the evolution time $t$ is normalized to the equilibrium relaxation time $\tau_R$ in Eq.~(\ref{eq-relaxation}) in order to take into account the leading coupling dependence.}
	\label{fig-OG-E}
\end{figure}

We first consider an over-occupied system with a relatively large scale separation $\left<p\right>_0/T_{\rm eq}$=0.2 at weak coupling $\lambda$=0.1, and investigate the evolutions of the spectra of quarks and gluons depicted in the top panel of Fig.~\ref{fig-OG-F}.
Starting from a large phase-space occupancy of soft gluons, the initial spectra undergo a quick memory loss at very early times and then gradually evolve into harder spectra through a direct energy cascade, pushing the low momentum gluons towards higher momenta.
In order to illustrate the self-similarity of this process, we follow previous works~\cite{Kurkela:2012hp,Schlichting:2019abc} and show re-scaled versions of the gluon spectra in the bottom panel of Fig.~\ref{fig-OG-F}. 
By re-scaling the phase-space distribution as $f_S(x)\simeq(t/t_0)^{-\alpha}f_g(\left<p\right>_0(t/t_0)^{-\beta}x,t)$, and plotting it against the re-scaled momentum variable $x=(t/t_0)^{\beta} p/\left<p\right>_0$, one indeed finds that in the relevant scaling window, which corresponds approximately to times $10^{-6} \le t\lambda^2T_{\rm eq}\le 10^{1}$ for this particular set of parameters, the spectra at different times overlap with each other to rather good accuracy, clearly indicating the self-similarity of the underlying process. Beside the gluons, all species of quarks/antiquarks are produced democratically over the course of the evolution from elastic $gg\rightarrow q\bar{q}$ conversions and inelastic splitting $g\rightarrow q\bar{q}$ processes. Generally, one finds that the quark/antiquark spectra follow the evolution of the gluon spectra, albeit due to their Fermi statistics the number of quarks/antiquarks in the system remains negligibly small compared to the overall abundance of gluons during the self-similar stage of the equilibration process.

Eventually for times $t\gtrsim 10^{2}/\lambda^2T_{\rm eq}$ the self-similar cascade in Fig.~\ref{fig-OG-F} stops as the occupancies of hard gluons fall below unity and the system subsequently approaches thermal equilibrium on time scales $\sim 10^{3}/\lambda^2T_{\rm eq}$ for the parameters chosen in Fig.~\ref{fig-OG-F}. It is interesting to point out, that due to the negligible abundance of quarks and antiquarks in the system, the evolution of the gluon spectra slightly overshoots the equilibrium temperature at times $t\gtrsim 10^{2}/\lambda^2T_{\rm eq}$, and subsequently relaxes back towards equilibrium as the correct equilibrium abundance of quarks and antiquarks is produced along the lines of our previous discussion of gluon dominated systems in Sec.~\ref{sec-evol-gluonquark}.

Next we will discuss the evolution of the average momentum $\left<p\right>(t)$,the screening mass square $m_D^2(t)$ and the effective temperature $T^{*}(t)$ summarized in Fig.~\ref{fig-OG-S}, where the upper panel shows the results for $\left<p\right>_0/T_{\rm eq}=0.2$, $\lambda=0.1$, i.e. the same parameters as in Fig.~\ref{fig-OG-F}, while the middle and bottom panels show the results for a smaller scale separation $\left<p\right>_0/T_{\rm eq}=1$, at larger values of coupling $\lambda=1, 10$. By comparing the evolutions of the various scales with the theoretical predicted power-law scaling (dashed line) in the turbulent regime (c.f. Eqns.~(\ref{eq-scale-p}), (\ref{eq-scale-mD}), (\ref{eq-scale-T})), one finds that the scaling behavior $\left<p\right>\propto t^{1/7}$, $T^{*}\propto t^{-3/7}$ and $m_D^2\propto t^{-2/7}$ associated with the turbulent energy transport towards the ultra-violet is indeed realized during intermediate times. Due to the large separation of scales for $\left<p\right>_0/T_{\rm eq}=0.2$, $\lambda=0.1$, the scaling window in the top panel of Fig.~\ref{fig-OG-S} extends over a significant period of time $10^{-6} \le t\lambda^2T_{\rm eq}\le 10^{1}$, consistent with the scaling of the gluon distribution observed in Fig.~\ref{fig-OG-F}. Even though the scaling window shrinks significantly for the smaller scale separations $\left<p\right>_0/T_{\rm eq}=1$ shown in the middle and bottom panels of Fig.~\ref{fig-OG-S}, it is remarkable that the same turbulent mechanism appears to be responsible for the energy transfer even for such moderately strongly coupled systems.

Even though a significant amount of time is spent to accomplish the turbulent energy transfer, the logarithmic representation in Fig.~\ref{fig-OG-S} spoils the fact, that it is in fact the ultimate approach towards equilibrium which requires the largest amount of time. Beyond the investigation of the dynamical evolutions of various scales, it is therefore useful to consider the evolutions of the ratios of different scales compared to their equilibrium values, as indicators of the equilibration progress. We present our results in Fig.~\ref{fig-OG-E}, where the upper panel shows the evolutions of the energy densities of gluons and quarks, approaching  their equilibrium limits around $t\simeq 1.5-2\tau_R$, similar to near-equilibrium systems shown in Fig.~\ref{fig-TMGQ-MU0E}.
The next two panels of Fig.~\ref{fig-OG-E} show the screening mass square evolutions of $m_D^2(t)$ and $m_Q^2(t)$, which rapidly decrease at early times, an eventually approach their equilibrium values at $t\simeq 1-1.5\tau_R$.  Similar observations also hold for the effective temperature $T^{*}(t)$ shown in the forth panel of Fig.~\ref{fig-OG-E}. Due to the delayed chemical equilibration of the system, the average momentum $\left<p\right>(t)$ shown in the bottom panel has a non-monotonic behavior, where the rapid increase at early times due to the direct energy cascade overshoots the equilibrium value, before $\left<p\right>(t)$'s gradual decrease at later times as energy is re-distributed between quarks and gluons, eventually approaching the equilibrium limit around $t\simeq 1.5-2\tau_R$. 

Since in Fig.~\ref{fig-OG-E} the ultimate approach towards equilibrium is mostly insensitive to the initial scale separations $\langle p \rangle_0/T_{\rm eq}$ and coupling strength $\lambda$ in Fig.~\ref{fig-OG-S} when expressed in units of the kinetic relaxation time $\tau_{R}$, we can estimate the  equilibration time of an over-occupied system as
\begin{eqnarray}
t_{\rm eq} \simeq 1-2 \tau_{R}
\end{eqnarray}
where, as usual, the exact numerical value depends the detailed criteria chosen to define the equilibration time.

\subsection{Equilibration of Under-occupied Systems}
\label{sec-evol-underoccupied}
Next we consider the opposite limit of an under-occupied system, where the energy density is initially carried by a small number $f_{0} \ll 1$ of high energetic particles, with average momentum $\langle p \rangle_{0} \gg T_{\rm eq}$. While there can be a large separation of scales, one finds that in contrast to over-occupied systems the final equilibrium temperate $T_{\rm eq}\sim f_0^{\frac{1}{4}}\left<p\right>_0\ll\left<p\right>_0$ is much smaller than the average initial momentum for under-occupied systems. Since the scale hierarchy is reversed, the thermalization process for an under-occupied system requires an energy transport from the ultra-violet to the infrared, which as we will discuss shortly will be accomplished via an inverse turbulent cascade of successive radiative emissions. While the qualitative features of this ``bottom-up'' thermalization mechanism have been established a long time ago~\cite {Baier:2000sb}, recent works in the context of thermalization and jet quenching studies~\cite{Blaizot:2013hx,Mehtar-Tani:2018zba,Schlichting:2020lef} have provided a more quantitative description of the different stages and clarified the relation to turbulence. Based on our effective kinetic description of QCD, we will extend previous findings in pure glue QCD~\cite{Kurkela:2014tea} to full QCD at zero and non-zero densities.

\begin{figure}
	\centering
	\includegraphics[width=0.48\textwidth]{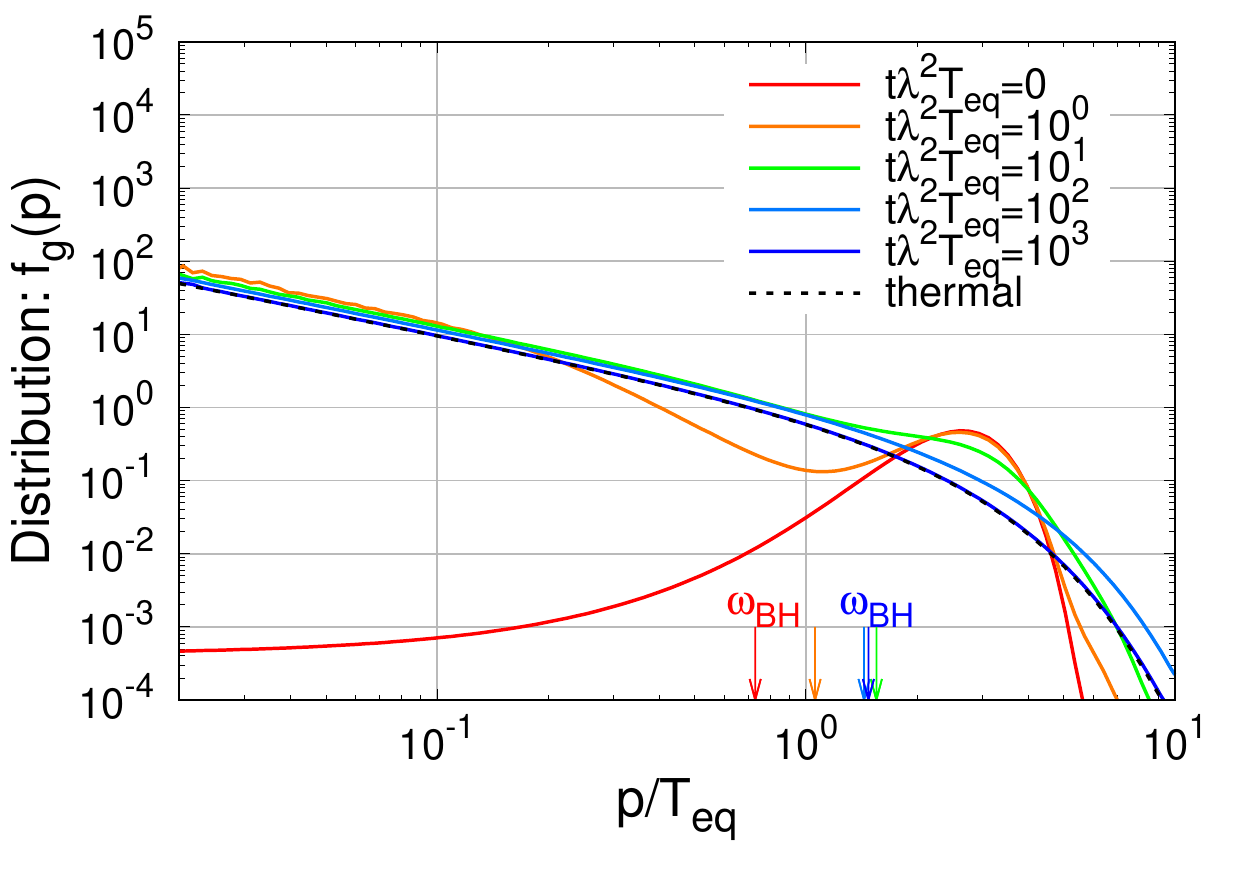}
	\hspace{\fill}
	\centering
	\includegraphics[width=0.48\textwidth]{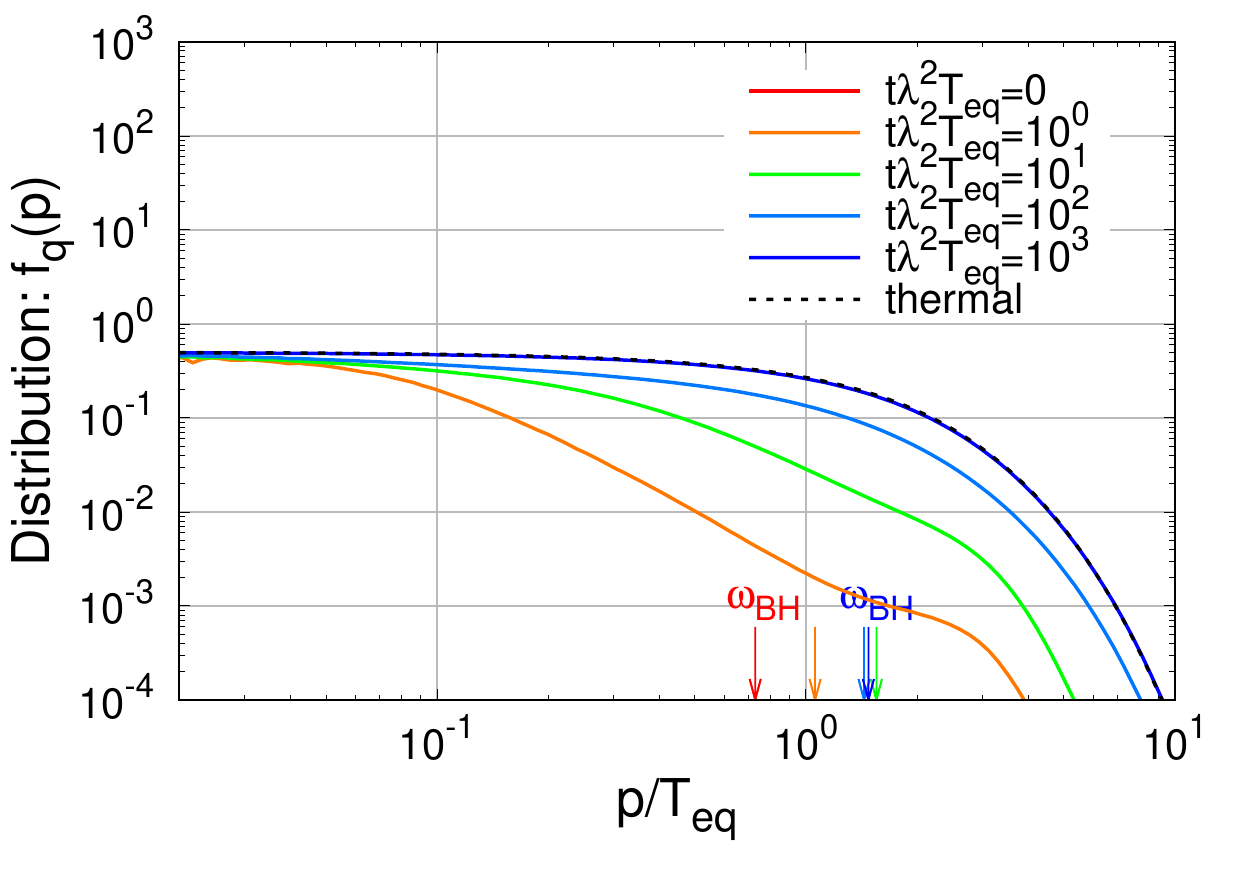}
	\caption{Evolution of the phase-space distributions of gluons (top) and quarks/anti-quarks (bottom) in an \emph{under-occupied gluon system} with $\left<p\right>_0/T_{\rm eq}$=3 at coupling $\lambda$=1. Dotted lines show the thermal equilibrium distributions. Vertical arrows mark the Bethe-Heitler frequencies $\omega_{BH}$.}
	\label{fig-UG-F3}
\end{figure}

\begin{figure}
	\centering
	\includegraphics[width=0.48\textwidth]{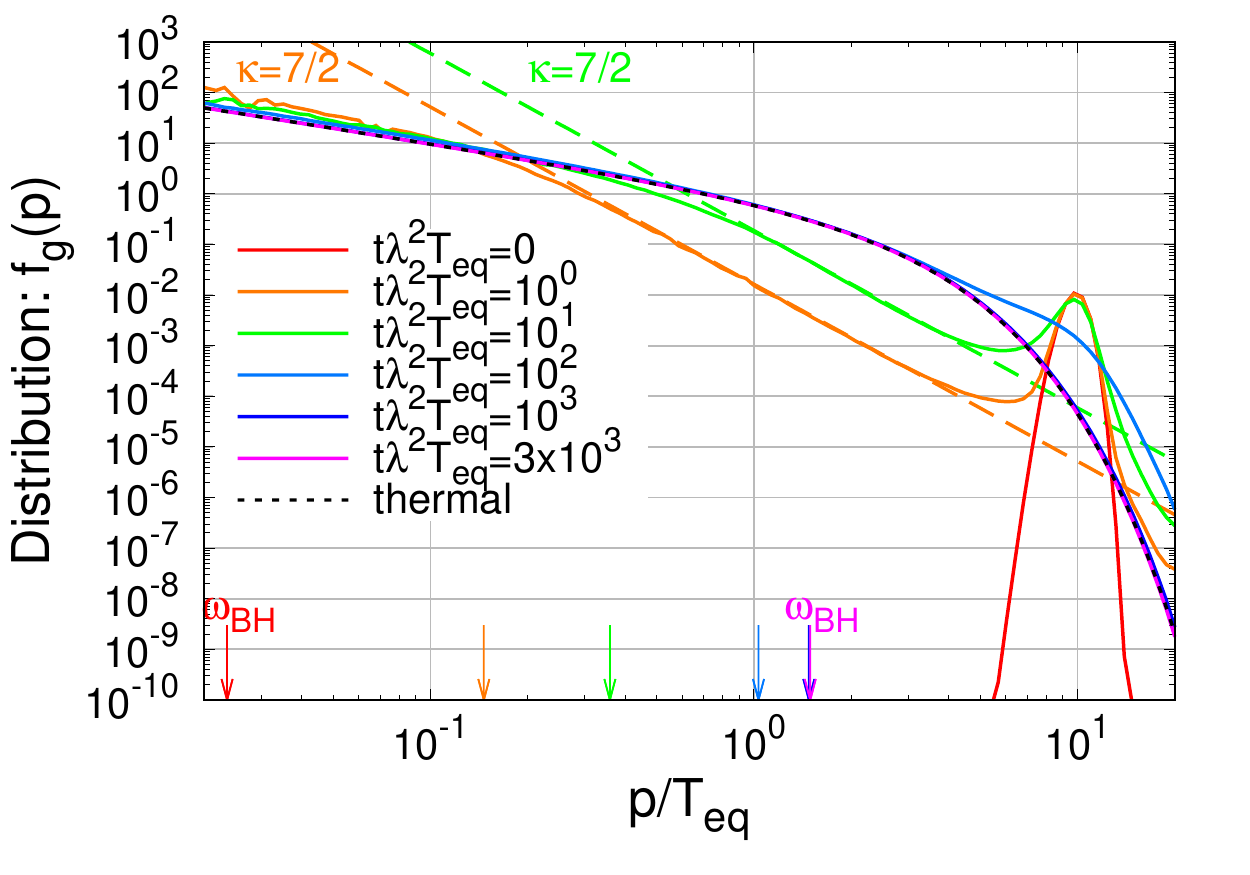}
	\hspace{\fill}
	\centering
	\includegraphics[width=0.48\textwidth]{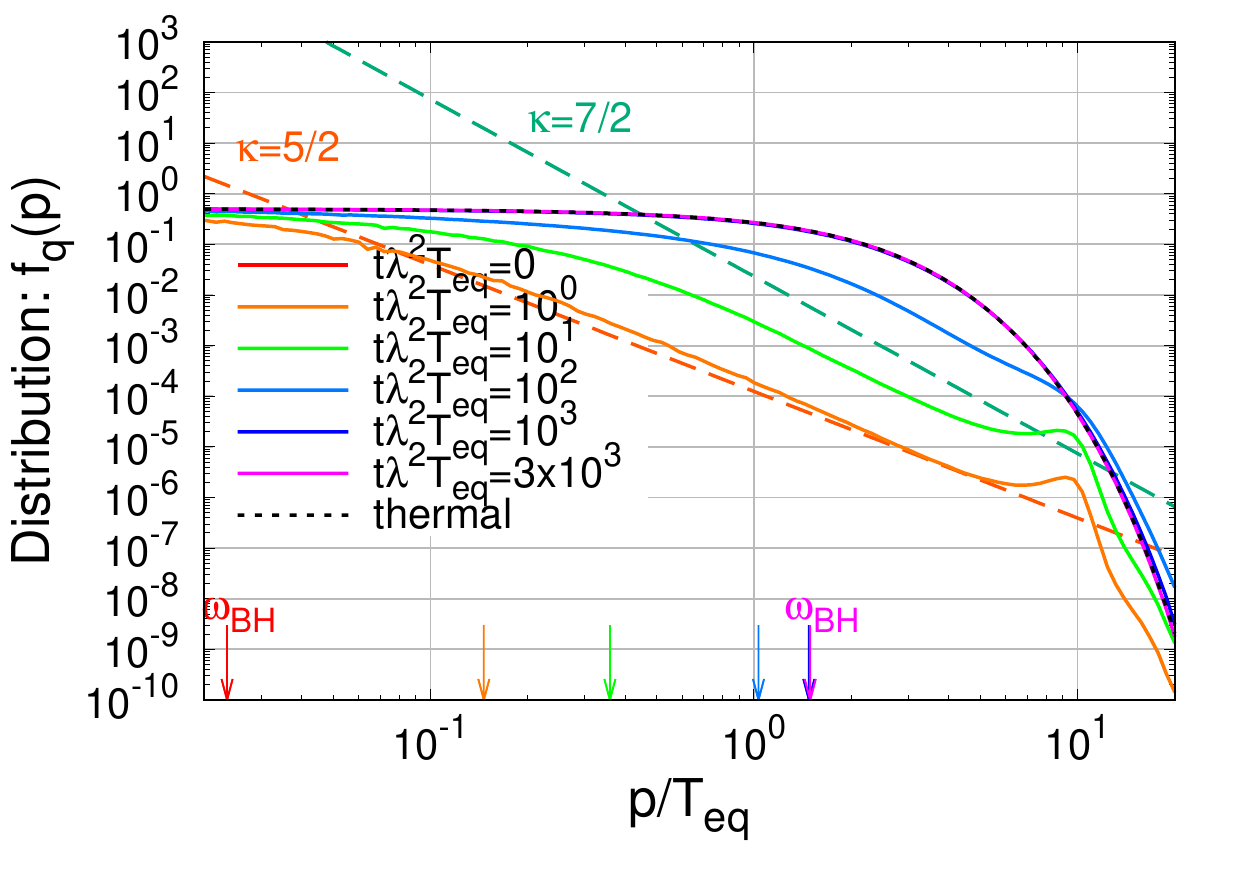}
	\caption{Evolution of the phase-space distributions of gluons (top) and quarks/anti-quarks (bottom) in an \emph{under-occupied gluon system} with $\left<p\right>_0/T_{\rm eq}$=30 at coupling $\lambda$=1. Dashed lines show the characteristic power law dependence of the single emission LPM spectra (orange) and the Kolmogorov Zakharov spectra (green). Dotted lines show the thermal equilibrium distributions. Vertical arrows mark the Bethe-Heitler frequencies $\omega_{BH}$.}
	\label{fig-UG-F10}
\end{figure}

\begin{figure}
	\centering
	\includegraphics[width=0.48\textwidth]{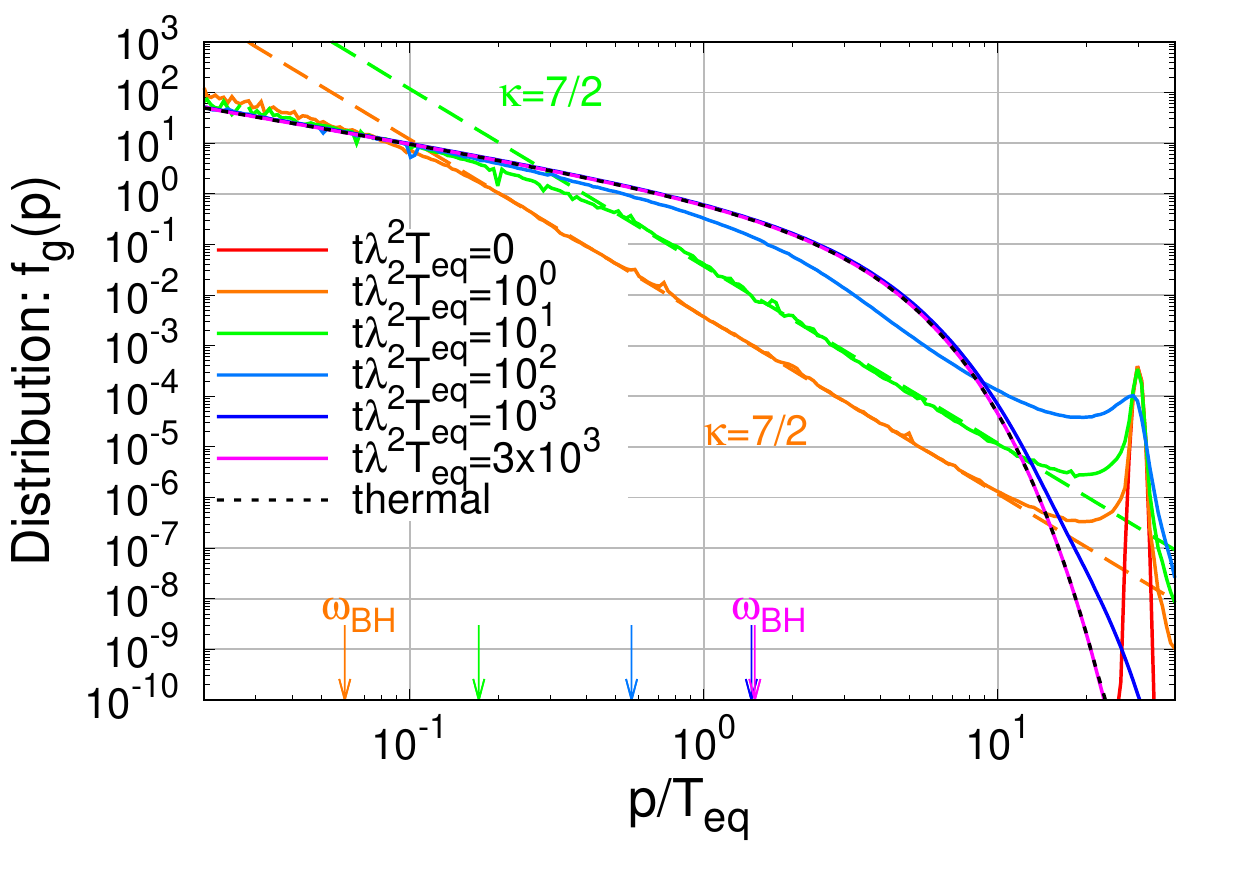}
	\hspace{\fill}
	\centering
	\includegraphics[width=0.48\textwidth]{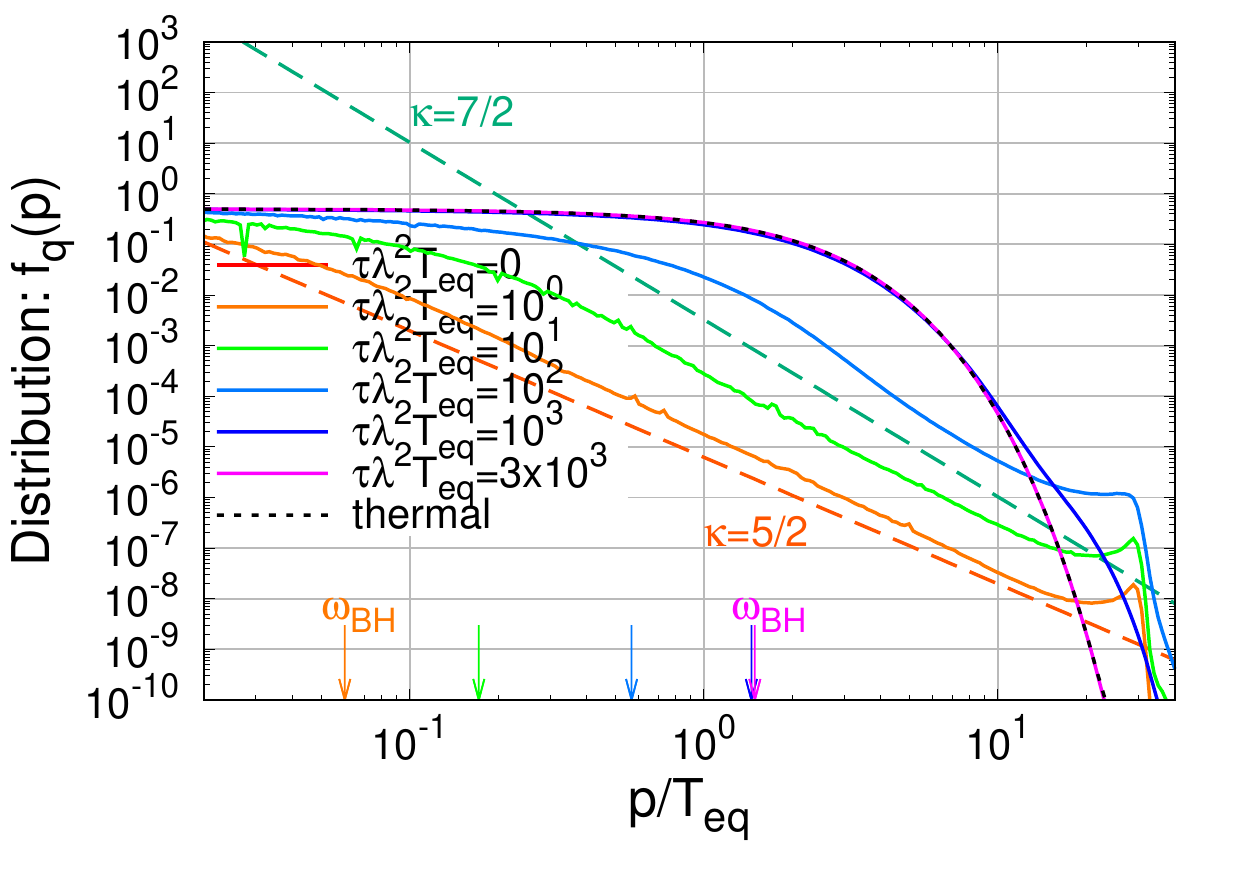}
	\caption{Evolution of the phase-space distributions of gluons (top) and quarks/anti-quarks (bottom) in an \emph{under-occupied gluon system} with $\left<p\right>_0/T_{\rm eq}$=30 at coupling $\lambda$=1. Dashed lines show the characteristic power law dependence of the single emission LPM spectra (orange) and the Kolmogorov Zakharov spectra (green). Dotted lines show the thermal equilibrium distributions. Vertical arrows mark the Bethe-Heitler frequencies $\omega_{BH}$.}
	\label{fig-UG-F30}
\end{figure}

\begin{figure}
		\centering
		\includegraphics[width=0.48\textwidth]{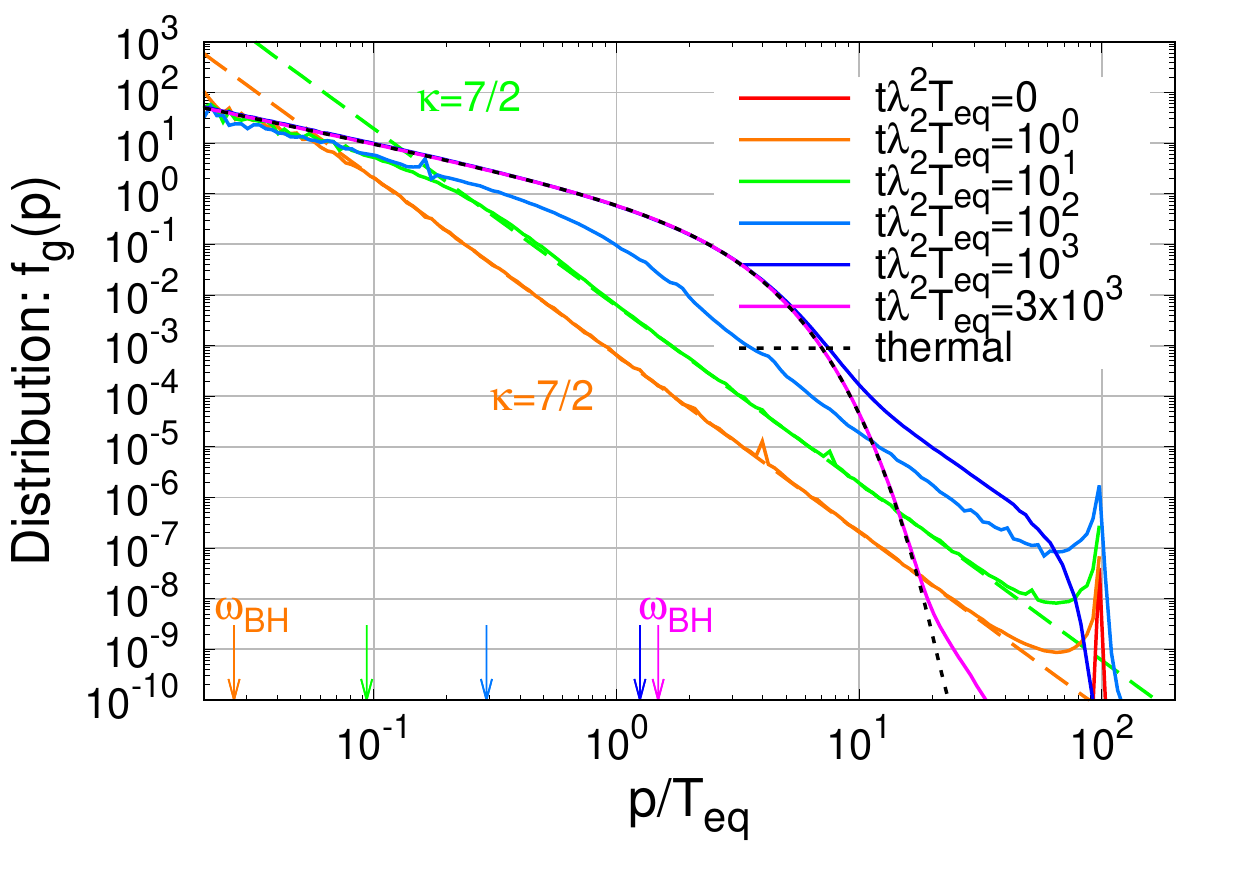}
	\hspace{\fill}
		\centering
		\includegraphics[width=0.48\textwidth]{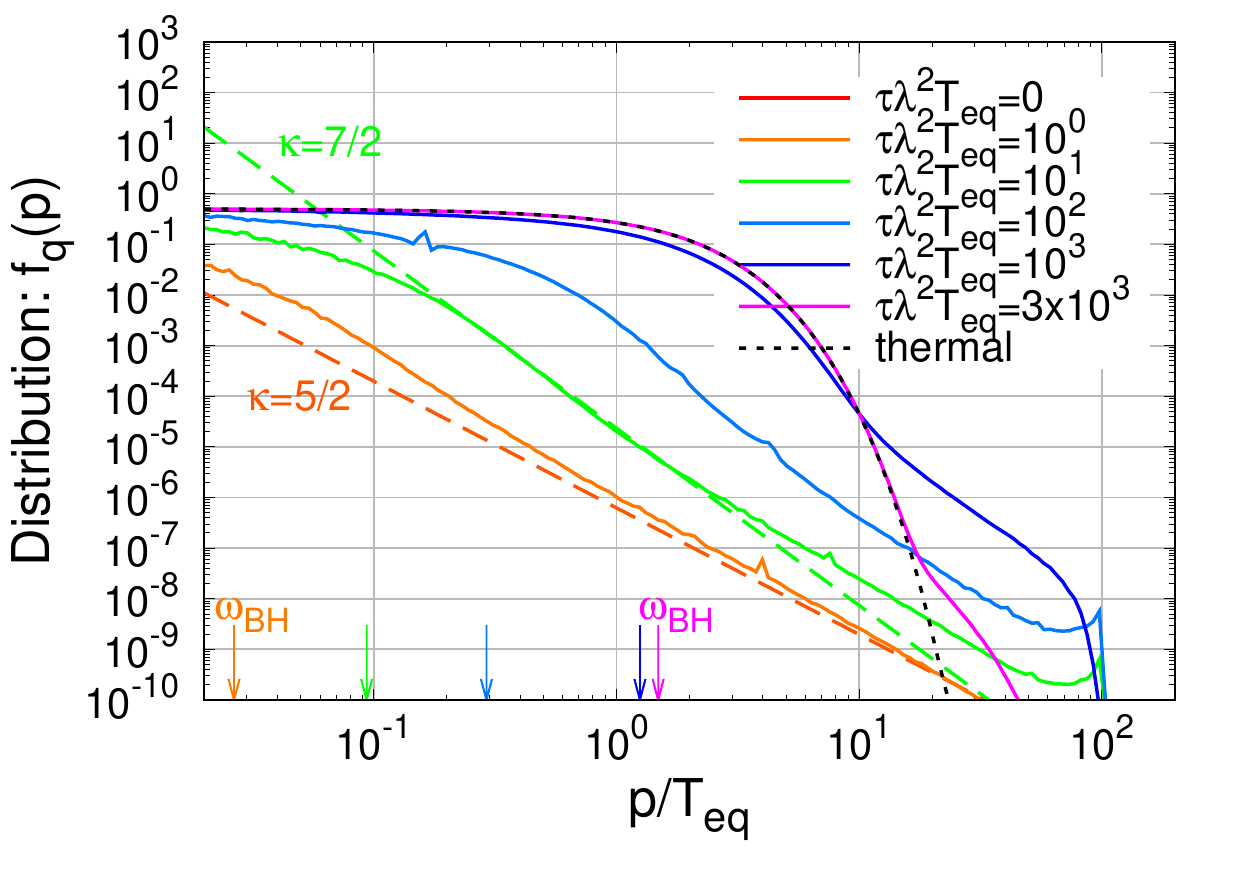}
	\caption{Evolution of the phase-space distributions of gluons (top) and quarks/anti-quarks (bottom) in an \emph{under-occupied gluon system} with $\left<p\right>_0/T_{\rm eq}$=100 at coupling $\lambda$=1. Dashed lines show the characteristic power law dependence of the single emission LPM spectra (orange) and the Kolmogorov Zakharov spectra (green). Dotted lines show the thermal equilibrium distributions. Vertical arrows mark the Bethe-Heitler frequencies $\omega_{BH}$.}
	\label{fig-UG-F100}
\end{figure}

\subsubsection{Theoretical Aspects}

Before we turn to our numerical results, we briefly recall the basic features of the bottom up thermalization in QCD plasmas following the discussion in~\cite{Schlichting:2019abc}. Starting from a dilute population of $f_0 \ll 1$ highly-energetic particles with $\langle p \rangle_0 \gg T_{\rm eq}$, elastic interactions between primary hard particles induce the emission of soft gluon radiation, which accumulates at low momenta. Due to the fact that elastic and inelastic interactions are more efficient at low momentum, the initially over-populated soft sector eventually thermalizes on a time scale $t \sim g^{-4} f_{0}^{-1/3} \langle p \rangle_{0}^{-1}$, before the highly-energetic primary particles have had sufficient time to decay. Even though at this time most of the energy is still carried by the hard primaries, the soft thermal bath begins to dominate screening and scattering, such that in the final stages of bottom-up equilibration, the few remaining hard particles loose their energy to the soft thermal bath, much like a jet loosing energy to a thermal medium~\cite{Baier:2000sb,Kurkela:2014tea,Schlichting:2019abc,Schlichting:2020lef}. 

Based on recent studies~\cite{Blaizot:2013hx,Mehtar-Tani:2018zba,Schlichting:2020lef}, the energy loss of hard primaries is accomplished by a turbulent inverse energy cascade, where the hard primary quarks/gluons, undergo successive splittings until the momenta of the radiated quanta becomes on the order of the temperature $T_{\rm soft}(t)$ of the soft thermal bath.
Specifically, at intermediate scales $T_{\rm soft}(t) \ll p \ll \langle p \rangle_{0}$, the distributions of quarks/antiquarks and gluons can be expected to feature the Kolmogorov-Zakharov spectra of weak-wave turbulence~\cite {Mehtar-Tani:2018zba,Schlichting:2020lef}
\begin{eqnarray}
f_{\rm KZ}(\vec{p},t) \propto \left(\frac{\left<p\right>_0}{p}\right)^{\frac{7}{2}}
\end{eqnarray}
which describe a scale-invariant energy flux from the ultra-violet $\sim \langle p \rangle_{0}$ to the infrared $\sim T_{\rm soft}(t)$, ensuring that the energy of the hard particles is deposited in the thermal medium without an accumulation of energy at intermediate scales.

Due to the energy loss of the hard primary particles, the temperature of the soft thermal bath increases until eventually the hard primaries have lost most of their energy to the thermal bath and the system approaches equilibrium. We note that due to the parametric suppression of inelastic rates for high-energy particles
\footnote{Since quasi-democratic $z\sim 1/2$ splittings dominate the turbulent energy transfer~\cite{Mehtar-Tani:2018zba,Schlichting:2019abc}, this can be seen by evaluating Eq.~(\ref{eq-HORate}) for $z\sim 1/2$}
$\Gamma_{\rm inel}^{\rm eq}(\langle p \rangle_0) \sim g^4 T_{\rm eq} \sqrt{\frac{T_{\rm eq}}{\langle p \rangle_0}}$, the energy loss of the hard primaries is slow compared to the equilibration of the soft sector, such that for sufficiently large scale separations  $\frac{\langle p \rangle_0}{T_{\rm eq}} \gg 1$ the thermalization of the system occurs on time scales $t\sim  g^{-4} T^{-1}_{\rm eq} 
\sqrt{\frac{\langle p \rangle_0}{T_{\rm eq}}} $, which can be significantly larger than the kinetic relaxation time $\tau_{R}\sim g^{-4} T^{-1}_{\rm eq}$.

\subsubsection{Bottom Up Thermalization of Quark-Gluon Plasma}
When considering the dynamics of under-occupied QCD plasmas, we need to specify the initial conditions for the momentum distribution and we can further distinguish different chemical compositions of the plasma.  We will limit our investigation to the following three cases, corresponding to (1) an initially under-occupied plasma of gluons, (2) an initially under-occupied plasma of quarks/antiquarks, and (3) an initially under-occupied plasma of quarks.

\begin{figure}
	\centering
	\includegraphics[width=0.87\textwidth]{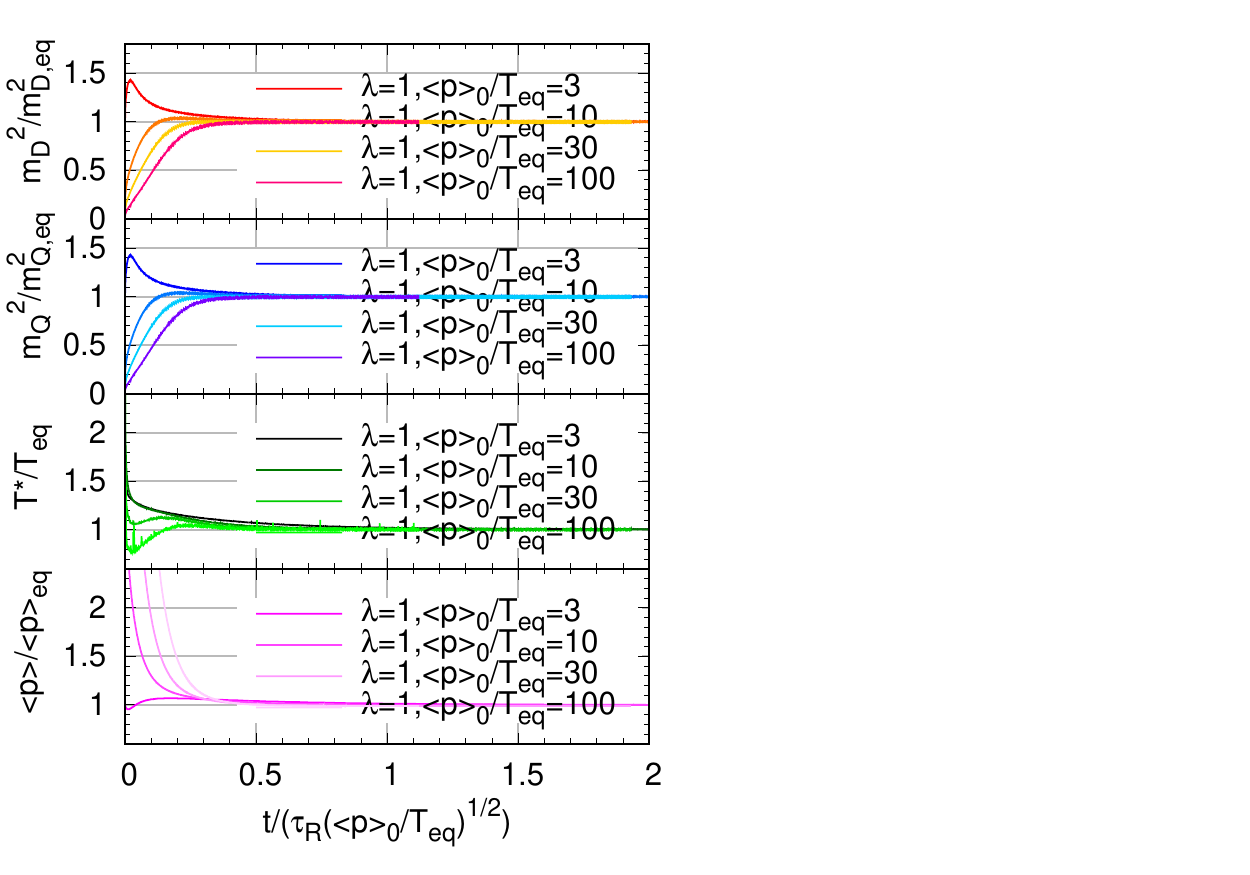}
	\caption{Evolution of the characteristic scales $m_D^2(t)$ (red), $m_Q^2(t)$ (blue), $T^{*}(t)$ (green) and $\left<p\right>$ (pink) in an \emph{under-occupied gluon system} at coupling strengths $\lambda=1$ for different scale separations $\langle p \rangle_{0}/T_{\rm eq}=3,10,30,100$ (darker to lighter colors). Scales are normalized to their respective equilibrium values, while the evolution time $t$ is normalized to $\tau_R\sqrt{\frac{p_0}{T_{\rm eq}}}$ in order to take into account the leading dependence on the initial energy $\langle p \rangle_{0}$.}
	\label{fig-UG-S}
\end{figure}

\begin{figure}
		\centering
		\includegraphics[width=0.48\textwidth]{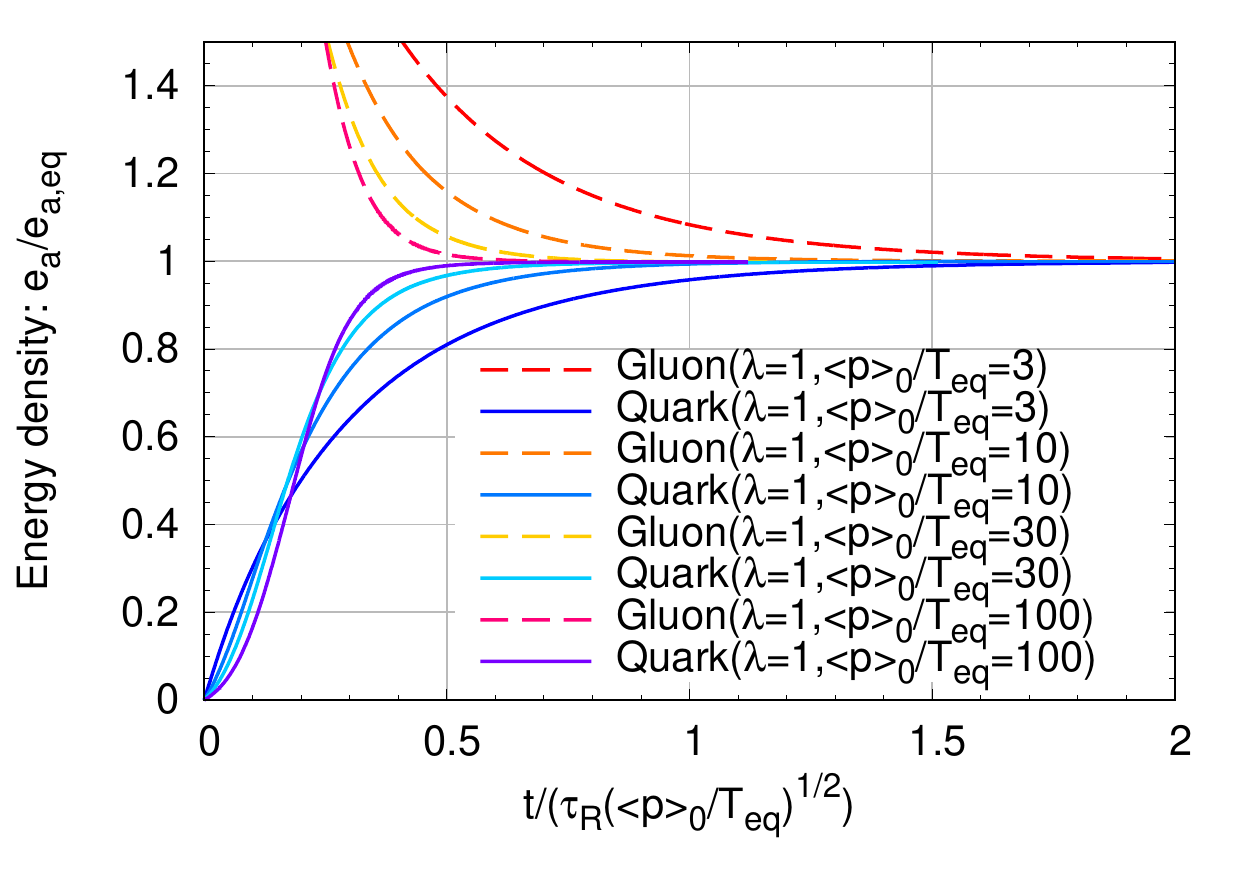}
		\centering
		\includegraphics[width=0.48\textwidth]{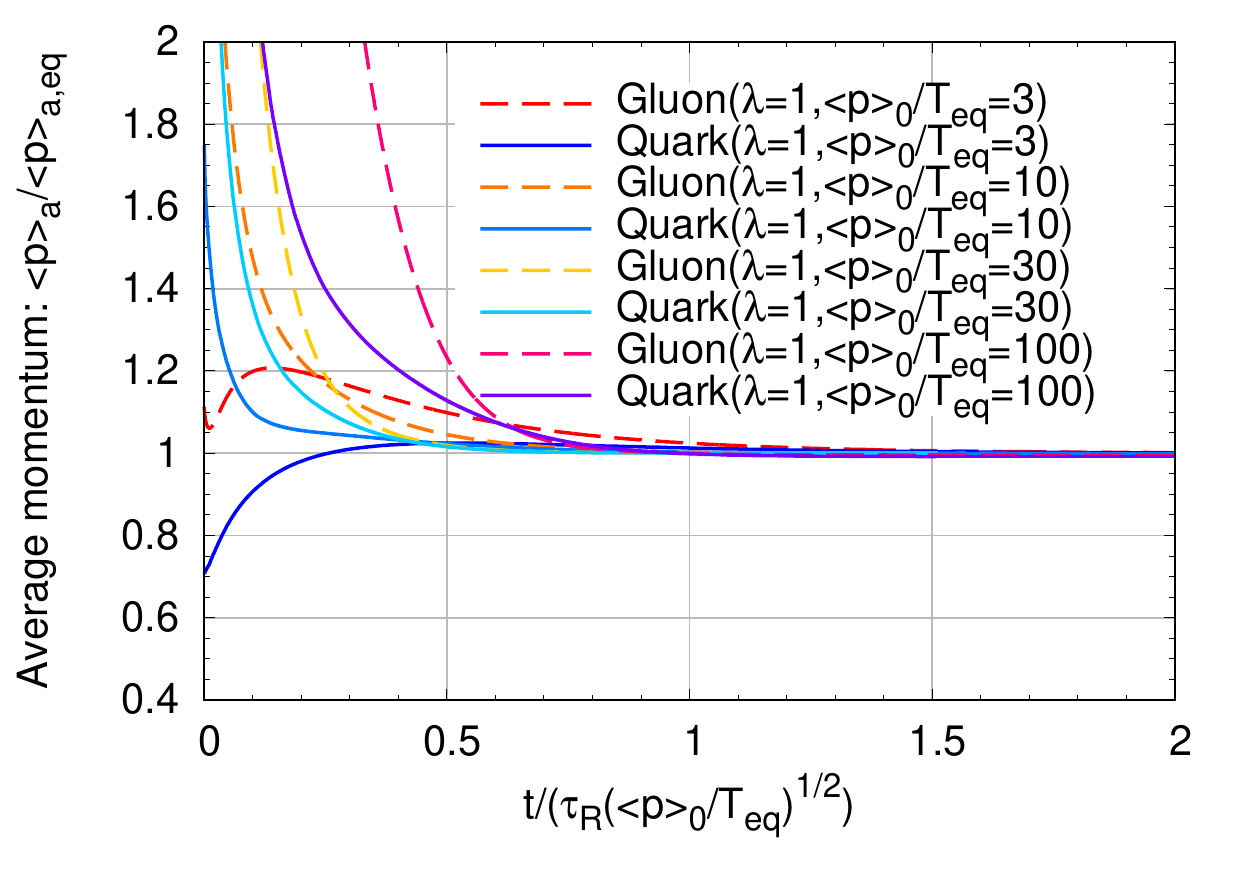}
	\caption{Evolution of the energy densities (top) and average momenta (bottom) of quarks (solid) and gluons (dashed) in an \emph{under-occupied gluon system} at coupling strengths $\lambda=1$ for different scale separation $\langle p \rangle_{0}/T_{\rm eq}=3,10,30,100$. Energy densities and average momenta are normalized to their respective equilibrium values, while the evolution time $t$ is normalized to $\tau_R\sqrt{\frac{p_0}{T_{\rm eq}}}$ in order to take into account the leading dependence on the initial energy $\langle p \rangle_{0}$.}	
	\label{fig-UG-E}
\end{figure}

We will employ the following initial conditions for an under-occupied plasma of gluons
\begin{eqnarray}
\label{eq-UO-INITI-G}
\nonumber
&&f_g(p,t=0)=\left(\frac{2\nu_q N_f}{\nu_g}\right) f_0 e^{-\frac{(p-p_0)^2}{Q^2}},\\
\nonumber
&&f_q(p,t=0)=0,\\
&&f_{\bar{q}}(p,t=0)=0\;,
\end{eqnarray}
while for an under-occupied plasma of quarks/antiquarks
\begin{eqnarray}
\label{eq-UO-INITI-QA}
\nonumber
&&f_g(p,t=0)=0,\\
\nonumber
&&f_q(p,t=0)=f_0e^{-\frac{(p-p_0)^2}{Q^2}},\\
&&f_{\bar{q}}(p,t=0)= f_0 e^{-\frac{(p-p_0)^2}{Q^2}},
\end{eqnarray}
and for an under-occupied plasma of quarks, the system is initialized as
\begin{eqnarray}
\label{eq-UO-INITI-Q}
\nonumber
&&f_g(p,t=0)=0,\\
\nonumber
&&f_q(p,t=0)=2 f_0 e^{-\frac{(p-p_0)^2}{Q^2}},\\
&&f_{\bar{q}}(p,t=0)=0,
\end{eqnarray}
where in all of the above relations the normalization 
\begin{eqnarray} 
f_0=\frac{2\pi^2e_{\rm eq}(T_{\rm eq},\mu_{\rm eq})}{9Q\left[\frac{2(p_0^2+Q^2)Q}{e^{p_0^2/Q^2}}+p_0\sqrt{\pi}(2p_0^2+3Q^2)(1+\mathrm{erf}(\frac{p_0}{Q}))\right]}
\end{eqnarray}
is chosen, such that all the systems have exactly the same energy density $e_{\rm eq}(T_{\rm eq},\mu_{\rm eq})$. By varying the parameter $p_0 \simeq \left<p\right>_0$, we can then adjust the separation of scales between the initial energy of the hard particles and the final equilibrium temperature. Since for $Q \ll p_0$, we do not expect a significant a sensitivity of our results to the parameter $Q$, that controls the initial width of the momentum distribution, we will employ $Q = T_{\rm eff}$ in the following.

We note that the under-occupied QCD plasmas of gluons in Eq.~(\ref{eq-UO-INITI-G}) and quarks/antiquarks in Eq.~(\ref{eq-UO-INITI-QA}), is charge neutral such that quarks and antiquarks of all light flavors ``$u,d,s$'' will be produced with equal abundancies. Notably this is not the case for the initial conditions in Eq.~(\ref{eq-UO-INITI-Q}), where the imbalance of quarks and antiquarks describes a system with a finite net-charge density, and we will for simplicity assume a degeneracy between ``$u,d,s$'' flavors.

\paragraph{Under-occupied gluons}

We start by analyzing the evolutions of under-occupied gluon systems in order to provide a direct and intuitive understanding of the bottom up thermalization scenario. The evolution of the momentum spectra of quarks and gluons during the thermalization process is presented in Figs.~\ref{fig-UG-F3}, \ref{fig-UG-F10}, \ref{fig-UG-F30} and \ref{fig-UG-F100} for weakly coupled plasmas $\lambda=1$ with different average initial momenta $\left<p\right>_0/T _{\rm eq}=3$ in Fig.~\ref{fig-UG-F3}, $\left<p\right>_0/T _{\rm eq}=10$ in Fig.~\ref{fig-UG-F10},  $\left<p\right>_0/T _{\rm eq}=30$ in Fig.~\ref{fig-UG-F30} and  $\left<p\right>_0/T _{\rm eq}=100$ in Fig.~\ref{fig-UG-F100}. Different panels show the evolutions of the gluon distributions $f_{g}(p)$ and quark/antiquark distributions $f_{q}(p)$, while different curves in each panel correspond to different evolution times $t\lambda T_{\rm eq}$ with vertical arrows marking the characteristic Bethe-Heitler frequency at each stage of the evolution.

\begin{figure}
		\centering
		\includegraphics[width=0.48\textwidth]{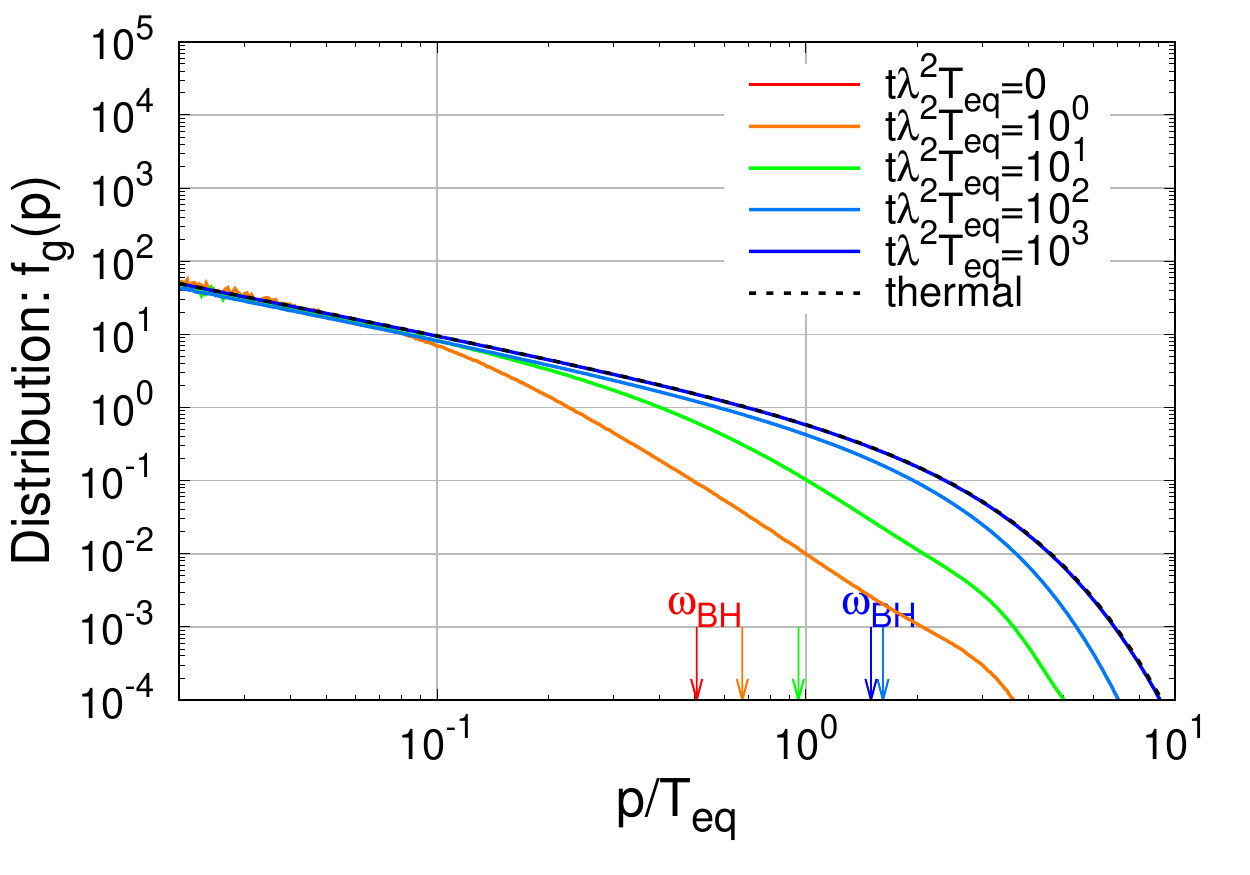}
		\centering
		\includegraphics[width=0.48\textwidth]{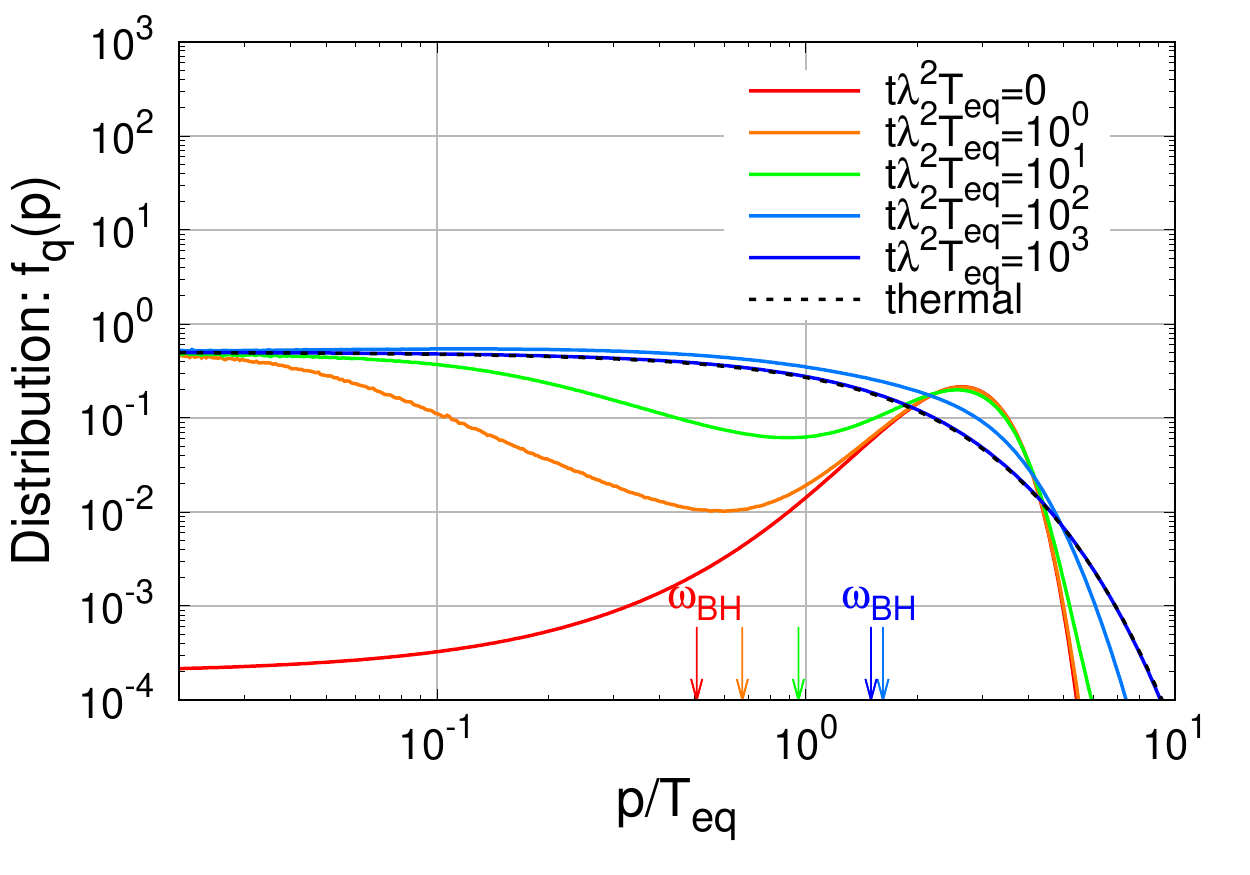}
	\caption{Evolution of the phase-space distributions of gluons (top) and quarks/anti-quarks (bottom) in an \emph{under-occupied quark/anti-quark system} with $\left<p\right>_0/T_{\rm eq}$=3 at coupling $\lambda$=1. Dotted lines show the thermal equilibrium distributions. Vertical arrows mark the Bethe-Heitler frequencies $\omega_{BH}$.}
	\label{fig-UQA-F3}
\end{figure}

By investigating the results for the larger scale separations $\left<p\right>_0/T _{\rm eq}=10$ in Fig.~\ref{fig-UG-F10},  $\left<p\right>_0/T _{\rm eq}=30$ in Fig.~\ref{fig-UG-F30} and  $\left<p\right>_0/T _{\rm eq}=100$ in Fig.~\ref{fig-UG-F100}, one clearly observes that soft radiation processes $g\rightarrow gg$ and $g\rightarrow q\bar{q}$ rapidly build up a large population of soft quarks and gluons with typical momenta $p \lesssim \omega_{BH}$. 
Even though at early times, such as e.g. $t\lambda^2T_{\rm eq}\ll 1$ in Fig.~\ref{fig-UG-F100}, the soft sector is over-occupied and thus highly gluon dominated, one finds that for sufficiently large scale separations, the over-occupation is depleted and the soft sector thermalizes before the hard primaries loose most of their energy to the soft thermal bath. 
Since at intermediate scales $\omega_{BH} \ll p \ll \left<p\right>_0$ the emission is in the LPM regime, the spectra of gluons and quarks initially feature a characteristic power law behavior $f_{g} \sim p^{-7/2}$, $f_{q} \sim p^{-5/2}$ for momenta $\omega_{BH} \ll p \ll \left<p\right>_0$, associated with the single emission spectra of the $g\rightarrow gg$ and $g\rightarrow q\bar{q}$ processes.
Subsequently, the energy of the hard primaries is transferred to the soft thermal bath, via an inverse turbulent cascade due to multiple successive $ g\to gg$, $g \to q\bar{q}$ and $q \to q g$ branchings, giving rise to the characteristic Kolmogorov-Zakharov spectrum $f_{g/q}\sim p^{-7/2}$ in both the gluon and quark sector. 
Since the energy injected into this cascade by the hard primaries at the scale $\sim \left<p\right>_0$, is transmitted all the way to the soft bath $\sim \omega_{BH}$ the temperature of the soft bath increases monotonically, as seen e.g. in Fig.~\ref{fig-UG-F100}, until eventually the hard primaries have lost nearly all of their energy and the system thermalizes. 
During the final stages of the approach towards equilibrium, a small number of hard primaries continues to loose energy giving rise to high momentum tails of the quark and gluon spectra seen for $t\lambda^2T_{\rm eq}=10^{3}$ in Figs.~\ref{fig-UG-F10}, \ref{fig-UG-F30}, \ref{fig-UG-F100}. Notably, the under-occupied system initially maintains a memory of the momentum distribution of hard primaries until the final stages of the thermalization process, which then closely resembles the mechanism of jet energy loss in a thermal medium~\cite{Schlichting:2020lef}.

Even for the smallest separation of scales $\left<p\right>_0/T_{\rm eq}$=3 shown in Fig.~\ref{fig-UG-F3}, some of the characteristic patterns of bottom up thermalization are still clearly visible, although in this case  radiative emissions occur in the Bethe-Heitler regime. 
Nevertheless, hard gluons with momenta $p \sim \left<p\right>_0$ still radiate soft gluons via $g\rightarrow gg$, leading to the formation of a soft thermal spectrum of gluons at low momenta. 
Even though quarks/antiquarks are also produced via $g\rightarrow q\bar{q}$ branching, one observes that the evolution in the quark sector is slightly slower than in the gluon sector, indicating once again that the energy transfer from gluons to quarks associated with the chemical equilibration of the system can cause a delay in the equilibration of the system.

\begin{figure}
	\centering
	\includegraphics[width=0.48\textwidth]{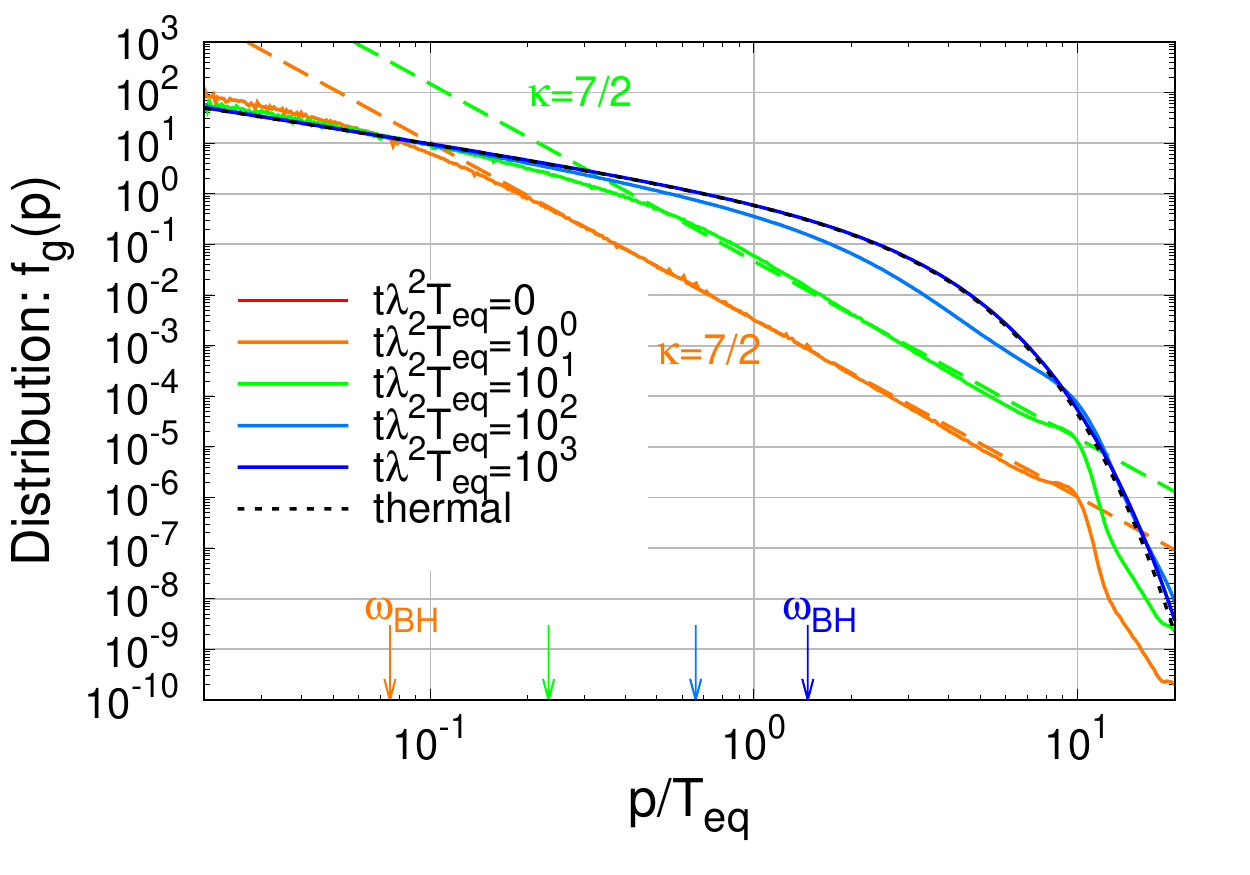}
	\centering
	\includegraphics[width=0.48\textwidth]{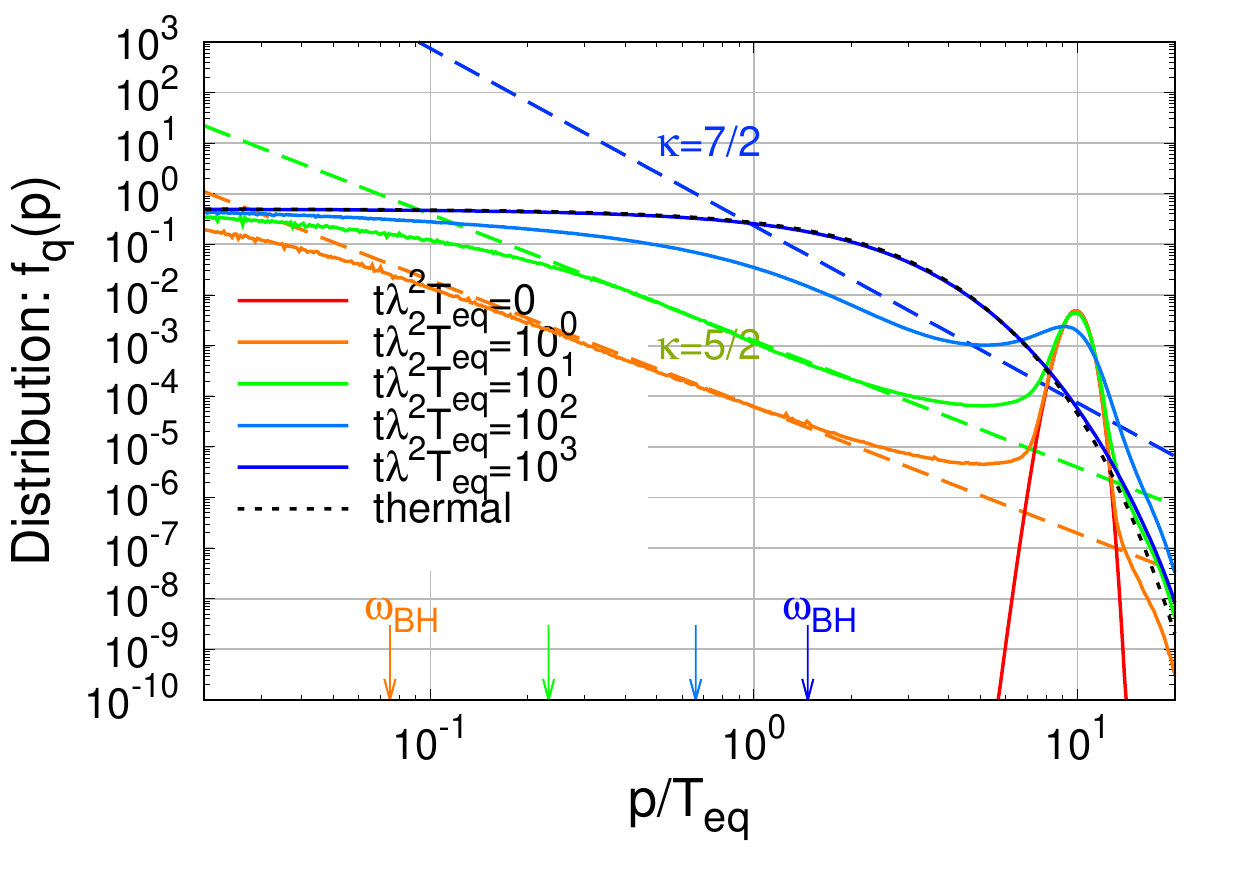}
    \caption{Evolution of the phase-space distributions of gluons (top) and quarks/anti-quarks (bottom) in an \emph{under-occupied quark/anti-quark system} with $\left<p\right>_0/T_{\rm eq}$=10 at coupling $\lambda$=1. Dashed lines show the characteristic power law dependence of the single emission LPM spectra ($\kappa=5/2,7/2$) and the Kolmogorov Zakharov spectra ($\kappa=7/2$). Dotted lines show the thermal equilibrium distributions. Vertical arrows mark the Bethe-Heitler frequencies $\omega_{BH}$.}
	\label{fig-UQA-F10}
\end{figure}

\begin{figure}
	\centering
	\includegraphics[width=0.48\textwidth]{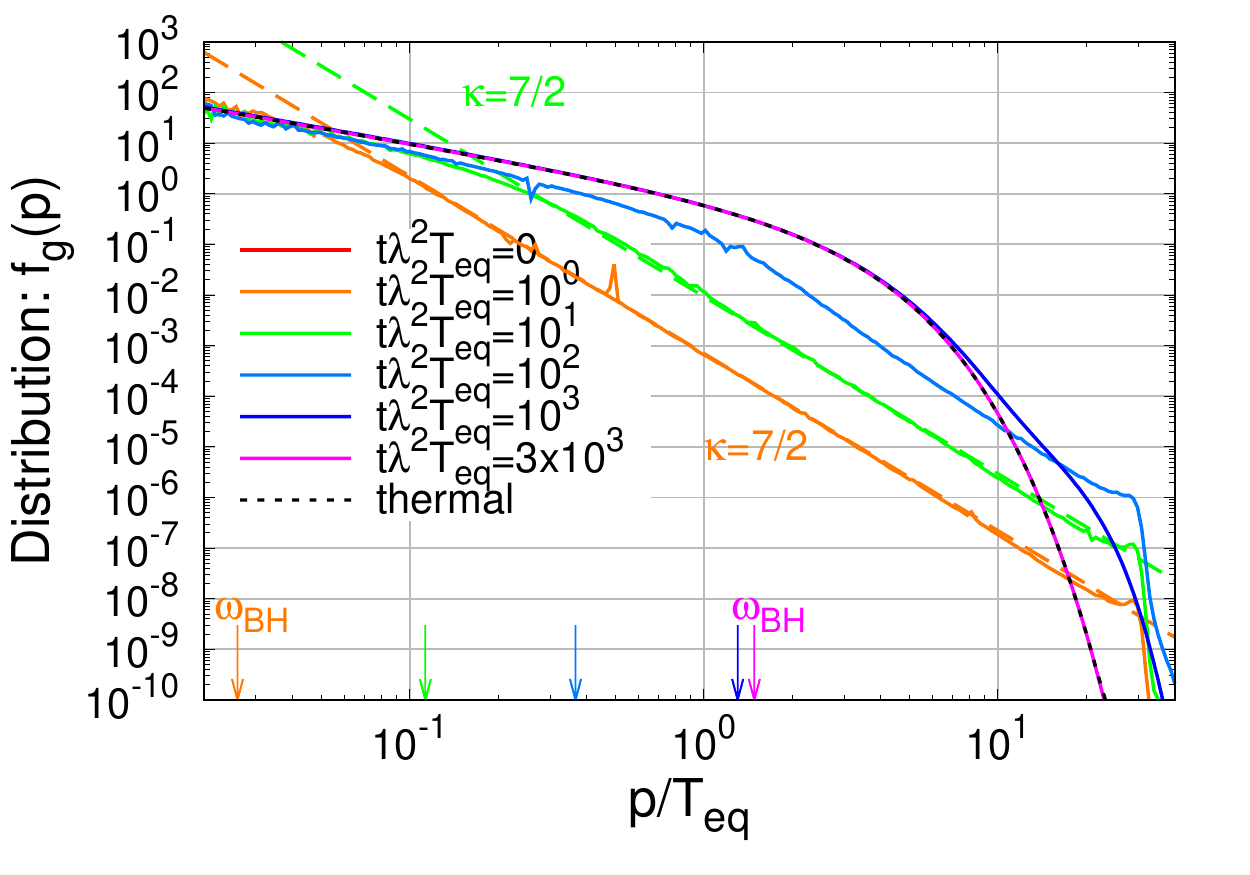}
	\centering
	\includegraphics[width=0.48\textwidth]{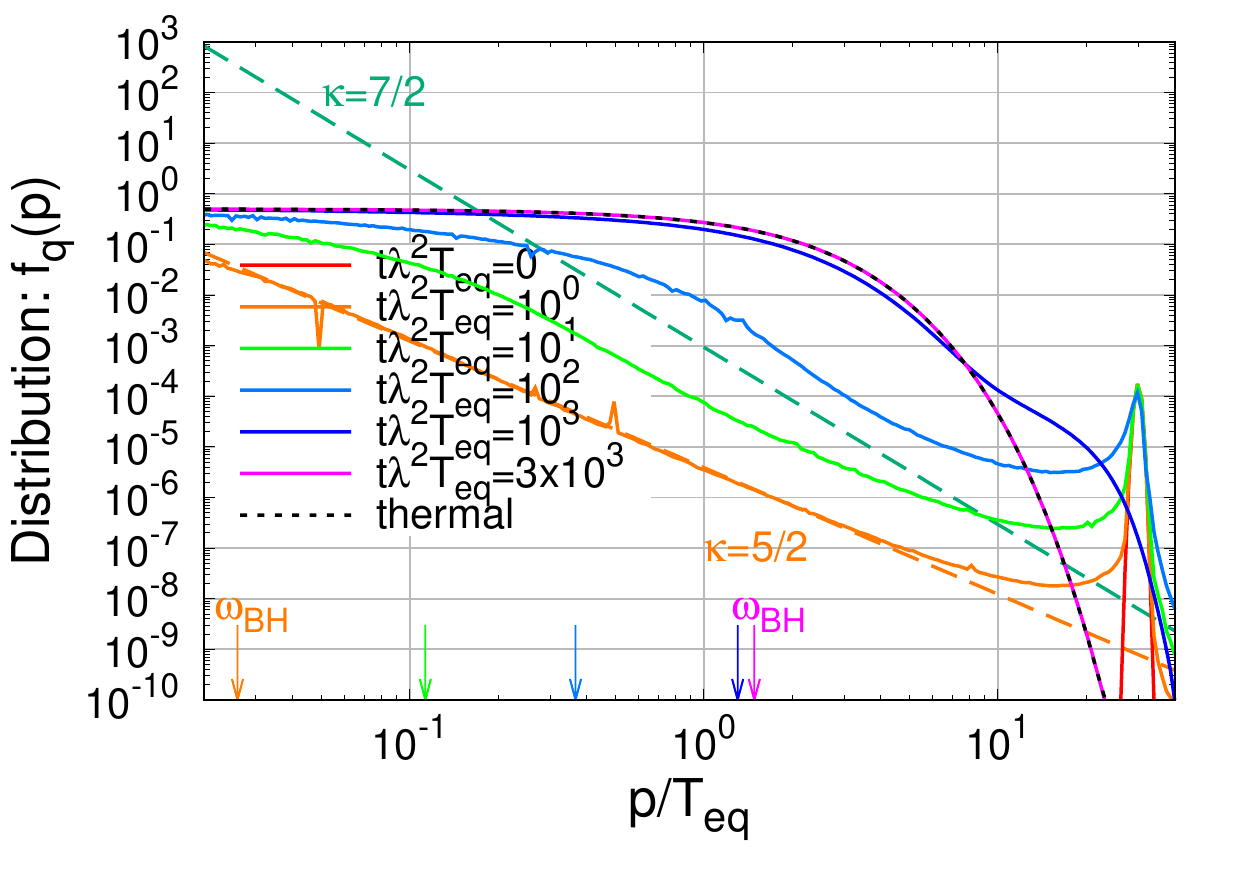}
    \caption{Evolution of the phase-space distributions of gluons (top) and quarks/anti-quarks (bottom) in an \emph{under-occupied quark/anti-quark system} with $\left<p\right>_0/T_{\rm eq}$=30 at coupling $\lambda$=1. Dashed lines show the characteristic power law dependence of the single emission LPM spectra ($\kappa=5/2,7/2$) and the Kolmogorov Zakharov spectra ($\kappa=7/2$). Dotted lines show the thermal equilibrium distributions. Vertical arrows mark the Bethe-Heitler frequencies $\omega_{BH}$.}
	\label{fig-UQA-F30}
\end{figure}

Now in order to compare the evolutions of the different systems, we again consider the evolutions of the characteristic dynamical scales $m_{D}^2, m_{Q}^2, T^{*}$ and $\left<p\right>$. 
Since in accordance with the discussion in Sec.~\ref{sec-evol-gluonquark} we anticipate that, for sufficiently large scale separations, the equilibration time of the system will be delayed by a factor  $\sqrt{\left<p\right>_0/T_{\rm eq}}$, relative to the equilibrium relaxation time $\tau_R$, we will consider normalizing the evolution time to $\tau_R\sqrt{\left<p\right>_0/T_{\rm eq}}$ when comparing the results for different average initial momenta $\left<p\right>_0/T_{\rm eq}$ in Figs.~\ref{fig-UG-S} and \ref{fig-UG-E}. Since the different scales $m_{D}^2,m_{Q}^2,T^{*}$ and $\left<p\right>$, exhibit different sensitivities to the hard and soft components of the plasma, their time evolutions are actually quite different. While for scale separations $\left<p\right>_0\gtrsim 10 T_{\rm eq}$, screening masses $m_{D}^2, m_{Q}^2$ are very quickly dominated by the soft thermal bath, and subsequently experience a strong rise as the soft bath heats up, the scale $T^{*}$ characterizing the strength of elastic interactions, receives significant contributions from the hard primaries at early times, before it is eventually dominated by the soft bath. Since the hard primaries carry most of the energy of the system until they eventually equilibrate, the average energy per particle $\left<p\right>$ is always dominated by the hard sector, and decreases monotonically over the course of the evolution. Besides the equilibration of the various scales, it is also interesting to consider the chemical equilibration of the system in Fig.~\ref{fig-UG-E}, where we present the energy fractions and average momenta separately for quarks and gluons. While for large scale separations, chemical equilibration in Fig.~\ref{fig-UG-E}  occurs on the same time scales as kinetic equilibration in Fig.~\ref{fig-UG-S}, one finds that for smaller scale separations the energy transfer from gluons to quarks requires additional time, delaying the equilibration of the system.

Generally, for scale separations $\left<p\right>_0/T _{\rm eq} \gtrsim 10 $, one finds that the scaling of the evolution time with $\sqrt{\left<p\right>_0/T_{\rm eq}}$, leads to comparable results for the equilibration time
\begin{eqnarray}
t_{\rm eq} \simeq 0.5-1.0 \tau_R\sqrt{\frac{\left<p\right>_0}{T_{\rm eq}}}\;,
\end{eqnarray}
albeit the curves for different $\left<p\right>_0/T_{\rm eq}$ in Figs.~\ref{fig-UG-S} and \ref{fig-UG-E} do not overlap completely, indicating that sub-leading corrections to this estimate still seem to be important for the scale separations considered in our study.

\paragraph{Under-occupied quarks and antiquarks}

\begin{figure}
	\centering
	\includegraphics[width=0.87\textwidth]{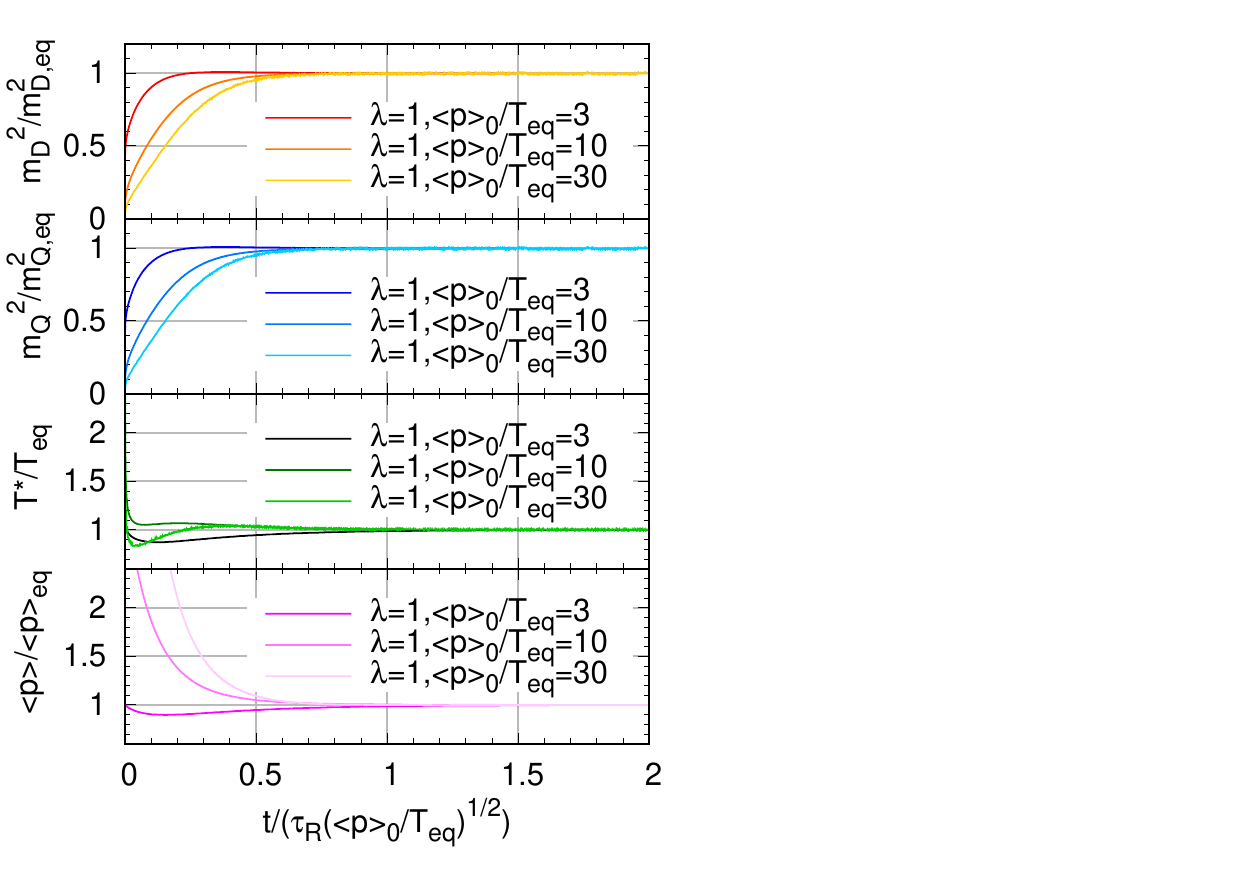}
	\caption{Evolution of the characteristic scales $m_D^2(t)$ (red), $m_Q^2(t)$ (blue), $T^{*}(t)$ (green) and $\left<p\right>$ (pink) in an \emph{under-occupied quark/anti-quark system} at coupling strengths $\lambda=1$ for different scale separations $\langle p \rangle_{0}/T_{\rm eq}=3,10,30,100$ (darker to lighter colors). Scales are normalized to their respective equilibrium values, while the evolution time $t$ is normalized to $\tau_R\sqrt{\frac{p_0}{T_{\rm eq}}}$ in order to take into account the leading dependence on the initial energy $\langle p \rangle_{0}$.}
	\label{fig-UQA-S}
\end{figure}

Similar to the under-occupied gluon systems, we will now consider charge neutral systems of under-occupied quarks/antiquarks. We proceed along the same lines and first investigate the evolution of the spectra for $\left<p\right>_0/T_{\rm eq}$=3, 10, 30 which are depicted in Figs.~\ref{fig-UQA-F3}, \ref{fig-UQA-F10} and \ref{fig-UQA-F30}. 
Generally, one finds that the thermalization processes follow essentially the same patterns as for the under-occupied gluon systems, with the inelastic production of soft quarks and gluons $q\rightarrow gq$ leading to the rapid build-up of the soft sector, before the hard primary quarks and antiquarks loose their energy via multiple successive branchings giving rise to the familiar
Kolmogorov-Zakharov spectra $f_{g/q}\sim p^{-7/2}$ at intermediate momentum scales $\omega_{\rm BH} \ll p \ll \left<p\right>_0$. Due to the radiative break-up of the hard primaries, the soft sector heats up, until the system eventually equilibrates when all of the hard primaries have had sufficient time to decay. While at early times, the hard components of the spectra ($p\sim \left<p\right>_0$) closely reflect the initial conditions, it is interesting to observe, that during the final approach towards equilibrium,  e.g. for $t\lambda^2 T_{\rm eq}=10^{3}$ in Fig.~\ref{fig-UG-F30} and Fig.~\ref{fig-UQA-F30}, the momentum distribution and chemistry of the remaining hard particles are significantly modified, and there is no longer a significant difference between under-occupied gluon systems and under-occupied quark/antiquark systems.

\begin{figure}
		\centering
		\includegraphics[width=0.48\textwidth]{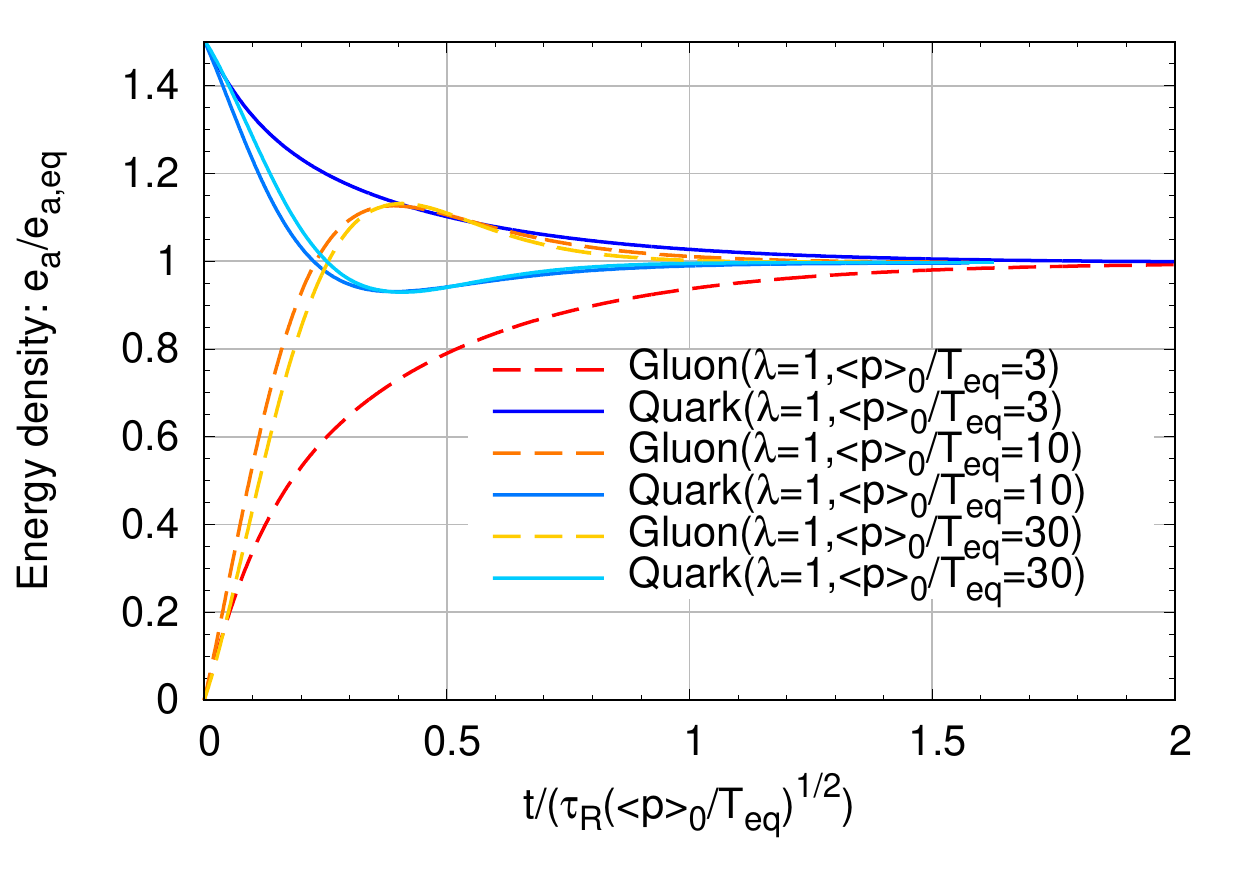}
		\centering
		\includegraphics[width=0.48\textwidth]{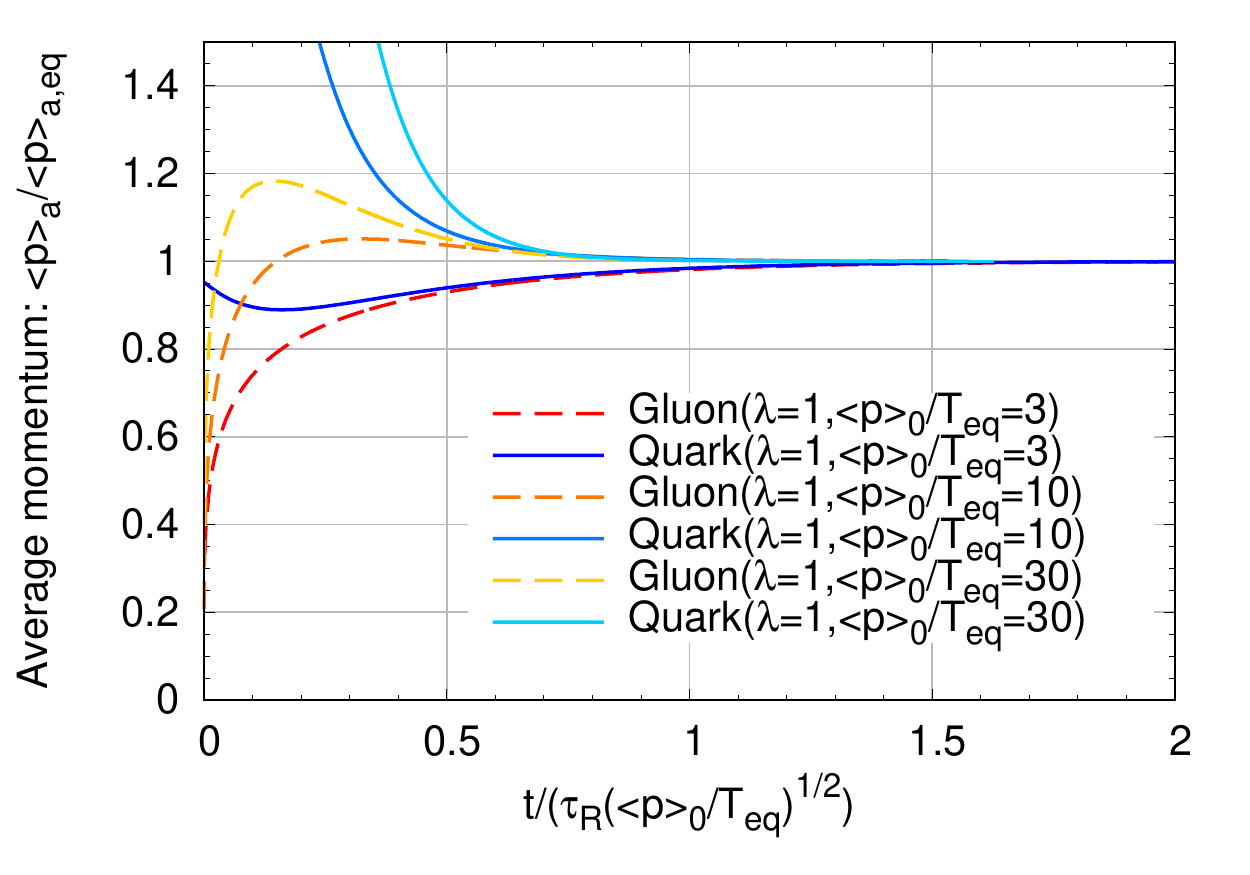}
	\caption{Evolution of the energy densities (top) and average momenta (bottom) of quarks (solid) and gluons (dashed) in an \emph{under-occupied quark/anti-quark system} at coupling strengths $\lambda=1$ for different scale separations $\langle p \rangle_{0}/T_{\rm eq}=3,10,30$. Energy densities and average momenta are normalized to their respective equilibrium values, while the evolution time $t$ is normalized to $\tau_R\sqrt{\frac{p_0}{T_{\rm eq}}}$ in order to take into account the leading dependence on the initial energy $\langle p \rangle_{0}$.}	
	\label{fig-UQA-E}
\end{figure}

By comparing the results for the evolutions of the dynamical scales $m_{D}^2,m_{Q}^2,T^{*}$ and $\left<p\right>$ in Fig.~\ref{fig-UQA-S} for the under-occupied quark/antiquark systems to the corresponding results for under-occupied gluons, one again observes essentially the same qualitative patterns. However, it is interesting to see, that for under-occupied systems of quarks and antiquarks, the approach towards equilibrium appears to occur on a somewhat larger time scale $\gtrsim 0.5 \tau_{R} \sqrt{\left<p\right>_0/T_{\rm eq}}$ as compared to under-occupied gluon systems, where by $0.5 \tau_{R} \sqrt{\left<p\right>_0/T_{\rm eq}}$ all the scales $m_{D}^2,m_{Q}^2,T^{*}$ and $\left<p\right>$ are already close to their respective equilibrium values. Based on our discussion in Sec.~\ref{subsec-inelastic-rate}, we believe that this discrepancy at intermediate times can be attributed to the different color factors in the inelastic interactions rates for the hard primary quarks/antiquarks and gluons, as discussed in detail in the context of jet quenching in~\cite{Schlichting:2020lef}. 
However, if one is concerned with the ultimate approach towards equilibrium, one should take into account the fact that at late times the quark/gluon composition is significantly modified, such that under-occupied systems of quarks and gluons can ultimately be expected to equilibrate at the same rate.

\begin{figure}[!t]
		\centering
		\includegraphics[width=0.48\textwidth]{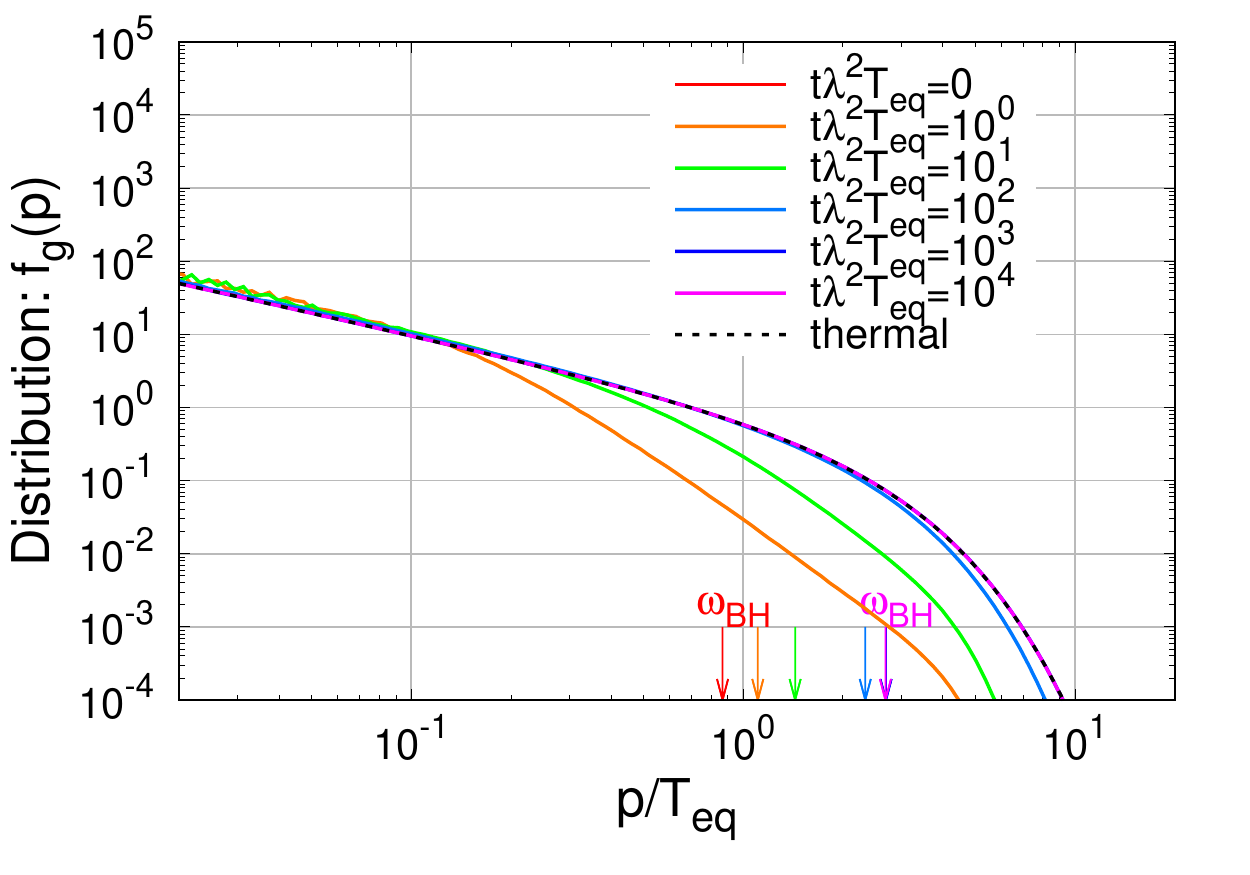}
		\centering
		\includegraphics[width=0.48\textwidth]{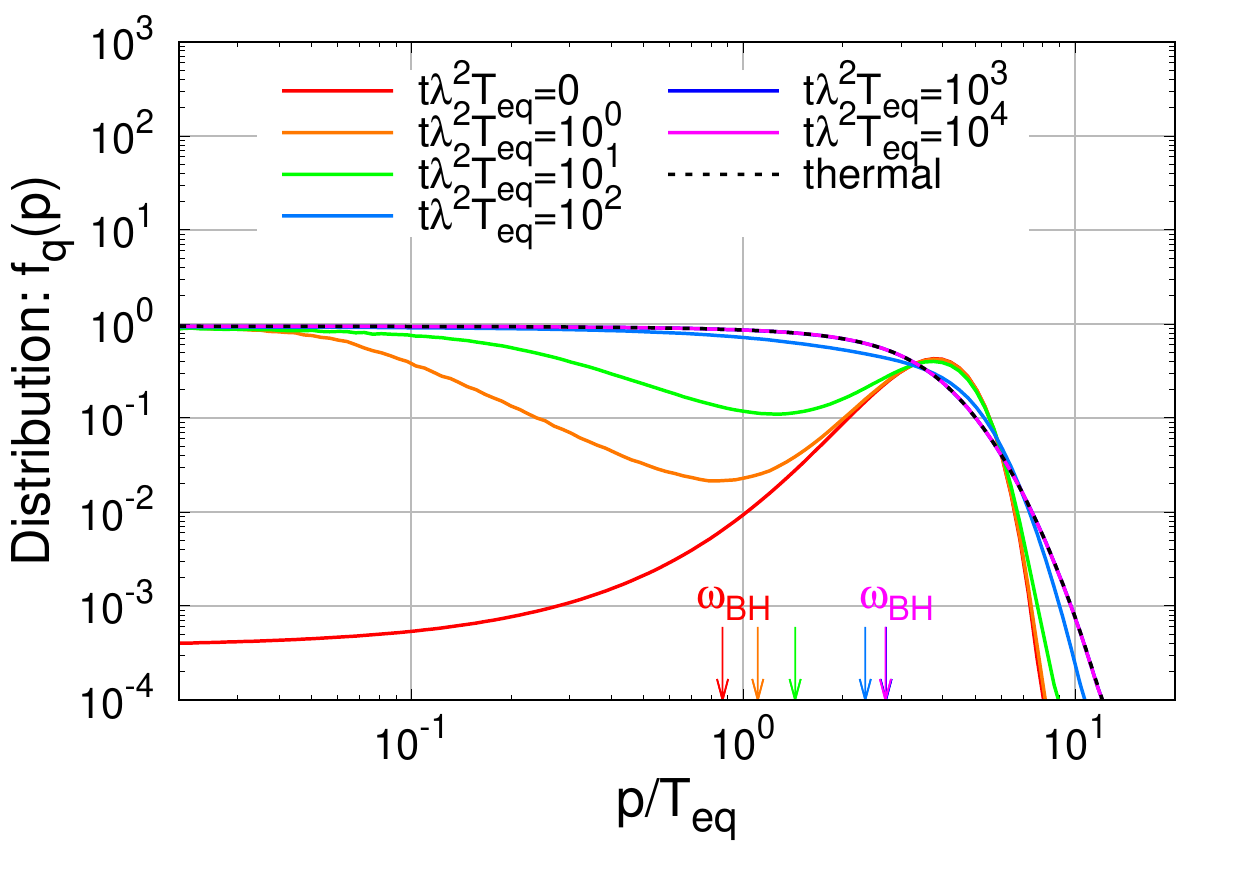}
		\centering
		\includegraphics[width=0.48\textwidth]{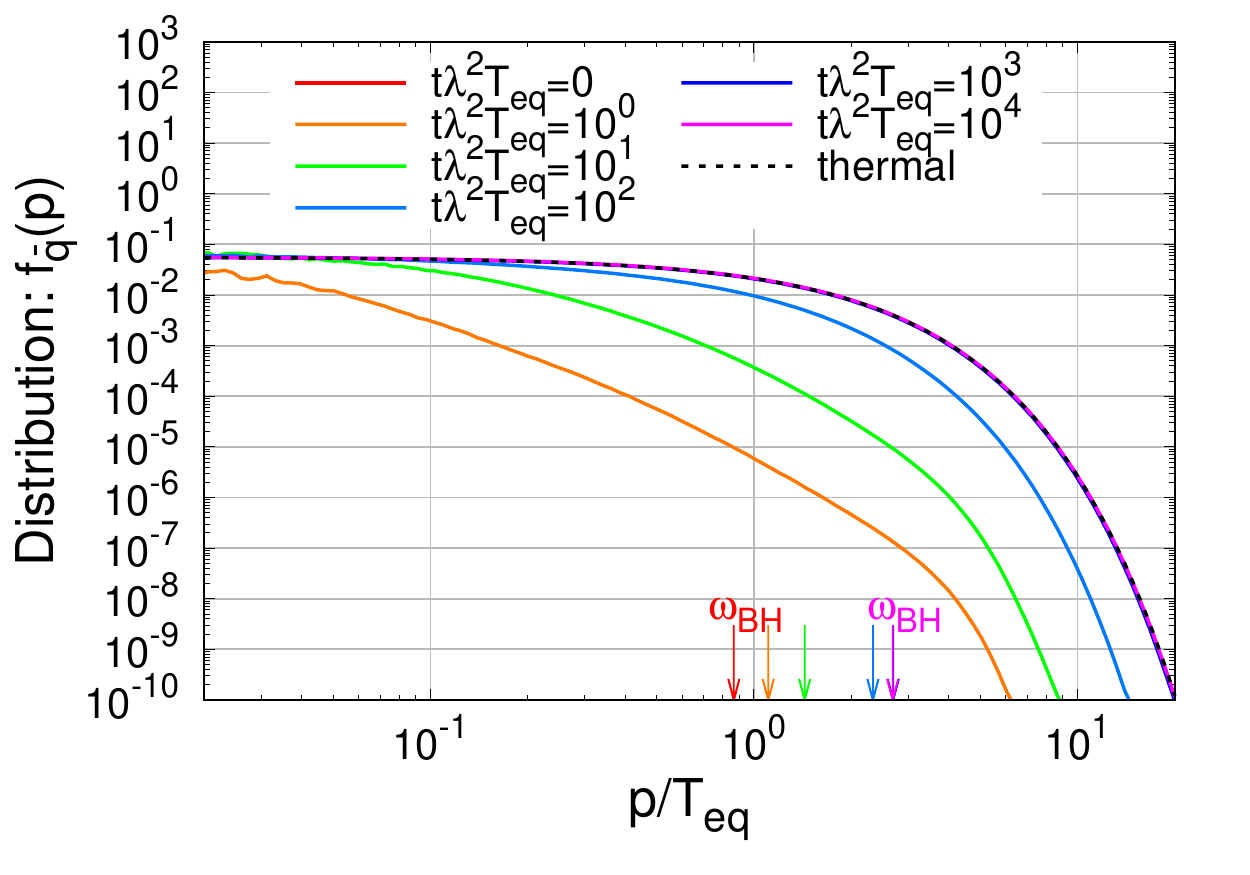}
	\caption{Evolution of the phase-space distributions of gluons (top) and quarks (middle) and anti-quarks (bottom) in an \emph{under-occupied quark system} with $\left<p\right>_0/T_{\rm eq}$=3 at coupling $\lambda$=1. Different flavors of quarks/anti-quarks are assumed to be identical and the equilibrium temperature is $T_{\rm eq}\simeq$0.70$T_{\rm eff}$. Dotted lines show the thermal equilibrium distributions. Vertical arrows mark the Bethe-Heitler frequencies $\omega_{BH}$.}
	\label{fig-UQ-F3}
\end{figure}

Next, in order to investigate the chemical equilibrations of the under-occupied quark/antiquarks systems, we present our simulation results for the evolutions of the energy fraction of quarks and gluons, and their average momenta in Fig.~\ref{fig-UQA-E}. Interestingly, one finds that in contrast to the behavior for under-occupied gluon systems in Fig.~\ref{fig-UG-E}, the energy fractions of quarks and gluons in the system exhibit a non-monotonic behavior. Even though initially all the energy is carried by the hard primary quarks and antiquarks, it turns out that for larger scale separations $\left<p\right>_0/T_{\rm eq}$=10, 30, gluons dominate the energy budget before the chemical equilibration of the system. 
By inspecting also the behaviors of the average momenta in the lower panel of Fig.~\ref{fig-UQA-E}, one finds that these gluons are typically soft, with the average momenta $\left<p\right>$ close to the equilibrium value. 
We believe that this behavior can be attributed to the fact that gluon radiation dominates the energy transfer from the hard to the soft sector, such that the soft thermal bath absorbs the energy pre-dominantly in form of gluons, before the energy is eventually re-distributed among quarks and gluons.

\begin{figure}[!t]
	\centering
	\includegraphics[width=0.48\textwidth]{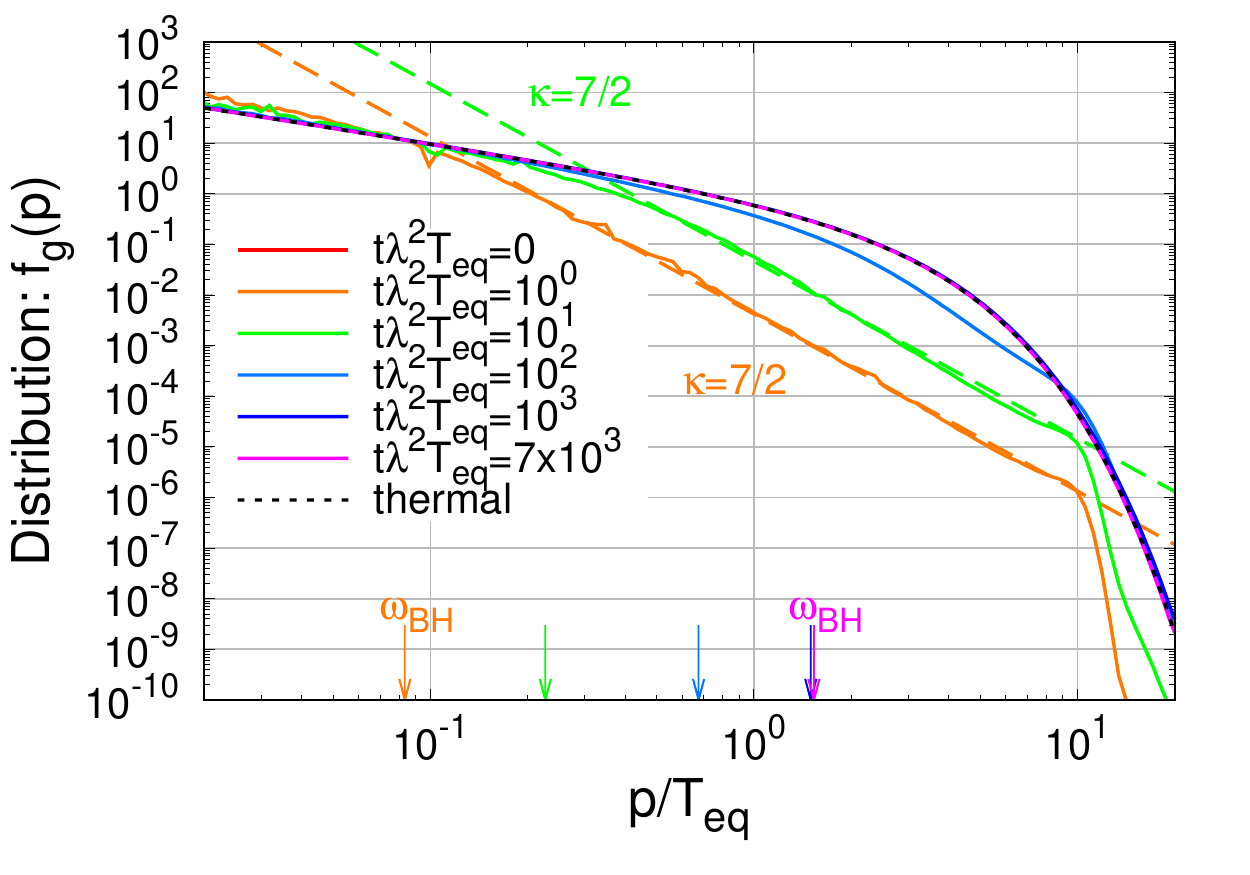}
	\centering
	\includegraphics[width=0.48\textwidth]{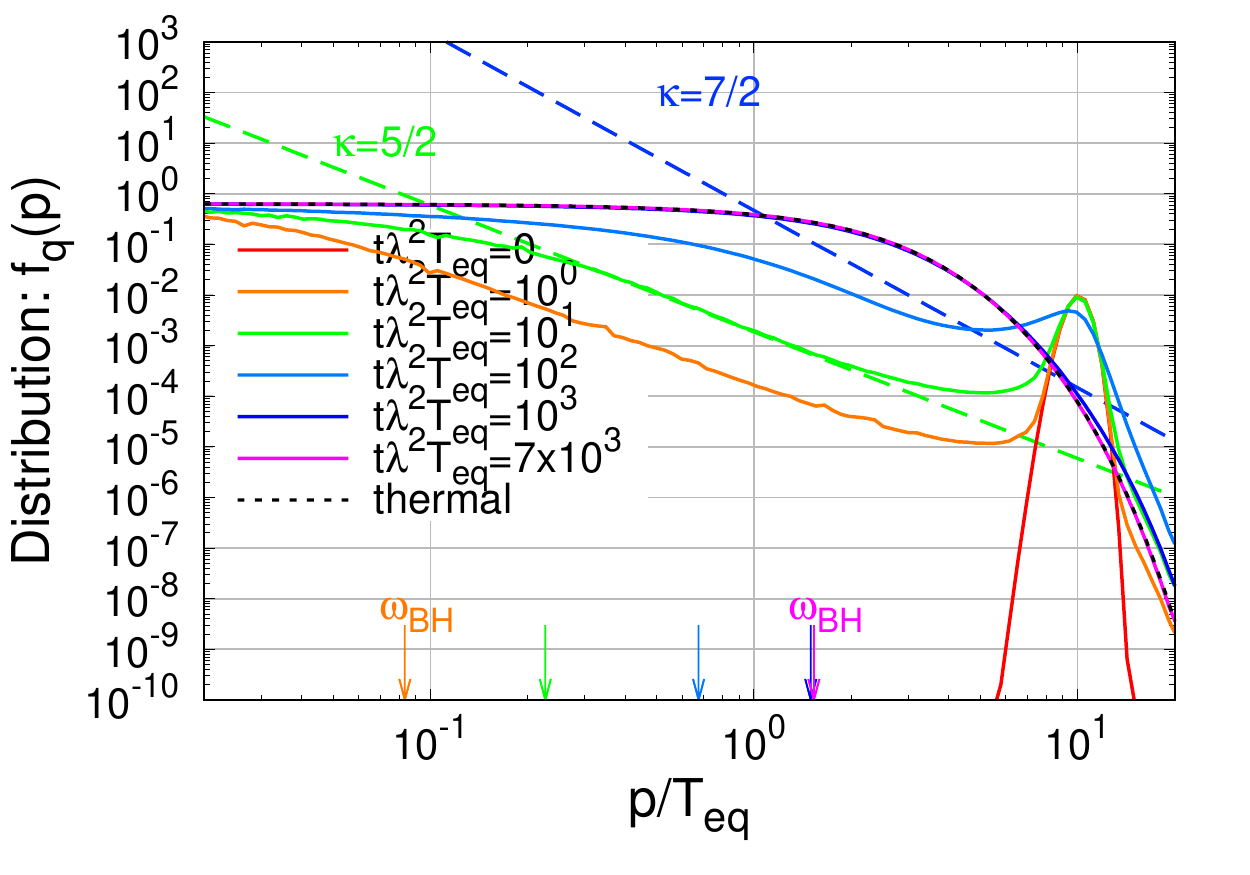}
	\centering
	\includegraphics[width=0.48\textwidth]{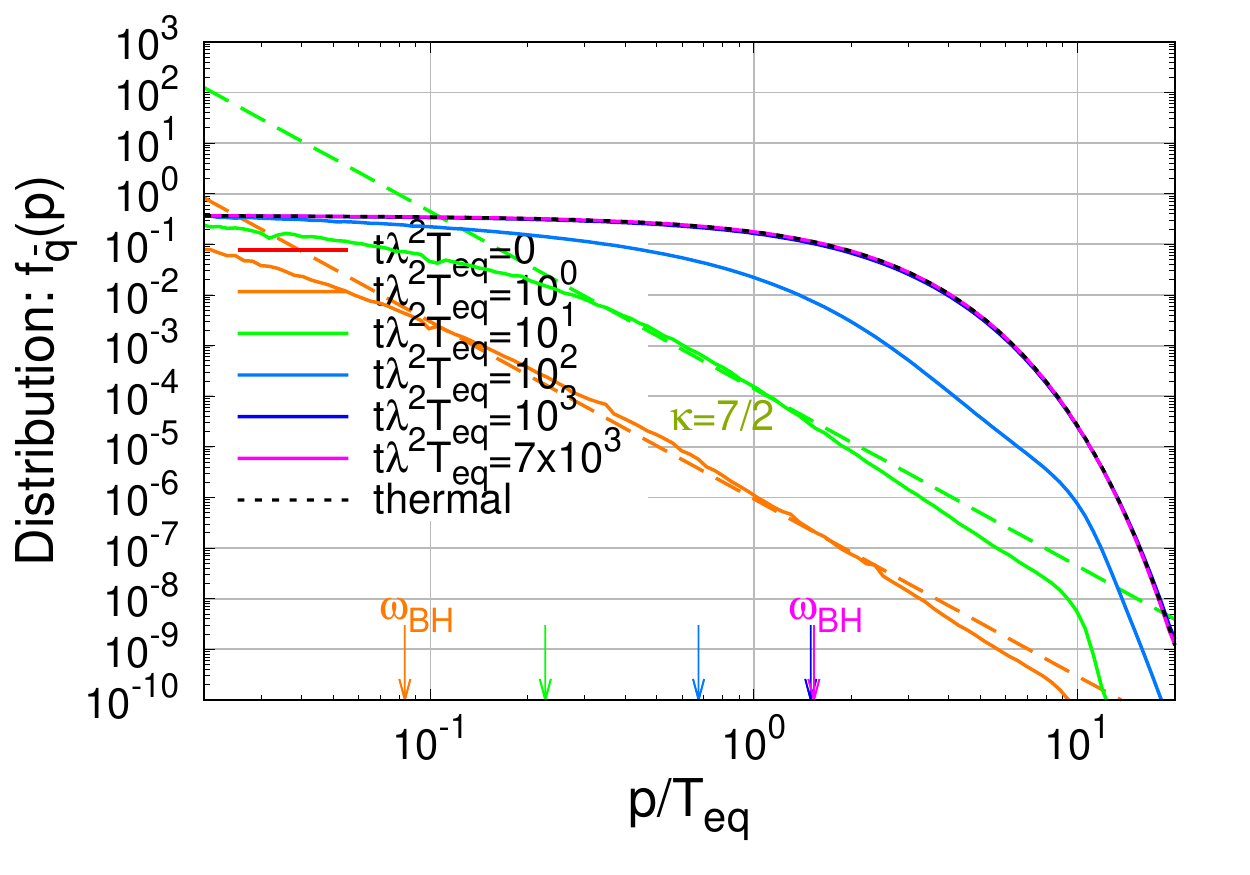}
	\caption{Evolution of the phase-space distributions of gluons (top) and quarks (middle) and anti-quarks (bottom) in an \emph{under-occupied quark system} with $\left<p\right>_0/T_{\rm eq}$=10 at coupling $\lambda$=1. Different flavors of quarks/anti-quarks are assumed to be identical and the equilibrium temperature is $T_{\rm eq}\simeq$0.98$T_{\rm eff}$. Dashed lines show the characteristic power law dependence of the single emission LPM spectra ($\kappa=5/2,7/2$) and the Kolmogorov Zakharov spectra ($\kappa=7/2$). Dotted lines show the thermal equilibrium distributions. Vertical arrows mark the Bethe-Heitler frequencies $\omega_{BH}$.}
	\label{fig-UQ-F10}
\end{figure}

\paragraph{Under-occupied quarks}

So far we have investigated the equilibrations of charge neutral systems of under-occupied gluons, quarks/antiquarks. 
Next we consider the equilibrations of under-occupied systems of quarks, which in accordance with Eq.~(\ref{eq-UO-INITI-Q}) carry non-zero densities of the conserved $u,d,s$ charges. 
Since in the presence of finite charge densities, the evolutions of quarks and antiquarks will be different, we first study the evolutions of spectra of gluons, quarks and antiquarks, which are depicted in Fig.~\ref{fig-UQ-F3} for $\left<p\right>_0/T_{\rm eff}=3$ and in Fig~\ref{fig-UQ-F10} for $\left<p\right>_0/T_{\rm eff}=10$. 
Evidently, the evolutions of the quark and gluon spectra in Figs.~\ref{fig-UQ-F3} and \ref{fig-UQ-F10} are very similar to the quark/antiquark spectra in Figs~\ref{fig-UQA-F3} and \ref{fig-UQA-F10} obtained in the zero density cases.
However, significant differences can be observed for the evolutions of the antiquarks, as for the under-occupied systems of quarks there are no antiquarks present in the initial conditions. 
Instead, the population of antiquarks observed at later times is produced via gluon splittings $g\rightarrow q\bar{q}$ and elastic $gg\rightarrow q\bar{q}$ conversions. 
Hence, the evolutions of the antiquark spectra closely follow the gluon spectra, as can be seen by comparing the upper and lower panels in Figs.~\ref{fig-UQ-F3} and \ref{fig-UQ-F10}.

\begin{figure}
		\centering
		\includegraphics[width=0.87\textwidth]{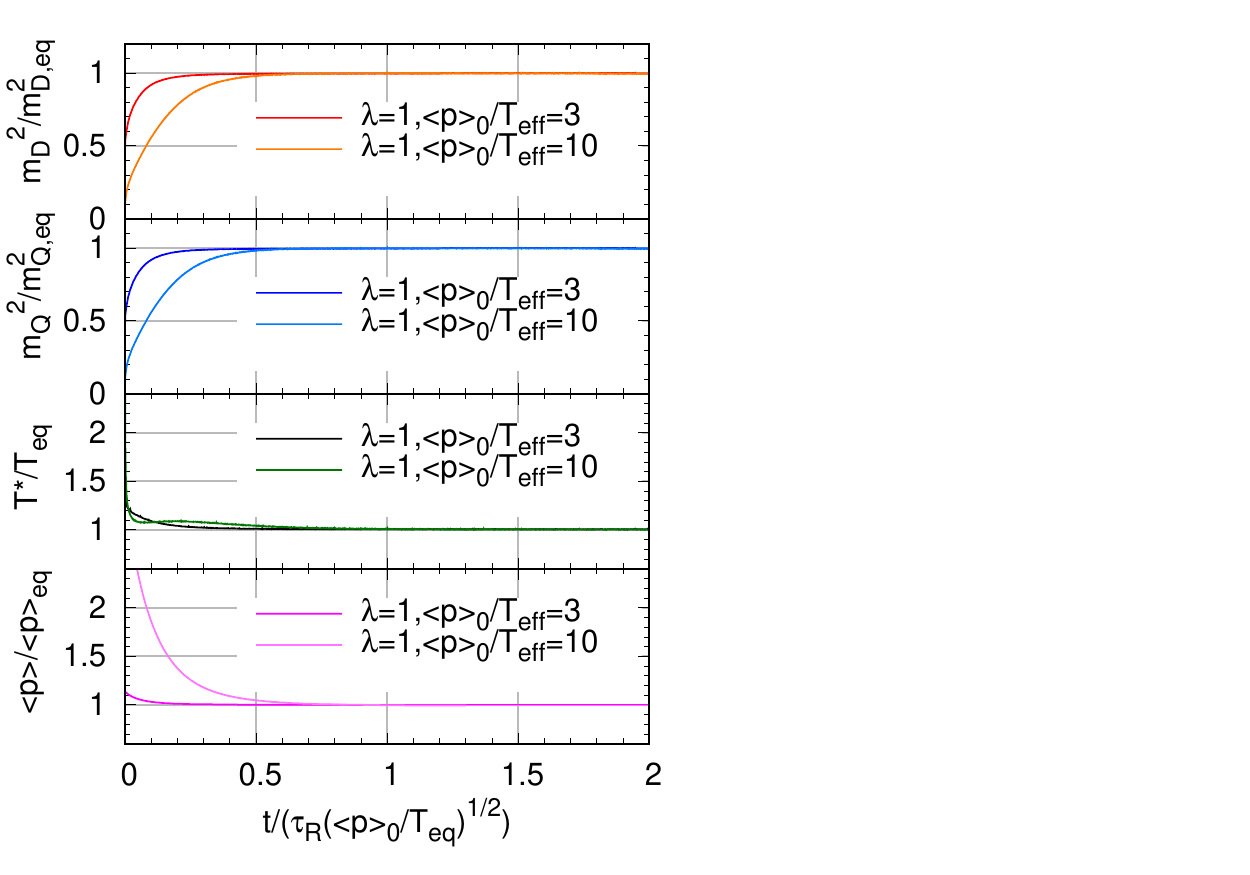}
	\caption{Evolution of the characteristic scales $m_D^2(t)$ (red), $m_Q^2(t)$ (blue), $T^{*}(t)$ (green) and $\left<p\right>$ (pink) in an \emph{under-occupied quark system} at coupling strengths $\lambda=1$ for different scale separation $\langle p \rangle_{0}/T_{\rm eq}=3,10$ (darker to lighter colors). Scales are normalized to their respective equilibrium values, while the evolution time $t$ is normalized to $\tau_R\sqrt{\frac{p_0}{T_{\rm eq}}}$ in order to take into account the leading dependence on the initial energy $\langle p \rangle_{0}$.}
	\label{fig-UQ-S}
\end{figure}

\begin{figure}
	\centering
	\includegraphics[width=0.48\textwidth]{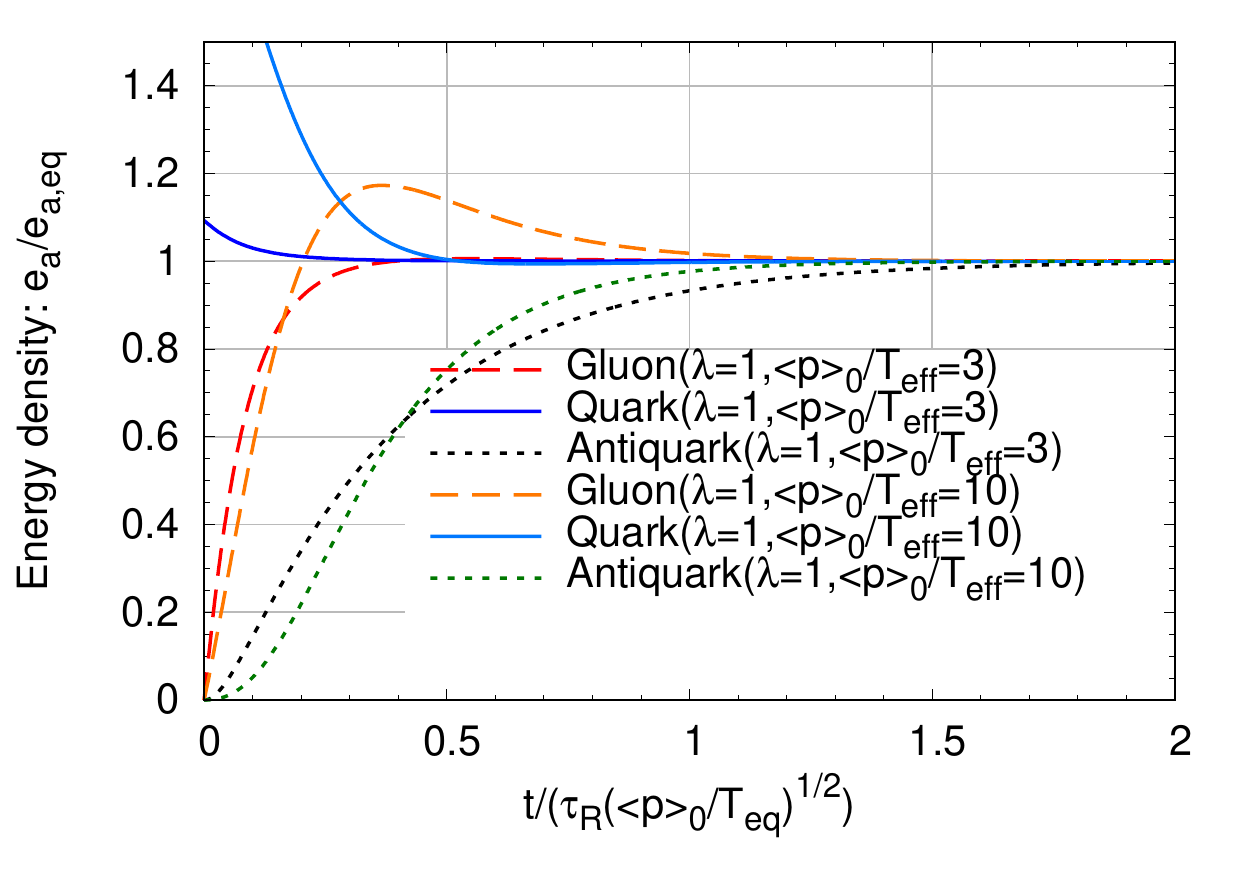}
	\centering
	\includegraphics[width=0.48\textwidth]{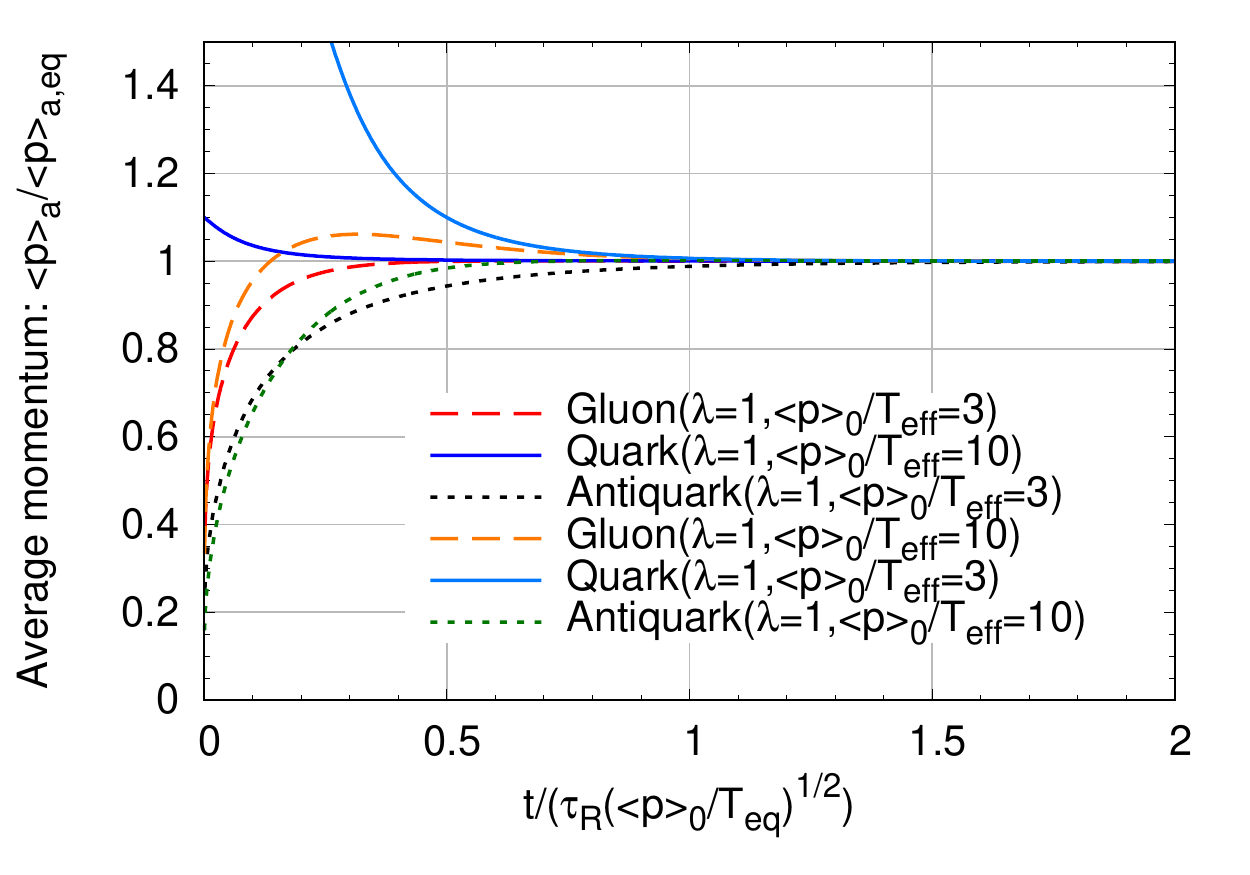}
		\caption{Evolution of the energy densities (top) and average momenta (bottom) of quarks (solid), anti-quarks (dotted) and gluons (dashed) in an \emph{under-occupied quark system} at coupling strengths $\lambda=1$ for different scale separation $\langle p \rangle_{0}/T_{\rm eq}=3,10,30$. Energy densities and average momenta are normalized to their respective equilibrium values, while the evolution time $t$ is normalized to $\tau_R\sqrt{\frac{p_0}{T_{\rm eq}}}$ in order to take into account the leading dependence on the initial energy $\langle p \rangle_{0}$.}	
	\label{fig-UQ-E}
\end{figure}

By comparing the evolutions of the characteristic scales for the zero and finite density systems in Figs.~\ref{fig-UQA-S} and \ref{fig-UQ-S}, one finds that the presence of the additional conserved charges does not significantly affect the kinetic equilibration of the system, in accordance with the finding that the evolutions of quark and gluon spectra are essentially unchanged. 
However, when considering the evolutions of the individual contributions of gluons, quarks and antiquarks to the energy densities in Figs.~\ref{fig-UQA-E} and \ref{fig-UQ-E}, one clearly observes that the chemical equilibration associated with the production of antiquarks requires a significant amount of time, with energy densities of gluons and antiquarks only approaching their equilibrium ratios for times $\gtrsim \tau_{R} \sqrt{\left<p\right>_0/T_{\rm eq}}$.
\section{Equilibration of longitudinally expanding plasmas}
\label{sec-evol-expansion}
So far we have discussed kinetic and chemical equilibrations for homogeneous and isotropic systems. Now in order to address the early time dynamics of high-energy heavy-ion collisions, we will focus on systems which are transversely homogeneous and longitudinally invariant under Lorentz boost, but can feature an anisotropy between longitudinal and transverse momenta. Denoting\begin{eqnarray}
\nonumber
&&x^{\mu} =\tau(\mathrm{cosh}(\eta),u_T^1,u_T^2,\mathrm{sinh}(\eta))
=(t,x_T^1,x_T^2,z)\\
\nonumber
&&p^{\mu} =p_T(\mathrm{cosh}(y),v_T^1,v_T^2,\mathrm{sinh}(y))
=(E,p_T^1,p_T^2,p_z)
\end{eqnarray}
where
\begin{eqnarray}
\nonumber
&&~~\tau=\sqrt{t^2-z^2},~~~~~~~\eta=\frac{1}{2}\mathrm{ln}\left(\frac{t+z}{t-z}\right)\\
&&p_T=\sqrt{E^2-p_z^2},~~~~~y=\frac{1}{2}\mathrm{ln}\left(\frac{E+p_z}{E-p_z}\right)
\end{eqnarray}
the phase-space distribution $f(x,p)$ for a longitudinally boost invariant and transversely homogeneous system can be conveniently expressed in the form $f(x,p)= f(\tau, p_T,p_{\|})$, where the variable $p_{\|}$ denotes the longitudinal momentum $p_{\|} = p_T \mathrm{sinh}(y-\eta)$ in the local rest-frame $u^{\mu}=(\cosh(\eta),0,0,\sinh(\eta))$ of the non-equilibrium plasma.

Since the system is homogeneous in transverse coordinates $\bf{x}_T$ and the longitudinal rapidity $\eta$, the resulting Boltzmann equation~\cite{Mueller:1999pi} 
\begin{eqnarray}
\label{eq-bolzmannExp}
&&\left[\frac{\partial}{\partial \tau} - \frac{p_{\|}}{\tau}\frac{\partial}{\partial p_{\|}} \right] f_a(\tau, p_T,p_{\|})=\\
&& \qquad -C^{{2\leftrightarrow2}}_a[f](\tau, p_T,p_{\|})-C^{{1\leftrightarrow2}}_a[f](\tau, p_T,p_{\|})\;, \nonumber 
\end{eqnarray}
takes essentially the same form as in Eq.~(\ref{eq-bolzmann}) with one additional term, that can be re-expressed in the form of an additional collision integral 
$-C^{{\rm exp}}_a[f]=\frac{p_{\|}}{\tau} \frac{\partial f_a(\tau, p_T,p_{\|})}{\partial p_{\|}}$ which characterizes the red-shift of the longitudinal momentum $p_{\|}$ due to the longitudinal expansion, and the discretized form of expansion term can be found in Sec.~\ref{chap-expansion}.

We note that in comparison to the previous discussion of homogenous and isotropic systems, there are two important physical differences when considering plasmas which are subject to a rapid longitudinal expansion. Due to the expansion, the system will on the one hand become more and more dilute over the course of the evolution, on the other hand the longitudinal expansion tends to reduce the longitudinal momenta in the local rest frame, thereby introducing an anisotropy which can persists on large time scales. We further note that momentum space anisotropic QCD plasmas are generally expected to be unstable~\cite{Mrowczynski:1993qm,Arnold:2003rq,Romatschke:2003ms,Romatschke:2004jh} due to the non-abelian analogue of the Weibel instability in electrodynamics~\cite{Weibel:1959zz}. While perturbative calculations of the one-loop self-energies in momentum space anisotropic plasmas suggest the presence of instabilities, and modifications to the thermalization scenario have been worked out~\cite{Bodeker:2005nv,Kurkela:2011ti}, sophisticated classical field simulations~\cite{Berges:2013eia,Berges:2013fga} point to the fact that plasma instabilities do not appear to play a dominant role in the non-equilibrium evolution of the system beyond very early times. Despite some recent progress~\cite{Hauksson:2020wsm}, it is currently not established how to properly include these effects into an effective kinetic description, and we will therefore follow previous works~\cite{Kurkela:2018wud,Kurkela:2018oqw,Mazeliauskas:2018yef} and neglect the effects of plasma instabilities, by resorting to the isotropic description of screening discussed in Sec.~\ref{subsec-elastic}.

While the kinetic evolution of homogeneous and isotropic systems is governed by the collision rates $\sim g^{4} T^{*}$ of elastic and inelastic collisions, the longitudinal expansion rate $\sim 1/{\tau}$ provides an additional time scale which dominates the kinetic evolution at early times. Even though the expansion rate is naively divergent in the limit $\tau \to 0$, it is important to point out that the effective kinetic description only becomes applicable on a time scale $\tau_0 \sim 1/Q_s$ corresponding to the formation time of hard particles in the system, where $Q_s$ is the typical momentum of quarks and gluons in the initial state. When describing the non-equilibrium evolution of long. expanding systems, we will therefore initialize the QCD kinetic simulations at a finite proper time $\tau_0 \sim 1/Q_s$ and consider the subsequent non-equilibrium evolution towards equilibrium.

With regards to the initial conditions for the phase-space distributions of quarks and gluons at initial time $\tau_0$, we follow previous works~\cite{Kurkela:2018oqw} and consider the following initial conditions, 
\begin{align}
&&f_{g}(\tau_0,p_{T},p_{\|})&=f^{0}_{g}
\frac{Q_0}{\sqrt{p_T^2+\xi_0^2p_{\|}^2}}e^{-\frac{2(p_T^2+\xi_0^2p_{\|}^2)}{3Q_0^2}}\;, \\
&&f_{q_{f}/\bar{q}_{f}}(\tau_0,p_{T},p_{\|})&=f^{0}_{q_{f}/\bar{q}_{f}} 
\frac{\sqrt{p_T^2+p_{\|}^2}}{\sqrt{p_T^2+\xi_0^2p_{\|}^2}}e^{-\frac{2(p_T^2+\xi_0^2p_{\|}^2)}{3Q_0^2}}\;,\nonumber
\end{align}
inspired by Color-Glass condensate calculations of the gluon spectra, where $Q_0=1.8 Q_s$ denotes the initial energy scale, $\xi_0=10$ characterizes the initial momentum space anisotropy and $f^{0}_{i}$ denotes the typical phase-space occupancy for the different particle species $i=g,u,\bar{u},d,\bar{d},s,\bar{s}$.
Since parton distributions at small $x$ are highly gluon dominated, gluons are copiously produced in high-energy collisions, and we will typically consider $f^{0}_{g} \sim 1/\alpha_s$. Since initial quarks can be produced either from the stopping of valence quarks, or via the production of quark anti-quark pairs, we consider
\begin{align}
\label{eq:QuarkRatios}
f^{0}_{u}&=\frac{7}{15} f_{val}^{0}+ \frac{1}{6} f_{split}^{0}\;, \qquad
f^{0}_{\bar{u}}=\frac{1}{6} f_{split}^{0}\;, \nonumber \\
f^{0}_{d}&=\frac{8}{15} f_{val}^{0}+ \frac{1}{6} f_{split}^{0}\;,  \qquad
f^{0}_{\bar{d}}=\frac{1}{6} f_{split}^{0}\;, \nonumber \\
f^{0}_{s}&=\frac{1}{6}f_{split}^{0}\;, \qquad \qquad \qquad 
f^{0}_{\bar{s}}=\frac{1}{6}f_{split}^{0}\;, 
\end{align}
where $f_{val}^{0}$ and $f_{split}^{0}$ represent the contributions from valence stopping and quark-anti-quark pair production, and we have implemented $(f_{u}-f_{\bar{u}})/(f_{d}-f_{\bar{d}})=7/8$ in Eq.~(\ref{eq:QuarkRatios}) to represent the different valence quark fractions taking into account a proton to neutron fraction $n_{p}/(n_{p}+n_{n})\approx 0.4$ in heavy nuclei. We present a compact summary of all the simulations performed in Tab.~\ref{tb-chemical}, where we list the corresponding initial conditions, and coupling strength $\lambda$, along with extracted values of the ratio $(\mu_{B}/T)_{\rm eq}$ at late times and the shear-viscosity $\eta T_{\rm eff}/(e+p)$ as discussed below.

\begin{center}
\begin{table}
\begin{tabular}{||c|c|c|c|c|c|c|c|c||}
\hline
$Q_s \tau_0$ & $f_g^0$ & $f_{val}^0$ & $f_{split}^0$ &
$\xi_0$ & $\lambda$ &
$\left(\frac{\eta T_{\rm eff}}{e+p}\right)$ & $(\frac{\mu_B}{T})_{\rm eq}$ & Figs\\ 
\hline \hline
5.5 & 1.068 & 0 & 0 & 
10 & 10 & 
1.00 & 0 & 
\ref{fig-EXP-pressure},
\ref{fig-EXP-MAP},
\ref{fig-EXP-E},
\ref{fig-EXP-S}\\ 
\hline
5.5 & 0.598 & 1.068 & 0 & 
10 & 10 & 
1.03 & 1.31 & 
\ref{fig-EXP-pressure},
\ref{fig-EXP-MAP},
\ref{fig-EXP-E},
\ref{fig-EXP-S}\\ 
\hline
12.5 & 0.363 & 1.602 & 0 & 
10 & 10 & 
1.08 & 2.38 & 
\ref{fig-EXP-pressure},
\ref{fig-EXP-MAP},
\ref{fig-EXP-E},
\ref{fig-EXP-S}\\ 
\hline
5.5 & 1.068 & 0 & 0 & 
10 & 5 & 
2.72 & 0 & 
\ref{fig-EXP-pressure},
\ref{fig-EXP-S} \\ 
\hline
5.5 & 0.598 & 1.068 & 0 & 
10 & 5 & 
2.76 & 0.97 & 
\ref{fig-EXP-pressure},
\ref{fig-EXP-S} \\ 
\hline
12.5 & 0.363 & 1.602 & 0 & 
10 & 5 & 
2.87 & 1.82 & 
\ref{fig-EXP-pressure},
\ref{fig-EXP-S}\\ 
\hline
5.5 & 0.950 & 0 & 0.269 & 
10 & 10 & 
1.00 & 0 & 
\ref{fig-EXP-E}\\ 
\hline
5.5 & 0.833 & 0 & 0.534 & 
10 & 10 & 
1.00 & 0 & 
\ref{fig-EXP-E} \\ 
\hline
5.5 & 0.598 & 0 & 1.068 & 
10 & 10 & 
1.00 & 0 & 
\ref{fig-EXP-E} \\ 
\hline
5.5 & 0.363 & 0 & 1.602 & 
10 & 10 & 
1.00 & 0 & 
\ref{fig-EXP-E} \\ 
\hline
5.5 & 0 & 0 & 2.427 & 
10 & 10 & 
1.00 & 0 & 
\ref{fig-EXP-E} \\ 
\hline
5.5 & 0.598 & 1.068 & 0.359 & 
10 & 10 & 
1.03 & 1.19 & 
\ref{fig-EXP-E} \\ 
\hline
5.5 & 0.598 & 1.068 & 1.077 & 
10 & 10 & 
1.02 & 1.01 & 
\ref{fig-EXP-E}\\ 
\hline
\end{tabular}
\caption{
Summary of the parameterization $\tau_0$, $f_{g}^0$, $f_{val}^0$, $f_{split}^0$, $\xi_0$, coupling $\lambda$
and extracted dimensionless coefficients $\eta T_{\rm eff}/(e+p)$ and baryon chemical potentials $(\mu_B/T)_{\rm eq}$.}
\label{tb-chemical}
\end{table}
\end{center}

\subsection{Early and late time behavior of $e$ and $\Delta n_{f}$}
Before we present the results of our QCD kinetic theory simulations, it is insightful to analyze the evolution of the energy-momentum tensor and conserved currents at early and late times. Due to the longitudinal expansion, the net-charge $\Delta n_{f}$ densities of all flavors are diluted according to
\begin{eqnarray}
\partial_{\tau} \Delta n_{f} = -\frac{\Delta n_{f}}{\tau}\;,
\end{eqnarray}
indicating that throughout the evolution the net-charge density per unit rapidity $\tau \Delta n_{f}(\tau)=(\tau \Delta n_{f})_{0}$ remains constant. Similarly, the energy density of a homogenous system undergoing a  boost-invariant longitudinal expansion decreases according to
\begin{eqnarray}
\partial_{\tau} e = - \frac{e+p_L}{\tau}\;,
\end{eqnarray}
where in addition to the trivial dilution, the second term on the right hand side characterizes the work performed against the longitudinal expansion~\cite{Bjorken:1982qr,Gyulassy:1983ub}, which is proportional to the longitudinal pressure $p_L=\tau^{2} T^{\eta\eta}$
given by
\begin{align}
p_L=\int \frac{d^2p_T}{(2\pi)^2} \frac{dp_{\|}}{(2\pi)} \frac{p_{\|}^{2}}{\sqrt{p_T^2+p_{\|}^2}} \sum_{i} \nu_{i} f_{i}(\tau,p_T,p_{\|})\;,
\end{align}
in QCD kinetic theory. Since at early times the system is rapidly expanding in the longitudinal direction, it is unable to maintain a sizeable longitudinal pressure. Early on, one therefore has $p_L \ll e$ , such that initially the energy per unit rapidity $\tau e(\tau)=(\tau e)_{0}$ remains approximately constant. Since initially $\tau \Delta n_{f}(\tau)=(\tau \Delta n_{f})_{0}$ and $\tau e(\tau)=(\tau e)_{0}$ are both constant, this further implies that, for finite density systems, the energy per baryon remains approximately constant at early times.

Evidently, this is sharp contrast to the behavior at asymptotically late times, where for an equilibrated QCD plasma the longitudinal pressure becomes $p_L=e/3$ such that $\tau^{4/3} e(\tau)=(\tau^{4/3} e(\tau))_{\rm \infty}$ approaches a constant and the energy per baryon decreases $\propto \tau^{-1/3}$. By considering the evolution of $e(\tau)$ along with the ratios of $\Delta n_{f}(\tau)/\Big(e(\tau)\Big)^{3/4}$ one then finds that the quantity $\tau^{1/3} T_{\rm eff}(\tau)=\left(\tau^{1/3} T_{\rm eff}\right)_{\rm \infty}$ as well as the ratios of the various chemical potentials to the temperature $\mu_{f,ldm}(\tau)/T_{ldm}(\tau)=(\mu_{f}/T)_{\rm \infty}$ become constant at asymptotically late times.

\subsection{Pressure isotropization and kinetic equilibration}
\begin{figure}
		\centering
		\includegraphics[width=0.48\textwidth]{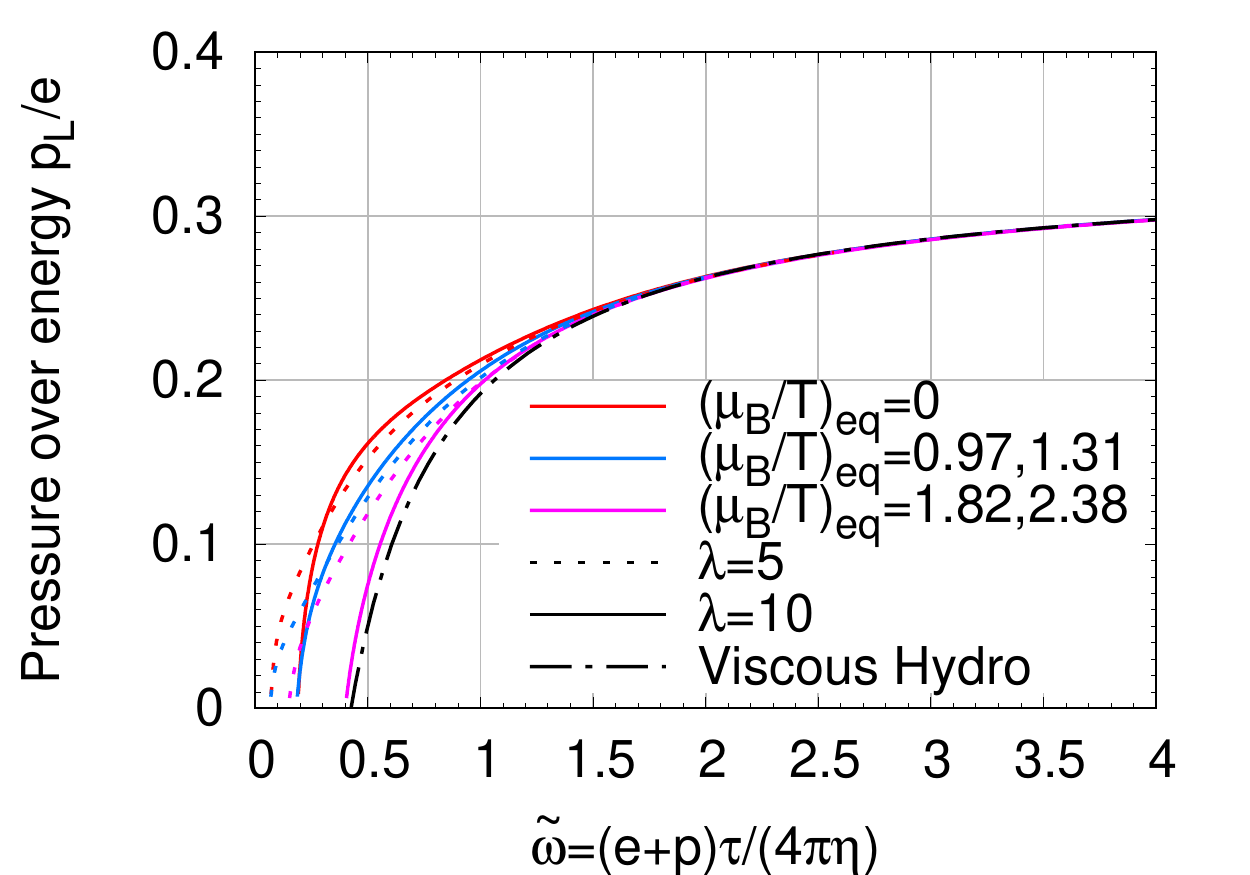}
	\caption{Non-equilibrium evolution of longitudinal pressure $p_L$ over energy density $e$ for different chemical potentials $(\mu_{B}/T)_{\rm eq}\approx 0,1,2$ (red,blue,pink) and coupling strength $\lambda=5,10$ (dotted,solid) as a function of the scaled time variable $\tilde{w}=(e+p)\tau/(4\pi \eta)$. Dashed dotted curve shows the asymptotic behavior of $p_L/e$ in viscous hydrodynamic.}
	\label{fig-EXP-pressure}
\end{figure}

We now turn to the presentation of our QCD kinetic theory results and first analyze the evolution of the bulk anisotropy, characterized by the ratio of the longitudinal pressure $p_L$ to the energy density $e$ shown in Fig.~\ref{fig-EXP-pressure}. Different curves in Fig.~\ref{fig-EXP-pressure} show the results for $p_L/e$ at zero and finite net-baryon density\footnote{Note that instead of characterizing the amount of net baryon density in terms of the initial energy per baryon $(e\tau)_0/(\Delta n_{B}\tau)_0$, the curves at different densities are labeled in terms of the asymptotic ratio of $(\mu_{B}/T)_{\rm eq}$ extracted from our simulations, and we refer to Tab.~\ref{tb-chemical} for the corresponding initial state parameters.} as a function of the scaling variable $\tilde{\omega}=(e+p)\tau/(4\pi \eta)$, which at zero net-baryon density ($\mu_{B}/T=0$) corresponds to the familiar expression $\tilde{\omega}=T\tau/(4\pi \eta/s)$ employed in previous works.

Starting from early times, where the system is dominated by the rapid longitudinal expansion and highly anisotropic $(p_L \ll e)$, the longitudinal pressure continuously rises as kinetic interactions become increasingly important. Despite the rapid increase of $p_L/e$ at early times, the system remains significantly anisotropic throughout the entire evolution shown in Fig.~\ref{fig-EXP-pressure} and only approaches an isotropic equilibrium state on much larger time scales.

Nevertheless, starting around $\tilde{\omega}\gtrsim 1$ the approach towards equilibrium is described by viscous hydrodynamics, where to leading order in the gradient expansion, the longitudinal pressure is given by
\begin{eqnarray}
\frac{p_L}{e}=\frac{1}{3}-\frac{16}{9}\frac{\eta}{(e+p)\tau}\;.
\end{eqnarray}
Expressing the non-equilibrium correction in terms of the dimensionless ratio $\eta T_{\rm eff}/(e+p)$ which at zero density reduces to the familiar $\eta/s$, the pressure evolution in hydrodynamics is the determined by
\begin{eqnarray}
\label{eq-visvoushydro}
\frac{p_L}{e}=\frac{1}{3}-\frac{4}{9\pi}\left(\frac{\eta T_{\rm eff}}{e+p}\right)\frac{4\pi}{\tau T_{\rm eff}}\;.
\end{eqnarray}
By analyzing the late time behavior of the different curves, we can then extract the transport coefficient  $\eta T_{\rm eff}/(e+p)$, who's values are indicated in Tab.~\ref{tb-chemical}. We note that although in principle $\eta T_{\rm eff}/(e+p)$ can exhibit a dependence on the chemical potentials $\mu_{f}/T$, we find that in the relevant range of $\mu_{B}/T$ for our simulations, the extracted values only differ by about ten percent, which a posteriori justifies its treatment as a constant when defining the scaling variable $\tilde{\omega}$ and extracting the values of $\eta T_{\rm eff}/(e+p)$ based on the late time behavior in Eq.~(\ref{eq-visvoushydro}).  When expressed in terms of the macroscopic scaling variable $\tilde{w}$, one also observes that the evolution of $p_L/e$ is rather insensitive to the microscopic coupling strength $\lambda=5,10$ in Fig.~\ref{fig-EXP-pressure}, as discussed in detail in \cite{Kurkela:2018oqw} for charge neutral plasmas.

By taking into account the (small) $\frac{\mu_{B}}{T}$ dependence of the transport coefficient $\eta T_{\rm eff}/(e+p)$, the results for $p_L/e$ in Fig.~\ref{fig-EXP-pressure} are presented such that they all exhibit the same hydrodynamic behavior in Eq.~(\ref{eq-visvoushydro}) at late times $\tilde{\omega} \gtrsim 1$ which is indicated by a black dashed line. However, by comparing the results for different net-baryon densities $(\mu_{B}/T)_{\rm eq}$, one clearly observes that at early times $\tilde{\omega}\lesssim 1$ the isotropization of the pressure proceeds more slowly for systems with a larger net baryon density. We will show shortly, that this feature can be understood by considering the fact that more baryon rich systems necessarily feature a larger abundance of quarks as compared to the initially gluon dominated zero density systems, which along with the less efficient isotropization of quark and anti-quark distributions leads to a slower build-up of the longitudinal pressure in the system.

\begin{figure}
	\centering
	\hspace*{-0.8in}
	\vspace*{-0.2in}
	\includegraphics[width=0.65\textwidth]{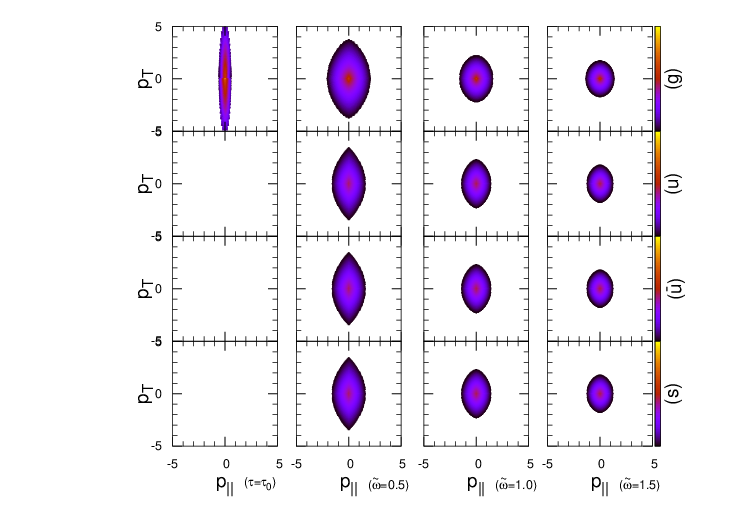}
	\centering
	\hspace*{-0.8in}
	\vspace*{-0.2in}
	\includegraphics[width=0.65\textwidth]{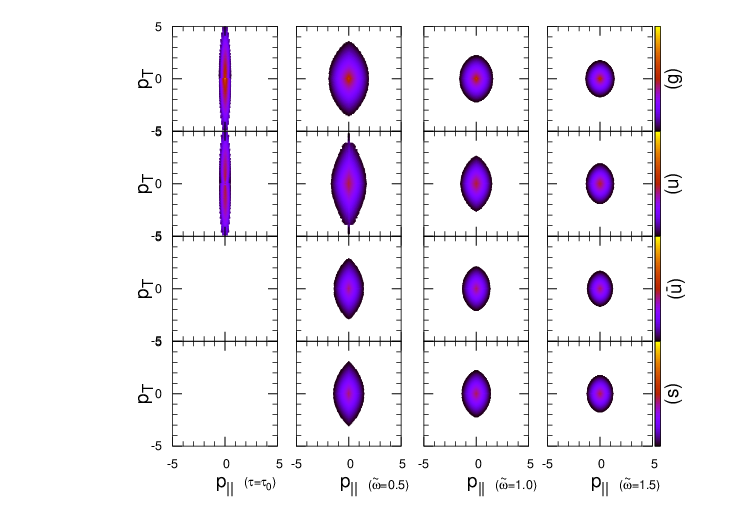}
	\centering
	\hspace*{-0.8in}
	\vspace*{-0.2in}
	\includegraphics[width=0.65\textwidth]{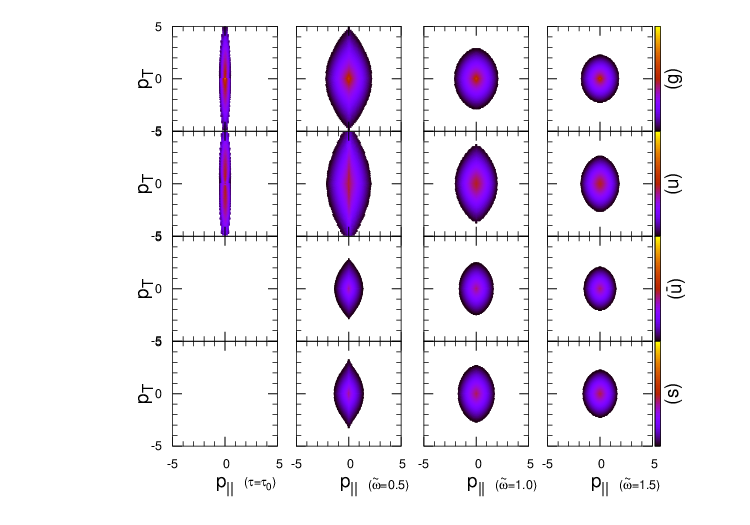}
	\caption{Evolution of the two-dimensional phase-space distributions $f(\tau,p_T,p_{\|})$ of gluons ($g$), up quarks ($u$), anti-up anti-quarks ($\bar{u}$) at times $\tau=\tau_0$, $\tilde{\omega}=0.5$, $\tilde{\omega}=1.0$, $\tilde{\omega}=1.5$ for different chemical potentials $(\mu_B/T)_{\rm eq}$=0 (top), 1.31 (middle), 2.38 (bottom) at coupling strength $\lambda$=10.}
	\label{fig-EXP-MAP}
\end{figure}

\subsection{Kinetic and chemical equilibration of light flavors}
Beyond the evolution of the pressure anisotropy, which provides an estimate of the range of applicability of a hydrodynamic description of the QGP, it is also insightful to consider the evolution of the phase-space densities of gluons, quarks and anti-quarks to scrutinize the underlying microscopic dynamics. Our results are compactly summarized in Fig.~\ref{fig-EXP-MAP}, where we present the evolution of the various distributions for three different values of the net baryon density $(\mu_{B}/T)_{\rm eq}=0$ (upper panel), $(\mu_{B}/T)_{\rm eq}=1.31$ (middle panel) and $(\mu_{B}/T)_{\rm eq}=2.38$ (lower panel). Different rows in each panel correspond to the distributions for different particle species, while different columns show the distributions at four different times, corresponding to the initial conditions in the first column, and $\tilde{\omega}=0.5,1.0,1.5$ in the second, third and fourth column. We focus on the evolution of the phase space distributions of gluons ($g$), up-quarks $(u)$, up-antiquarks ($\bar{u}$) and strange quarks ($s$), noting that the distributions of strange and anti-strange quarks are identical $f_{s}=f_{\bar{s}}$, and that up and down quark distributions exhibit essentially the same features.

Starting from the behavior at zero net-baryon density $(\mu_{B}/T)_{\rm eq}=0$ depicted in the top panel, where we assume that there are initially no quarks present in the system, one finds that quark/antiquarks of all flavors are democratically produced, and naturally inherit the anisotropy of the gluon distribution. However, the quark/antiquark distributions at intermediate stages of the evolution $\tilde{\omega}=0.5,1.0$ exhibit a larger degree of anisotropy as compared to the gluon distribution, indicating the slower isotropization of quarks/antiquarks. By considering the underlying microscopic processes in the bottom up scenario~\cite{Baier:2000sb}, one expects the isotropization of the gluon distribution to be driven by the radiative decay of hard gluons due to collinear $g\to gg$ and $g \to q\bar{q}$ processes, followed by $gg\rightarrow gg$, $g q\to g q$ and $g \bar{q}\to g \bar{q}$ elastic scatterings which isotropize the momentum distribution of soft gluons, whereas quarks/anti-quarks are pre-dominantly produced via collinear $g\to q\bar{q}$ splittings and to a lesser extent by $gg \to q\bar{q}$ elastic conversions, with the subsequent isotropization of soft quarks/anti-quarks due to $qg\rightarrow qg$ $\bar{q}g\rightarrow \bar{q}g$,$qq\rightarrow qq$,$\bar{q}\bar{q}\rightarrow \bar{q}\bar{q}$ and $q\bar{q}\to q\bar{q}$ elastic scattering processes. Based on the different color factors for the elastic scattering processes involving quarks and gluons, e.g. $|\mathcal{M}_{gg}^{gg}|^2 \propto C_{A}^2$ and $|M_{gq}^{gq}|^2 \propto C_{F}C_{A}$ (see Tab.~\ref{tb-pQCD}), it is then natural to expect a faster isotropization of the gluon distribution.

When considering the evolution of the phase-space distributions at finite net-baryon density, shown in the central and bottom panels of Fig.~\ref{fig-EXP-MAP} for $(\mu_{B}/T)_{\rm eq}=1.31$ and $2.38$, one finds that the overall behavior of the phase-space distributions at different times is rather similar to the zero density case. However, at finite density, the non-zero values of the conserved $u$ and $d$ charges lead to an overabundance of up and down quarks as compared to anti-quarks of the same flavor. Since at larger net-baryon density, $u$ and $d$ quarks carry a significant fraction of the initial energy, the larger degree of anisotropy of the quark distribution then manifests itself at the level of the bulk anisotropy $p_L/e$, seen in Fig.~\ref{fig-EXP-pressure}. 

\begin{figure}
	\centering
	\includegraphics[width=0.48\textwidth]{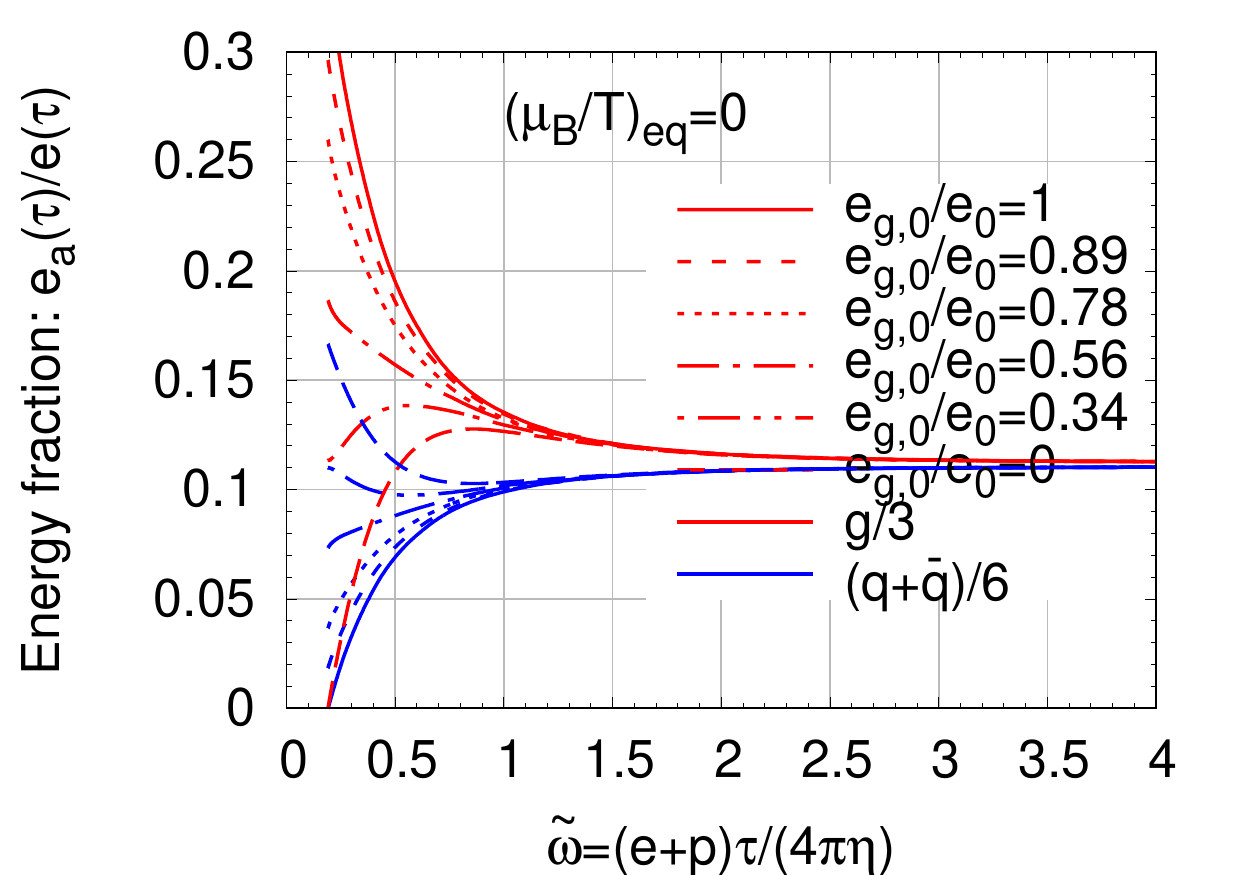}
	\hspace{\fill}
	\centering
	\includegraphics[width=0.48\textwidth]{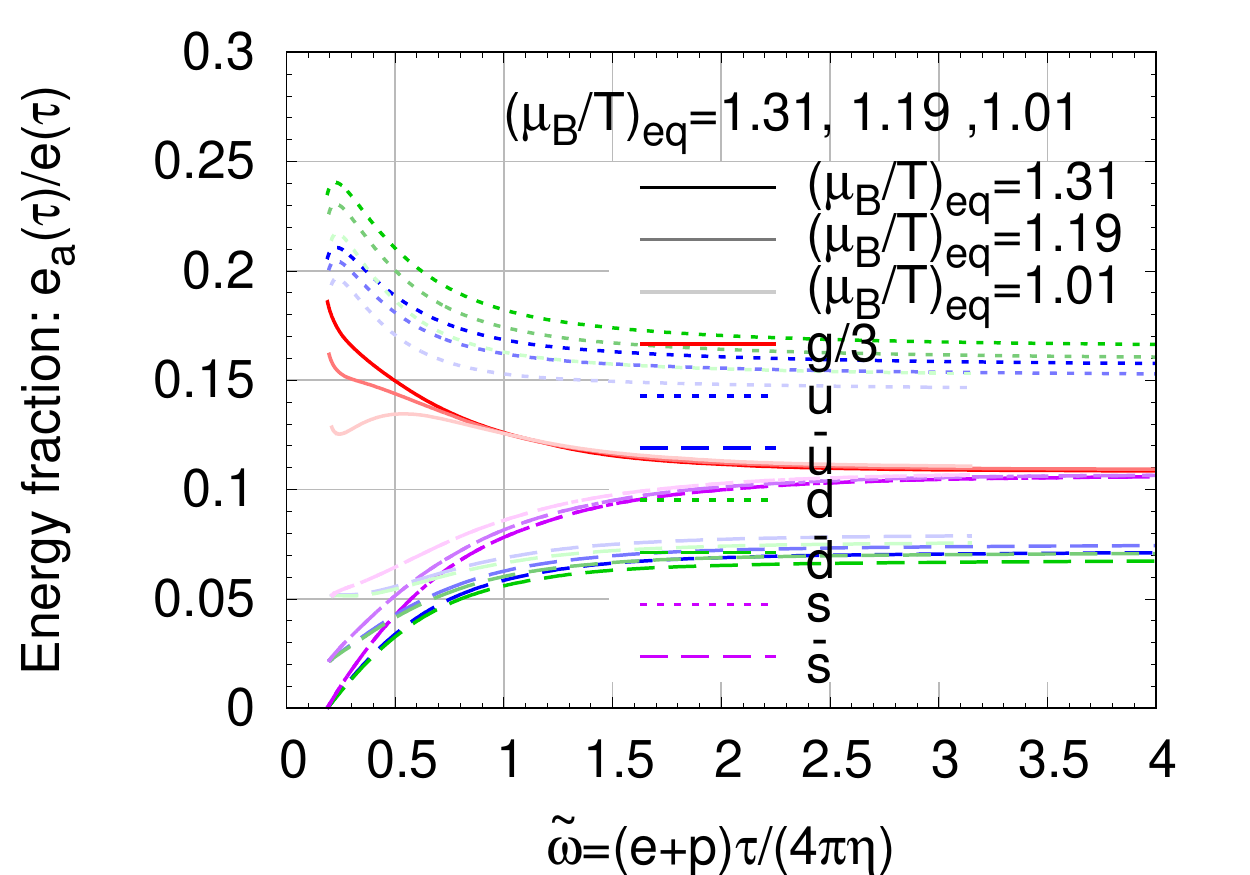}
	\centering
	\includegraphics[width=0.48\textwidth]{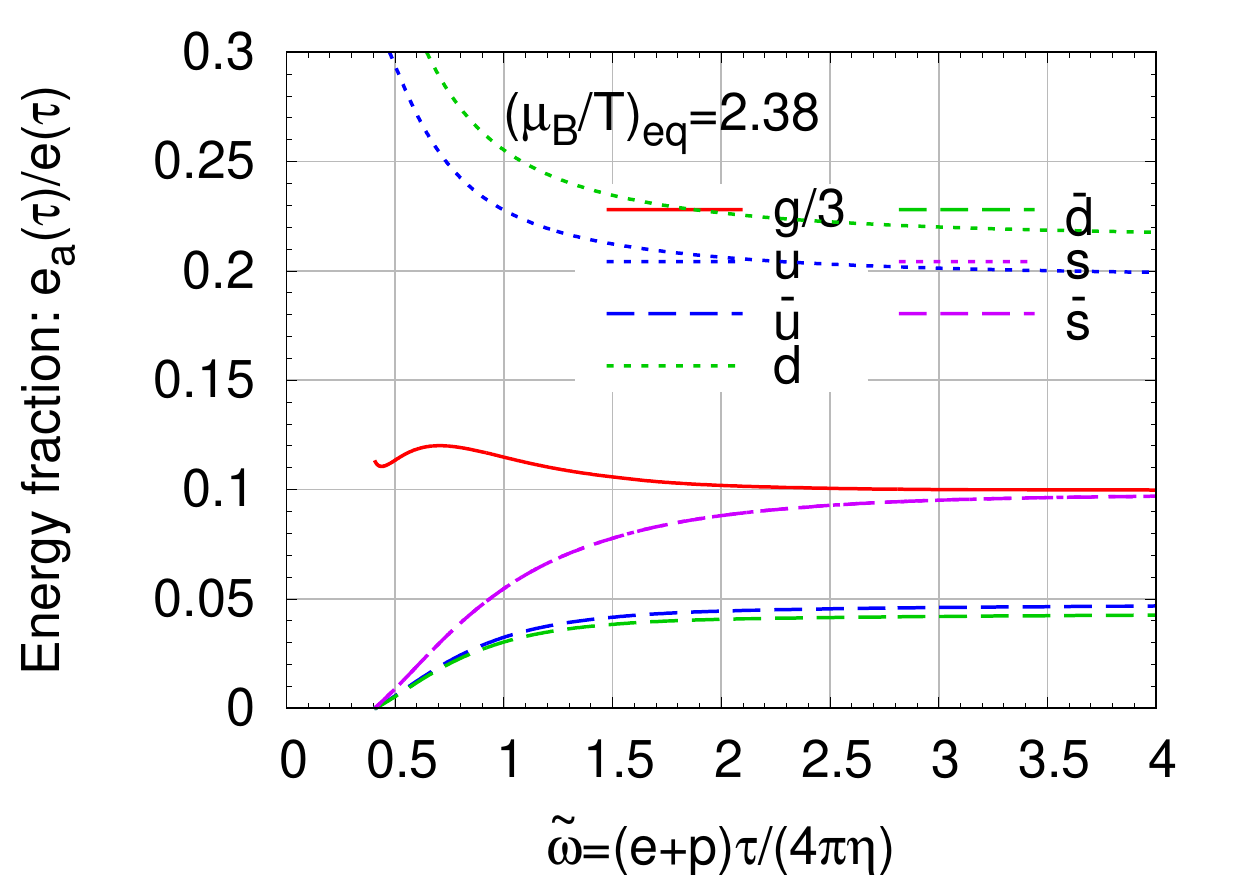}
	\caption{
	Evolution of fraction of energy carried by gluons ($g$), up ($u$), anti-up ($\bar{u}$), down ($d$), anti-down ($\bar{d}$), strange $(s)$ and anti-strange ($\bar{s}$) quarks as a function of the scaled time variable $\tilde{w}=(e+p)\tau/(4\pi \eta)$. Different panels show the results for different chemical potentials $(\mu_B/T)_{\rm eq}$=0 (top), 1.31, 1.19, 1.01 (middle), 2.38 (bottom). Different curves of same or similar color in the top and middle panels correspond show results for variations of the initial QGP chemistry (c.f. Tab.~\ref{tb-chemical}).}
	\label{fig-EXP-E}
\end{figure}

Besides the dynamics of the up and down flavors, it is also interesting to compare the evolution of the strange quark distribution $(f_s)$ at zero and finite density. While at zero density strange quarks can be efficiently produced via inelastic $g \to q\bar{q}$ and elastic conversions $gg \to q\bar{q}$ conversion, the direct production of $s\bar{s}$-pairs from $u$ and $d$ quarks is only possible through quark/anti-quark annihilation $q\bar{q} \to q\bar{q}$, which at finite density is suppressed due to the lack of anti-quarks. By comparing the results for $f_{s}$ in the upper and lower panels of Fig.~\ref{fig-EXP-MAP}, one therefore finds that the strangeness production at finite density is delayed until $\tilde{\omega}\sim 1$, when strangeness is efficiently produced from inelastic $g \to q\bar{q}$ and elastic conversions $gg \to q\bar{q}$ conversions.

Next in order to further analyze the chemical composition of the QGP, we follow \cite{Kurkela:2018wud} and investigate the fraction of energy $e_{a}(\tau)/e(\tau)$ carried by each individual species $a$ during the non-equilibrium evolution. Our results for this quantity, $e_{a}(\tau)/e_{\rm total}(\tau)$ are presented in Fig.~\ref{fig-EXP-E} as a function of the scaling variable $\tilde{\omega}$. Different panels in Fig.~\ref{fig-EXP-E} show the results for different net-baryon densities, with $(\mu_{B}/T)_{\rm eq}=0$ in the top panel, $(\mu_{B}/T)_{\rm eq}=1.31,1.19,1.01$ in the central panel and  $(\mu_{B}/T)_{\rm eq}=2.38$ in the bottom panel, while different solid, dashed and dotted curves in each panel correspond to the result obtained by varying the chemical composition of the initial state (see Tab.~\ref{tb-chemical}). Starting with the evolution at zero net-baryon density, we find that for gluon dominated initial conditions $(e_{g,0}/e_{0}=1)$ a large part of the initial energy of gluons is rapidly transferred to quarks and anti-quarks of all flavors. Similarly for quark/anti-quark dominated initial conditions at zero density $(e_{g,0}/e_{0}=0)$, a rapid energy transfer from the quark to the gluon sector occurs, effectively resulting in a memory loss of the initial QGP chemistry on time scales $\tilde{w}\sim 1$.  Eventually, for $\tilde{w}\gtrsim 0.5$ the zero density plasma becomes gluon dominated, before relaxing towards chemical equilibrium on time scales $\tilde{\omega} \sim 1-2$. Clearly, the situation is different at moderate or large net baryon density shown in the bottom panel of Fig.~\ref{fig-EXP-E}, where $u$ and $d$ quarks carry the dominant fraction of the energy density throughout the evolution. Due to the fact that multiple quark/anti-quark species contribute different amounts, one observes that the evolution of the chemistry of the QGP at moderate and large net-baryon density is significantly more complicated, and the approach towards equilibrium occurs on somewhat larger time scales $\tilde{\omega}\sim 1.5-2.5$, due to the less efficient production of anti-quarks ($\bar{u},\bar{d}$) and strangeness ($s,\bar{s}$).

\begin{figure}
	\centering
	\includegraphics[width=0.7\textwidth]{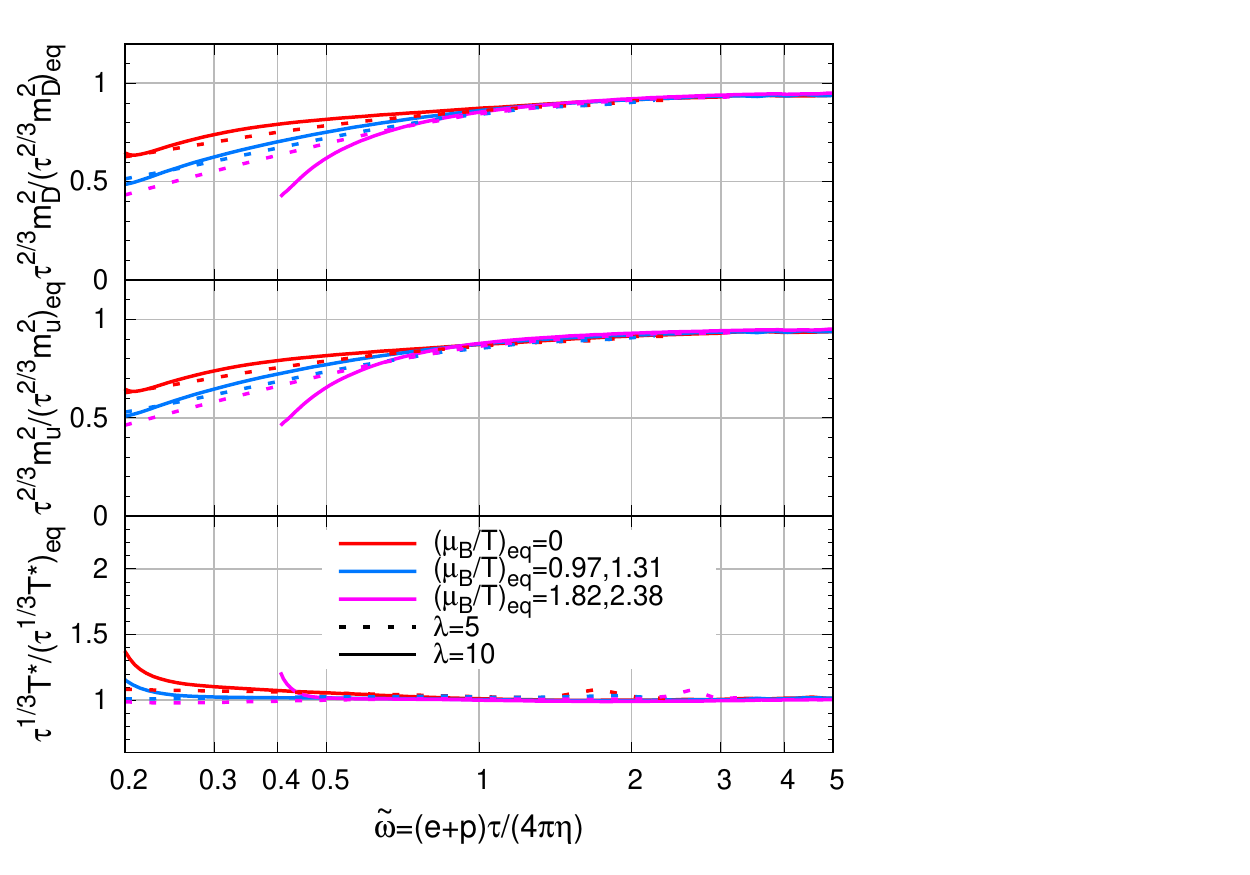}
	\hspace{\fill}
	\caption{Evolution of the characteristic scales $\tau^{2/3} m_D^2(\tau)$ (top), $\tau^{2/3}m_u^2(\tau)$ (middle), $\tau^{1/3}T^{*}(\tau)$ (bottom) normalized to their asymptotic equilibrium values, as a function of the scaled time variable $\tilde{w}=\frac{(e+p)\tau}{4\pi \eta}$ for different chemical potentials $(\mu_B/T)_{\rm eq}$=0, 1.31, 2.38 for $\lambda$=10 and $(\mu_B/T)_{\rm eq}$=0, 0.97, 1.82 for $\lambda$=5.}
	\label{fig-EXP-S}
\end{figure}

We conclude our discussion of equilibration in longitudinally expanding QCD plasmas, by considering once again the evolution of the characteristic scales $m_{D}^2$, $m_{Q,u}^2$ and $T^{*}$ that govern the rates of elastic and inelastic interactions in the plasma.  
The time evolution of these quantities is presented in Fig.~\ref{fig-EXP-S}, where in order to account for the continuous expansion of the system we have normalized the respective quantities as $\tau^{\frac{2}{3}} m_{D}^2/(\tau^{\frac{2}{3}} m_{D}^2)_{\rm eq}$, $\tau^{\frac{2}{3}} m_{Q}^2/(\tau^{\frac{2}{3}} m_{Q}^2)_{\rm eq}$ and $\tau^{\frac{1}{3}} T^{*}/(\tau^{\frac{1}{3}} T^{*})_{\rm eq}$ such that for $\tilde{\omega} \gg 1$ all ratios approach unity. By comparing the evolution of the different curves, we find that simulation results at different coupling strength $\lambda=5,10$ are in good overall agreement when expressing the evolution in terms of the scaling variable $\tilde{w}$. While the effective temperature $T^{*}$ relaxes towards its equilibrium value on time scales $\tilde{w} \sim 1$, the screening masses $m_{D}^2$,$m_{u}^{2}$ for gluons and (up-) quarks only approach their equilibrium values at asymptotically late times, indicating residual deviations from local thermal equilibrium on the order of $10\%$.

\section{Conclusions}
~\label{sec:conclusions}

We developed a QCD kinetic description of the light flavor QCD degrees of freedom to study near and far-from equilibrium dynamics of the Quark Gluon Plasma (QGP) at zero and finite density of the conserved baryon number, electric charge and strangeness. Based on numerical solution of the kinetic equations, including all leading order elastic and in-elastic interactions between gluons, quarks and anti-quarks, we exposed the general features of kinetic and chemical equilibration of non-equilibrium QCD plasmas in the perturbative regime at (asymptotically) high energies.

Generally, we find that, albeit the energy transfer between quark and gluon degrees of freedom can take a significant time, kinetic and chemical equilibrations of QCD plasmas occur roughly on the same time scale. 
By performing detailed investigations of the evolution of the spectra and collision rates, we further established a microscopic understanding of different equilibration processes in QCD plasmas, which generalizes earlier results obtained in pure glue QCD~\cite{Kurkela:2014tea,Kurkela:2015qoa,Kurkela:2018vqr} and QCD at zero density~\cite{Kurkela:2018oqw,Kurkela:2018xxd,Mazeliauskas:2018yef}.  

Specifically, for over-occupied systems, which initially feature a large number of low energy gluons, we find that the thermalization process proceeds via a self-similar turbulent cascade, before eventually reaching equilibrium on a time scale $\sim 4\pi \eta/s/T_{\rm eq}$. Conversely, for under-occupied systems, which initially feature a small number of high energy quarks or gluons, thermalization is achieved via the bottom-up scenario, with a number of interesting features regarding the role of quark and gluon degrees of freedom.

Studies of the equilibration of the QGP in a longitudinally expanding system provide the basis for a realistic matching of the initial state in heavy-ion collisions to initial conditions for the subsequent hydrodynamic evolution. By analyzing the macroscopic evolution of the energy momentum tensor and the microscopic evolution of the phase-space distributions of quarks and gluons, we found that viscous hydrodynamics typically becomes applicable on time scales where $(e+p)\tau/(4\pi\eta)\sim 1$; however, isotropization and strangeness production proceed more slowly for finite density systems, and we refer to our companion paper~\cite{Du:2020pre} for further discussions of phenomenological consequences.

We finally note that the numerical framework to solve the QCD kinetic equations presented in this paper could be extended in several regards, e.g. by including heavy flavor degrees of freedom or electroweak interactions, to study a variety of aspects regarding the early time dynamics of high-energy heavy-ion collisions and the thermalization of the early universe.
\section*{Acknowledgment}
We thank Giuliano Giacalone, Aleksi Kurkela, Aleksas Mazeliauskas, Jean-Francois Paquet, Ismail Soudi, Derek Teaney for discussions and collaboration on related projects. We are especially grateful to Aleksas Mazeliauskas for fruitful exchanges on the numerical implementation of QCD kinetic theory. This work is supported by the Deutsche Forschungsgemeinschaft (DFG, German Research Foundation) – project number 315477589 – TRR 211. The authors also gratefully acknowledge computing time provided by the Paderborn Center for Parallel Computing (PC2) and National Energy Research Scientific Computing Center under US Department of Energy.

\appendix
\section{Weight Function Discretization}
\label{sec-algorithm}
\subsection{Weighted Integral: Discretization}
We discretize the Boltzmann equation Eq.~(\ref{eq-bolzmann}) with the {\it weighted integral} of a function $\mathcal{F}(\vec{p})$ (particle phase-space distribution $\mathcal{F}(\vec{p})$=$f(\vec{p})$ or collision integral $\mathcal{F}(\vec{p})$=$C[f](\vec{p})$) transforming from the continuous domain $\Omega_{p,\theta,\phi}=\{(p,\theta,\phi)|p\in \mathbb{R}^{+};\theta\in[0,\pi]; \phi\in[0,2\pi)\}$ to a discretized domain $\mathbb{N}^3_{p,\theta,\phi}=\{(i_{p},j_\theta,k_\phi)|i_{p}=0,\dots,N_p-1; j_{\theta}=0,\cdots,N_\theta-1; k_{\phi}=0,\cdots,N_\phi-1
\}$. 

The weighted integral of distribution function $f(\vec{p})$ reads
\begin{eqnarray}
\label{eq-fton}
n(i_{p},j_{\theta},k_{\phi})
=\int \frac{d^3p}{(2\pi)^3}
w^{(p)}_{i}(p) w^{(\theta)}_{j}(\theta) w^{(\phi)}_{k}(\phi)f(\vec{p})
\end{eqnarray}
where $w^{(p)}_{i}(p)$,$w^{(\theta)}_{j}(\theta)$ and $w^{(\phi)}_{k}(\phi)$ are weight functions satisfying a {\it completeness relation}
\begin{eqnarray}
\label{eq-wCompleteness}
\sum_{i\in \mathbb{N}_x}w_{i}^{(x)}(x)=\chi_{\Omega_x}(x)
=\left\{
\begin{matrix}
1~&~,~x\in \Omega_x\\
0~&~,~x\notin \Omega_x
\end{matrix}
\right.
\end{eqnarray}
with $\chi_{\Omega_x}(x):\Omega_x\rightarrow\mathbb{Z}_2$ the indicator function on domain $\Omega_x$.
The completeness relation ensures the summation of weighted integral $n(i_{p},j_{\theta},k_{\phi})$ to be the total number of specific particle
\begin{eqnarray}
\sum_{(i_{p},j_\theta,k_\phi)}n(i_{p},j_{\theta},k_{\phi})
=\int \frac{d^3p}{(2\pi)^3}f(\vec{p}).
\end{eqnarray}

The weight functions satisfying completeness relation Eq.~(\ref{eq-wCompleteness}) can be achieved by decomposing into two parts
\begin{eqnarray}
w_{i}^{(p)}(p)=w_{i}^{L}(p) + w_{i}^{R}(p)
\end{eqnarray}
with left and right sided weights
\begin{eqnarray}
\nonumber
&&w_{i}^{L}(p)
=\left\{
\begin{matrix} \qquad ~S_{(p)}(p)|_{p_{i}}^{p_{i+1}} \chi_{[p_{i},p_{i+1}]}(p) & ,0\leq i \leq N_{p}-2\\
0 & ,i=N_{p}-1  \end{matrix} \right.\\
\nonumber
&&w_{i}^{R}(p)
=\left\{
\begin{matrix} \left(1-S_{(p)}(p)|_{p_{i-1}}^{p_{i}} \right) \chi_{[p_{i-1},p_{i}]}(p) & ,1\leq i \leq N_{p}-1\\
0 & ,i=0\end{matrix}\right.\\
\end{eqnarray}.

The {\it spectral weight} $S_{(p)}(p)|_{p_{i}}^{p_{i+1}}$ needs to be constructed in a form
\begin{eqnarray}
S_{(p)}(p)|_{p_{i}}^{p_{i+1}}=\frac{y(p_{i+1})-y(p)}{y(p_{i+1})-y(p_{i})},
\end{eqnarray}
with $y(p)$ an arbitrary function of $p$ so that $S_{(p)}(p_{i})|_{p_{i}}^{p_{i+1}}=1$, $S_{(p)}(p_{i+1})|_{p_{i}}^{p_{i+1}}=0$ and Eq.~(\ref{eq-wCompleteness}) satisfied.

\subsection{Sum Rules}
Indeed, the above function $y(p\in[p_{i},p_{i+1}])$ can be reversely expressed with the left and right weights
\begin{eqnarray}
\label{eq-sumDomain}
y(p\in[p_{i},p_{i+1}])=
w_{i}^{L}(p)y(p_{i})+w_{i+1}^{R}(p)y(p_{i+1})
\end{eqnarray}
which yields a {\it sum rule}
\begin{eqnarray}
\label{eq-wSumRuleI}
\sum_{i=0}^{N_p-1}w_{i}^{(p)}(p)y(p_{i})=y(p)
\end{eqnarray}
for $p\in [p_{\rm min},p_{\rm max}]$.

Specifically, we work with a properly choice of the functions for $p$, ${\rm cos}(\theta)$ and $\phi$
\begin{eqnarray}
\label{eq-LinearWeight}
y_{p}(p)=p,
~y_{\theta}(\theta)={\rm cos}(\theta),
~y_{\phi}(\phi)=1.
\end{eqnarray}
that provides the way to evaluate energy and longitudinal momentum of the particle in discretized form, following the definition of weighted integral Eq.~(\ref{eq-fton}), completeness relation Eq.~(\ref{eq-wCompleteness}) and sum rule Eq.~(\ref{eq-wSumRuleI})
\begin{eqnarray}
\label{eq-conservation}
&&\sum_{(i_p,j_\theta,k_\phi)}
p_{i}n(i_{p},j_{\theta},k_{\phi})
=\int\frac{d^3p}{(2\pi)^3}p f(\vec{p})\\
\nonumber
&&\sum_{(i_p,j_\theta,k_\phi)}
p_{i}{\rm cos}(\theta_j)n(i_{p},j_{\theta},k_{\phi})
=\int\frac{d^3p}{(2\pi)^3}p{\rm cos}(\theta)f(\vec{p})
\end{eqnarray}
with 
$(p,\theta,\phi)\in \Omega_{p,\theta,\phi}$.
Their weighted functions are
\begin{eqnarray}
\label{eq-LinearWeightF}
\nonumber
w_{i}^{(p)}(p)&=&\left[
\frac{p_{i+1}-p}{p_{i+1}-p_{i}} \chi_{[p_{i},p_{i+1}]}(p)
+\frac{p-p_{i-1}}{p_{i}-p_{i-1}} \chi_{[p_{i-1},p_{i}]}(p)\right]\\
\nonumber
w_{j}^{(\theta)}(\theta)&=&\left[
\frac{{\rm cos}(\theta_{j+1})-{\rm cos}(\theta)}{{\rm cos}(\theta_{j+1})-{\rm cos}(\theta_{j})} \chi_{[\theta_{j},\theta_{j+1}]}(\theta)\right.\\
\nonumber
&&\left.+\frac{{\rm cos}(\theta)-{\rm cos}(\theta_{j-1})}{{\rm cos}(\theta_{j})-{\rm cos}(\theta_{j-1})} \chi_{[\theta_{j-1},\theta_{j}]}(\theta)\right]\\
w_{k}^{(\phi)}(\phi)&=&
\chi_{[\phi_{k},\phi_{k+1}]}(\phi).
\end{eqnarray}

\subsection{Weighted Sum: Continuation}
We Taylor expand $f(\vec{p})$ around the points $(p_i,\theta_j,\phi_k)$ and use the nearest grid points to reconstruct the phase-space distribution at any specific coordinates $(p,\theta,\phi)$ from its discretized form. Assuming 
\begin{eqnarray}
p^{l}f(\vec{p})\approx {\rm const.}=\left<p^{l}f(\vec{p})\right>_{(i_{p},j_{\theta},k_{\phi})}
\end{eqnarray}
holds for a neighborhood $\vec{p}\in U_{(i_{p},j_{\theta},k_{\phi})}$ around $(p_i,\theta_j,\phi_k)$, the weighted integral Eq.({\ref{eq-fton}}) reads
\begin{eqnarray}
\nonumber
&&n(i_{p},j_{\theta},k_{\phi})
\approx\left<p^{l}f(\vec{p})\right>_{(i_{p},j_{\theta},k_{\phi})}\int\frac{d^3p}{(2\pi)^3}\frac{1}{p^l}W_{(i_p,j_\theta,k_\phi)}(\vec{p})\\
\end{eqnarray}
with $W_{(i_p,j_\theta,k_\phi)}(\vec{p})=w^{(p)}_{i}(p)w^{(\theta)}_{j}(\theta)w_{k}^{(\phi)}(\phi)$.
Defining the {\it modified weighted volume} (MWV)
\begin{eqnarray}
A_{(i_{p},j_{\theta},k_{\phi})}^{(2-l)}=\int\frac{d^3p}{p^l}W_{(i_p,j_\theta,k_\phi)}(\vec{p})
\end{eqnarray}
we may evaluate the constant for $\vec{p}\in U_{(i_{p},j_{\theta},k_{\phi})}$ via
\begin{eqnarray}
\left<p^{l}f(\vec{p})\right>_{(i_{p},j_{\theta},k_{\phi})}\approx \frac{(2\pi)^3n(i_{p},j_{\theta},k_{\phi})}{A_{(i_{p},j_{\theta},k_{\phi})}^{(2-l)}}.
\label{eq-WeightConst}
\end{eqnarray}

The azimuthal symmetric weight functions in Eq.~(\ref{eq-LinearWeightF}) gives $A_{(i_{p},j_{\theta},k_{\phi})}^{(2-l)}
=2\pi A_{(i_{p})}^{(2-l)}A_{(j_{\theta})}$ and the modified weighted volumes are listed in Table~\ref{tb-MWV}.

In order to evaluate the distribution between points in the grids, we perform interpolation similar to Eq.~(\ref{eq-sumDomain}) for azimuthal symmetric case so that 
\begin{eqnarray}
f(\vec{p}\in\Omega_{p})
=\sum_{(\alpha,\beta)\in\mathbb{N}^2_{\vec{p}}}
w_{\alpha}^{(p)}(p)w_{\beta}^{(\theta)}(\theta)\frac{(2\pi)^2n(\alpha,\beta)}{p^{l}A_{(\alpha)}^{(2-l)}A_{(\beta)}}
\label{eq-ntof}
\end{eqnarray}
where the sub-domain cubic $\Omega_{\vec{p}}=[p_{L},p_{R}]\times[\theta_{L},\theta_{R}]\times[\phi_{L},\phi_{R}]\subset\Omega$
and the corresponding nearest grids in discretized sub-domain $\mathbb{N}^2_{\vec{p}}=\{(i_{P},j_{\Theta})|P=p_L,p_R;\Theta=\theta_L,\theta_R;i_{p_R}-i_{p_L}=j_{\theta_R}-j_{\theta_L}=1\}$.

\begin{widetext}
\begin{center}
\begin{table}
\begin{tabular}{|c|c|c|}
\hline
MWV & formula & index range\\ 
\hline
&$(p_{i+1}-p_{i})(3p_{i}^2+2p_{i}p_{i+1}+p_{i+1}^2)/12$& $i=0$ \\ 
$A_{(i_{p})}^{(2)}$ & $(p_{i}-p_{i-1})(3p_{i}^2+2p_{i}p_{i-1}+p_{i-1}^2)/12$& $i=N_{p}-1$ \\ 
&$\left[(p_{i+1}-p_{i})(3p_{i}^2+2p_{i}p_{i+1}+p_{i+1}^2)+(p_{i}-p_{i-1})(3p_{i}^2+2p_{i}p_{i-1}+p_{i-1}^2)\right]/12$& $i\in[1,N_{p}-2]$ \\ 
\hline
&$(p_{i+1}-p_{i})(2p_{i}+p_{i+1})/6$& $i=0$ \\ 
$A_{(i_{p})}^{(1)}$ & $(p_{i}-p_{i-1})(2p_{i}+p_{i-1})/6$& $i=N_{p}-1$ \\ 
&$(p_{i+1}-p_{i-1})(p_{i+1}+p_{i}+p_{i-1})/6$& $i\in[1,N_{p}-2]$ \\ 
\hline
&$(p_{i+1}-p_{i})/2$& $i=0$ \\ 
$A_{(i_{p})}^{(0)}$ & $(p_{i}-p_{i-1})/2$& $i=N_{p}-1$ \\ 
&$(p_{i+1}-p_{i-1})/2$& $i\in[1,N_{p}-2]$ \\ 
\hline
&$\left[{\rm cos}(\theta_{j+1})-{\rm cos}(\theta_{j})\right]/2$& $i=0$ \\ 
$A_{(j_{\theta})}$ & $\left[{\rm cos}(\theta_{j})-{\rm cos}(\theta_{j-1})\right]/2$& $i=N_{p}-1$ \\ 
&$\left[{\rm cos}(\theta_{j+1})-{\rm cos}(\theta_{j-1})\right]/2$& $i\in[1,N_{p}-2]$ \\
\hline
\end{tabular}
\caption{Modified weighted volume (MWV) with choice of weight functions from Eq.~(\ref{eq-LinearWeightF})}
\label{tb-MWV}
\end{table}
\end{center}
\end{widetext}

\section{Discretization of Collision Integrals}
\label{sec-discretization}
\subsection{Elastic Collision Integrals}
\label{sec-discretization-elastic}
\subsubsection{Discretization and Efficient Samplings}
\label{sec-de-form}
The elastic collision integral for particle ``a'' with momentum $\vec{p}_1$ in process $a,b \rightarrow c,d$ ($p_{1},p_{2}\leftrightarrow p_{3},p_{4}$) reads:
\begin{eqnarray}
\label{eq-cint-elastic-app}
\nonumber
&&C^{{2\leftrightarrow2}}_a[f](\vec{p}_1)
=\frac{1}{2 \nu_{a}}\frac{1}{2 E_{p_1}}{\sum_{cd}}
\int d\Pi_{2\leftrightarrow2}\\
&&\times|\mathcal{M}_{cd}^{ab}(\vec{p}_1,\vec{p}_2|\vec{p}_3,\vec{p}_4)|^2F_{cd}^{ab}(\vec{p}_1,\vec{p}_2|\vec{p}_3,\vec{p}_4),
\end{eqnarray}
with measure
\begin{eqnarray}
\nonumber
&&d\Pi_{2\leftrightarrow2} = \frac{d^3p_2}{(2\pi)^3} \frac{1}{2E_{p_2}} \frac{d^3p_3}{(2\pi)^3} \frac{1}{2E_{p_3}} \frac{d^3p_4}{(2\pi)^3} \frac{1}{2E_{p_4}}\\
&&\times(2\pi)^4 \delta^{(4)}({p}_1+{p}_2-{p}_3-{p}_4),
\end{eqnarray}
$|\mathcal{M}_{cd}^{ab}(\vec{p}_1,\vec{p}_2|\vec{p}_3,\vec{p}_4)|^2$ is the matrix element square for process ``$a,b\leftrightarrow c,d$'' summed over spin and color for all particles, and $F_{cd}^{ab}(\vec{p}_1,\vec{p}_2|\vec{p}_3,\vec{p}_4)$ describes the statistical factor for ``$a,b \leftrightarrow c,d$''.

The discretized form of collision integral Eq.~(\ref{eq-cint-elastic-app}) follows the the transformation according to Eq.~(\ref{eq-fton})
\begin{eqnarray}
\label{eq-elastic-moment}
&&C^{{2\leftrightarrow2}}_a[n](i_{p},j_{\theta},k_{\phi})\\
\nonumber
&&=\int\frac{d^2p_1}{(2\pi)^3}W_{(i_{p},j_{\theta},k_{\phi})}(\vec{p}_1)C^{{2\leftrightarrow2}}_a[f](\vec{p}_1)\\
\nonumber
&&=\frac{1}{2 \nu_{a}}{\sum_{cd}}
\int d\Omega_{2\leftrightarrow2}
W_{(i_{p},j_{\theta},k_{\phi})}(\vec{p}_1)
Q_{cd}^{ab}(\vec{p}_1,\vec{p}_2|\vec{p}_3,\vec{p}_4)
\end{eqnarray}
with measure
\begin{eqnarray}
\label{eq-elastic-moment-measure}
\nonumber
&&d\Omega_{2\leftrightarrow2}=\frac{d^3p_1}{(2\pi)^3} \frac{1}{2E_{p_1}} \frac{d^3p_2}{(2\pi)^3} \frac{1}{2E_{p_2}} \frac{d^3p_3}{(2\pi)^3} \frac{1}{2E_{p_3}} \frac{d^3p_4}{(2\pi)^3} \frac{1}{2E_{p_4}}\\
&&\times(2\pi)^4\delta^{(4)}({p}_1+{p}_2-{p}_3-{p}_4).
\end{eqnarray}
and a $Q$-Factor 
\begin{equation}
Q_{cd}^{ab}(\vec{p}_1,\vec{p}_2|\vec{p}_3,\vec{p}_4)
=|\mathcal{M}_{cd}^{ab}(\vec{p}_1,\vec{p}_2|\vec{p}_3,\vec{p}_4)|^2F_{cd}^{ab}(\vec{p}_1,\vec{p}_2|\vec{p}_3,\vec{p}_4)
\end{equation}
which has the following symmetries
\begin{eqnarray}
\label{eq-elastic-factor-sym}
\nonumber
&&Q_{dc}^{ba}(\vec{p}_2,\vec{p}_1|\vec{p}_4,\vec{p}_3)=
Q_{cd}^{ba}(\vec{p}_2,\vec{p}_1|\vec{p}_3,\vec{p}_4)\\
&&=
Q_{cd}^{ab}(\vec{p}_1,\vec{p}_2|\vec{p}_3,\vec{p}_4)=
-Q_{ab}^{cd}(\vec{p}_3,\vec{p}_4|\vec{p}_1,\vec{p}_2)
\end{eqnarray}

One good feature of utilizing weight function algorithm is that the phase-space measure $d\Omega_{2\leftrightarrow2}$ in Eq.~(\ref{eq-elastic-moment-measure}) is invariant under arbitrary exchanges between $\vec{p}_1, \vec{p}_2, \vec{p}_3, \vec{p}_4$, meaning we can maximize Mont-Carlo samplings expanding the terms with those exchanges. 
One sampling of $\vec{p}_1, \vec{p}_2, \vec{p}_3, \vec{p}_4$ can be freely used by all combinations, increasing the efficiency of numerical calculation and statistical accuracy.
Indeed, denoting $W_1=W_{(i_{p},j_{\theta},k_{\phi})}(\vec{p}_1)$,  $Q_{cd}^{ab}(12|34)=Q_{dc}^{ba}(\vec{p}_2,\vec{p}_1|\vec{p}_4,\vec{p}_3)$, the discretized form of collision integral Eq.~(\ref{eq-elastic-moment}) can be expanded according to the symmetry as
\begin{widetext}
\begin{equation}
\label{eq-elastic-moment-8}
\boxed{
\begin{aligned}
&C^{{2\leftrightarrow2}}_a[n](i_{p},j_{\theta},k_{\phi})
=\frac{1}{2 \nu_{a}}{\sum_{cd}}
\int d\Omega_{2\leftrightarrow2}
W_1Q_{cd}^{ab}(12|34)\\
&=\frac{1}{2 \nu_{a}}{\sum_{cd}}
\int d\Omega_{2\leftrightarrow2}\frac{1}{8}\left[
W_1(Q_{cd}^{ab}(12|34)+Q_{cd}^{ab}(12|43))+W_2(Q_{cd}^{ab}(21|43)+Q_{cd}^{ab}(21|34))\right.\\
&\left.+W_3(Q_{cd}^{ab}(34|12)+Q_{cd}^{ab}(34|21))+W_4(Q_{cd}^{ab}(43|21)+Q_{cd}^{ab}(43|12))\right],\\
\end{aligned}
}
\end{equation}
\end{widetext}

As the energy density and longitudinal momentum flux can be directly evaluated from the discretized form in Eq.(\ref{eq-conservation}), energy and longitudinal momentum conservation can be exactly fulfilled by the discretized form of collision integral, as a derivative of distributions. 

We take the most complicated process $q_1\bar{q}_1 \leftrightarrow q_2\bar{q}_2$ as an example, other processes follow.
According to Eq.~(\ref{eq-elastic-moment-8}), the discretization forms read:\\
(1)~For quark $q_1$, note that $Q_{\bar{q}_2q_2}^{q_1\bar{q}_1}(12|34)=Q_{q_2\bar{q}_2}^{q_1\bar{q}_1}(12|43)$
\begin{align}
&C^{{2\leftrightarrow2}}_{q_1}[n](i_{p},j_{\theta},k_{\phi})\\
\nonumber
&=\frac{1}{2 \nu_{q}}
\int d\Omega_{2\leftrightarrow2}
W_1(Q_{q_2\bar{q}_2}^{q_1\bar{q}_1}(12|34)+Q_{\bar{q}_2q_2}^{q_1\bar{q}_1}(12|34))\\	
\nonumber
&=\frac{1}{\nu_{q}}
\int d\Omega_{2\leftrightarrow2}\frac{1}{8}\left[
W_1(Q_{q_2\bar{q}_2}^{q_1\bar{q}_1}(12|34)+Q_{q_2\bar{q}_2}^{q_1\bar{q}_1}(12|43))\right.\\
\nonumber
&\left.+W_2(Q_{q_2\bar{q}_2}^{q_1\bar{q}_1}(21|43)+Q_{q_2\bar{q}_2}^{q_1\bar{q}_1}(21|34))\right.\\
\nonumber
&\left.+W_3(Q_{q_2\bar{q}_2}^{q_1\bar{q}_1}(34|12)
+Q_{q_2\bar{q}_2}^{q_1\bar{q}_1}(34|21))\right.\\
\nonumber
&\left.+W_4(Q_{q_2\bar{q}_2}^{q_1\bar{q}_1}(43|21)+Q_{q_2\bar{q}_2}^{q_1\bar{q}_1}(43|12))\right]
\end{align}
(2)~For antiquark $\bar{q}_1$, note that $Q_{\bar{q}_2q_2}^{\bar{q}_1q_1}(12|34)=Q_{q_2\bar{q}_2}^{q_1\bar{q}_1}(21|43)$, $Q_{q_2\bar{q}_2}^{\bar{q}_1q_1}(12|34)=Q_{q_2\bar{q}_2}^{q_1\bar{q}_1}(21|34)$
\begin{align}
&C^{{2\leftrightarrow2}}_{\bar{q}_1}[n](i_{p},j_{\theta},k_{\phi})\\
\nonumber
&=\frac{1}{2 \nu_{q}}
\int d\Omega_{2\leftrightarrow2}
W_1(Q_{\bar{q}_2q_2}^{\bar{q}_1q_1}(12|34)+Q_{q_2\bar{q}_2}^{\bar{q}_1q_1}(12|34))\\	
\nonumber
&=\frac{1}{\nu_{q}}
\int d\Omega_{2\leftrightarrow2}\frac{1}{8}\left[
W_1(Q_{q_2\bar{q}_2}^{q_1\bar{q}_1}(21|43)+Q_{q_2\bar{q}_2}^{q_1\bar{q}_1}(21|34))\right.\\
\nonumber
&\left.+W_2(Q_{q_2\bar{q}_2}^{q_1\bar{q}_1}(12|34)+Q_{q_2\bar{q}_2}^{q_1\bar{q}_1}(12|43))\right.\\
\nonumber
&\left.+W_3(Q_{q_2\bar{q}_2}^{q_1\bar{q}_1}(43|21)+Q_{q_2\bar{q}_2}^{q_1\bar{q}_1}(43|12))\right.\\
\nonumber
&\left.+W_4(Q_{q_2\bar{q}_2}^{q_1\bar{q}_1}(34|12)+Q_{q_2\bar{q}_2}^{q_1\bar{q}_1}(34|21))\right]
\end{align}
(3)~For quark $q_2$, note that $Q_{q_1\bar{q}_1}^{q_2\bar{q}_2}(12|34)=-Q_{q_2\bar{q}_2}^{q_1\bar{q}_1}(34|12)$, $Q_{\bar{q}_1q_1}^{q_2\bar{q}_2}(12|34)=-Q_{q_2\bar{q}_2}^{q_1\bar{q}_1}(43|12)$
\begin{align}
&C^{{2\leftrightarrow2}}_{q_2}[n](i_{p},j_{\theta},k_{\phi})\\
\nonumber	
&=\frac{1}{2 \nu_{q}}
\int d\Omega_{2\leftrightarrow2}
W_1(Q_{q_1\bar{q}_1}^{q_2\bar{q}_2}(12|34)+Q_{\bar{q}_1q_1}^{q_2\bar{q}_2}(12|34))\\	
\nonumber
&=\frac{1}{\nu_{q}}
\int d\Omega_{2\leftrightarrow2}\frac{1}{8}\left[
-W_1(Q_{q_2\bar{q}_2}^{q_1\bar{q}_1}(34|12)+Q_{q_2\bar{q}_2}^{q_1\bar{q}_1}(43|12))\right.\\
\nonumber
&\left.-W_2(Q_{q_2\bar{q}_2}^{q_1\bar{q}_1}(43|21)+Q_{q_2\bar{q}_2}^{q_1\bar{q}_1}(34|21))\right.\\
\nonumber
&\left.-W_3(Q_{q_2\bar{q}_2}^{q_1\bar{q}_1}(12|34)+Q_{q_2\bar{q}_2}^{q_1\bar{q}_1}(21|34))\right.\\
\nonumber
&\left.-W_4(Q_{q_2\bar{q}_2}^{q_1\bar{q}_1}(21|43)+Q_{q_2\bar{q}_2}^{q_1\bar{q}_1}(12|43))\right]
\end{align}
(4)~For antiquark $\bar{q}_2$, note that $Q_{\bar{q}_1q_1}^{\bar{q}_2q_2}(12|34)=-Q_{q_2\bar{q}_2}^{q_1\bar{q}_1}(43|21)$, $Q_{q_1\bar{q}_1}^{\bar{q}_2q_2}(12|34)=-Q_{q_2\bar{q}_2}^{q_1\bar{q}_1}(34|21)$
\begin{align}
&C^{{2\leftrightarrow2}}_{\bar{q}_2}[n](i_{p},j_{\theta},k_{\phi})\\
\nonumber
&=\frac{1}{2 \nu_{q}}
\int d\Omega_{2\leftrightarrow2}
W_1(Q_{\bar{q}_1q_1}^{\bar{q}_2q_2}(12|34)+Q_{q_1\bar{q}_1}^{\bar{q}_2q_2}(12|34))\\	
\nonumber
&=\frac{1}{\nu_{q}}
\int d\Omega_{2\leftrightarrow2}\frac{1}{8}\left[
-W_1(Q_{q_2\bar{q}_2}^{q_1\bar{q}_1}(43|21)+Q_{q_2\bar{q}_2}^{q_1\bar{q}_1}(34|21))\right.\\
\nonumber
&\left.-W_2(Q_{q_2\bar{q}_2}^{q_1\bar{q}_1}(34|12)+Q_{q_2\bar{q}_2}^{q_1\bar{q}_1}(43|12))\right.\\
\nonumber
&\left.-W_3(Q_{q_2\bar{q}_2}^{q_1\bar{q}_1}(21|43)+Q_{q_2\bar{q}_2}^{q_1\bar{q}_1}(12|43))\right.\\
\nonumber
&\left.-W_4(Q_{q_2\bar{q}_2}^{q_1\bar{q}_1}(12|34)+Q_{q_2\bar{q}_2}^{q_1\bar{q}_1}(21|34))\right]
\end{align}

We have following conservation laws automatically proved by the discretized collision integral from the completeness relation Eq.~(\ref{eq-wCompleteness}) and sum rule Eq.~(\ref{eq-wSumRuleI}).\\
(1)~Charge conservation from completeness relation:
\begin{align}
\nonumber
&\sum_{(i_{p},j_{\theta},k_{\phi})}\left[\nu_{q}C^{{2\leftrightarrow2}}_{q_1}[n](i_{p},j_{\theta},k_{\phi})-\nu_{q}C^{{2\leftrightarrow2}}_{\bar{q}_1}[n](i_{p},j_{\theta},k_{\phi})\right]\\
\nonumber
&=\sum_{(i_{p},j_{\theta},k_{\phi})}
\int d\Omega_{2\leftrightarrow2}\frac{1}{8}
\left[(W_1-W_2)(
Q_{q_2\bar{q}_2}^{q_1\bar{q}_1}(12|34)
+Q_{q_2\bar{q}_2}^{q_1\bar{q}_1}(12|43)\right.\\
\nonumber
&\left.
-Q_{q_2\bar{q}_2}^{q_1\bar{q}_1}(21|34)
-Q_{q_2\bar{q}_2}^{q_1\bar{q}_1}(21|43))\right.\\
\nonumber
&\left.
+(W_3-W_4)(Q_{q_2\bar{q}_2}^{q_1\bar{q}_1}(34|12)
+Q_{q_2\bar{q}_2}^{q_1\bar{q}_1}(34|21)\right.\\
\nonumber
&\left.
-Q_{q_2\bar{q}_2}^{q_1\bar{q}_1}(43|12)
-Q_{q_2\bar{q}_2}^{q_1\bar{q}_1}(43|21))\right]\\
\nonumber
&=\int d\Omega_{2\leftrightarrow2}\frac{1}{8}
\left[(1-1)(
Q_{q_2\bar{q}_2}^{q_1\bar{q}_1}(12|34)
+Q_{q_2\bar{q}_2}^{q_1\bar{q}_1}(12|43)\right.\\
\nonumber
&\left.
-Q_{q_2\bar{q}_2}^{q_1\bar{q}_1}(21|34)
-Q_{q_2\bar{q}_2}^{q_1\bar{q}_1}(21|43))\right.\\
\nonumber
&\left.
+(1-1)(Q_{q_2\bar{q}_2}^{q_1\bar{q}_1}(34|12)
+Q_{q_2\bar{q}_2}^{q_1\bar{q}_1}(34|21)\right.\\
&\left.
-Q_{q_2\bar{q}_2}^{q_1\bar{q}_1}(43|12)
-Q_{q_2\bar{q}_2}^{q_1\bar{q}_1}(43|21))\right]=0
\end{align}
(2)~Total number conservation from completeness relation:
\begin{align}
\nonumber
&\sum_{(i_{p},j_{\theta},k_{\phi})}\left[\nu_{q}C^{{2\leftrightarrow2}}_{q_1}[n](i_{p},j_{\theta},k_{\phi})+\nu_{q}C^{{2\leftrightarrow2}}_{\bar{q}_1}[n](i_{p},j_{\theta},k_{\phi})\right.\\
\nonumber
&\left.+\nu_{q}C^{{2\leftrightarrow2}}_{q_2}[n](i_{p},j_{\theta},k_{\phi})+\nu_{q}C^{{2\leftrightarrow2}}_{\bar{q}_2}[n](i_{p},j_{\theta},k_{\phi})\right]\\
\nonumber
&=\sum_{(i_{p},j_{\theta},k_{\phi})}
\int d\Omega_{2\leftrightarrow2}\frac{1}{8}
(W_1+W_2-W_3-W_4)\\
\nonumber
&\times\left(
Q_{q_2\bar{q}_2}^{q_1\bar{q}_1}(12|34)
+Q_{q_2\bar{q}_2}^{q_1\bar{q}_1}(12|43)
+Q_{q_2\bar{q}_2}^{q_1\bar{q}_1}(21|43)
+Q_{q_2\bar{q}_2}^{q_1\bar{q}_1}(21|34)\right.\\
\nonumber
&\left.
-Q_{q_2\bar{q}_2}^{q_1\bar{q}_1}(34|12)
-Q_{q_2\bar{q}_2}^{q_1\bar{q}_1}(43|12)
-Q_{q_2\bar{q}_2}^{q_1\bar{q}_1}(43|21)
-Q_{q_2\bar{q}_2}^{q_1\bar{q}_1}(34|21)\right)\\
\nonumber
&=\int d\Omega_{2\leftrightarrow2}\frac{1}{8}
(1+1-1-1)\left(
Q_{q_2\bar{q}_2}^{q_1\bar{q}_1}(12|34)
+Q_{q_2\bar{q}_2}^{q_1\bar{q}_1}(12|43)\right.\\
\nonumber
&\left.
+Q_{q_2\bar{q}_2}^{q_1\bar{q}_1}(21|43)
+Q_{q_2\bar{q}_2}^{q_1\bar{q}_1}(21|34)
-Q_{q_2\bar{q}_2}^{q_1\bar{q}_1}(34|12)
-Q_{q_2\bar{q}_2}^{q_1\bar{q}_1}(43|12)\right.\\
&\left.
-Q_{q_2\bar{q}_2}^{q_1\bar{q}_1}(43|21)
-Q_{q_2\bar{q}_2}^{q_1\bar{q}_1}(34|21)\right)=0
\end{align}
(3)~Energy momentum conservation from sum rule, denoting $P_{ij}$=$(p_i,p_i\rm{cos}(\theta_j))$ with $(i,j)\in \mathbb{N}^2_{p,\theta}$, $P_{s}$=$(p_s,p_s\rm{cos}(\theta_s))$ with $s=1,2,3,4$, and notice that $W_s=W_{(i_{p},j_{\theta},k_{\phi})}(\vec{p}_s)$:
\begin{align}
\nonumber
&\sum_{(i_{p},j_{\theta},k_{\phi})}P_{ij}\left[\nu_{q}C^{{2\leftrightarrow2}}_{q_1}[n](i_{p},j_{\theta},k_{\phi})+\nu_{q}C^{{2\leftrightarrow2}}_{\bar{q}_1}[n](i_{p},j_{\theta},k_{\phi})\right.\\
\nonumber
&\left.+\nu_{q}C^{{2\leftrightarrow2}}_{q_2}[n](i_{p},j_{\theta},k_{\phi})+\nu_{q}C^{{2\leftrightarrow2}}_{\bar{q}_2}[n](i_{p},j_{\theta},k_{\phi})\right]\\
\nonumber
&=\sum_{(i_{p},j_{\theta},k_{\phi})}P_{ij}
\int d\Omega_{2\leftrightarrow2}\frac{1}{8}
(W_1+W_2-W_3-W_4)\times\\
\nonumber
&\left(
Q_{q_2\bar{q}_2}^{q_1\bar{q}_1}(12|34)
+Q_{q_2\bar{q}_2}^{q_1\bar{q}_1}(12|43)
+Q_{q_2\bar{q}_2}^{q_1\bar{q}_1}(21|43)
+Q_{q_2\bar{q}_2}^{q_1\bar{q}_1}(21|34)\right.\\
\nonumber
&\left.
-Q_{q_2\bar{q}_2}^{q_1\bar{q}_1}(34|12)
-Q_{q_2\bar{q}_2}^{q_1\bar{q}_1}(43|12)
-Q_{q_2\bar{q}_2}^{q_1\bar{q}_1}(43|21)
-Q_{q_2\bar{q}_2}^{q_1\bar{q}_1}(34|21)\right)\\
\nonumber
&=\int d\Omega_{2\leftrightarrow2}\frac{1}{8}
(P_1+P_2-P_3-P_4)\left(
Q_{q_2\bar{q}_2}^{q_1\bar{q}_1}(12|34)\right.\\
\nonumber
&\left.
+Q_{q_2\bar{q}_2}^{q_1\bar{q}_1}(12|43)
+Q_{q_2\bar{q}_2}^{q_1\bar{q}_1}(21|43)
+Q_{q_2\bar{q}_2}^{q_1\bar{q}_1}(21|34)
-Q_{q_2\bar{q}_2}^{q_1\bar{q}_1}(34|12)\right.\\
&\left.
-Q_{q_2\bar{q}_2}^{q_1\bar{q}_1}(43|12)
-Q_{q_2\bar{q}_2}^{q_1\bar{q}_1}(43|21)
-Q_{q_2\bar{q}_2}^{q_1\bar{q}_1}(34|21)\right)=0
\end{align}

\subsubsection{Phase-Space Integration}
\label{sec-phasespace}
Evaluation of the phase-space integrals can be achieved by expressing $p_{3}=p_{1}+q$ and $p_{4}=p_{2}-q$ to eliminate the momentum conservation constraint
\begin{eqnarray}
&&\int d\Omega_{2\leftrightarrow2} = \int\frac{d^3p_1}{(2\pi)^3}\int\frac{d^3p_2}{(2\pi)^3} \int\frac{d^3q}{(2\pi)^3} \\
\nonumber
&&\times\frac{1}{16 E_{p_1} E_{p_{2}} E_{p_{3}} E_{p_{4}}}(2\pi) \delta(E_{p_1} +E_{p_{2}} -E_{p_{3}}- E_{p_{4}})
\end{eqnarray}
with 
\begin{eqnarray}
E_{p_1}=p_1\;, \qquad E_{p_3}=\sqrt{p_1^2+q^2+2p_1 q \cos(\theta_{1q})}\;,  \nonumber \\
E_{p_2}=p_2\;, \qquad E_{p_4}=\sqrt{p_2^2+q^2-2p_2 q \cos(\theta_{2q})}\;.
\end{eqnarray}
Similarly, to eliminate the energy conservation constraint we follow the standard trick to parameterize the integral in terms of the energy transfer $\omega$
\begin{eqnarray}
&&(2\pi) \delta(E_{p_1} +E_{p_{2}} -E_{p_{3}}- E_{p_{4}}) \\
\nonumber
&&= (2\pi) \int d\omega~\delta(E_{p_1} +\omega -E_{p_{3}})~\delta(E_{p_2} - \omega -E_{p_{4}})
\end{eqnarray}
By expressing the arguments of the $\delta$ functions as
\begin{eqnarray}
\nonumber
p_1 +\omega = \sqrt{p_1^2+q^2+2p_1 q \cos(\theta_{1q})}\;,  \\
p_2 -\omega = \sqrt{p_2^2+q^2-2p_2 q \cos(\theta_{2q})}\;,
\end{eqnarray}
this can be re-cast into a constraint for the angles
\begin{eqnarray}
\cos(\theta_{1q})= \frac{w}{q} + \frac{w^2-q^2}{2p_1q}\;,
\cos(\theta_{2q})= \frac{w}{q} - \frac{w^2-q^2}{2p_2q}\;,
\end{eqnarray}
which has a valid solution for
\begin{eqnarray}
|\omega| < q\;, \qquad  p_{1}>\frac{q-\omega}{2}\;, \qquad p_{2} > \frac{q+\omega}{2}\;.
\end{eqnarray}
Evaluating the Jacobian of this transformation as
\begin{eqnarray}
\nonumber
\frac{\partial}{\partial \cos(\theta_{1q})}  \left( p_1 +\omega - \sqrt{p_1^2+q^2+2p_1 q \cos(\theta_{1q})}  \right) = \frac{-p_1 q}{E_{p_3}}\;,\\
\nonumber
\frac{\partial}{\partial \cos(\theta_{2q})}  \left( p_2 - \omega - \sqrt{p_2^2+q^2-2p_2 q \cos(\theta_{2q})} \right) = \frac{+p_2 q}{E_{p_4}}\;,\\
\end{eqnarray}
the phase-space integral can then be re-cast into the form
\begin{eqnarray}
&&\int d\Omega_{2\leftrightarrow2}= (2\pi)~\int\frac{d^3p_1}{(2\pi)^3}\int\frac{d^3p_2}{(2\pi)^3} \int\frac{d^3q}{(2\pi)^3}\\
\nonumber
&& \int d\omega  \frac{1}{16 p_1^2 p_2^2 q^2} \theta(q-|\omega|) \theta\left(p_1-\frac{q-\omega}{2}\right) \theta\left(p_2-\frac{q+\omega}{2}\right) \\
\nonumber
&&\delta\left(\frac{w}{q} + \frac{w^2-q^2}{2p_1q} - \cos(\theta_{1q})\right) \delta\left(\frac{w}{q} - \frac{w^2-q^2}{2p_2q} - \cos(\theta_{2q})\right)\;,
\end{eqnarray}

Clearly the most straightforward way to implement the constraints is to perform the $q$ and $\omega$ integrations prior to the $p_1$ and $p_2$ integrations, such that the vectors $p_1$ and $p_2$ can be parameterized in terms of spherical coordinates in a right-handed orthonormal system spanned by the unit vectors
\begin{eqnarray}
\nonumber
&&\vec{e}_{1}=\vec{e}_{q}=\frac{\vec{q}}{|\vec{q}|}, ~~~~~
\vec{e}_{2}=\frac{\vec{e}_{n}-\cos(\theta_{q}) \vec{e}_{q}}{|\vec{e}_{n}-\cos(\theta_{q}) \vec{e}_{q}|}, \\
&&\vec{e}_{3}=\frac{\vec{e}_{q}\times \left(\vec{e}_{n}-\cos(\theta_{q}) \vec{e}_{q}\right)}{|\vec{e}_{q} \times \left(\vec{e}_{n}-\cos(\theta_{q}) \vec{e}_{q}\right)|},
\end{eqnarray}
where $\vec{e}_{n}$ denotes the preferred axis of the coordinate system, such that
\begin{eqnarray}
\nonumber
&&\vec{p}_{1}= p_{1} \left[ \cos(\theta_{1q}) \vec{e}_{q} + \sin(\theta_{1q}) \cos(\phi_{1}) \frac{\vec{e}_{n}-\cos(\theta_{q}) \vec{e}_{q}}{|\vec{e}_{n}-\cos(\theta_{q}) \vec{e}_{q}|}\right.\\
\nonumber
&&\left.+ \sin(\theta_{1q}) \sin(\phi_{1})  \frac{\vec{e}_{q}\times \left(\vec{e}_{n}-\cos(\theta_{q}) \vec{e}_{q}\right)}{|\vec{e}_{q} \times \left(\vec{e}_{n}-\cos(\theta_{q}) \vec{e}_{q}\right)|} \right],\\
\nonumber
&&\vec{p}_{2}= p_{2} \left[ \cos(\theta_{2q}) \vec{e}_{q} + \sin(\theta_{2q}) \cos(\phi_{2}) \frac{\vec{e}_{n}-\cos(\theta_{q}) \vec{e}_{q}}{|\vec{e}_{n}-\cos(\theta_{q}) \vec{e}_{q}|} \right.\\
&&\left.+  \sin(\theta_{2q}) \sin(\phi_{2})  \frac{\vec{e}_{q}\times \left(\vec{e}_{n}-\cos(\theta_{q}) \vec{e}_{q}\right)}{|\vec{e}_{q} \times \left(\vec{e}_{n}-\cos(\theta_{q}) \vec{e}_{q}\right)|} \right]\;, \nonumber \\
\end{eqnarray}
which allows for a straightforward evaluation of the constraints on $\cos(\theta_{1q})$ and $\cos(\theta_{2q})$ yielding
\begin{eqnarray}\nonumber
&&\int d\Omega_{2\leftrightarrow2}=\frac{1}{16 (2\pi)^8}\int_{0}^{\infty} dq \int_{-1}^{1} d\cos(\theta_q) \int_{0}^{2\pi} d\phi_{q} \int_{-q}^{q} d{\omega}\\ &&\times\int_{\text{max}(\frac{q-\omega}{2},0)}^{\infty} dp_{1} \int_{\text{max}(\frac{q+\omega}{2},0)}^{\infty} dp_{2} \int_{0}^{2\pi} d\phi_{1} \int_{0}^{2\pi} d\phi_{2}.
\end{eqnarray}
Specifically in terms of these coordinates the angles of $p_1$ and $p_2$ with respect to to the anisotropy direction are given by
\begin{eqnarray}
\nonumber
\cos(\theta_1)&=&\cos(\theta_{1q})\cos(\theta_{q})+\sin(\theta_{1q}) \sin(\theta_{q}) \cos(\phi_{1}),\\
\nonumber
\cos(\theta_2)&=&\cos(\theta_{2q})\cos(\theta_{q})+\sin(\theta_{2q}) \sin(\theta_{q}) \cos(\phi_{2}).\\
\end{eqnarray}
Note that if we wish to impose constraints on the magnitudes of $p_1^{\rm min}<p_1<p_1^{\rm max}$  and $p_2^{\rm min}<p_2<p_2^{\rm max}$, then the corresponding phase-space constraints take the form
\begin{eqnarray}
\nonumber
&&\text{max}\left(\frac{q-\omega}{2},p_1^{\rm min}\right) <p_1 < p_1^{\rm max},\\
&&\text{max}\left(\frac{q+\omega}{2},p_2^{\rm min}\right) <p_2 < p_2^{\rm max}.
\end{eqnarray}

\subsection{Inelastic Collision Integrals}
\label{sec-discretization-inelastic}
\subsubsection{Collinear Radiation}
\label{subsec-collinear}
The inelastic collision integral for particle ``a'' with momentum $\vec{p}_1$ in splitting process $a \rightarrow b,c$ ($p_{1}\leftrightarrow p_{2},p_{3}$) and joining process $a,b \rightarrow c$ ($p_{1},p_{2}\leftrightarrow p_{3}$) reads:

\begin{eqnarray}
\label{eq-cint-inelastic-app}
&&C^{{1\leftrightarrow2}}_a[f](\vec{p}_1)\\
\nonumber
&&=\frac{1}{2 \nu_{a}}\frac{1}{2 E_{p_1}}\sum_{bc}\int d\Pi_{1\leftrightarrow 2}^{a\leftrightarrow bc}
|\mathcal{M}_{bc}^{a}(\vec{p}_1|\vec{p}_2,\vec{p}_3)|^2F_{bc}^{a}(\vec{p}_1|\vec{p}_2,\vec{p}_3)\\
\nonumber
&&+\frac{1}{\nu_{a}}\frac{1}{2 E_{p_1}}\int d\Pi_{1\leftrightarrow 2}^{ab\leftrightarrow c}
|\mathcal{M}_{c}^{ab}(\vec{p}_1,\vec{p}_2|\vec{p}_3)|^2F_{c}^{ab}(\vec{p}_1,\vec{p}_2|\vec{p}_3)\\
\nonumber
&&=\frac{1}{2 \nu_{a}}\frac{1}{2 E_{p_1}}\sum_{bc}\int d\Pi_{1\leftrightarrow 2}^{a\leftrightarrow bc}
|\mathcal{M}_{bc}^{a}(\vec{p}_1|\vec{p}_2,\vec{p}_3)|^2F_{bc}^{a}(\vec{p}_1|\vec{p}_2,\vec{p}_3)\\
\nonumber
&&-\frac{1}{\nu_{a}}\frac{1}{2 E_{p_1}}\int d\Pi_{1\leftrightarrow 2}^{ab\leftrightarrow c}
|\mathcal{M}_{ab}^{c}(\vec{p}_3|\vec{p}_1,\vec{p}_2)|^2F_{ab}^{c}(\vec{p}_3|\vec{p}_1,\vec{p}_2)
\end{eqnarray}
with measure
\begin{eqnarray}
\nonumber
&&\int d\Pi_{1\leftrightarrow2}^{a\leftrightarrow bc}= \int\frac{d^3p_2}{(2\pi)^3} \frac{1}{2E_{p_2}} \int\frac{d^3p_3}{(2\pi)^3} \frac{1}{2E_{p_3}}\\
\nonumber
&&\times(2\pi)^4\delta^{(4)}({p}_1-{p}_2-{p}_3).\\
\nonumber
&&\int d\Pi_{1\leftrightarrow2}^{ab\leftrightarrow c}= \int\frac{d^3p_2}{(2\pi)^3} \frac{1}{2E_{p_2}} \int\frac{d^3p_3}{(2\pi)^3} \frac{1}{2E_{p_3}}\\
&&\times(2\pi)^4\delta^{(4)}({p}_1+{p}_2-{p}_3)
\end{eqnarray}
$|\mathcal{M}_{bc}^{a}(\vec{p}_1|\vec{p}_2,\vec{p}_3)|^2$ and $|\mathcal{M}_{c}^{ab}(\vec{p}_1,\vec{p}_2|\vec{p}_3)|^2$ are the matrix element squares for process ``$a\leftrightarrow b,c$'' and ``$a,b\leftrightarrow c$''.
$F_{cd}^{a}(\vec{p}_1|\vec{p}_3,\vec{p}_4)$ and $F_{c}^{ab}(\vec{p}_1,\vec{p}_2|\vec{p}_3)$
are the statistical factors.
Similarly, we define $Q$-Factors
\begin{eqnarray}
\nonumber
	\begin{aligned}
	&Q_{bc}^{a}(1|23)=Q_{bc}^{a}(\vec{p}_1|\vec{p}_2,\vec{p}_3)=|\mathcal{M}_{bc}^{a}(\vec{p}_1|\vec{p}_2,\vec{p}_3)|^2F_{bc}^{a}(\vec{p}_1|\vec{p}_2,\vec{p}_3)\\
	&Q_{c}^{ab}(12|3)=Q_{c}^{ab}(\vec{p}_1,\vec{p}_2|\vec{p}_3)=|\mathcal{M}_{c}^{ab}(\vec{p}_1,\vec{p}_2|\vec{p}_3)|^2F_{c}^{ab}(\vec{p}_1,\vec{p}_2|\vec{p}_3)
	\end{aligned}\\
\end{eqnarray}
which have the following symmetries
\begin{eqnarray}
\label{eq-inelastic-factor-sym}
Q_{bc}^{a}(\vec{p}_1|\vec{p}_2,\vec{p}_3) =
Q_{cb}^{a}(\vec{p}_1|\vec{p}_3,\vec{p}_2) = 
-Q_{a}^{bc}(\vec{p}_2,\vec{p}_3|\vec{p}_1).
\end{eqnarray}

As was suggested by AMY~\cite{Arnold:2002zm}, in a nearly collinear inelastic scattering, after integration all soft kick particles from the medium, and assuming massless kinematics, the collision integral in Eq.(\ref{eq-cint-inelastic-app}) can be recast into 1D collinear process (we denote $p$=$|\vec{p}|$$\approx$$p_{\rm \|}$)
\begin{eqnarray}
\label{eq-ccint-inelastic}
&&C^{{1\leftrightarrow2}}_a[f]({p}_1)\\
\nonumber
&&=\frac{(2\pi)^3}{2\nu_{a}p_1^2}\sum_{bc}\int_0^{\infty} dp_2dp_3
\delta(p_1-p_2-p_3)\\
\nonumber
&&\times\gamma_{bc}^{a}(p_1|p_2,p_3)F_{bc}^{a}(p_1|p_2,p_3)\\
\nonumber
&&-\frac{(2\pi)^3}{\nu_{a}p_1^2}\int_0^{\infty} dp_2dp_3
\delta(p_1+p_2-p_3)\\
\nonumber
&&\times\gamma_{ab}^{c}(p_3|p_1,p_2)F_{ab}^{c}(p_3|p_1,p_2)
\end{eqnarray}
where in the statistical distribution, we only put the momentum components of $p_2$=$\vec{p}_2\cdot{\vec{p}_1}/|\vec{p}_1|$ and $p_3$=$\vec{p}_3\cdot{\vec{p}_1}/|\vec{p}_1|$ which is in parallel to the $p_1$, while $p_1$=$|\vec{p}_1|$ is its full momentum. 

we parameterize the phase-space for inelastic rates by setting $p_1=p$, $p_2=zp$, $p_3=\bar{z}p$ in the first term
\begin{eqnarray}
\nonumber
&&\int_{0}^{\infty}dp_2dp_3\delta(p_1-p_2-p_3)\gamma^{a}_{bc}(p_1|p_2,p_3)\\
&&= p\int_{0}^{1}dzd\bar{z}\delta(1-z-\bar{z})\gamma^{a}_{bc}\Big(p | zp,\bar{z}p\Big)\;,
\end{eqnarray}
and similarly by setting $p_1=p$, $p_3=p_1/z$, $p_2=\bar{z}p_3$ in the second term
\begin{eqnarray}
\nonumber
&&\int_{0}^{\infty}dp_2dp_3~\delta(p_1+p_2-p_3)\gamma^{c}_{ab}(p_3|p_1,p_2)\\
&&=p\int_{0}^{1}dzd\bar{z}\frac{1}{z^2}\delta(1-z-\bar{z})\gamma^{c}_{ab}\Big(\frac{p}{z}|p,\frac{\bar{z}}{z}p\Big),
\end{eqnarray}
such that upon defining the effective rate
\begin{eqnarray}
\frac{d\Gamma_{bc}^{a}}{dz}\left(p,z\right)=\frac{(2\pi)^3}{\nu_{a}p}\gamma^{a}_{bc}\Big(p|zp,\bar{z}p\Big)
\end{eqnarray}
we arrive at
\begin{eqnarray}
&&C^{{1\leftrightarrow2}}_a[f](p)\\
\nonumber
&&=\frac{(2\pi)^3}{2\nu_{a}p}\sum_{bc}\int_0^1dzd\bar{z}
\delta(1-z-\bar{z})\\
\nonumber
&&\times\gamma_{bc}^{a}\big(p|zp,\bar{z}p\big)F_{bc}^{a}(p|zp,\bar{z}p)\\
\nonumber
&-&\frac{(2\pi)^3}{\nu_{a}p}\int_0^1dzd\bar{z}\frac{1}{z^2}
\delta(1-z-\bar{z})\\
\nonumber
&&\times\gamma_{ab}^{c}\big(\frac{p}{z}|p,\frac{\bar{z}}{z}p\big)F_{ab}^{c}(\frac{p}{z}|p,\frac{\bar{z}}{z}p)\\
\nonumber
&=&\frac{1}{2\nu_{a}}\int_0^1dz\sum_{bc}\left[
\frac{d\Gamma_{bc}^{a}}{dz}\big(p_1,z\big)\nu_{a}F_{bc}^{a}(p|zp,\bar{z}p)\right.\\
\nonumber
&&\left.-\frac{2}{z^3}
\frac{d\Gamma_{ab}^{c}}{dz}\big(\frac{p}{z},z\big)\nu_{c}F_{ab}^{c}(\frac{p}{z}|p,\frac{\bar{z}}{z}p)\right]
\end{eqnarray}
where $\bar{z}=1-z$. We may further symmetrize this expression by splitting the second term into two terms integrating over $z$ and $1-z$ such that
\begin{eqnarray}
\label{eq-ccint-inelastic-cast-app}
&&C^{{1\leftrightarrow2}}_a[f](p)=\frac{1}{2\nu_{a}}\int_0^1dz
\left[
\sum_{bc}\frac{d\Gamma_{bc}^{a}}{dz}\big(p,z\big)\nu_{a}F_{bc}^{a}(p|zp,\bar{z}p)\right.\\
\nonumber
&&\left.-
\frac{1}{z^3}\frac{d\Gamma_{ab}^{c}}{dz}\big(\frac{p}{z},z\big)\nu_{c}F_{ab}^{c}(\frac{p}{z}|p,\frac{\bar{z}}{z}p)-
\frac{1}{\bar{z}^3}\frac{d\Gamma_{ab}^{c}}{dz}\big(\frac{p}{\bar{z}},\bar{z}\big)\nu_{c}F_{ab}^{c}(\frac{p}{\bar{z}}|p,\frac{z}{\bar{z}}p)
\right]
\end{eqnarray}

\subsubsection{Discretization of Collinear Form}

The discretized form of collinear collision integral in Eq.~(\ref{eq-ccint-inelastic-cast-app}) follows the transformation in Eq.~(\ref{eq-fton}):
\begin{eqnarray}
\nonumber
&&C^{{1\leftrightarrow2}}_a[n](i_{p},j_{\theta},k_{\phi})=
\int\frac{d^3p}{(2\pi)^3}W_{(i_{p},j_{\theta},k_{\phi})}(\vec{p})C^{{1\leftrightarrow2}}_a[f](p)\\
\nonumber
&&=\frac{1}{4\pi^2}\int_0^{\infty}p^2dp\int_{-1}^{+1}d{\rm cos(\theta)}W_{(i_{p},j_{\theta},k_{\phi})}(\vec{p})C^{{1\leftrightarrow2}}_a[f](p)\\
\nonumber
&&
=\frac{1}{8\pi^2\nu_{a}}\int_0^{\infty}p^2dp\int_{-1}^{+1}d{\rm cos(\theta)}
\int_0^1dzW_{(i_{p},j_{\theta},k_{\phi})}(\vec{p})\\
\nonumber
&&\times\left[
\sum_{bc}\frac{d\Gamma_{bc}^{a}}{dz}\big(p,z\big)\nu_{a}F_{bc}^{a}(p|zp,\bar{z}p)\right.\\
\nonumber
&&\left.-
\frac{1}{z^3}\frac{d\Gamma_{ab}^{c}}{dz}\big(\frac{p}{z},z\big)\nu_{c}F_{ab}^{c}(\frac{p}{z}|p,\frac{\bar{z}}{z}p)\right.\\
\nonumber
&&\left.-
\frac{1}{\bar{z}^3}\frac{d\Gamma_{ab}^{c}}{dz}\big(\frac{p}{\bar{z}},\bar{z}\big)\nu_{c}F_{ab}^{c}(\frac{p}{\bar{z}}|p,\frac{z}{\bar{z}}p)
\right]\\
\nonumber
&&
=\frac{1}{8\pi^2\nu_{a}}\int_0^{\infty}p^2dp\int_{-1}^{+1}d{\rm cos(\theta)}
\int_0^1dz\\
\nonumber
&&\times
\left[
\sum_{bc}\frac{d\Gamma_{bc}^{a}}{dz}\big(p,z\big)\nu_{a}F_{bc}^{a}(p|zp,\bar{z}p)W_{(i_{p},j_{\theta},k_{\phi})}(p)\right.
\\
\nonumber
&&\left.-
\frac{d\Gamma_{ab}^{c}}{dz}\big(p,z\big)\nu_{c}F_{ab}^{c}(p|zp,\bar{z}p)W_{(i_{p},j_{\theta},k_{\phi})}(zp)\right.\\
&&\left.-
\frac{d\Gamma_{ab}^{c}}{dz}\big(p,\bar{z}\big)\nu_{c}F_{ab}^{c}(p|\bar{z}p,zp)W_{(i_{p},j_{\theta},k_{\phi})}(\bar{z}p)
\right]
\end{eqnarray}
where we used a redefinition of the integrals $\int \frac{d^3p}{(2\pi)^3} \frac{1}{z^3}=\int \frac{d^3p'}{(2\pi)^3}$ (and similarly for $\bar{z}$) to bring all terms into the same form.

Note that
\begin{eqnarray}
\nonumber
&&\frac{d\Gamma_{ab}^{c}}{dz}\big(p,\bar{z}\big)=
\frac{d\Gamma_{ba}^{c}}{dz}\big(p,z\big),\\ 
&&F_{ab}^{c}(p|\bar{z}p,zp)=
F_{ba}^{c}(p|zp,\bar{z}p).
\end{eqnarray}

Define a $\Delta$-Factor for inelastic collision
\begin{eqnarray}
\Delta_{ab}^{c}(p,z)=\frac{d\Gamma_{ab}^{c}}{dz}\big(p,z\big)\nu_{c}F_{ab}^{c}(p|zp,\bar{z}p)
\end{eqnarray}
which has the following symmetry
\begin{eqnarray}
\Delta_{ab}^{c}(p,z)=\Delta_{ba}^{c}(p,\bar{z})
\end{eqnarray}

With a refinement of the measure and weight in the integral $d\Pi$=$\frac{1}{8\pi^2}\int_0^{\infty}p^2dp\int_{-1}^{+1}d{\rm cos(\theta)}\int_0^1dz$ and $W(p)$=$W_{(i_{p},j_{\theta},k_{\phi})}(p)$, we can rewrite the form into
\begin{widetext}
\begin{eqnarray}
\label{eq-inelastic-cfinal}
\boxed{C^{{1\leftrightarrow2}}_a[n](i_{p},j_{\theta},k_{\phi})
=\frac{1}{\nu_{a}}\int d\Pi\left[\sum_{bc}\Delta_{bc}^{a}(p,z)W(p)-\Delta_{ab}^{c}(p,z)W(zp)-\Delta_{ba}^{c}(p,z)W(\bar{z}p)\right]}
\end{eqnarray}
\end{widetext}

For $b\neq c$, we have the following useful equivalent form of Eq.(\ref{eq-inelastic-cfinal}) for convenience:
\begin{eqnarray}
\nonumber
&&C^{{1\leftrightarrow2}}_a[n](i_{p},j_{\theta},k_{\phi})=\\
\nonumber
&&=\frac{1}{\nu_{a}}\int d\Pi\left[
\left(\Delta_{bc}^{a}(p,z)+\Delta_{cb}^{a}(p,z)\right)W(p)\right.\\
\nonumber
&&\left.-
\Delta_{ab}^{c}(p,z)W(zp)-
\Delta_{ba}^{c}(p,z)W(\bar{z}p)
\right]\\
\nonumber
&&=\frac{1}{\nu_{a}}\int d\Pi\left[
\left(\Delta_{bc}^{a}(p,z)+\Delta_{bc}^{a}(p,\bar{z})\right)W(p)\right.\\
&&\left.-
\Delta_{ab}^{c}(p,z)W(zp)-
\Delta_{ab}^{c}(p,\bar{z})W(\bar{z}p)
\right]
\end{eqnarray}

For $b=c$, we have
\begin{eqnarray}
\nonumber
&&C^{{1\leftrightarrow2}}_a[n](i_{p},j_{\theta},k_{\phi})=\\
\nonumber
&&=\frac{1}{\nu_{a}}\int d\Pi\left[
\Delta_{cc}^{a}(p,z)W(p)\right.\\
\nonumber
&&\left.-
\Delta_{ac}^{c}(p,z)W(zp)-
\Delta_{ca}^{c}(p,z)W(\bar{z}p)
\right]\\
\nonumber
&&=\frac{1}{\nu_{a}}\int d\Pi\left[
\Delta_{cc}^{a}(p,z)W(p)\right.\\
&&\left.-
\Delta_{ac}^{c}(p,z)W(zp)-
\Delta_{ac}^{c}(p,\bar{z})W(\bar{z}p)
\right]
\end{eqnarray}
where we use the fact that for any function $f(p,z)$, we have
\begin{eqnarray}
\int_0^1dzf(p,z)=\int_0^1dzf(p,1-z)
\end{eqnarray}

Now we list the discretized collision integrals for all inelastic processes and prove the exact conservation taking the most complicated process $g\leftrightarrow q\bar{q}$ for example.
\paragraph{Process $g\leftrightarrow q\bar{q}$}
The available collision integrals are listed below\\

(1)~For gluon $\gl$ $(\qu\qb=u\bar{u},d\bar{d},s\bar{s})$
\begin{eqnarray}
&&C^{{1\leftrightarrow2}}_{\gl}[n](i_{p},j_{\theta},k_{\phi})\\
\nonumber
&&=
\frac{1}{\nu_{\gl}}\int d\Pi\left[
(\Delta_{\qu\qb}^{\gl}(p,z)
+\Delta_{\qb\qu}^{\gl}(p,z))W(p)\right]
\end{eqnarray}

(2)~For quark $\qu$ $(\qu=u,d,s)$
\begin{eqnarray}
&&C^{{1\leftrightarrow2}}_{\qu}[n](i_{p},j_{\theta},k_{\phi})\\
\nonumber
&&=
\frac{1}{\nu_{\qu}}\int d\Pi\left[-
\Delta_{\qu\qb}^{\gl}(p,z)W(zp)-
\Delta_{\qb\qu}^{\gl}(p,z)W(\bar{z}p)\right]
\end{eqnarray}

(3)~For antiquark $\qb$ $(\qb=\bar{u},\bar{d},\bar{s})$
\begin{eqnarray}
&&C^{{1\leftrightarrow2}}_{\qb}[n](i_{p},j_{\theta},k_{\phi})\\
\nonumber
&&=
\frac{1}{\nu_{\qb}}\int d\Pi\left[-
\Delta_{\qb\qu}^{\gl}(p,z)W(zp)-
\Delta_{\qu\qb}^{\gl}(p,z)W(\bar{z}p)\right]
\end{eqnarray}
Again, we denote $P_{ij}$=$(p_i,p_i\rm{cos}(\theta_j))=p_i\Theta_j$ with $(i,j)\in \mathbb{N}^2_{p,\theta}$.
Since collinear collision has the same scattering angle for in and out particles, the particle $a$, $b$, $c$ has energy and longitudinal momentum $P=(p,p\rm{cos}(\theta))$, $zP=(zp,zp\rm{cos}(\theta))$, $\bar{z}P=(\bar{z}p,\bar{z}p\rm{cos}(\theta))$. The collision integral has the following conservation laws automatically proved by the discretized collision integral from the completeness relation Eq.~(\ref{eq-wCompleteness}) and sum rule Eq.~(\ref{eq-wSumRuleI}).\\

(1)~Charge conservation for $\gl\rightarrow\qu\qb$ $(\qu\qb=u\bar{u},d\bar{d},s\bar{s})$
\begin{eqnarray}
\nonumber
&&\sum_{(i_{p},j_{\theta},k_{\phi})}\left[
\nu_{\qu}C^{{1\leftrightarrow2}}_{\qu}[n]-
\nu_{\qb}C^{{1\leftrightarrow2}}_{\qb}[n]
\right]\\
\nonumber
&&=\sum_{(i_{p},j_{\theta},k_{\phi})}\int d\Pi\left[\left(\Delta_{\qb\qu}^{\gl}(p,z)-
\Delta_{\qu\qb}^{\gl}(p,z)\right)\left(W(zp)-W(\bar{z}p)\right)\right]\\
&&=\int d\Pi\left[\left(\Delta_{\qb\qu}^{\gl}(p,z)-
\Delta_{\qu\qb}^{\gl}(p,z)\right)\left(1-1\right)\right]
=0
\end{eqnarray}

(2)~Energy momentum conservation for $\gl\rightarrow\qu\qb$ $(\qu\qb=u\bar{u},d\bar{d},s\bar{s})$
\begin{eqnarray}
\nonumber
&&\sum_{(i_{p},j_{\theta},k_{\phi})}P_{ij}\left[
\nu_{\gl}C^{{1\leftrightarrow2}}_{\gl}[n]+
\nu_{\qu} C^{{1\leftrightarrow2}}_{\qu}[n]+
\nu_{\qb} C^{{1\leftrightarrow2}}_{\qb}[n]
\right]\\
\nonumber
&&\sum_{(i_{p},j_{\theta},k_{\phi})}P_{ij}
\int d\Pi\left[
(\Delta_{\qu\qb}^{\gl}(p,z)
+\Delta_{\qb\qu}^{\gl}(p,z))\right.\\
&&\left.\times\left(W(p)-W(zp)-W(\bar{z}p)\right)\right]\\
\nonumber
&&=\int d\Pi\left[
(\Delta_{\qu\qb}^{\gl}(p,z)
+\Delta_{\qb\qu}^{\gl}(p,z))
\left(P-zP-\bar{z}P\right)\right]=0
\end{eqnarray}

\paragraph{Process $g\leftrightarrow gg$}
The gluon collision integral is
\begin{eqnarray}
&&C^{{1\leftrightarrow2}}_{\gl}[n](i_{p},j_{\theta},k_{\phi})\\
\nonumber
&&=\frac{1}{\nu_{\gl}}\int d\Pi\left[
\Delta_{\gl\gl}^{\gl}(p,z)(W(p)-W(zp)-W(\bar{z}p))\right]
\end{eqnarray}

\paragraph{Process $q\leftrightarrow gq$}
The available collision integrals for partons are listed below\\

(1)~For gluon $\gl$ $(\qu=u,d,s,\bar{u},\bar{d},\bar{s})$
\begin{eqnarray}
&&C^{{1\leftrightarrow2}}_{\gl}[n](i_{p},j_{\theta},k_{\phi})\\
\nonumber
&&=
\frac{1}{\nu_{\gl}}\int d\Pi\left[-
\Delta_{\gl\qu}^{\qu}(p,z)W(zp)-
\Delta_{\qu\gl}^{\qu}(p,z)W(\bar{z}p)\right]
\end{eqnarray}

(2)~For quark/anti-quark $\qu$ $(\qu=u,d,s,\bar{u},\bar{d},\bar{s})$
\begin{eqnarray}
&&C^{{1\leftrightarrow2}}_{\qu}[n](i_{p},j_{\theta},k_{\phi})\\
\nonumber
&&=
\frac{1}{\nu_{\qu}}\int d\Pi\left[
(\Delta_{\qu\gl}^{\qu}(p,z)
+\Delta_{\gl\qu}^{\qu}(p,z))W(p)\right.\\
\nonumber
&&\left.-
\Delta_{\qu\gl}^{\qu}(p,z)W(zp)-
\Delta_{\gl\qu}^{\qu}(p,z)W(\bar{z}p)\right]
\end{eqnarray}

\subsection{Longitudinal Expansion Integrals}
\label{chap-expansion}
In a longitudinally expanding system, there is an additional contribution to the collision integral which can be expressed in the form of a collision integral (see Eq.(\ref{eq-bolzmannExp}))
\begin{eqnarray}
&&C^{\rm z-exp}_a[f](\vec{p},t)=-\frac{p_{\|}}{t} \frac{\partial f_a(\vec{p},t)}{\partial {p_{\|}}}\\
\nonumber
&&=-\frac{p\cos(\theta)}{t}\left[\cos(\theta)\frac{\partial f_a(\vec{p},t)}{\partial p}+\frac{\sin^2(\theta)}{p}\frac{\partial f_a(\vec{p},t)}{\partial\cos(\theta)}\right].
\end{eqnarray}

Thus the discretized form of the collisions integral for azimuthal-isotropic medium can be evaluated as
\begin{eqnarray}
&&C^{\rm z-exp}_a[n](i_{p},j_{\theta})\\
\nonumber
&&=-\frac{1}{(2\pi)^2} \int_{0}^{\infty} dp~p^2~\int_{-1}^{1} d\cos(\theta)
w_{i}^{(p)}(p)w_{j}^{(\theta)}(\theta)\\
\nonumber
&&\frac{p\cos(\theta)}{t}\left[\cos(\theta)\frac{\partial f_a(\vec{p},t)}{\partial p}+\frac{\sin^2(\theta)}{p}\frac{\partial f_a(\vec{p},t)}{\partial\cos(\theta)}\right].
\end{eqnarray}

Note from Eq.~(\ref{eq-ntof}), we can re-express the distribution $f(\vec{p},t)$ in continuous domain in terms of a summation of its neighbouring discretized forms
\begin{eqnarray}
\nonumber
&&f(\vec{p},t)
=\frac{1}{p^{l}}\sum_{\alpha,\beta}
w_{\alpha}^{(p)}(p)w_{\beta}^{(\theta)}(\theta)\frac{(2\pi)^2}{A_{(\alpha)}^{(2-l)}A_{(\beta)}}
n(\alpha,\beta),\\
&&((\alpha,\beta)\in\mathbb{N}^2_{p,\theta}).
\end{eqnarray}

We approximate the $f(\vec{p}\in U_{(i_{p},j_{\theta})})$
with $f(\vec{p}\in U_{p})$ where
$U_{(i_{p},j_{\theta})}
=(p_{i-1},p_{i+1})\times(\theta_{j-1},\theta_{j+1})$
and 
$U_{p}
=(p_{i},p_{i+1})\times(\theta_{j},\theta_{j+1}).$
Then we can construct the collision term in discretized grids with derivative of weight functions
\begin{widetext}
\begin{eqnarray}
\boxed{
\begin{aligned}
\nonumber
&C^{\rm z-exp}[n](i_{p},j_{\theta})=M^{\rm z-exp}(_{i_{p},j_{\theta}}^{\alpha,\beta})n(\alpha,\beta)\\
\nonumber
&=-\sum_{(\alpha,\beta)\in\mathbb{N}_{p,\theta}^2}\int\frac{d^3p}{(2\pi)^3}\left(\frac{(2\pi)^2p\cos(\theta)}{tA_{(\alpha)}^{(2-l)}A_{(\beta)}}\right)
w_{i}^{(p)}(p)w_{j}^{(\theta)}(\theta)\left[\cos(\theta)\frac{\partial}{\partial p}\left(\frac{w_{\alpha}^{(p)}}{p^{l}}\right)w_{\beta}^{(\theta)}(\theta)
+\frac{\sin^2(\theta)}{p}\left(\frac{w_{\alpha}^{(p)}}{p^{l}}\right)\frac{\partial w_{\beta}^{(\theta)}(\theta)}{\partial\cos(\theta)}\right]n(\alpha,\beta)\\
\nonumber
&{\rm with~derivatives}\\
\nonumber
&\frac{\partial}{\partial p}\left(\frac{w_{\alpha}^{(p)}(p)}{p^l}\right)=
\frac{1}{p^{l+1}}\left[
\frac{(l-1)p-l p_{\alpha+1}}{p_{\alpha+1}-p_{\alpha}} \chi_{[p_{\alpha},p_{\alpha+1}]}(p)
+\frac{l p_{\alpha-1}-(l-1)p}{p_{\alpha}-p_{\alpha-1}} \chi_{[p_{\alpha-1},p_{\alpha}]}(p)\right]\\
\nonumber
&\frac{\partial w_{\beta}^{(\theta)}(\theta)}{\partial {\rm cos}(\theta)}=\left[
\frac{-1}{{\rm cos}(\theta_{\beta+1})-{\rm cos}(\theta_{\beta})} \chi_{[\theta_{\beta},\theta_{\beta+1}]}(\theta)
+\frac{1}{{\rm cos}(\theta_{\beta})-{\rm cos}(\theta_{\beta-1})} \chi_{[\theta_{\beta-1},\theta_{\beta}]}(\theta)\right]\\
\end{aligned}
}\\
\end{eqnarray}
\end{widetext}
The integrated out expression is analytic but lengthy, and we don't list it here.
\section{Evolution Algorithm}
\label{sec-algorithm3}
\subsection{Grid Setup}
\label{sec-grids}
We discussed the discretized algorithm in Sec.~\ref{sec-algorithm} and Sec.~\ref{sec-discretization} without specifying the choice of grids distribution in the phase space since the algorithm is generally independent of the choice. However, since the particle distributions are generally accumulated at low-momentum and eliminated at high-momentum near equilibrium, an exponential distribution of grid points along momentum $p$ is preferred. Similarly, the system is close to isotropic near equilibrium, thus a linear distribution of grid points along angle $\theta$ is preferred. With those considerations, we setup the grids as
\begin{eqnarray}
&&p_i=p_{\rm min}\left(\frac{p_{\rm max}}{p_{\rm min}}\right)^{\frac{i}{N_{p}-1}},~~~~~~~i=0,...,N_{p}-1;\\
&&\theta_j=\frac{j}{N_{\theta}-1}(\theta_{\rm max}-\theta_{\rm min}),~~~j=0,...,N_{\theta}-1.
\end{eqnarray}
with $p_{\rm min}$ chosen as nonzero small value to avoid singularity at the initial point and $\theta_{\rm min}=0$, $\theta_{\rm max}=\pi$. The choice of $p_{\rm max}$ and maximal number of intervals depend on the specific systems we want to inspect. For under-occupied systems with high momentum jets, their values should increase.

We take $N_{p}$=64, $N_{\rm \theta}$=64, for isotropic systems and $N_{p}$=256, $N_{\rm \theta}$=64, for anisotropic systems where higher accuracy is needed.

\subsection{Inelastic Rate Interpolation}
\label{sec-ratesinterpolation}
The effective inelastic rates $\frac{d\Gamma^{a}_{bc}}{dz}(p,z)$ are also depending on screening masses $m_D$, $m_{Q f}$ which are evaluated every time step during the evolution, hence we need to evaluate the effective inelastic rates dynamically. The calculation of the effective inelastic rate is numerically intensive thus we do not calculate the rate for every time step. 
Instead, we first calculate the rates as functions of $\omega_{\rm BH}=\frac{m_D^2}{g^2 T^{*}}$, $x_{\rm DQf}=\frac{m_D^2}{m_{Qf}^2}$ for each inelastic process and each quark flavor. Then we set up 2D local rate grids on scales $\omega_{\rm BH}$, $x_{\rm DQf}$ with grid boundary varied by $\pm 20\%$ of their central values. Since the scales $m_D^2$, $m_{Qf}^2$, $T^{*}$ are changing modestly (see Sec.~\ref{sec-adaptive}), the rates in the upcoming steps can be interpolated from the grids as long as the upcoming scales $\omega_{\rm BH}$, $x_{\rm DQf}$ are within the grid squares. 
Once either $\omega_{\rm BH}$, $x_{\rm DQf}$ evolves outside of the grid squares, we recalculate the inelastic rates based on the current scales.

\subsection{Monte-Carlo Sampling}
\label{sec-mcsampling}
We perform Monte-Carlo integration of collision integrals for both elastic and inelastic processes.

For the elastic samplings, we first sample $q$, then $-q\le\omega\le q$ and finally $\frac{q-\omega}{2}\le p_1$, $\frac{q+\omega}{2}\le p_2$ according to the discussions in Sec.~\ref{sec-phasespace}. 
The samplings of angles ${\rm cos}(\theta_{q})$, $\phi_{q}$ together with $\phi_{1}$, $\phi_{2}$ help us determine the values for $p_3$, $p_4$.
With each of set of samplings for momenta $p_1$, $p_2$, $p_3$, $p_4$, we calculate the discretized collision integral according to Eq.~(\ref{eq-elastic-moment-8}) which simultaneously updates the gain and loss terms of all the processes, which by virtue of the sum rules ensures exact energy and particle number conservation as was discussed in Sec.~\ref{sec-de-form}.
Similarly, the evaluation of inelastic collision integrals are performed by sampling $p$, $z$ and the angle with respect to the longitudinal direction ${\rm cos}(\theta)$ according to Eq.~(\ref{eq-inelastic-cfinal}) which also simultaneously updates the gain and loss terms.

The summation of all relevant processes and all samplings for collision integrals provides the total collision integral in the Boltzmann equation for specific particle.

The sampling numbers are chosen to be $N_{\rm sample, elastic}$=512 for each specific elastic process and $N_{\rm sample, inelastic}$=256 for each specific inelastic process. 

\subsection{Adaptive time step}
\label{sec-adaptive}
Evolving the particle distributions in the discretized domain, we need an adaptive time step size $\Delta t$ to perform a stable increment for each distinct step
\begin{eqnarray}
&&\Delta n_a(i_{p},j_{\theta},k_{\phi},t)=-\left[C^{{2\leftrightarrow2}}_a[n](i_{p},j_{\theta},k_{\phi},t)\right.\\
\nonumber
&&\left.+C^{{1\leftrightarrow2}}_a[n](i_{p},j_{\theta},k_{\phi},t)
+C^{\rm z-exp}_a[n](i_{p},j_{\theta},k_{\phi},t)\right]\Delta t.
\end{eqnarray}
In order to do that, we need to make sure essential physics scales are not changing rapidly in each step. Common scales into considerations are total number density $n$, total energy density $e$ and total longitudinal pressure $p_L$, Debye screening mass square $m_D^2$ quark screening mass square $m_{Q f}^2$, effective temperature $T^{*}$. Some other scales may also be considered. However, more scales will not only increase the stability, but also slow down the evolution with a shorter resulting time step $\Delta t$. 
According to their expressions listed in Sec.~\ref{sec-theory-thermodynamics}, their relative changing rate can be approximated by
\begin{widetext}
\begin{eqnarray}
\nonumber
&&\frac{\partial_t n}{n}=\frac{\int d^3p\left[\nu_g C_A \partial_t f_g+\nu_q C_F \sum_f\left(\partial_t f_{q}+\partial_t f_{\bar{q}}\right)\right]}
{\int d^3p\left[\nu_g C_A f_g+\nu_q C_F \sum_f\left(f_{q}+f_{\bar{q}}\right)\right]}\\
\nonumber
&&\frac{\partial_t e}{e}=\frac{\int pd^3p\left[\nu_g C_A \partial_t f_g+\nu_q C_F \sum_f\left(\partial_t f_{q}+\partial_t f_{\bar{q}}\right)\right]}
{\int pd^3p\left[\nu_g C_A f_g+\nu_q C_F \sum_f\left(f_{q}+f_{\bar{q}}\right)\right]}\\
\nonumber
&&\frac{\partial_t p_L}{p_L}=\frac{\int p{\rm cos}^2(\theta)d^3p\left[\nu_g C_A \partial_t f_g+\nu_q C_F \sum_f\left(\partial_t f_{q}+\partial_t f_{\bar{q}}\right)\right]}
{\int p{\rm cos}^2(\theta)d^3p\left[\nu_g C_A f_g+\nu_q C_F \sum_f\left(f_{q}+f_{\bar{q}}\right)\right]}\\
\nonumber
&&\frac{\partial_t m_D^2}{m_D^2}=\frac{\int\frac{d^3p}{2p}\left[\nu_g C_A \partial_t f_g+\nu_q C_F \sum_f\left(\partial_t f_{q}+\partial_t f_{\bar{q}}\right)\right]}
{\int\frac{d^3p}{2p}\left[\nu_g C_A f_g+\nu_q C_F \sum_f\left(f_{q}+f_{\bar{q}}\right)\right]}\\
\nonumber
&&\frac{\partial_t m_{Q f}^2}{m_{Q f}^2}=\frac{\int\frac{d^3p}{2p}\left[2\partial_t f_G+\left(\partial_t f_{Q f}+\partial_t f_{\bar{Q} f}\right)\right]}
{\int\frac{d^3p}{2p}\left[2 f_G+\left(f_{Q f}+f_{\bar{Q} f}\right)\right]}\\
&&\frac{\partial_t m_D^2T^{*}}{m_D^2T^{*}}=\frac{\int d^3p\left[\nu_g C_A (\partial_t f_g+2f_g df_g)
+\nu_q C_F\sum_{f}\left(\partial_t f_q-2f_q \partial_t f_q+\partial_t f_{\bar{q}}-2f_{\bar{q}}\partial_t f_{\bar{q}}\right)\right]}
{\int d^3p\left[\nu_g C_A f_g(1+f_g)+\nu_q C_F \sum_f\left((f_{q}(1-f_{q})+f_{\bar{q}}(1-f_{\bar{q}})\right)\right]}.
\end{eqnarray}
\end{widetext}

In those above formula the time derivative of distribution $\partial_t$ can be evaluated according to Eq.~(\ref{eq-ntof})
\begin{eqnarray}
\nonumber
&&\partial_t f(\vec{p},t)
=\frac{1}{p^{l}}\sum_{\alpha,\beta}
w_{\alpha}^{(p)}(p)w_{\beta}^{(\theta)}(\theta)\frac{(2\pi)^2}{A_{(\alpha)}^{(2-l)}A_{(\beta)}}
\frac{\Delta n(\alpha,\beta)}{\Delta t},\\
\nonumber
&&=-\frac{1}{p^{l}}\sum_{\alpha,\beta}
w_{\alpha}^{(p)}(p)w_{\beta}^{(\theta)}(\theta)\frac{(2\pi)^2}{A_{(\alpha)}^{(2-l)}A_{(\beta)}}
\sum C^{\rm process}_a(\alpha,\beta),\\
&&((\alpha,\beta)\in\mathbb{N}^2_{p,\theta}).
\end{eqnarray}
We control the change of those scales for each time step less than 5\% via a primary time step size goal
\begin{eqnarray}
\nonumber
\Delta t_{\rm goal}=\frac{0.05}{{\rm max}\{|\frac{\partial_t n}{n}|,|\frac{\partial_t e}{e}|,|\frac{\partial_t p_L}{p_L}|,|\frac{\partial_t m_D^2}{m_D^2}|,|\frac{\partial_t m_{Q f}^2}{m_{Q f}^2}|,|\frac{\partial_t m_D^2T^{*}}{m_D^2T^{*}}|\}}\\
\end{eqnarray}
In order to have a smooth change of the time step size, we ultimately choose time step size for step $i$ as 
\begin{eqnarray}
\Delta t_i=\left(\Delta t_{i-1}^3\Delta t_{\rm goal}\right)^\frac{1}{4}
\end{eqnarray}

\section*{References}
\bibliography{bib/ref}

\end{document}